\documentclass[letterpaper, 11pt, openany,twoside]{book}
\input{mydefs.sty}

\renewcommand{\thesisdepartmentname}{Mathematics Institute}    
\renewcommand{\thesissubmission}{Submitted to the University of Warwick in partial fulfilment \\[-10pt] of the requirements for admission to the degree of}

\renewcommand{\thesisauthor}{David Philip Sanders}    

\renewcommand{\thesismonth}{January}     

\renewcommand{\thesisyear}{2005}      

\renewcommand{\thesistitle}{Deterministic Diffusion in Periodic Billiard Models}     

\renewcommand{\thesisauthorpreviousdegrees}{}

 \renewcommand{\thesisdegree}{Doctor of Philosophy in Mathematics}  

\begin{document}
\pagenumbering{roman}


\thesistitlecolourpage           

\startonright

\normalsize \tableofcontents




\addcontentsline{toc}{chapter}{Acknowledgements}
\chapter*{Acknowledgements}

{ 
I would firstly  like to thank my
PhD supervisor, Robert MacKay, for his comments, suggestions and
encouragement during the last four years.  I have learnt a great
deal from him about how to approach research.

I thank Pierre Gaspard for introducing me to billiard models and
for suggesting to look at 3D billiards, Hern\'an Larralde for
discussions about Maxwell distributions, and Leonid Bunimovich,
Nikolai Chernov, Carl Dettmann, Martin Henk, Rainer Klages, Greg
Pavliotis,  Florian Theil and S\'ebastien Viscardy for useful
conversations.  I would also like to thank my examiners, Andrew
Stuart and Stephen Wiggins, for their probing questions and
comments.

I thank \mbox{EPSRC} for a PhD studentship, and the Mathematics
Institute for providing me with enough teaching to stay just about
financially afloat after the EPSRC money ran out.

I used extensively computing facilities provided by the University
of Warwick Centre for Scientific Computing; thanks to Matt Ismail
and Jaroslaw Zachwieja for assistance with their use, and to Colm
Connaughton for getting me started.  I also thank Harald Harders,
Theo Hopman and
 Ethan Merritt for help with \texttt{gnuplot} via e-mail, and Billy
 Donegan for useful discussions on computing issues.


Many thanks to my colleagues in the Mathematics Institute for
discussions about maths and many other things; and to all my
friends for musical and other interludes.

None of this would have been possible without  my family,  who
have always been there to support me.

Finally, I would not have come anywhere close to finishing this
thesis without Gaby's love and support. }





\addcontentsline{toc}{chapter}{Declaration}
\chapter*{Declaration}        

I declare that this thesis is my own work and has not been
submitted for a degree at any other university.

\noindent
\revision{
%
\emph{Material between square brackets and in red consists of comments and
corrections added in August 2008.} \\[10pt]
\noindent Most of the contents of \chapref{chap:fine-structure},
together with some of \secref{sec:polyg-normal-diffn}, 
were published as \cite{SandersFineStructurePRE2005}.\\[10pt]
\noindent An extended version of the remainder of \chapref{chap:polygonal} was
published jointly with
Hern\'an Larralde
\cite{SandersLarraldeNormalAnomalousDiffusionPolygonalPRE2006}. \\[10pt]
\noindent The numerical results in \ref{chap:3dmodel} are incorrect; a corrected
version has been submitted for publication \cite{SandersNormalDiffusion3D}.
}





\addcontentsline{toc}{chapter}{Summary}
\chapter*{Summary}


\vspace*{-20pt}

We investigate statistical properties of several classes of
periodic billiard models which can be regarded as diffusive. We
begin by motivating the study of such models in
\chapref{chap:introduction} and reviewing how statistical
properties arise in \chapref{chap:stat-props}.

In \chapref{chap:geom-dependence} we consider a periodic Lorentz
gas satisfying a geometrical condition, for which diffusion has
been rigorously proved. We discuss how to estimate diffusion
coefficients from numerical data and then study their geometry
dependence, finding a qualitative change in the shape of curves as
one parameter is varied.  We discuss the application of a random
walk approximation of the diffusion coefficient and a related
Green--Kubo formula. We also consider the effect on the diffusion
coefficient of reducing the geometrical symmetry.

In \chapref{chap:fine-structure} we study the shape of position
and displacement distributions, which converge to a normal
distribution by the central limit theorem. We find a fine-scale
oscillation in the densities which prevents them from converging
pointwise to Gaussian densities, and relate this to the geometry
of the billiard domain, giving an analytical expression for the
fine-structure function. We provide strong evidence that, when
demodulated, the densities converge uniformly to Gaussians,
strengthening the standard central limit theorem, and we find an
upper bound on the rate of this convergence. We further consider
the effect of a non-constant distribution of particle speeds,
showing that the
 limiting position distributions can be non-Gaussian.

\chapref{chap:polygonal} treats polygonal billiard channels, where
few rigorous results are known.  We provide numerical evidence
that normal diffusion can
 occur, and that the central limit theorem can be
satisfied.  We develop a  picture of how normal diffusion can fail
if there are parallel scatterers, and we characterise the
resulting anomalous diffusion, as well as the crossover from
normal to anomalous diffusion as such a geometrical configuration
is approached.

In \chapref{chap:3dmodel} we extend our methods to a
three-dimensional periodic Lorentz gas.  We present a model with
overlapping scatterers exhibiting normal diffusion in a certain
regime. Outside this regime we provide evidence that the type of
holes present in the structure strongly influences the statistical
properties, and show that normal diffusion may be a possibility
 even in the presence of cylindrical holes.

We finish in \chapref{chap:conclusions} with conclusions and some
directions for future research.



\thispagestyle{empty}
\begin{center}
\vspace*{\stretch{1}}
 {\LARGE \emph{Para Gaby}}
\vspace*{\stretch{3}}
\end{center}

\startonright

\pagestyle{fancy} \pagenumbering{arabic} \setcounter{page}{1}





\graphicspath{{figs/}}

\chapter{Introduction}
\label{chap:introduction}

\section{Motivation: dynamics of fluids and statistical mechanics}

The dynamics of fluids is comparatively well
understood\footnote{The major exception to this is the class of
turbulent phenomena.} at the length scales of everyday experience:
we call this the
\defn{macroscopic level}.
At this level, the dynamics is described by partial differential
equations modelling the behaviour of continua.  But we can also
think of fluids as made up of a vast number of \defn{microscopic
particles} which are orders of magnitude smaller than typical
macroscopic lengths, and whose interactions we again understand
well. The question then arises to relate these two levels of
description of the same substance. Since many states of the
microscopic system correspond to one state of the macroscopic
system (specified by a few macroscopic variables such as
temperature and pressure), the relation must be of a statistical
nature: we seek to relate \emph{averages} over microscopic states
to macroscopic phenomena, where there is a separation of length
scales between microscopic and macroscopic.

\subsection{Hard-sphere fluids}

At a microscopic level, the simplest physical\footnote{Many
seemingly less-physical microscopic models satisfying certain
conservation laws give rise to fluid behaviour at a macroscopic
level.  A particularly good example is that of lattice-gas
cellular automata, where particles jump between neighbouring
points of regular lattices and obey certain microscopic
conservation laws: see e.g.\ \cite{BoonRivet}.} microscopic
picture of a fluid is a collection of a very large number of
identical atoms or molecules moving through empty space and
undergoing collisions with each other.  Restricting to a classical
(as opposed to a quantum-mechanical) description, we consider a
Hamiltonian system with inter-particle interactions  described by
a short-range potential.

The simplest potential, giving the most naive picture of a  fluid,
is the \emph{hard-sphere} potential, which jumps from $0$ to
$\infty$ at a distance $a$; the dynamics then corresponds to
spheres of radius $a$ in free motion which undergo elastic
collisions when they meet. Simulations of a relatively small
number of hard spheres (\defn{hard-sphere fluids}) show behaviour
which,  in certain regimes, resembles that of real fluids. The
hope is that studying this relatively simple system should give
insight into the relation between microscopic dynamics and
macroscopic fluid dynamics.



\subsection{Transport properties}

One of the key aims of (non-equilibrium) statistical mechanics  is
to relate the microscopic properties of a fluid (such as a
knowledge of the interaction potential referred to above) to a set
of coefficients which appear in the macroscopic equations of
motion for the fluid, and describe the \defn{transport} (motion in
space) of conserved quantities; they are therefore known as
transport coefficients.  These include coefficients describing
diffusion (transport of mass), viscosity (transport of momentum)
and heat conduction (transport of energy in the form of
heat)\footnote{These are the only transport coefficients for a
\defn{simple} fluid, i.e.\ a fluid consisting of one component.}.

\subsection{Diffusion and Brownian motion}

In this thesis we discuss only diffusion in detail.  This is one
of the most fundamental phenomena in fluids and refers to the
spreading out of matter which is initially confined to a small
subregion of a system. This is the process by which concentration
gradients in the system are smoothed out, and hence one of the
ways in which a system returns to equilibrium.

In seminal work   published in his \emph{annus mirabilis} 1905
\cite{Einstein} (see also \cite[Chap.~1]{Gardiner}), Einstein
related diffusion to the   motion of a large particle subjected to
repeated impacts with molecules of the surrounding fluid.
Regarding these impacts as \emph{random}, due to the complexity of
the presumed true microscopic dynamics of the huge number of fluid
molecules, he showed that certain plausible assumptions led to an
equation describing the probability distribution of the position
of the particle; this equation is identical to the classical
diffusion equation.  The seemingly random motion of such particles
had been observed in 1827 by Brown, a botanist, studying the
motion of a pollen grain in water, and is hence called
\defn{Brownian motion}.  For a recent experiment of relevance to
this thesis see \cite{GaspardBrownian}.

Einstein's derivation effectively models the motion of the
particle as a \emph{stochastic process}, namely a random walk; it
is thus a \defn{mesoscale} model intermediate between the true
underlying microscopic dynamics and the observed macroscopic
behaviour.


\section{Billiard models}

\subsection{Toy models exhibiting diffusive properties}

Recently it has been realised that it is possible to study the
statistical properties of  some
 simple deterministic dynamical systems  at the level of the  full microscopic
dynamics, and that these resemble to some extent those of
diffusion.  When this is the case, we talk about
\defn{deterministic diffusion}.  As discussed extensively in \chapref{chap:stat-props},
 there is a hierarchy of  related statistical properties which may `look diffusive'.  The concept of
deterministic diffusion thus actually consists of several levels.

Such systems can be regarded as `toy' models to understand
transport processes in more realistic systems \cite{DorfBook}.
Examples include classes of uniformly hyperbolic one-dimensional
(1D) maps (see e.g.\ \cite{KlagesD99} and references therein) and
multibaker models \cite{GaspBook}.  Often, rigorous results are not
available, but numerical results and analytical arguments indicate
that diffusion occurs, for example in Hamiltonian systems such as
the standard map \cite{LichtenbergLieberman}.

\subsection{Billiard models}

\begin{figure}[tp]
\centering
\includegraphics*{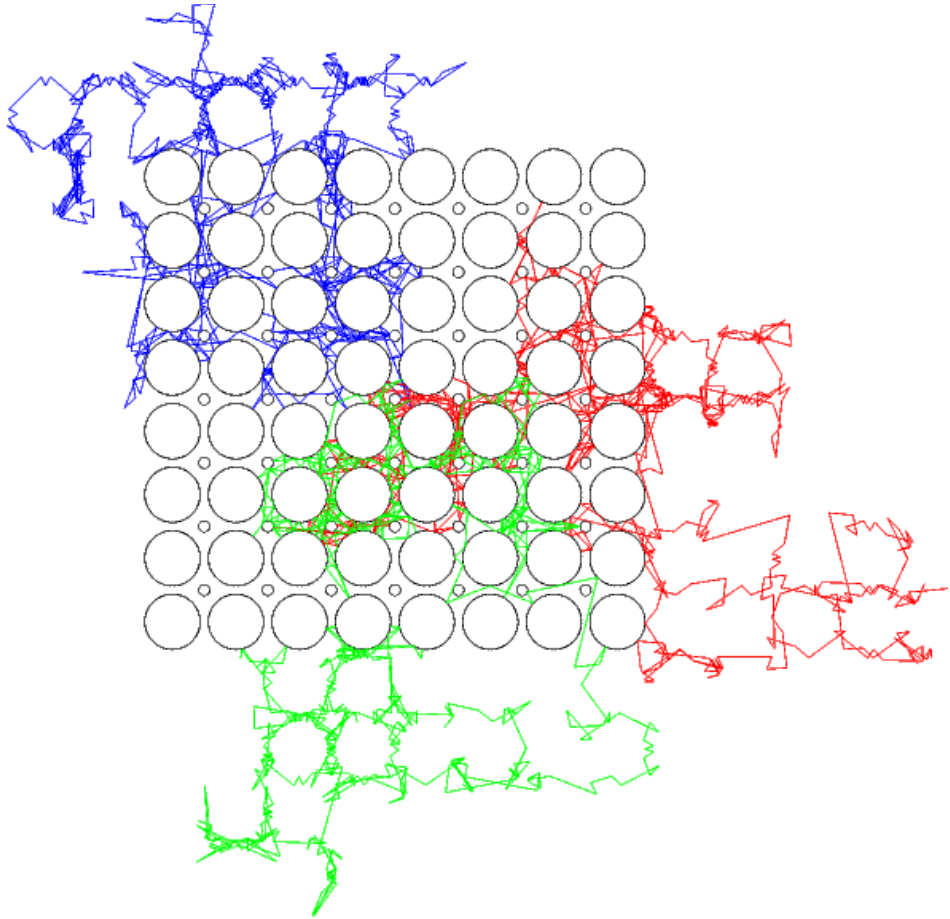}
\caption{\label{fig:lorentz-gas-traj} Sample trajectories  in a
periodic Lorentz gas.  Each trajectory emanates from one point in
the central unit cell, with only the initial velocities being
different.  The lattice of scatterers extends throughout space,
but for clarity only a portion is shown.}
\end{figure}

Billiard models constitute an important class of such models.
Here, non-interacting point particles in free motion undergo
elastic collisions with an array of fixed scatterers. Such models
were introduced by Lorentz \cite{Lorentz}, who modelled electron
flow through an amorphous metal by point particles moving through
a random array of hard spheres;  such a model (with hard discs or
spheres as the scatterers) is now termed a \defn{Lorentz gas}.
Several example trajectories of a \emph{periodic} Lorentz gas  are
shown in \figref{fig:lorentz-gas-traj}.

Lorentz gases have been used  to model neutron transport in dense
media \cite{CaseZweifel} and can be viewed as modelling the flow
of a dilute gas of light particles through a gas of heavy
particles, in the limit where the ratio of masses of the light
particles to those of the heavy particles goes to infinity
\cite{ChapmanCowling}. The Lorentz gas is also a basic model in
kinetic theory, since certain questions are simpler  to answer  in
this context \cite{DorfmanInSzasz, Hauge}.

\subsection{Hard-sphere fluids as billiard models}
\label{subsec:hard-spheres-as-billiards}

\begin{figure}
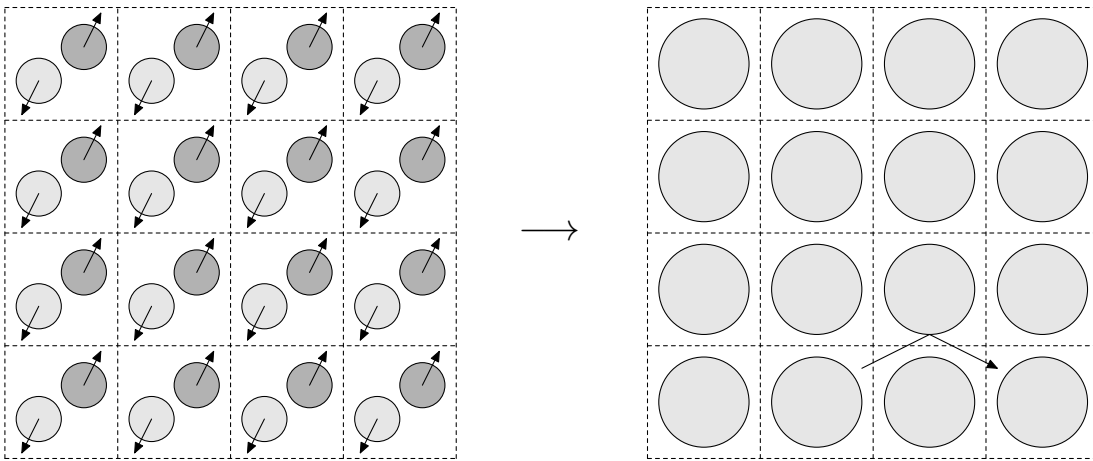

\centering
\begin{minipage}[c]{0.4\textwidth}
    \centering \includegraphics{periodic-fluid.1}
    \end{minipage}
\qquad 
{\Large $\longrightarrow$} \qquad
\begin{minipage}[c]{0.4\textwidth}
    \centering \includegraphics{periodic-fluid.2}
    \end{minipage}
\caption{\label{fig:periodic-fluid} Reduction of a 2-disc periodic fluid to a periodic Lorentz gas.}
\end{figure}

Furthermore, a hard-sphere fluid can be regarded as a billiard
model in a high-dimensional phase space.  The simplest example of
this is a 2-disc periodic fluid consisting of two discs on a
torus, or equivalently two discs in each copy of a
periodically-repeated unit cell, as shown on the left of
\figref{fig:periodic-fluid}. Call the discs A and B, with radii
$r_A$ and $r_B$.  Suppose we stand at the centre of disc A and
look at the motion of disc B bouncing off disc A.  We see the same
dynamics as we would if a point particle located at the centre of
B were colliding with a fixed disc of radius $r_A+r_B$ located at
the centre of disc A. In this way the dynamics of the periodic
fluid is equivalent to the dynamics of the periodic Lorentz gas
model shown on the right of \figref{fig:periodic-fluid}.

More generally, any hard-sphere fluid in a torus can be regarded
as a billiard in a high-dimensional phase space, as follows.
Consider
 $N$ hard spheres in a periodic box $Q \subset \R^d$, with
positions $\q_i \in Q$, velocities $\v_i \in \R^d$, masses $m_i$
and radii $r_i$.  The vector $(\q_1, \ldots, \q_N, \v_1, \ldots,
\v_N) \in  \M'$ then specifies the instantaneous state of the
system, where $\M' \defeq Q^N \times \R^{Nd}$ is the phase space
of the dynamical system.

The total momentum is conserved, so we can change to a frame of
reference in which the centre of mass is fixed at $\vect{0}$ and
the total momentum is $\vect{0}$, allowing us to restrict
attention to a reduced phase space $\M$ of lower dimension
\cite{SzaszClassicalBilliardBalls}. Furthermore, regions of $\M$
corresponding to configurations where the hard spheres overlap are
not allowed; these excluded regions are high-dimensional cylinders
in $\M$. The dynamics of the hard-sphere fluid then corresponds
exactly to free motion in $\M$, together with elastic collisions
on these cylinders, i.e.\ to
billiard dynamics in the phase space $\M$  \cite{SzaszClassicalBilliardBalls}.

\subsection{Transport in billiards}

Billiards can be regarded as the simplest physical systems in
which diffusion, understood as the large-scale transport of mass
through the system, can occur: as pointed out  in \cite{BunRev},
all that is required for diffusion to be possible is some mass
which can move through the system, which is exactly the situation
in billiards. Other transport processes have also been studied in
billiards, e.g.\ electrical conduction \cite{ChernovElecCondCMP,
ChernovElecPRL}, heat conduction \cite{AlonsoLorentzChannel} and
viscosity \cite{BunSpohn, BunRev,ViscGaspI}.

 In this thesis we consider only
diffusion in \emph{periodic} billiard models, where the scatterers
form a periodic array.  In this case the dynamics in the extended
(\defn{unfolded}) system can be obtained by looking at the
dynamics on a single unit cell with perodic boundaries, i.e.\ a
torus with scatterers removed, and keeping track of which lattice
cell particles are in. The region $Q$ exterior to the scatterers
on the torus is then called the
\defn{billiard domain}. Since the particles are non-interacting,
it is usual to set all velocities to $1$ by a geometrical
rescaling, although in \secref{sec:maxwell-vel-distn} we discuss
the effect of
 a Gaussian velocity distribution.

\section{Classical diffusion equation}
\label{sec:diffn-eqn}

Since we wish to  describe dynamical systems as  diffusive if
their statistical behaviour looks like diffusion, we briefly
review some features of the classical diffusion equation (also
called the heat equation). This is  a partial differential
equation which models the empirically observed
\defn{diffusion} (flow / transport) of matter from regions of high
concentration  to regions of low concentration, and is one of the
classical field equations of macroscopic physics.  We begin by
deriving the diffusion equation via a macroscopic balance
equation.

\subsection{Balance equations}

Let $\rho$ be the density (per unit volume) of an extensive
quantity.  Then $\rho(\r, t) \from \R^d \times \Rplus \to \R$ is a
scalar time-dependent field depending on position $\r\in \R^d$ in
space and time $t$, where $d$ is the number of spatial dimensions.
Examples of such fields are the mass density of a substance,
 the heat content, the
concentration of a chemical species, and the $x$ component of the
momentum.  We assume that $\rho$ is sufficiently smooth that we
can differentiate with respect to space and time.

Consider a fixed region $V$ (a `volume') in the space $\R^d$, with
outward unit normal vector $\n$ and boundary $S \defeq \boundary  V$.
 The total amount of substance (or in general of the extensive quantity of which
 $\rho$ is the density) inside
$V$ at time $t$ is $\int_V \rho(\r, t) \d \r$; this is the mass
inside $V$ if $\rho$ is the mass density of a gas, for example.

This amount of $\rho$ inside $V$ can change over time by exactly
two (local) mechanisms: $\rho$ can be created or destroyed inside
$V$, with source strength $\sigma$, the amount created per unit
volume, per unit time; or $\rho$ can flow across the boundary of
the volume $V$, with flux vector $\J$ per unit area, per unit
time. This neglects non-local effects such as radiation.

Thus
\begin{equation}\label{}
   \frac{\mathrm{d}}{\mathrm{d}t} \int_V \rho(\r, t) \d \r =
\int_V \sigma(\r, t) \d \r - \int_{\boundary V} \J \cdot \n \d S.
\end{equation}
Since $V$ is fixed, we can move the time derivative inside the
integral.  The divergence theorem then gives
\begin{equation}\label{eq:integrated-balance}
    \int_V \frac{\partial \rho}{\partial t}(\r, t) \d \r =
\int_V \sigma(\r, t) \d \r - \int_V \nabla \cdot \J \d \r,
\end{equation}
for any fixed volume $V$.  This implies that\footnote{Taking all
terms to one side gives an equation of the form $\int_V f(\r,t)
\d\r = 0$.  If $f$ is not (almost) everywhere $0$, then it must be
 positive (without loss of generality) on some region $V^{+}$ of non-zero
volume.  Then $\int_{V^{+}} f(\r,t) \d\r > 0$, contradicting the
assumption that the integral equal zero over \emph{any} volume.}
\begin{equation} \label{eq:diffsource} 
\frac{\partial \rho}{\partial t}(\r, t) = \sigma(\r, t) - \nabla
\cdot \J(\r,t).
\end{equation}

The equation \eqref{eq:diffsource} is termed a \defn{balance
equation} \cite{KondepudiPrigogine}.  It forms the basis for
deriving macroscopic evolution equations for any scalar field, and
hence also vector and tensor fields  by considering components. In
order to derive such equations, we must specify $\sigma$ and $\J$
in terms of other known quantities using \defn{constitutive
equations}, which model the `constitution' (behaviour) of a
substance, to get a
\defn{closed} system of equations: see e.g.\
\cite{KondepudiPrigogine}.

\subsection{Derivation of the (anisotropic) diffusion equation}

We now specialise to the case where $\rho$ is the concentration,
or mass density, of a substance which is diffusing. If the mass of
the substance is (locally) conserved, we have $\sigma=0$.%
\footnote{This is not the case, for example, if we have several
reacting chemical species.  The source $\sigma_A$ of
species A then contains terms describing the creation and
destruction of A in the reactions (the \emph{total} mass, however,
being conserved). We then obtain a set of
\emph{reaction--diffusion} equations describing the
spatio-temporal evolution of the concentration fields.  Such
equations can yield  interesting spatial patterns
\cite{KondepudiPrigogine}.} We must now specify $\J$, a question
which is considered at length in non-equilibrium thermodynamics
\cite{DeGrootMazur}.  We consider only the (simplest) case, when
\defn{Fick's law} is obeyed, so that the flux $\J$ is a linear
function of the concentration gradient, $\nabla \rho$. (This can
be viewed as a first-order \emph{approximation} under the
assumption of local thermodynamic equilibrium and slowly varying
concentration profile \cite{DeGrootMazur,KondepudiPrigogine}.)
Hence
\begin{equation}\label{eq:constitutive-diffusion}
    \J = -\tilde{\tD} \cdot \nabla \rho,
\end{equation}
where $\tilde{\tD}$ is a second rank tensor called the
\defn{diffusion tensor}.  The minus sign accounts for the
empirical fact that matter diffuses from regions of high
concentration to regions of low concentration. This is an example
of a constitutive equation.

Assuming that $\tilde{\tD}$ is \emph{independent of position}, we
have
\begin{equation}
\nabla \cdot \J = - \sum_{i,j} \tilde{D}_{i j} \, \partial_{i j}
\rho = - \sum_{i,j} D_{i j} \, \partial_{i j} \rho,
\end{equation}
 where $\tilde{D}_{i j}$ are the components of the tensor
$\tilde{\tD}$ with respect to a Cartesian coordinate system, and
$\partial_i \rho \defeq \frac{\partial \rho}{\partial r_i}$. We
have defined the (symmetric) \defn{diffusion tensor} $\tD$ as the
symmetric part of $\tilde{\tD}$, \ie $\tD =
\frac{1}{2}(\tilde{\tD} + \tilde{\tD} \transp)$, with components
$D_{i j}$. The antisymmetric part does not play any role, since the
matrix of second partial derivatives of $\rho$ is symmetric.
Substituting in \eqref{eq:diffsource} gives the
\defn{(anisotropic) diffusion equation}
\begin{equation}\label{eq:diffusion}
\frac{\partial \rho}{\partial t} = \sum_{i,j} D_{i j} \,
\partial_{i j} \rho.
\end{equation}
We can also write this independently of coordinate system as
\begin{equation}\label{eq:constitutive-no-components}
\frac{\partial \rho}{\partial t} = \nabla \cdot (\tD \cdot \nabla
\rho).
\end{equation}



\subsection{Solution of the diffusion equation in an unbounded domain}

We consider the diffusion equation in one
dimension, with $D$ independent of space:
\begin{equation}\label{eq:diffusion-eqn}
    \pd{\rho(t;x)}{t} = D \, \pd{^2 \rho(t;x)}{x^2}.
\end{equation}

We  solve equation \eqref{eq:diffusion-eqn}, a linear partial
differential equation with constant coefficients,  using a
standard Green function method.  For $k \in \R$ we define the
Fourier transform of $\rho$ at time $t$ by
\begin{equation}\label{eq:defn-fourier-transf}
    \rh{t;k} \defeq \int_{x=-\infty}^{\infty} \e^{-\i kx}\, \rho(t;x)
    \d x.
\end{equation}
Fourier transforming \eqref{eq:diffusion-eqn} gives
\begin{equation}\label{eq:diff-eqn-ft}
    \pd{\rh{t;k}}{t} = -D \, k^2 \, \rh{t;k},
\end{equation}
an ordinary differential equation for $\rh{t;k}$, the solution of
which is
\begin{equation}\label{eq:diff-eqn-ft-soln}
    \rh{t;k} = \rh{0;k} \, \e^{-D\,k^2\,t}.
\end{equation}
Taking the inverse transform of this product gives a convolution:
\begin{align}\label{eq:diff-eqn-soln}
    \rho_t(x) = \rho(t;x) &= \frac{1}{2\pi} \int_{k=-\infty}^{\infty} \e^{+ \i kx} \, \rh{t;k}  \d k \\
              &= [\rho_0 \conv G^t](x),
\end{align}
where $\rho_0(x) \defeq \rho(0;x)$ is the initial concentration
distribution and $G^t$ is the \defn{Green function} (or
\defn{propagator})  for the diffusion equation on an unbounded
$1$-dimensional domain.  This Green function is given by the
\defn{Gaussian}
\begin{equation}\label{eq:green-fn-diffn-eqn}
    G^t(x) \defeq G(t;x) \defeq \frac{1}{\nsqrt{4\pi \, D \, t}}
    \, \exp \lt[-\frac{x^2}{4 D \, t} \rt]
\end{equation}
with mean $0$ and variance
\begin{equation}\label{eq:green-fn-variance}
    \var{G^t} \defeq \int_{-\infty}^{\infty} x^2 \, G^t(x)
     \d x = 2 D\, t
\end{equation}
at time $t$, and Fourier transform
\begin{equation}\label{}
\hat{G}(t; k) = \e^{-D k^2 t}.
\end{equation}
%
We recall that the convolution operation is defined by
\begin{equation}\label{eq:defn-convolution}
    (u \conv v)(x) \defeq \int_{y=-\infty}^{\infty} u(x-y) \, v(y)
     \d y.
\end{equation}

\paragraph*{Multidimensional diffusion equation}
In the multidimensional case, since $\tD$ is a symmetric tensor it
is possible to choose an orthogonal coordinate system in which
$\tD$ is represented by a \emph{diagonal} matrix.  In these new
coordinates, the solution is a product of solutions of the 1D
equation.  Reverting to the original coordinates gives a
multi-dimensional Gaussian \cite{Grimmett} for a Dirac-delta
initial condition.

\subsection{Solutions of the diffusion equation as probability
densities}

Henceforth we regard the diffusion equation as describing the
evolution of probability density functions representing
probability distributions, as follows.

Let the initial condition at time $t=0$ be $\rho_0$. Physically we
are interested in non-negative $\rz$ with finite mass, i.e.\
\begin{equation}\label{}
\int_{y=-\infty}^{\infty} \rz(y) \rd y < \infty.
\end{equation}
By normalising if necessary we can instead assume that
\begin{equation}\label{}
\int_{y=-\infty}^{\infty} \rz(y) \rd y = 1.
\end{equation}
It then follows that $\int_{-\infty}^{\infty} \rho_t(y) \d y = 1$
and that the solution remains non-negative for all times $t$, so
that we can regard the diffusion equation as describing the time
evolution of
\defn{probability densities}.


\subsection{Calculation of moments from Fourier transform}

We can calculate \defn{moments} of a distribution directly from
its Fourier transform as follows (see \eg \cite{ResiboisDeLeener,
Balescu}).

Differentiate \eqref{eq:defn-fourier-transf} with respect to $k$
to get
\begin{equation}\label{eq:diff-four-transf}
    \pd{\rh{t;k}}{k} = \int_{-\infty}^{\infty} -\i \, x \, \e^{-\i \, k \, x} \,
    \rho(t;x) \d x,
\end{equation}
and hence
\begin{equation}\label{eq:first-moment-from-ft}
\lt. \pd{\rh{t;k}}{k} \evalat_{k=0} =  \int_{-\infty}^{\infty} -\i \,x\,\,%
 \rho(t;x) \d x \eqdef -\i \, \meanat{x}{t}.
\end{equation}
Here we denote by $\meanat{f(x)}{t}$ the mean of the function
$f(x)$ with respect to the distribution at time $t$. Similarly,
higher moments can be obtained by differentiating repeatedly with
respect to $k$:
\begin{equation}\label{eq:higher-moments-from-ft}
    \lt. \pd{^m\rh{t;k}}{k^m} \evalat_{k=0} =  \int_{-\infty}^{\infty} (-\i)^m \,x^m\,\,
 \rho(t;x) \d x = (-\i)^m \, \meanat{x^m}{t}.
\end{equation}

In a similar way, in the full multi-dimensional, anisotropic
diffusion equation \eqref{eq:diffusion} for $\rho(t;\x)$ in $n$
dimensions, we use the multi-dimensional Fourier transform
\begin{equation}\label{eq:multi-dim-ft}
\rh{t;\k} \defeq \int_{-\infty}^{\infty} \! \cdots
\int_{-\infty}^{\infty} \e^{-\i \, \k \cdot \x} \rho(t;\x) \d x_1
\cdots \d x_n,
\end{equation}
where $\k \cdot \x \defeq \sum_{i=1}^n k_i \, x_i$. Then we have,
for example,
\begin{equation}\label{eq:multi-dim-means-from-ft}
\left. \meanat{x_i \, x_j}{t} = - \frac{\partial^2
\rh{t;\k}}{\partial k_i \,
\partial k_j} \evalat_{\k=\vect{0}},
\end{equation}
with higher moments calculated in an analogous way.

\subsection{Asymptotic behaviour of moments of solutions of the diffusion equation}
\label{subsec:asymp-second-moment-diffn}

We assume that $\rho_0$ decays sufficiently fast at infinity for
the relevant integrals to exist. Physically relevant sufficient
conditions for this are, for example, that the initial condition
has compact support, \ie it vanishes outside a finite interval, or
that the initial condition is exponentially localised, in the
sense that
\begin{equation}\label{}
\rz(y) \le \e^{-K\modulus{y}}
\end{equation}
for some constant $K>0$.

 In this section we denote partial differentiation with respect
to $k$ by a subscript $k$, so that $\rhat_k \defeq
\partial \rhat/\partial k$.

\paragraph{First moment}
We first calculate the time dependence of the first moment (\ie
the centre of mass) of the solution of the diffusion equation.
 Differentiating \eqref{eq:diff-eqn-ft-soln} (the solution of the
diffusion equation expressed in Fourier transforms) once with
respect to $k$, we have
\begin{equation}\label{}
\rhat_k(t;k) = \e^{-D\,k^2\,t} \lt\{ -2k \, D \,t  \rzhat(k) +
\rzhat_k(k) \rt\},
\end{equation}
so that using equation \eqref{eq:first-moment-from-ft} twice, we
obtain
\begin{equation}\label{}
\meanat{x}{t} = \i \, \rhat_k(t;0) = \i \, \rzhat_k(0) = \i \,
\rhat_k(0;0) = \meanat{x}{0}.
\end{equation}
Hence the mean position (centre of mass) is constant.  So the
second moment is the first non-trivial one.

\paragraph{Second moment}
To simplify the calculation of the time evolution of the second
moment, we now change to a coordinate system with the origin
located at the centre of mass, \ie we assume without loss of
generality that $\meanat{x}{0} = 0$.

Differentiating \eqref{eq:diff-eqn-ft-soln}  again with respect to
$k$ gives the second derivative
\begin{equation}\label{eq:second-deriv-of-rho-hat}
\rhat_{kk}(t;k) = \e^{-D\,k^2\,t}  \lt\{ \rzhat_{kk}(k) + 2
\rzhat_k(k).(-2kD\,t) + \rzhat(k)\lt[ -2D\,t + (-2kD\,t)^2 \rt]
\rt\},
\end{equation}
so that
\begin{equation}\label{}
\meanat{x^2}{t} = -\rhat_{kk}(t;0) = -\rzhat_{kk}(0) + 2D\,t \,
\rzhat(0) = \meanat{x^2}{0} + 2D\,t.
              \label{eq:msd-behaviour-for-classical-diffn}
\end{equation}
(Note that $\rzhat(0)= \int \e^{0.kx} \, \rho_0(x) \d x = 1$.)
Since $\rho_0$ was assumed to have compact support, $\rzhat$ is
infinitely differentiable, by regularity results for Fourier
transforms \cite{Katznelson}. Hence $\meanat{x^2}{t}$ is a
straight line with slope $2D$, which does not pass through the
origin unless the initial condition is a Dirac delta (since the
initial variance is $0$ only if the initial distribution is
concentrated on a single point); the variance hence grows
\emph{asymptotically} linearly.

From the above result we can derive a relation between the
diffusion coefficient and the rate of growth of the variance.
Dividing \eqref{eq:msd-behaviour-for-classical-diffn} by $t$ and
taking the limit gives
\begin{equation}\label{eq:einstein-relation-classical-diffn}
D = \lim_{t \to \infty} \frac{\meanat{x^2}{t}}{2t}.
\end{equation}
This result is known as the \defn{Einstein relation}, since it was
first obtained by Einstein \cite{Einstein}.
 We
also have
\begin{equation}\label{}
D = \lim_{t\to\infty} \frac{1}{2} \ddt{}\meanat{x^2}{t},
\end{equation}
but note that the first limit can exist when the second one does
not, for example if $\meanat{x^2}{t} = \sin \, t$. Similarly in
the multi-dimensional case we find
\begin{equation}\label{eq:einstein-multidim}
D_{ij} = \lim_{t \to \infty} \frac{\meanat{x_i \, x_j}{t}}{2t}.
\end{equation}

If the system has enough  symmetry
(\secref{sec:reducing-symmetry}), then we have
 $D_{ii} = D$ for all $i$,
 with all other
components equal to $0$, and
\begin{equation}\label{}
D = \lim_{t \to \infty} \frac{\meanat{\x^2}{t}}{2dt},
\end{equation}
where $d$ is the spatial dimension and $\x^2 \defeq \x \cdot \x =
\sum_{i=1}^{d} x_i^2$. We remark that this reduction of the
diffusion tensor to a multiple of the identity tensor also occurs
if the system is
\defn{isotropic}, i.e.\ has the same properties in \emph{any} direction
in space, but isotropy is a stronger condition than necessary for
this reduction to happen.  

\subsection{Convergence of solutions of the diffusion equation to Gaussians}

If the initial condition for the diffusion equation is $\rho_0$ at
time $t=0$, then the solution at time $t$ is given by the
convolution
\begin{equation}\label{}
\rho_t(x) = \int_{y=-\infty}^{\infty} \rz(y) \frac{1}{\nsqrt{4\pi
Dt}} \, \e^{-(x-y)^2/4Dt} \d y.
\end{equation}

We are interested in the shape of the distribution for long times.
In \appref{app:conv-solns-diffn-eqn} we show that for suitable
initial data the solution converges to a Gaussian when
appropriately rescaled. This convergence requires the limiting
function to be non-degenerate.  Since $\rho_t$ tends to $0$
pointwise as $t\to\infty$, we must first rescale $\rho_t$.  We
know from the above argument or from dimensional considerations
that $X^2 \sim T$, where $X$ is a typical lengthscale and $T$ is a
timescale. Hence we rescale $x$ by $\nsqrt{t}$, putting
\begin{equation}\label{}
\trho_t(x) \defeq \trho(t;x) \defeq \nsqrt{t} \,.\, \rho(t;x
\nsqrt{t}).
\end{equation}
The first factor of $\nsqrt{t}$ is to normalise the integral of
$\trho_t$
 to $1$.

In \chapref{chap:fine-structure} we study the convergence to a
limiting normal distribution of rescaled distribution functions in
the context of billiards. For comparison, in
\appref{app:conv-solns-diffn-eqn} we consider the convergence of
rescaled solutions of the diffusion equation to the limiting
Gaussian.  We show that
  $\trho_t(x)$ converges
pointwise to a Gaussian with variance $2D$:
\begin{equation}\label{}
\trho_t(x) \stackrel{t \to \infty}{\longrightarrow} \gauss{2D}(x),
\end{equation}
where
\begin{equation}\label{}
\gauss{2D}(x) \defeq \frac{1}{\nsqrt{4 \pi D}} \, \e^{-x^2/4D}
\end{equation}
is the Gaussian density with mean $0$ and variance $2D$.
Furthermore the convergence is uniform with
\begin{equation}\label{}
\modulus{\trho_t(x) - \gauss{2D}(x)} \le \frac{C}{t},
\end{equation}
for all $x$ and some constant $C$ which is independent of $x$.

\subsection{Numerical confirmation of $\bigO{t^{-1}}$ decay}

We numerically confirm the $\bigO{t^{-1}}$ decay found in
\appref{app:conv-solns-diffn-eqn} by considering particular
initial distributions. The first initial condition we consider is
\begin{equation}\label{}
\rho_1(x) \defeq \textfrac{1}{3} \delta(x-2) + \textfrac{2}{3}
\delta(x+1),
\end{equation}
where $\delta(x-x_0)$ is the Dirac delta function at position
$x_0$.  The diffusion equation with this initial condition is
analytically soluble, with solution
\begin{equation}\label{}
\rho_t(x) = \textfrac{1}{3} G_t(x-2) + \textfrac{2}{3} G_t(x+1),
\end{equation}
so that the solution is smooth and rapidly decaying for any $t>0$
and hence fits into the class for which we can prove the
$\bigO{t^{-1}}$ convergence.

\begin{figure}
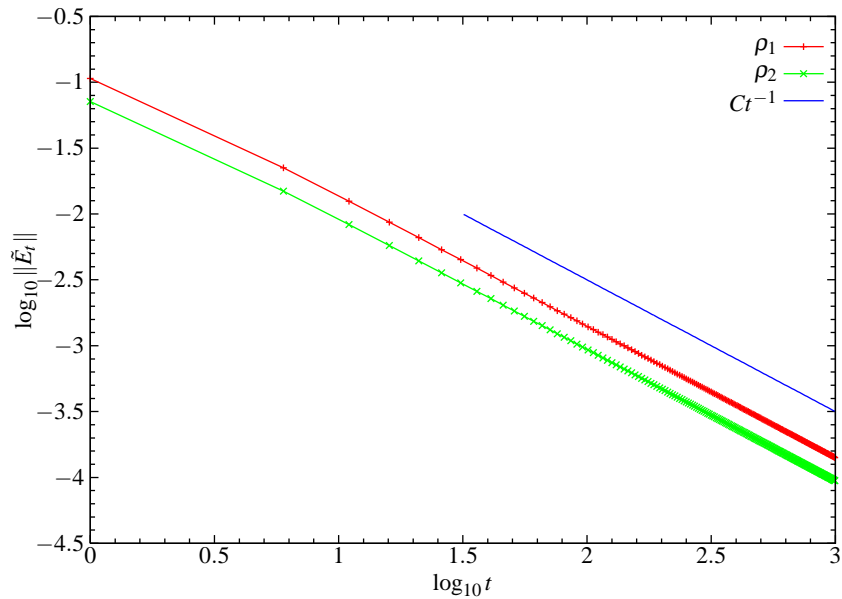

  \centrefig{rate-conv-diffn-eqn.eps}
  \caption{\label{fig:rate-conv-diffn-eqn} Uniform distance $\tilde{E}_t$ (defined
  in \appref{app:conv-solns-diffn-eqn}) from the limiting Gaussian of
  the rescaled
  solution of the diffusion equation with initial conditions
  $\rho_1(x) = \textfrac{1}{3} G_t(x-2) + \textfrac{2}{3} G_t(x+1)$ and
  $\rho_2(x) = \frac{1}{2K} \indic{[-K,K]}(x)$.}
\end{figure}

To verify the convergence numerically, we calculate $\norm{\rho_t
- \gauss{2D}} = \sup_{x \in \R} \modulus{\rho_t(x) -
\gauss{2D}(x)}$, setting $D=1$ by rescaling time.  Plotting the
logarithm of this distance against the logarithm of time gives a
straight line with slope $-1$, confirming the $t^{-1}$ decay, as
shown in \figref{fig:rate-conv-diffn-eqn}.

Another initial condition for which the equation is exactly
soluble is
\begin{equation}\label{}
\rho_2(x) = \frac{1}{2K} \indic{[-K,K]}(x),
\end{equation}
with solution \cite{Crank}
\begin{equation}\label{}
\rho_t(x) = \frac{1}{4K} \lt[ \erf \lt(\frac{x+K}{2 \sqrt{Dt}}
\rt) - \erf \lt(\frac{x-K}{2 \sqrt{Dt}} \rt) \rt],
\end{equation}
where
\begin{equation}\label{}
\erf(x) \defeq \frac{2}{\surd{\pi}} \int_0^x \e^{-t^2/2} \rd t
\end{equation}
is a special function called the \defn{error function}.
Again the numerical $\bigO{t^{-1}}$ decay rate, as shown in
\figref{fig:rate-conv-diffn-eqn}, confirms the analytical result.

\subsection{Hallmarks of diffusion}

 We conclude from the above remarks that hallmarks of diffusion are the following
 features of the
\emph{asymptotic} behaviour as $t\to\infty$:
\begin{itemize}
\item mean squared displacement growing asymptotically
    linearly in time $t$, with constant of proportionality
    $2D$, where $D$ is the diffusion coefficient; and

\item convergence of the $\sqrt{t}$-rescaled position distribution
to a non-degenerate Gaussian.
\end{itemize}
Here by `non-degenerate' we mean that the limiting Gaussian has
non-zero variance, or in higher dimensions that the covariance
matrix is positive definite.

These are the statistical features which we will use to
characterise a process as diffusive, together with a third dealing
with convergence of rescaled paths to Brownian motion.

We remark that in general inhomogeneous physical systems, the rate
of diffusion characterised by the diffusion coefficient can vary
over the system.  In such a case, we can say that we have
diffusion if Fick's law holds locally, \ie if the flux of the
diffusing quantity is locally proportional to its  gradient
\cite{Crank}.

\section{Statistical properties}


Many dynamical systems, including important classes of billiards,
are
\defn{chaotic}, in the sense that trajectories emanating from
nearby initial conditions separate fast as time evolves.  Any
physical measurement has a limited precision, so that if we repeat
an experiment several times then we cannot be sure that the
initial condition is identical each time, but only that it lies
within a certain tolerance.  Averaging over the results of the
experiments thus corresponds to averaging over a small ball of
initial conditions as it spreads out over possible states of the
system. This argument motivates the need for introducing
probabilistic notions, namely ergodic and statistical properties,
to study deterministic dynamical systems which are chaotic; see
\chapref{chap:stat-props}.

One of the main reasons for the interest in billiard models is the
possibility of obtaining rigorous results on their ergodic and
statistical properties: see e.g.\ \cite{BS,BSC}.  The techniques
are most highly developed for \defn{dispersing} billiards such as
the Lorentz gas, where the curved, convex scatterers cause nearby
trajectories to separate exponentially fast.  In fact, Lyapunov
exponents (which measure the rate of separation) exist
and are non-zero almost everywhere for the
billiard map, so that the system is
\defn{hyperbolic} (chaotic).  Due to the Hamiltonian nature of
the system, the Lyapunov exponents come in positive--negative
pairs, so that at least one Lyapunov exponent is positive almost
everywhere \cite{CherMark, GaspBook}; further, the
Kolmogorov--Sinai entropy, which measures the rate of generation of information
in time, is positive \cite{GaspBook}. These are
standard indicators of the chaotic nature of the system
\cite{EckmannRuelle}.

 Hard-sphere fluids are only
\defn{semi-dispersive}, due to the flat, neutral directions along
the cylinder axis in the many-particle phase space; this makes
their rigorous analysis much harder: see e.g.\ \cite{BalintCylind}
for a review. Nonetheless, a rigorous proof of the celebrated
Boltzmann ergodic hypothesis has recently been achieved for hard
discs ($d=2$) and spheres ($d=3$): see references in
\secref{subsec:billiards-erg-stat-props}.

The main focus of this thesis is \defn{deterministic diffusion}, a
statistical property of certain dynamical systems, including
certain classes of billiards. A definition often used in the
physical literature is that a system is \defn{diffusive} if the
mean squared displacement grows proportionally to time $t$,
asymptotically as $t \to \infty$. However, there are stronger
properties which are also characteristic of diffusion, which a
given system may or may not possess: (i) a \defn{central limit
theorem} may be satisfied, i.e.\ rescaled distributions converge
to a normal distribution as $t \to \infty$; and (ii) the rescaled
dynamics may `look like' Brownian motion.  See
\chapref{chap:stat-props} for details.

Recently there has been much interest in the question of which
microscopic features are \emph{necessary} for a system to exhibit
strong ergodic and statistical properties.  The proofs of these properties for
dispersive billiards depend crucially on the fact that they are hyperbolic,
but numerical evidence has been given that systems with weaker chaotic
properties may also show strong statistical properties, for
example the polygonal billiard channels studied in
\chapref{chap:polygonal}.


\section{Overview of thesis}

\bchapref{chap:stat-props} reviews concepts from dynamical
systems, probability theory and ergodic theory which we require to
discuss deterministic diffusion. Chapters
\ref{chap:geom-dependence}--\ref{chap:3dmodel} present our results
on deterministic diffusion in three types of billiard model: a 2D
periodic Lorentz gas with a two-dimensional parameter space
(Chaps.~\ref{chap:geom-dependence} and \ref{chap:fine-structure});
several classes of polygonal billiard channel
(\chapref{chap:polygonal}); and 3D periodic Lorentz gases
(\chapref{chap:3dmodel}).  Conclusions are presented in
\chapref{chap:conclusions}, followed by several appendices giving
technical details of results needed in the main text;  we expect
that many of these results are already known, but in several cases
we were unable to find suitable references.

The discussion in each chapter can be thought of in terms of three
related threads. The first is mathematical: to what extent do
rigorous results hold beyond their immediate range of
applicability, and can we understand more precisely the
statistical behaviour described by
 those results? The second is physical:
how do geometrical features affect the dynamics and statistical
properties of the system?  The third is statistical: how can we
best estimate statistical properties from numerical data?



\startonright





\graphicspath{{figs/}}

\chapter{Statistical properties of dynamical systems}
\label{chap:stat-props}

In this chapter we review in what sense deterministic dynamical
systems `look like' stochastic processes. We make a (somewhat
arbitrary) distinction between
\defn{ergodic} properties and
\defn{statistical} properties: the former deal with general theorems
referring to large classes of observables, whereas the latter hold
only for observables with a certain degree of smoothness, are
related to rates of convergence in ergodic theorems, and are
of physical relevance.

\section{Dynamical systems and stochastic
processes}

\subsection{Invariant measures}

Let $\fl^t \from \M \to \M$ be the flow of a continuous ($t \in
\R$) or discrete ($t \in \Z$) dynamical system with phase space
$\M$. Measures $\mu$ are defined on some $\sigma$-algebra $\Balg$
on $\M$.  Our dynamical systems will be defined on metric spaces,
so that we can take, for example, $\Balg$ to be the Borel
measurable subsets of $\M$.  Then any physically relevant subset
of $\M$ will be in $\Balg$, and from now on we will usually refer
to subsets of $\M$ without explicitly mentioning $\Balg$.

Suppose that we start the evolution of the dynamical system with a
distribution of initial conditions described by the measure
$\mu_0$. This distribution will evolve in time to the measure
$\mu_t$ given by
\begin{equation}\label{}
\mu_t \defeq (\fl^t)\lowerstar(\mu_0); \quad \mu_t(A) \defeq
\mu_0(\fl^{-t}(A));
\end{equation}
here the right hand side defines the meaning of the
\defn{push-forward} $(\fl^t)\lowerstar$.

A particularly simple and interesting case occurs if the measure
is preserved by the system, or is invariant, as follows. We say
that $\mu
\defeq \mu_0$ is
\defn{invariant} with respect to the flow if, for all $t$,
$(\fl^t)\lowerstar(\mu) = \mu$, or equivalently if
\begin{equation}\label{}
\mu(\fl^{-t}(A)) = \mu(A), \quad \text{for all } A \in \Balg.
\end{equation}
Here we also require that $\fl^t$ be \defn{measurable} with
respect to $\Balg$, i.e.\ such that $\fl^{-t}(A) \in \Balg$ for
all  $A \in \Balg$. In the discrete time case, where the dynamics
is given by a map $T \from M \to M$, it is enough to have
$\mu(T^{-1}(A)) = \mu(A)$  for all measurable sets $A$.

\paragraph{Natural invariant measures}

Some systems possess \emph{natural} invariant measures if the
dynamics preserves some structure.  The main class of interest to
us is \defn{Hamiltonian} systems, which preserve Liouville
measure: see \secref{sec:billiards-rigorous-results} in the case
of billiards and \cite[Chap.~5]{KatokHasselblatt} for other
examples. We are then most interested in statistical properties
with respect to these measures.

\subsection{Dynamical systems with invariant measures as
 stationary stochastic processes}

Fix an \defn{observable} $f \from \M \to \R$ (that is, a quantity that we
could in principle measure when the system is in different states
in phase space) and look at $X_t \defeq f \comp \fl^t \from \M \to
\R$. Given a measure $\mu \defeq \mu_0$ on $\M$, we can regard
$(X_t)$ as a collection of random variables indexed by time, so
that we have a
\defn{stochastic process} \cite{Grimmett}.

If the measure $\mu$ is invariant, then the process $(X_t)$ is
\defn{stationary}, which means that for all $n$, all $t_1, \ldots,
t_n$ and all $h>0$, the families
\begin{equation}\label{}
(X_{t_1}, \ldots, X_{t_n}) \quad \text{and} \quad (X_{t_1+h},
\ldots, X_{t_n+h})
\end{equation}
have the same joint distribution as each other \cite{Grimmett}.
The proof is as follows.  For two times $t$ and $s$ and two sets
$A, B \in \Balg$, we have
\begin{align}\label{}
& \prob{f \comp \fl^t \in A, f \comp \fl^s \in B}
= \mu\left(\fl^{-t}(f\inv(A)) \cap \fl^{-s}(f\inv(B))\right), \\
\intertext{which by the measure-preserving property of $\Phi$ is
equal to}
 &=
\mu\left(\fl^{-r}\lt(\fl^{-t}(f\inv(A)) \cap
\fl^{-s}(f\inv(B))\rt) \rt)  =\prob{f \comp \fl^{t+r} \in A, f
\comp \fl^{s+r} \in B}.
\end{align}
An analogous argument holds for $n$ times and $n$ sets; hence the
random variables $\lt(X_t = f \comp \Phi^t\rt)$ form a stationary
stochastic process.

\section{Ergodic properties}

Ergodic properties with respect to an invariant measure can be
thought of loosely as a measure-theoretic description of
chaoticity.  There is a hierarchy of increasingly strong
properties; the ideas originated in statistical mechanics: see
e.g.\ \cite{DorfBook}.  A general reference is, for example,
\cite{Cornfeld}.

\subsection{Ergodicity}

In the billiard systems we study there is a natural invariant
measure $\mu$, namely Liouville measure (see
\secref{sec:billiards-rigorous-results}), which we regard as fixed
in the following.

Suppose that the invariant measure $\mu$ is finite, i.e.\ $\mu(\M)
< \infty$. Then we can normalise $\mu$ to get a \defn{probability
measure} with $\mu(\M)=1$.
 The
flow $\fl^t$ is
\defn{ergodic} with respect to  $\mu$ if\footnote{We continue to suppress the necessity of
having $A \in \Balg$.} for all $A \subset \M$, we have
\begin{equation}\label{}
\fl^{-t}(A) = A \quad \Rightarrow \quad \: \mu(A)=0 \thickspace
\text{ or } \thickspace \mu(A) = 1,
\end{equation}
i.e.\ any set $A$ which is invariant under the flow has measure
$0$, so that it is trivial from the point of view of measure
theory, or has measure $1$, so that it covers the whole space
except for a set of measure $0$.


\subsection{Birkhoff ergodic theorem}

Let $\mu$ be an invariant probability measure for the dynamical
system $\fl^t$, and let $f \from \M \to \R$ be integrable.  Then
Birkhoff's ergodic theorem states (see e.g.\ \cite{Cornfeld}) that
the limit
\begin{equation}\label{}
\bar{f}(x) \defeq \lim_{T \to \infty} \frac{1}{T} \int_0^T (f
\comp \fl^t)(x) \rd t
\end{equation}
exists for almost all $x \in \M$ with respect to the measure
$\mu$, and we call $\bar{f}(x)$ the \defn{time average} of $f$.
Furthermore, if also the system is ergodic, then
\begin{equation}\label{}
\bar{f}(x) = \mean{f}_{\mu} \defeq \int_\M f \rd \mu \quad
\text{a.e.},
\end{equation}
so that time averages are almost everywhere (a.e.) equal to the
\defn{space average} $\mean{f}$ with respect to the ergodic invariant
measure $\mu$.


The motivation for this theorem came originally from Boltzmann,
who had the idea that a sufficiently complicated dynamical system
should explore the whole accessible phase space, an idea known as
the \defn{ergodic hypothesis}.  This has recently been proved for
hard-sphere fluids in a series of papers
\cite{Simanyi2DErgHypNoExcep, SimanyiErgHypTypHardBall,
 SimanyiErgHypTypHardDisc, SimanyiSzaszCompleteHyp}.

\subsection{Mixing}

The flow $\fl^t$ is \defn{mixing} with respect to the invariant
measure $\mu$ if for any sets $A, B \subset \M$, we have
\begin{equation}\label{}
\mu(A \cap \fl^t(B)) \stackrel{t \to \infty}{\longrightarrow}
\mu(A) \, \mu(B).
\end{equation}
Mixing implies ergodicity.  An important interpretation for us is
that mixing is equivalent to weak convergence of densities: see
\appref{app:conv-proj-densities}.

There is also a notion of \defn{weak-mixing}, intermediate in
strength between ergodicity and mixing.  For discrete time systems
we say that a transformation $T$ preserving a measure $\mu$ is
\defn{weak-mixing} if for any $A,B \subset \M$, we have
\begin{equation}\label{}
\lim_{n\to\infty} \frac{1}{n} \sum_{i=0}^{n-1}\modulus{\mu(T^{-i}
A \cap B) - \mu(A) \mu(B)} = 0.
\end{equation}
For comparison, there is a characterisation of ergodicity in
similar terms, stating that $T$ is ergodic if for any $A,B \subset
\M$ we have
\begin{equation}\label{}
\lim_{n\to\infty} \frac{1}{n} \sum_{i=0}^{n-1} \mu(T^{-i} A \cap
B) = \mu(A) \mu(B).
\end{equation}
See \cite{Walters, Cornfeld} for detailed comparisons of the
different notions, and \cite{Cornfeld} for formulations for
continuous-time systems.

\subsection{K-systems}

K-systems have very strong ergodic properties: in particular their
Kolmogorov--Sinai entropy is positive.  See e.g.\
\cite[p.~71]{SinaiTopicsErgTheory} for the definition.  K-systems
are multiply mixing (a generalisation of mixing), mixing and
ergodic.

\section{Statistical properties: probabilistic limit theorems}
\label{sec:prob-limit-thms}

We now turn to statistical properties, reviewing the application
of results from the theory of stationary stochastic processes to
the context of dynamical systems.

Diffusion in billiards concerns the statistical behaviour of the
particle positions. Denoting the position at time $t$ by $\x_t$
and restricting attention to the first component $x_t$, we can
write
\begin{equation}\label{eq:x-as-integral}
x_t = \int_0^t v_1(s) \rd s + x_0 = \int_0^t f \comp \Phi^s(\cdot)
\rd s + x_0 ,
\end{equation}
where $f = v_1$, the first velocity component.  This expresses
$x_t$ solely in terms of functions defined on the torus, so that
in a sense we have reduced a spatially extended problem (spreading
out over an infinite lattice) to a problem on the torus.  Equation
\eqref{eq:x-as-integral} shows that the displacement $\Dx_t \defeq
x_t-x_0$ is in some sense a more natural observable than the
position $x_t$ in this context.

In the above we are regarding $v_1 \from \M \to \R$ as the
observable
\begin{equation}\label{}
v_1(\omega) = v_1(q,v) = v_1
\end{equation}
which returns the first component of the velocity of the initial
condition, so that $v_1 \comp \Phi^t(\omega)$ is the first
velocity component at time $t$ starting from the initial condition
$\omega=(q,v) \in \M$.

More generally, we can consider integrals of the form in
\eqref{eq:x-as-integral} over other observables $f \from \M \to
\R$; these are important in the study of other transport
processes, for example \cite{BunRev}.  The question we wish to
answer then concerns the distribution of \defn{accumulation
functions} of the form \cite{CherYoung}
\begin{equation}\label{eq:accumulation-function}
S_t(\cdot) \defeq \int_0^t f \comp \Phi^s(\cdot) \rd s,
\end{equation}
in particular in the limit as $t \to \infty$.

 The  integral in
\eqref{eq:accumulation-function} is a continuous-time version of a
Birkhoff sum
\begin{equation}\label{}
S_n = f + f\comp T + f \comp T^2 + \cdots + f \comp T^{n-1} = X_1
+ \cdots + X_n,
\end{equation}
where the $X_n \defeq f \comp T^{n-1}$ are stationary random
variables.  We are thus interested in statistical properties, for
example means and shapes of limiting distributions, of sums and
integrals of stationary random variables.

Intuitively, if the correlations between the $X_i$ decay
sufficiently fast, then they are asymptotically independent, and
we can hope that the classical limit theorems for independent and
identically distributed random variables can be extended to the
stationary case. The Bernstein method, summarised below, is a
rigorous justification for this.

\subsection{Averages}
The simplest statistical properties to study are \defn{averages}
of observables. If $f\from \M \to \R$ is an observable on the
phase space of the dynamical system, then we denote the
\defn{mean} of the observable $f$ by
\begin{equation}\label{}
\mean{f} \defeq \Exp{\mu}{f} = \int_{\M} f \rd \mu = \int_{\M}
f(\omega) \rd \mu(\omega).
\end{equation}
Here, $\mu$ is the distribution of initial conditions $\omega \in
\M$ and $\Exp{\mu}{f}$ denotes the expectation of the random
variable $f$ with respect to the probability measure $\mu$.

We will be interested in the evolution over time of such averages
when the observable $f$ involves the flow $\Phi^t$.  A key role is
played by the \defn{mean squared displacement} at time $t$,
denoted
\begin{equation}\label{eq:defn-msd}
\mean{\Delta x^2}_t \defeq \mean{(\Delta x_t)^2}_\mu = \mean{(x_t
- x_0)^2}_\mu = \left\langle  \left( \int_0^t v_1 \comp
\Phi^s(\cdot) \rd s \right)^2 \right \rangle_\mu.
\end{equation}
In the final expression in \eqref{eq:defn-msd}, the dot denotes
the variable $\omega \in \M$ over which the average
$\mean{\cdot}_\mu$ is taken.  We will usually think of this mean
squared displacement as a function of time $t$, so that it is
convenient to use the notation $\mean{\Delta x^2}_t$, which makes
this dependence explicit.

The physical interpretation of the above definition is as an
average over initial conditions of the time-dependent observable.
 An alternative point of view is to regard $\mean{\Delta
x^2}_t$ as an average over the evolved probability distribution
$\mu_t$ at time $t$ starting from a distribution $\mu_0$ at time
$0$; the average is then over a fixed observable at points
determined by evolving the initial conditions in time.

\subsection{Central limit theorem: independent, identically distributed case}
\label{subsec:clt-iid}

Let $(X_i)$ be independent and identically distributed
(i.i.d.) random variables with mean $0$ and variance $\sigma^2 <
\infty$. We are interested in statistical properties of the
accumulation function
\begin{equation}
S_n \defeq X_1 + \cdots + X_n,
\end{equation}
as $n \to \infty$. We have
\begin{equation}\label{}
\E{S_n} = 0; \quad \var{S_n} = n \var{X},
\end{equation}
so that if we normalise by $\sqrt{n}$, setting $\tilde{S}_n
\defeq S_n / \sqrt{n}$, then we get
\begin{equation}\label{}
\var{\tilde{S}_n} = \var{\frac{S_n}{\sqrt{n}} } =
\frac{\var{S_n}}{n} = \sigma^2.
\end{equation}
Since the variances of the $\tilde{S}_n$ are now independent of
$n$, we may hope that the distributions converge in shape to some
limiting distribution.

With the above conditions, the classical \defn{central limit
theorem} (see e.g.\ \cite{FellerII,Durrett}) states that
\begin{equation}\label{}
\frac{S_n}{\sqrt{n}} \distconv Z_{\sigma^2}, \quad \text{i.e.}
\quad \prob{\frac{S_n}{\sqrt{n}}  \leq x } \longrightarrow
\frac{1}{\sqrt{2 \pi \sigma^2}} \int_{-\infty}^x \exp \left(
\frac{-s^2}{2 \sigma^2} \right) \rd s,
\end{equation}
so that the rescaled accumulation functions converge \defn{in
distribution} to a \defn{normal distribution}.

We recall the definition of this notion of convergence
\cite{Grimmett, Billingsley}. Let $X_n$ be a sequence of
real-valued random variables with distribution functions $F_n$, so
that $F_n(x)
\defeq \prob{X_n \le x}$.  We say that the sequence $X_n$
\defn{converges in distribution} to the random variable $X$, with distribution function
$F(x) \defeq \prob{X\le x}$, written $X_n \distconv X$,  if
$F_n(x) \to F(x)$ at all $x$ where $F(x)$ is continuous.  There is
a generalisation of this concept which applies in much more
general situations: see \appref{app:weak-conv-measures}.

\subsection{Central limit theorem: stationary case}

If now the $X_n$ are no longer independent but they are
stationary, we have
\begin{equation}
\Var[S_n] = n \, C(0) + 2 \sum_{j=1}^{n-1} (n-j) \, C(j)
\end{equation}
where the \defn{autocorrelation function} of $f$ is
\begin{equation}\label{}
C(n) \defeq \E{X_0 \, X_n} = \mean{f . (f \comp T^n)} -
\mean{f}^2,
\end{equation}
 and $\mean{\cdot} = \E{\cdot}$ both denote averages
 (expectations) over the invariant measure.
It follows that
\begin{equation}\label{}
\frac{\Var[S_n]}{n} \stackrel{n \to \infty}{\longrightarrow}
\sigma^2 \defeq C(0) + 2 \sum_{j=1}^\infty C(j),
\end{equation}
\emph{provided $\sum_j j \, C_j$ exists}.  A sufficient condition
for this\footnote{Note that $C_j = \littleo{j^{-2}}$ is not
sufficient, since $\sum_j j \, C_j$ is divergent for $C_j
= j^{-2}$. On the other hand, $C_j = \littleo{j^{-2-\epsilon}}$ is
not necessary either, since e.g.\ for  $C_j = j^{-2} (\log
j)^{-p}$ the sum $\sum_j j \, C_j$ is convergent for $p>1$
\cite{RudinPrinciples}.} is that $C_j = \littleo{j^{-2-\epsilon}}$
for some $\epsilon>0$.

Now we may again hope/expect that
\begin{equation}\label{}
\frac{S_n}{\sqrt{n}} \distconv Z_{\sigma^2}.
\end{equation}
We can also regard the central limit theorem as describing the
distribution of  `fluctuations' around the mean in the Birkhoff
ergodic theorem.  (Note that the Birkhoff ergodic theorem is a
version of the strong law of large numbers for stationary
processes \cite{Durrett}.)

\subsection{Central limit theorem for stationary processes: Bernstein method}

The Bernstein method can be used to prove results of central limit
theorem (CLT) type for stationary processes, based on the idea
that if $\E{X_0 \, X_n}$ decays sufficiently fast then the $X_i$
are
\defn{asymptotically independent} and we can reduce to the case of
independent random variables.

\begin{figure}
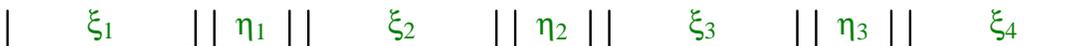

 \centrefig[1]{bernstein-method.1}
 \caption{\label{fig:bernstein-method} Partition for the Bernstein method.}

\end{figure}

The method works as follows (see e.g.\ \cite[Chap.~7]{Durrett} and
\cite{Billingsley}). Split the sum $S_n$ up into alternate blocks
$\xi_j$ of length $p$ and $\eta_j$ of length $q$, so that $\xi_1 =
X_1 + \cdots + X_p$, $\eta_1 = X_{p+1} + \cdots + X_{p+q}$, $\xi_2
= X_{p+q+1} + \cdots X_{2p+q}$, etc.; see
\figref{fig:bernstein-method}.

If  $q$ is sufficiently large then the $\xi_j$ are almost
independent; on the other hand, if $q$ is small compared to $p$
then the sum $\sum_j \eta_j$ of the small blocks is small compared
to the total size of the sum $S_n$.  In a suitable limit we can
thus reduce to the independent case, with an error which tends to
$0$.

\subsection{Rate of mixing}

For the above to work, we need to show for a given stationary
sequence that the correlations $C_n$ decay fast enough, i.e.\ that
we have a fast enough rate of mixing. There are various sufficient
`mixing conditions' for the central limit theorem to hold: see
e.g.\ \cite{Durrett, Billingsley, DavidsonStochLimitTheory} in the
context of probability theory, and \cite{DenkerCLT, LiveraniCLT}
in the specific context of dynamical systems.

\subsection{Functional central limit theorem}

The central limit theorem quoted above deals with one-dimensional
distributions.  There is a multi-dimensional version, where
multi-dimensional distributions converge to multi-dimensional
normal distributions. If also a further technical condition,
called
\defn{tightness}, is satisfied, then the probability distribution
on the space of continuous paths which is induced by the dynamical
system converges weakly to a continuous-time stochastic process.
If further the correlations behave correctly, then the limit will
be the particular case of Brownian motion.  Such a result is known
as a \defn{functional central limit theorem} (i.e.\ a version of
the central limit theorem for paths represented by functions).
This is also called the \defn{weak invariance principle}.

For further details in the
 context of diffusion see \secref{sec:defn-determ-diffn} and
\appref{app:weak-conv-measures}.

\subsection{Almost-sure invariance principle}

The strongest type of statistical property is known as an
\defn{almost-sure invariance principle}, since it says that a
stochastic process can be written, almost surely, as the sum of a
Brownian motion on a suitable space, together with an error which
can be bounded in a precise way: see e.g.\ \cite{PhilippStout,
DenkerPhilipp, MelbourneStatLimit}.

\subsection{Transition from discrete-time to continuous-time}
\label{subsec:green-kubo-disc-cont}

In billiards, we are interested in statistical properties of the
physical continuous-time dynamics.  Currently these are not
directly available, due to the lack of information on decay of
correlations in continuous time; rather they are derived from the
results on the billiard map, using the fact that the flow is a
suspension flow over that map.

We refer to \appref{app:suspension-flows} for the definition of
suspension flows and for a statement of the key theorem relating
the central limit theorem for the flow to that of the map. We give
here a derivation of the Green--Kubo formula from the Einstein
formula for the diffusion coefficient in continuous time,
 valid \emph{only} if the velocity
autocorrelation function decays sufficiently fast, which for
billiards has not yet been proved \cite{CherYoung}. References
containing similar derivations include
\cite{vanBeijeren,DetRev,BoonYip, ResiboisDeLeener}. We remark
that recent results\footnote{\revision{Stretched
exponential bounds on correlations for H\"older-continuous observables
for the 2D periodic Lorentz gas were recently proved in
\cite{ChernovBoundsCorrelationsBilliardFlowsJSP2007}.}} have shown the
exponential decay of
correlations in continuous time for H\"older observables of
Anosov geodesic flows \cite{Dolgopyat, ChernovDecayCorrFlows}
and contact flows \cite{LiveraniContactFlows}.

We take as a starting point the Einstein formula
\begin{equation}\label{eq:einstein}
  D = \lim_{t \to \infty} \frac{1}{2t} \msd_t.
\end{equation}
Using $\D x(t) = x(t) - x(t_0) = \int_{t'=0}^{t} v(t') \d t'$, we
have
\begin{equation}\label{eq:time_translation_inv}
\msd_t = \int_{t'=0}^t \int_{t''=0}^t \vcf{t'}{t''} \d t' \d t''
=\int_{t'=0}^t \int_{\tau=t'-t}^{t'} \vcf{\tau}{0} \d t' \d \tau.
\end{equation}
Here we have used the fact that the averages are invariant with
respect to time translation, due to the stationarity of the
stochastic process; we also performed a change of variables from
$t''$ to $\tau
\defeq t'-t''$.

We now change the order of integration, obtaining
\begin{equation}\label{eq:split_integration_region}
\msd_t = \int_{\tau=-t}^{0} \int_{t'=0}^{t+\tau} \vcf{\tau}{0} \d
t' \d \tau %
+ \int_{\tau=0}^t \int_{t'=\tau}^t \vcf{\tau}{0} \d t' \d \tau.
\end{equation}
Hence
\begin{equation}\label{eq:recombine}
\msd_t=2 \int_{\tau=0}^t (t-\tau) \vcf{\tau}{0} \d \tau.
\end{equation}

We can now define a
\defn{finite-time diffusion coefficient} $D(t)$, an estimate
for the (infinite-time) diffusion coefficient based on the
information available up to time $t$.  There are two possible ways
to do this.  Using the Einstein definition, we have
\begin{equation}\label{eq:finite_time_diffn_coeff}
  D(t) \defeq \frac{\msd_t}{2t} = \int_{\tau=0}^t
  \left(1-\frac{\tau}{t}\right) C(\tau) \d \tau,
\end{equation}
where $C(\tau) \defeq \vcf{\tau}{0}$ is the velocity
autocorrelation function. Hence the diffusion coefficient
\begin{equation}\label{eq:diffn_coeff}
  D \defeq \lim_{t \to \infty} D(t) = \int_0^\infty C(\tau) \rd
  \tau
\end{equation}
exists if: (i) $\int_{\tau=0}^{\infty} C(\tau) \d \tau < \infty$,
i.e. $C$ is integrable; and (ii) $\frac{1}{t} \int_{\tau=0}^t \tau
\, C(\tau) \d \tau \to 0$ as $t \to \infty$.  In fact (ii) follows
from (i) by integrating by parts, so that a necessary and
sufficient condition for the existence of the diffusion
coefficient is that $C(t)$ is integrable.

The equation \eqref{eq:diffn_coeff}, relating the   time integral
of the velocity autocorrelation function to the transport
coefficient $D$, is a
\defn{Green--Kubo relation}
\cite{DorfBook}; Green--Kubo formulae can be found for all
transport coefficients \cite{EvansMorriss, GaspBook}, e.g.\ via
linear response theory \cite{EvansMorriss}, expressing the
transport coefficient as a time integral of the autocorrelation
function of the flux of the quantity being transported.


Another possible definition of a finite-time diffusion coefficient
$\tilD$, which is more closely related to the numerical method we
shall use later \cite{Balescu, AllenTildesley}, is to define
$\tilD$ using the \defn{local slope} of the mean squared
displacement $\msd_t$:
\begin{equation}\label{eq:finite-time-diff-coeff-deriv-defn}
    \tilD(t) \defeq \frac{1}{2} \frac{\d}{\d t}  \msd_t.
\end{equation}
From \eqref{eq:recombine}, we have
\begin{equation}\label{eq:msd-expanded}
\msd_t=2t \int_{\tau=0}^t C(\tau) \d \tau - 2 \int_{\tau=0}^t \tau
\, C(\tau) \d \tau,
\end{equation}
so that the fundamental theorem of calculus gives
\begin{equation}\label{eq:finite-time-diff-coeff-evaluation}
\tilD(t) = \int_{\tau=0}^t C(\tau) \d \tau + t \, C(t) - t \, C(t)
=\int_{\tau=0}^t C(\tau) \d \tau.
\end{equation}
This definition avoids the slowly-decaying $1/t$ tail of the first
definition \cite{Balescu}. Further, if we have a bound on the rate
of decay of correlations then we can use
\eqref{eq:finite-time-diff-coeff-evaluation}
 to estimate the error $\modulus{D-\tilD(t)}$ resulting from using the finite-time
 diffusion coefficient $\tilD(t)$ instead of its infinite-time
 limit $D$.
For example, if correlations decay exponentially as
\begin{equation}\label{eq:exp-decay-corr-result}
\modulus{C(t)} \le K \, e^{-\alpha \, t},
\end{equation}
then we can estimate
\begin{equation}\label{eq:discrepancy-diff-coeff-finite-one}
\modulus{D-\tilD(t)} = \modulus{\int_{\tau=t}^{\infty} C(\tau) \rd
\tau} \le \int_{\tau=t}^{\infty} K \, e^{-\alpha \, \tau} \rd \tau
=  \frac{K}{\alpha} \, e^{-\alpha \, t}.
\end{equation}
Hence this difference tends very rapidly to zero, as seen in
numerical simulations.  However, obtaining the constants $K$ and
$\alpha$, even numerically, is difficult or impossible.

If, on the other hand, correlations decay only algebraically as
\begin{equation}\label{}
\modulus{C(t)} \le K t^{-1-\epsilon},
\end{equation}
then
\begin{equation}\label{}
\modulus{D-\tilD(t)} \le \modulus{\int_{s=t}^\infty C(s) \rd s} =
\frac{1}{\epsilon} \, t^{-\epsilon}.
\end{equation}
This error tends to zero very slowly for small $\epsilon$, so that
the finite-time estimation method for any practical time has an
additional uncertainty built in.


\section{Definition of deterministic diffusion}
\label{sec:defn-determ-diffn}

There is no single definition of deterministic diffusion; instead,
we say that a dynamical system is diffusive if one or more of the
following hierarchy of statistical properties holds.  The idea is
that the statistical properties behave to some extent like those
of solutions of the diffusion equation described in
\secref{sec:diffn-eqn}, or those of Brownian motion (see below).
We use terminology suitable for diffusion in billiards, although
these properties are also relevant for more abstract dynamical
systems.

\begin{enumerate}[\quad (a)]
\item the mean squared displacement  grows linearly asymptotically;

\item the position or
displacement distribution, rescaled by $\sqrt{t}$, converges in
distribution to a non-degenerate Gaussian, i.e.\ a central limit
theorem holds; and

\item the whole process,
suitably rescaled, converges in distribution to a Wiener process.
\end{enumerate}

Properties (a) and (b) are based on the corresponding properties
of the diffusion equation, although  property (b) will in general
only hold at the level of \emph{weak} convergence, whereas in the
classical diffusion equation there is pointwise (and even uniform)
convergence of the \emph{densities}. Property (c) says that the
process, when suitably rescaled, looks like Brownian motion, which
can be thought of as a probabilistic model of diffusion. We have
the implications\footnote{\revision{In fact, as was pointed out to me by Ian
Melbourne, (b) does \emph{not} imply (a) in general. However, in billiards,
they tend to go together -- to know which scaling factor to use in the central
limit theorem, it is necessary to calculate the mean squared displacement.}} (c)
$\Rightarrow$ (b) $\Rightarrow$ (a), so that
(c) is the strongest property. We now give precise versions of
these statements.

\paragraph{(a) Asymptotic linearity of mean squared displacement}

 The limit
\begin{equation}\label{}
2D \defeq \lim_{t\to\infty} \frac{1}{t} \mean{\Dx^2}_t
\end{equation}
exists, so that the mean squared displacement  $\meanat{\Dx^2}{t}
\defeq \mean{\left[\Dx(t)\right]^2}$ (the variance of the
displacement distribution) grows asymptotically linearly in time:
\begin{equation}\label{}
\meanat{\Dx^2}{t} \sim 2Dt \quad \text{as }t \to \infty,
\end{equation}
where $D$ is the \defn{diffusion coefficient}. In $d \ge 2$
dimensions, letting $\Dx_i(t) \defeq x_i(t) - x_i(0)$ be the $i$th component of
the displacement, we have
\begin{equation}\label{}
\meanat{\Dx_i \, \Dx_j}{t} \sim 2 D_{ij} t,
\end{equation}
where the $D_{ij}$ are components of a symmetric diffusion tensor.

\paragraph{(b) Central limit theorem: convergence to normal distribution}

Scale the displacement distribution by $\sqrt{t}$, so that the
variance of the rescaled distribution is bounded. Then this
distribution converges
\defn{weakly}, or \defn{in distribution}, to a normally
distributed random variable $\z$ \cite{GaspNic, CherYoung}:
\begin{equation}\label{}
    \frac{\x(t)-\x(0)}{\sqrt{t}} \distconv
  \z, \qquad \text{as } t \to \infty.
\end{equation}
In the $1$-dimensional case, this means that
\begin{equation}\label{eq:CLT-defn}
\lim_{t \to \infty} \P\left(\frac{x_t-x_0}{\sqrt{t}} < u\right) =
\frac{1}{\sigma \sqrt{2 \pi}} \int_{s=-\infty}^u
\e^{-s^2/2\sigma^2} \rd s,
\end{equation}
where $\P(\cdot)$ denotes the probability of the event inside the
parentheses with respect to the initial distribution of the random
variable $x_0$, and $\sigma^2$ is the variance of the limiting
normal distribution. In $d \ge 2$ dimensions, this is replaced by
similar statements about probabilities of $d$-dimensional sets.
This is the
\defn{central limit theorem} for the random variable $\Dx$.
From (a) we know that in 1D, the variance of the limiting normal
distribution is $\sigma^2 = 2D$; in $d \ge 2$ dimensions, the
covariance matrix of $\z$ is given by the matrix $(2 D_{ij})$
\cite{BS, DettCoh2}.

\paragraph{(c) Functional central limit theorem: convergence of path
distribution to Brownian motion} Consider the following rescaling
of the whole path of the process:
\begin{equation}\label{}
    \tilde{\x}_t(s) := \frac{\x(st) - \x(0)}{\sqrt{t}},
\end{equation}
where $s \in [0,1]$.  The scale $\sqrt{t}$ is the `natural' scale
coming from (b). This rescaling `squashes' the entire path of the
process onto the interval $[0,1]$, so that we can compare paths at
different times. In fact, we compare the induced probability
measures on the space of continuous functions $[0,1] \to \R^d$.

We say that the process satisfies a
\defn{functional central limit theorem} if the probability
distribution of the rescaled
\defn{paths} of the process converges weakly to Wiener measure (Brownian
motion) $\B$ with covariance matrix as in (b), as $t \to \infty$:
\begin{equation} \label{eq:functional-CLT}
    \tilde{\x}_t \distconv \B \quad \text{as }
    t \to \infty,
\end{equation}
 This  is known as a \defn{functional central limit theorem}, or
\defn{weak invariance principle} \cite{CherYoung}.

This makes precise the sense in which  $\tilde{\x}_t$ looks like
Brownian motion on long length and time scales. In the following
sections we discuss in more detail the meaning of the above
statement, by showing how diffusion can be regarded as a
stochastic process and in what sense these rescaled processes
converge to diffusion processes.

\subsection{Diffusion as a stochastic process}
\label{subsec:diffn-as-stoch-proc}
 As recalled in \secref{sec:diffn-eqn}, diffusion is described classically by the diffusion
equation
\begin{equation}\label{eq:diffn-eqn}
\pd{\rho(t, \r)}{t} = D \,   \nabla^2 \rho(t, \r).
\end{equation}
We would like a microscopic model which gives behaviour on a
macroscopic level consistent with this equation. Following
Einstein and Wiener (see e.g.\ \cite{Gardiner}), we look for a
stochastic process $B_t$, determined by the probability density
$p(\x,t)$ of a particle being at position $\x$ at time $t$ given
that it started at $\x=\vect{0}$ at time $t=0$. Based on physical
reasoning, we impose that the sample paths of the process should
be continuous, and that displacements $B_{t+h}-B_t$ should be
independent of the history of the particle motion up to time $t$.

Under certain technical conditions, $p(\x,t)$ then satisfies the
equation
\begin{equation}\label{eq:Fokker-Planck}
\pd{p}{t}  + \pd{}{x_i} \left[A_i \, p - \frac{1}{2} \sum_j
\pd{}{x_j} \left(B_{ij}\, p \right) \right] = 0,
\end{equation}
which is known as \defn{Kolmogorov's forward equation} or the
\defn{Fokker--Planck equation}
\cite{Gardiner}. The  \defn{drift vector} $\vect{A}(\x,t)$ and the
\defn{diffusion tensor} $\tens{B}(\x,t)$
  give the mean and variance, respectively, of  infinitesimal displacements
at position $\x$ and time $t$ \cite{Gardiner}.

If the system is \defn{homogeneous}, then $\vect{A}$ and
$\tens{B}$ are independent of $\x$ and $t$.  If the system is also
sufficiently symmetric, then the drift is zero and the diffusion
tensor is a multiple of the identity tensor. The stochastic
process is then
\defn{Brownian motion},
and the Fokker--Planck equation  \eqref{eq:Fokker-Planck} reduces
to the diffusion equation \eqref{eq:diffn-eqn}.  (We remark that
the factor $1/2$ in \eqref{eq:Fokker-Planck}
 occurs naturally in the probabilistic setting, and often leads to a
discrepancy between the probability and physics literature.)  A
general diffusion process, however, can be
\defn{inhomogeneous} in both space and time.

Brownian motion is defined as follows \cite{Durrett}. A standard
one-dimensional \defn{Brownian motion} (also called the
\defn{Wiener process}) is a real-valued stochastic process $B_t$, $t\ge 0$,
such that:
\begin{enumerate}
\item For all $n$, if $t_0 < t_1 < \ldots < t_n$, then the
\defn{increments} $B(t_0), B(t_1)-B(t_0), \ldots,
B(t_n)-B(t_{n-1})$ are independent random variables.

\item If $t > s \ge 0$, then $B(t)-B(s)$ is a normal random
variable with mean $0$ and variance $t-s$, so that
\begin{equation}\label{}
\P(B(t)-B(s) \in A) = \int_A \frac{1}{\sqrt{2 \pi (t-s)}} \, \exp
\left[ -\frac{x^2}{2 (t-s)} \right] \rd x.
\end{equation}

\item With probability $1$, the function $t \mapsto B(t)$ is
continuous.

\end{enumerate}

Standard $d$-dimensional Brownian motion is then the vector
process $\B(t)
\defeq (B_1(t), \ldots, B_d(t))$, where each $B_i$ is an
independent standard 1D Brownian motion.

Another approach to  study such diffusion processes is via
stochastic differential equations (SDEs) \cite{Gardiner}. The
above process corresponds to solutions of the SDE
\begin{equation}\label{}
\rd \x = \vect{A}(\x,t) \rd t + \vect{\sigma}(\x,t) \rd \B(t),
\end{equation}
where $\B$ is an $n$-dimensional Brownian motion and
$\vect{\sigma} \vect{\sigma}\transp = \tens{B}$.  This form makes
more explicit the infinitesimal increments referred to above, and
is now in a form which is more suitable for numerical simulation.

\subsection{Weak convergence to Brownian motion}
\label{subsec:weak-conv-to-brownian-non-rigorous}

We must make precise what we mean by saying that
\begin{equation}
\label{}
    \tilde{\x}_t \weakconv \B \quad \text{as }
    t \to \infty.
\end{equation}

The rigorous definition of this notion of weak convergence is
given in \appref{app:weak-conv-measures}; see \cite{Billingsley}.
Necessary and sufficient conditions are \cite{Billingsley} (1)
that the
 finite-dimensional distributions of the process
$\tx_t$ converge to those of Brownian motion; and (2) that the
class of induced measures  on path space  is
\defn{tight}: see \appref{app:weak-conv-measures} for the definition.

Property (1) means that for any $n$, any times $s_1 < \cdots <
s_n$, and any reasonable sets $D_1, \ldots, D_n$ in $\R^d$, we
have
\begin{equation}\label{eq:defn-fclt}
\prob{\tx_t(s_1) \in D_1, \ldots, \tx_t(s_n) \in D_n} \stackrel{t
\to \infty}{\longrightarrow} \prob{\B(s_1) \in D_1, \ldots,
\B(s_n) \in D_n}.
\end{equation}
This is called the
\defn{multi-dimensional central limit theorem} in
\cite{CherLimit}. The right-hand side can be expressed as a
multi-dimensional integral over Gaussians, as follows. We see
that\footnote{The probability that $\B(t_1)=\x_1$ exactly is  $0$,
but the argument can be made rigorous.}
\begin{align}\label{}
\condprob{\B(t_2) = \x_2}{\B(t_1)=\x_1} &= \condprob{\B(t_2)-\B(t_1) = \x_2-\x_1}{\B(t_1)=\x_1}\\
&= \prob{\B(t_2)-\B(t_1) = \x_2-\x_1} \\
&= p(\x_2-\x_1, t_2-t_1),
\end{align}
since by the definition of Brownian motion $\B$, the events
$\B(t_2)-\B(t_1) = \x_2-\x_1$ and $\B(t_1)=\x_1$ are independent,
 and $\B(t_2)-\B(t_1)$ is normal with mean
$0$ and variance $2 D_{ij}(t_2-t_1)$; here $p(\x,t)$ is the
Gaussian probability density function with that mean and variance.
Integrating over all $\x_1 \in D_1$ and $\x_2 \in D_2$, we have
\begin{equation}\label{}
\prob{\B(t_1) \in D_1, \B(t_2) \in D_2} =  \int_{D_1} \rd \x_1
\int_{D_2} \rd \x_2 \, p(\x_1, t_1) \, p(\x_2-\x_1, t_2-t_1),
\end{equation}
with a similar expression for the right hand side of
\eqref{eq:defn-fclt}.

Since $\tx(s_i) \in D_i$ if and only if $\x(t s_i)-\x(0) \in
\sqrt{t} D_i$, the multidimensional central limit theorem becomes
\begin{multline}\label{}
\prob{\x(\lambda^2 t_1) \in \lambda D_1,\x(\lambda^2 t_2) \in
\lambda D_2, \ldots, \x(\lambda^2 t_n) \in \lambda D_n} \\
\stackrel{\lambda \to \infty}{\longrightarrow} \int_{D_1} \rd \x_1
\int_{D_2} \rd \x_2 \cdots \int_{D_n} \rd \x_n \,p(\x_1, t_1) \,
p(\x_2-\x_1, t_2-t_1) \cdots p(\x_n-\x_{n-1}, t_n-t_{n-1}),
\end{multline}
setting $\lambda \defeq \sqrt{t}$ and then renaming the times
$s_i$ as $t_i$. This relation was used in \cite{DettCoh2} as the
definition of a diffusive process, although the functional central
limit theorem is a stronger statement which also requires
tightness.  We remark that another name for the functional central
limit theorem is the weak (or Donsker) invariance principle
\cite{CherYoung}.

In general, we may have convergence to a more general diffusion
process than Brownian motion. For example, in
\cite{DurrMechanicalBM} it was shown that the motion of a large
particle embedded in an infinite ideal gas converges to an
Ornstein--Uhlenbeck process.
 In the case of the periodic Lorentz gas, however, we have
time-independent dynamics defined on a system which is
space-homogeneous and symmetric on a large scale;
 the only possible limiting diffusion process which satisfies these conditions
 is Brownian motion.

\subsection{Further remarks}

An important part of the above limit theorems is proving that the
limiting distribution is non-degenerate, i.e.\ that the diffusion
tensor is positive definite \cite{BunRev}. In general, we can
consider the above limit theorems for an arbitrary observable $f$.

We will mostly consider $1$-dimensional projections (marginals) of
the above distributions; we can equivalently regard this as
studying the limit theorems for $f$ being one component of $\D\x$,
which follow from the multi-dimensional results stated above.

\subsection{Discussion of definitions}

Property (c)
 makes precise in what sense a dynamical system looks like Brownian motion
 when correctly
rescaled.  This is the strongest, and so in some sense the best,
property with which we could define deterministic diffusion (i.e.\
a dynamical system is diffusive when it satisfies property (c)).
However, there are very few physically relevant systems which have
been proved to satisfy (c). Interest in the periodic Lorentz gas
comes in large part from the fact that it is one of the only such
systems; another is the triple linkage \cite{HuntMackay}.  As
mentioned above, in general we may instead have convergence to
some other diffusion process.

The multi-dimensional central limit theorem part of (c) was
studied in \cite{DettCoh2}, where both Lorentz gases and
wind--tree models were found to obey it, tested for certain sets
$D_i$ and certain values of $n$.  However, as stated in
\cite{DettCoh2}, (c) is difficult to investigate numerically, and
the results in that paper seem to be the best that we can expect.

Property (b), the central limit theorem, has been shown for large
classes of observables $f$ in many dynamical systems (see e.g.\
\cite{CherLimit} and references therein),
 but
again they are not often physically relevant systems.  Property
(b) was used in \cite{GaspNic} as the definition of a diffusive
system, but does not seem to have been applied in the physics
literature; this is the approach taken in
\chapref{chap:fine-structure}.

Most emphasis in the physical literature is placed on property
(a): many authors define a system to be diffusive if only property
(a) is verified (numerically), e.g.\
\cite{KlagesD00,AlonsoPolyg,DettCoh1}.  Many types of system are
diffusive in this sense, including 1D maps \cite{KlagesD99},
Lorentz gases (periodic and random) \cite{BS, KlagesD00, DettCoh1}
and Ehrenfest wind--tree models, both periodic \cite{AlonsoPolyg}
and random \cite{DettCoh1}.

The implication (c) $\Rightarrow$ (b) always holds, and (a) and (b)
usually go together. The reverse implications
 (a) $\Rightarrow$ (b) and (b) $\Rightarrow$ (c) hold only under
certain additional
 conditions: see e.g.\ \cite{CherLimit}.
It is thus of interest to ask if dynamical systems showing
property (a) also show properties (b) and (c). For example, in
\cite{CohenKong} a \emph{disordered} lattice-gas wind--tree model
was reported as having an asymptotically linear mean squared
displacement, but a non-Gaussian distribution function, i.e.\ (a)
but not (b). However, disorder can lead to trapping effects which
cannot occur in periodic systems \cite{AlonsoPolyg}; we are not
aware of a \emph{periodic} (and hence ordered) billiard model with
a unit-speed velocity distribution which shows (a) but not (b),
although in \secref{sec:maxwell-vel-distn} we show that this can
occur with a Maxwellian velocity distribution.






\subsection{Approach via cumulants}

Several papers have approached property (b) by studying higher
order cumulants of the distribution,  e.g.\ \cite{AlonsoPolyg,
DettCoh1}.  A normal distribution has all cumulants higher than
the second equal to zero. It thus seems that we need all cumulants
of the rescaled displacement distribution to tend to zero as $t
\to \infty$ for the distribution to approach a normal
distribution; this was stated in \cite[p.~790]{DettCoh1}, for
example.

In fact, the weak type of convergence occurring in the central
limit theorem (convergence in distribution) does not in general
require higher moments and cumulants even to exist: a faster rate
of decay of correlations is necessary to ensure the existence of
higher cumulants than is required for the central limit theorem to
hold: see \cite{vanBeijeren, CherDett}.

Higher cumulants are instead related to large deviations,
describing behaviour in the tails of the distribution \cite[Sec.\
7.3.6]{GaspBook}. We show in \chapref{chap:polygonal} that a class
of polygonal billiard channels which was found in
\cite{AlonsoPolyg} to have rapidly-growing cumulants
 nonetheless
appears to satisfy the central limit theorem.

\section{Rigorous results on billiards}
\label{sec:billiards-rigorous-results}


In general, a billiard table is a Riemannian manifold with a
piecewise smooth boundary.  The dynamics is given by geodesic flow
away from the boundary and specular reflection (elastic collision)
at the boundary. Here we restrict
attention to two-dimensional periodic Lorentz gases, where
non-interacting point particles in free motion on a torus-shaped
billiard table undergo elastic collisions with a set of fixed
scatterers.   We give only enough detail for our purposes, and
refer to \cite{Tabachnikov}, \cite[Chap.~6]{Cornfeld} and
\cite[Chap.~4]{CherMark} for further information\footnote{\revision{An
excellent recent reference on chaotic billiards is
\cite{ChernovMarkarianChaoticBilliardsAMS2006}.}}.


\subsection{Continuous-time dynamics}

\begin{figure}
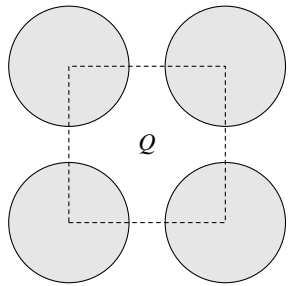

\centrefig{square_geom.2} \caption{\label{fig:bill-domain} Example
of a billiard domain $Q$ with scatterers removed.}
\end{figure}

We denote by $Q$ the \defn{billiard domain}, i.e.\ the available
region where particles can move, given by
\begin{equation}\label{}
Q \defeq \T^2 \setminus \bigcup_{i} \Disc_i,
\end{equation}
where the scatterers $\Disc_i$ are non-intersecting discs in the
case of the periodic Lorentz gas.  We  visualise the torus $\T^2$
as a square $[0,1)^2$ with periodic boundary conditions, as in
\figref{fig:bill-domain}.

Each particle moves in a straight line within $Q$ at constant
velocity $\v$ until it hits the boundary $\boundary Q$ at a point
$\q$ on the boundary $\boundary \Disc_i$ of scatterer $i$. It then
undergoes an
\defn{elastic collision}, giving a new velocity
\begin{equation}\label{}
\v' = \v - 2(\v\cdot \n) \, \n
\end{equation}
after collision, where $\n \defeq \n(\q)$ is the outward normal
vector at $\q$ to the scatterer $\Disc_i$. The particle continues
its motion by alternation of straight line motion and elastic
collisions with the boundary.

The above prescription defines a continuous-time dynamical system
given by a flow
\begin{gather}\label{}
\tilde{\flow}^t \from \Mt \to \Mt;\\
 \tilde{\flow}^t \from
(\q_0,\v_0) \mapsto (\q_t,\v_t),
\end{gather}
where $\Mt \defeq \tang Q$ is the tangent bundle to $Q$, i.e.\ the
set
\begin{equation}\label{}
\Mt \defeq \theset{ (\q, \v) \colon \q \in Q, \v \in
\tangsp{\q}{Q} },
\end{equation}
where $\tangsp{\q}{Q}$ is the tangent space to $Q$ at the point
$\q$.


Since the speed $\norm{\v}$ is preserved, it is customary to take
all particles with unit speed\footnote{We consider non-constant
speed distributions in \secref{sec:maxwell-vel-distn}.}.  We can
thus restrict attention to the new phase space $\M$ given by the
unit tangent bundle\footnote{The identification of the unit
tangent bundle $\tang_1 Q$ with the direct product $Q \times S^1$
is not in general valid, but it does hold for the simple situation
we consider.}
\begin{equation}\label{}
\M \defeq \unittang Q = Q \times S^1,
\end{equation}
where
\begin{equation}\label{}
S^1 \defeq \theset{\v \in \R^2 : \norm{\v}=1},
\end{equation}
and the flow
 $\Phi^t \from \M \to \M$, which we call the \defn{billiard flow}.
We can parametrise the space $\M$ with coordinates $(x,y,\theta)$,
where $(x,y) \in [0,1)^2$ are position coordinates and $\theta \in
[0,2\pi)$ is the angle of the velocity vector.

We remark that for billiard tables with corners, where two smooth
parts of the boundary touch at a non-zero angle, the trajectory
cannot be continued if it exactly hits a corner.  However, the set
of all such trajectories has measure $0$.

\subsection{Billiard map: Poincar\'e section}

As is often the case in dynamical systems, it turns out to be
technically easier to study a Poincar\'e section of the flow.  A
natural cross-section is given by
\begin{equation}\label{}
M \defeq \theset{(\q,\v) \in \M \colon \q \in \boundary Q,
\thickspace \v \cdot \n(\q) > 0},
\end{equation}
i.e.\ the set of vectors whose base points lie on a scatterer and
which point into the interior of the billiard domain.  This
cross-section $M$ forms a two-dimensional surface in the
three-dimensional phase space $\M$.  One way of parametrising $M$
is to take as coordinates the arc length $r$ around the scatterer
of the point of collision (from some reference point), and the
angle $\varphi$ between the velocity vector and the normal vector
\cite{CherMark}. Alternatively we can take the arc length around
the scatterer and the sine of $\varphi$, giving canonically
conjugate
\defn{Birkhoff coordinates} \cite[Chap.~5]{GaspBook}.  In general
we must also keep track of a label $i$ denoting which scatterer
was hit. The flow $\flow^t$ then \defn{induces} an invertible map
$\map \from M \to M$ which maps one collision into the next,
called the
\defn{billiard map}.

The billiard flow $\flow^t$ can now be regarded as a
\defn{suspension flow} over $T$, under the free path length function
$\freepath \from M \to \Rplus$, where $\freepath(\q,\v)$ is the
length of the collision-free part of the trajectory emanating from
the initial condition $(\q,\v) \in M$.  Suspension flows are
treated in detail in \appref{app:suspension-flows}.

\subsection{Measures}

Since the billiard flow $\flow^t$ is Hamiltonian, it preserves a
natural invariant measure $\mu$, called \defn{Liouville measure},
given by\footnote{This is a useful shorthand for the statement
\begin{equation}\label{}
\mu(A) = \int_A \rd \mu  = c_{\mu} \int_{(x,y,\theta)\in A} \rd x
\rd y \rd \theta \qquad \text{for all } A \in \Balg.
\end{equation}%
}
\begin{equation}\label{}
\rd \mu \defeq c_{\mu} \rd x \rd y \rd \theta,
\end{equation}
where $\rd x$ etc.\ denote Lebesgue measure in the respective
coordinate, and $c_{\mu}$ is a normalising constant chosen so that
$\mu(\M)=1$. This measure induces a measure $\nu$ on $M$ which is
invariant under $T$, given by \cite{CherMark}
\begin{equation}\label{}
\rd \nu \defeq c_{\nu} \cos \varphi \rd r \rd \varphi.
\end{equation}
For a 2D billiard (i.e.\ one for which the dynamics is restricted
to a plane), we have
\begin{equation}\label{}
c_{\mu} = \frac{1}{2 \pi \modulus{Q}}
\end{equation}
and
\begin{equation}\label{}
c_{\nu} = \frac{1}{2 \modulus{\boundary Q}}.
\end{equation}

\subsection{Ergodic and statistical properties}
\label{subsec:billiards-erg-stat-props}

The ergodic and statistical properties of billiards have been the
object of intense scrutiny since the seminal work of Sinai
\cite{Sinai70}, where the K-property of the periodic 2-disc fluid
was proved.  All of the proofs depend on the \defn{hyperbolicity}
of scattering billiards, i.e.\ the existence of stable and
unstable directions; for details see
\cite{ChernovMarkarianChaoticBilliardsAMS2006, CherMark, GaspBook,
Tabachnikov} and the references cited below.  A few of the key
rigorous results for dispersing and semi-dispersing billiards are:

\begin{itemize}
\item 2D periodic Lorentz gas models for which a geometrical \defn{finite horizon}
condition holds (\secref{subsec:finite-horizon}) satisfy the
central limit theorem and functional central limit theorem if the
scatterers are disjoint and piecewise $C^3$ smooth \cite{BS, BSC};

\item higher-dimensional Lorentz gases with the same conditions also satisfy the CLT and
FCLT \cite{Chernov94}, but there are issues with the proofs in
higher dimensions that need addressing\footnote{N.~I.~Chernov,
private communication.} after the work of \cite{BalintAsterisque,
BalintAnnHenriPoinc};

\item velocity autocorrelation functions of the billiard map decay
exponentially \cite{Young, Cher99}; and

\item hard ball fluids are ergodic, mixing and K-systems: see
\cite{Simanyi2DErgHypNoExcep} and references therein.

\end{itemize}
There is an excellent collection of reviews in \cite{Szasz}.

\section{Numerical evaluation of statistical quantities}
\label{sec:monte-carlo}

To evaluate numerically statistical quantities such as the mean
squared displacement, we use a simple Monte Carlo method.  We take
a large sample $(\x\up{i}_0,\v\up{i}_0)_{i=1}^N$ of size $N$ of
initial conditions chosen uniformly with respect to Liouville
measure in one unit cell using a random number generator:  the
positions $\x_0$ are uniform with respect to Lebesgue measure in
the billiard domain $Q$, and the velocities $\v_0$ are uniform in
the unit circle $S^1$, i.e.\ with angles between $0$ and $2 \pi$,
and unit speeds. These evolve after time $t$ to
$(\x\up{i}(t),\v\up{i}(t))_{i=1}^N$; the distribution of this
ensemble then gives an approximation to that of $(\x(t), \v(t))$.

We use the Mersenne Twister random number generator
\cite{MersenneTwister}, which seems to be considered to be one of
the best-performing generators; we have also done some comparisons
with a standard \texttt{rand} implementation.  Since our random
numbers determine only the initial conditions, and since the
system is strongly mixing, we do not expect the choice of random
number generator to be  important, provided the initial conditions
are sufficiently well distributed through the phase space, so that
no small regions with important dynamical effects  are omitted.

We denote averages over the initial conditions, or equivalently
expectations with respect to the distribution of $(\x_0, \v_0)$,
by $\langle \cdot \rangle$.  Approximations of such averages can
now be evaluated via  \cite{NR}
\begin{equation}\label{}
\mean{f(\x_0,\v_0)} = \lim_{N \to \infty} \frac{1}{N}\sum_{i=1}^N
f(\x_0\up{i},\v_0\up{i}).
\end{equation}
The infinite sample size limit, although unobtainable in practice,
reflects the expectation that larger $N$ will give a better
approximation.

\paragraph{Lyapunov instability}

The Lyapunov instability in dispersing billiards implies that
numerical simulations of trajectories will not be accurate beyond
a time where the uncertainty in the initial condition has grown
exponentially to $\bigO{1}$, so that some authors do not allow any
data obtained beyond that time \cite{GarrGall}. However, most
physics papers implicitly reject this argument by presenting
long-time results from simulations, and we shall also do this,
whilst being careful about
 other sources of error in statistical calculations.  One partial
 justification for this comes from \defn{shadowing}
\cite{ChernovMarkarianChaoticBilliardsAMS2006}: in certain
 hyperbolic systems, it can be proved that there is a true
 trajectory close to a simulated trajectory.  The difficulty in
 billiards comes from the discontinuous nature of the dynamics;
 nonetheless, some results of shadowing type were proved in
 \cite{TroubetzkoyShadowing}.  We are not aware of any papers
 which investigate shadowing in particular billiard models,
 but early numerical investigations of the validity of numerical estimates of
 ergodic properties for uniformly hyperbolic
 systems can be found in \cite{Benettin1, Benettin2}.



\startonright




\graphicspath{{figs/}}

\chapter[Geometry-dependence of diffusion coefficint]{Geometry-dependence of the diffusion coefficient in a 2D
periodic Lorentz gas model}

\label{chap:geom-dependence}

\section{Two-dimensional periodic Lorentz gas model}
\label{sec:2d-lorentz-gas-model}


A \defn{periodic Lorentz gas} is a periodic billiard with disjoint
circular scatterers (discs).  The term is also applied to other
periodic billiards of dispersing type with scatterers of more
general shape: see e.g.\ \cite{HarayamaKG02}.

In this chapter and the following one we study aspects of
deterministic diffusion in a particular 2D periodic Lorentz gas
model introduced below: here we consider only the mean squared
displacement and the diffusion coefficient, while in
\chapref{chap:fine-structure} we examine the shape of
distributions.

\subsection{Finite horizon condition} \label{subsec:finite-horizon}

Periodic Lorentz gases were shown in \cite{BS, BSC} to be
diffusive
 (\secref{sec:defn-determ-diffn}), provided they satisfy the
\defn{finite horizon} condition:
  there is an upper bound on the free path length between collisions.
 If this is not the case, so that a particle can travel infinitely far
without colliding with any scatterers (the billiard has an
\defn{infinite horizon}), then \defn{corridors} exist, which allow
for fast-propagating orbits.  This leads to
\defn{super-diffusive} behaviour, in the sense that the  mean squared
displacement grows like $t \log t$ (which is faster than $t$), as
has recently been proved in
\cite{SzaszVarjuLimitLawsRecurrencePlanarLorentzInfiniteHorizonJSP2007}, after
heuristic and numerical arguments
\cite{FriedmanMartin84, GeiselAnomDiff} and analytical 
evidence \cite{Bleher} were
previously given.  There is a more detailed discussion of
statistical behaviour in the infinite-horizon regime in
\secref{sec:stat-behaviour-inf-horiz}.

We are mainly interested in normal diffusion, so that we wish to
restrict attention to values of the geometrical parameters for
which there is a finite horizon.  The simplest periodic
two-dimensional lattice\footnote{We consistently use the term
`lattice' in the sense of physics to mean any periodic structure.
Mathematicians instead use the term in the narrower sense of the
set of points in $\R^d$ of the form $\sum_{i=1}^d a_i \vect{e}_i$,
where the $\vect{e}_i$ are basis vectors of $\R^d$ and  $a_i \in
\Z$; these are called
\defn{Bravais lattices} in physics \cite{AshcroftMermin}.} is the
square lattice.  To allow the possibility of diffusion in a square
lattice, the scatterers must not touch or overlap: if they do, we
obtain a localised regime where any particle is trapped in a
bounded region of space, so that the diffusion coefficient is
necessarily $0$. However, non-touching scatterers in a square
lattice always imply an infinite horizon: trajectories parallel to
the lattice axes and passing sufficiently close to the centre of a
unit cell never collide.

For this reason a periodic Lorentz gas on a \emph{triangular}
lattice is often used, where there is a finite horizon when the
non-touching discs are sufficiently close together: see e.g.\
\cite{MZ, KlagesD00}.  We instead elect to retain a square lattice
model, blocking all possible corridors by adding an additional
disc at the centre of each cell.

\subsection{Model studied}

\label{subsec:lorentz-model-studied}

\begin{figure}
\centering
 \subfigure[]{\label{fig:2dlattice}
\includegraphics{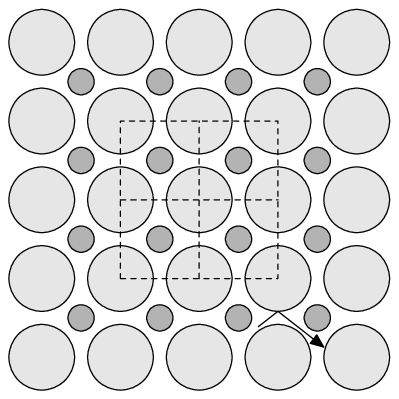}}
\qquad \qquad \subfigure[]{\label{fig:geom2d}
\includegraphics{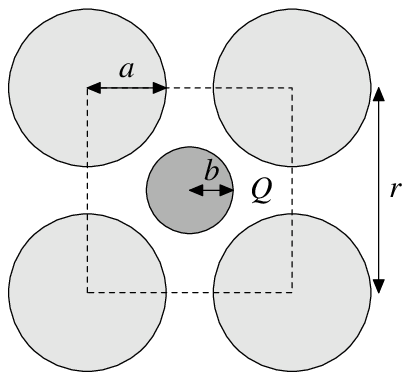}}
\caption{ (a) Part of the infinite system, constructed from two
square lattices of discs, shown in different shades of grey; the
dashed lines indicate several unit cells. (b) A single unit cell,
defining the geometrical parameters.  The billiard domain is the
area $Q$ exterior to the discs.}
\end{figure}

The model which we focus on, previously studied in
\cite{GarrGall,Garrido}, consists of two inter-penetrating square
lattices of discs; they have the same lattice spacing $r$, and
radii $a$ and $b$, respectively, and are arranged such that each
unit cell of the $a$-lattice contains a $b$-disc at its center.
Part of the infinite system is shown in \figref{fig:2dlattice} and
the geometrical parameters are defined in \figref{fig:geom2d}.

We were led to this model  as a cross-section of the 3D model
presented in \chapref{chap:3dmodel}, independently of \cite{GarrGall,
Garrido},
although as described above, it is a natural model to consider.
The same model has also been studied in \cite{BonettoThermoI,
BonettoThermoII} in a different context. The two-dimensional
parameter space of this model allows us to
 vary independently, and hence study the effect of, two physical quantities:
 the size of exits of each unit cell, and the
 accessible area of a unit cell.  This is not possible in the
 standard triangular Lorentz gas.


\subsection{Length scale}
\label{subsec:length-scale}

Since the speed is conserved, scaling the geometry by a constant
factor does not alter the essentials of the dynamics, just the
time-scale on which it occurs.  We may hence use dimensionless
units such that the speed is equal to $1$ and we are free to
choose a length scale.

In this thesis we use one of two choices of length scale: in
analytical calculations it is often convenient to fix $a=1$ and
vary $r$ and $b$; while in numerical calculations we often fix
$r=1$ and vary $a$ and $b$. Where necessary we distinguish the
latter case using tildes, writing
\begin{equation}\label{}
\ta\defeq a/r; \qquad \tb \defeq b/r.
\end{equation}

\subsection{Parameter space}
\label{subsec:2d-parameter-space}

Garrido \cite{Garrido} derived the `phase diagram'\ of the model,
showing  the regions in parameter space corresponding to the
various localised, finite horizon and infinite-horizon regimes.
Here we correct this diagram by including the following features
that were incorrectly treated in \cite{Garrido}: (i) the system is
symmetrical under interchange of $\ta$ and $\tb$; and (ii) the
finite-horizon regime is smaller than was found in \cite{Garrido}.
We fix $r=1$ but omit the tildes on $\ta$ and $\tb$. The
derivation, which reduces to geometrical considerations, is useful
since it provides pointers for the geometrically more involved
calculation for the 3D model in \chapref{chap:3dmodel}.

\begin{figure}[p]
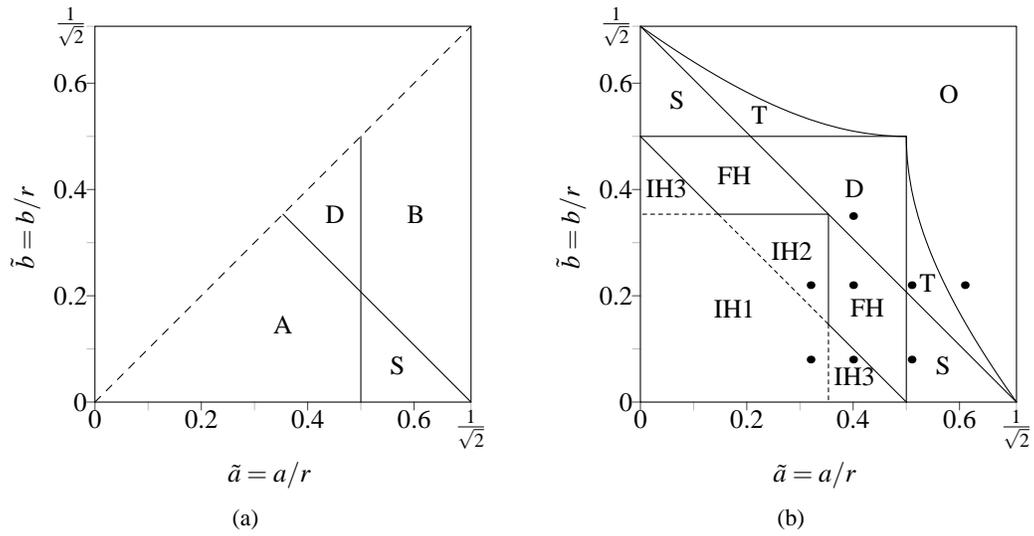

  \centering
\subfigure[]{\label{fig:squarephase-init}\includegraphics{square-phase-diag.2}}
  \qquad
\subfigure[]{\label{fig:square-phase-diag}\includegraphics{square-phase-diag.1}}
  \caption{(a) Four initial regions. (b) Full phase diagram; dots indicate
parameter values for
Figs.~\ref{fig:non-localised-configs}~and~\ref{fig:localised-configs}.}

\end{figure}

If we interchange the radii $a$ and $b$, then the resulting
structure is the same as the initial one, except for a translation
by the vector $(\texthalf, \texthalf)$.  Hence the statistical
properties are the same, so that the whole phase diagram is
reflection symmetric in the line $a=b$.  In the following we hence
restrict attention to the triangular region $a \ge b$ shown in
\figref{fig:squarephase-init}, and then complete the diagram by
reflection.

The main features of the geometry are (i) whether the $a$ discs
touch or overlap with each other; and (ii) whether the $b$ discs
overlap with the $a$ discs. The respective boundaries in parameter
space are given by the straight lines with equations
\begin{equation}\label{eq:phase-diag-lines}
   a=1/2 \quad \text{and} \quad b=\frac{1}{\sqrt{2}}-a,
\end{equation}
respectively. These divide the region $a>b$  into $4$ sub-regions
A, B, D (\defn{diamond}) and S (\defn{star}), shown in
\figref{fig:squarephase-init}.

In the complete phase diagram, shown in
\figref{fig:square-phase-diag}, we have the following
subdivisions:
\begin{itemize}
\item region A into FH (\defn{finite horizon}) and IH
(\defn{infinite horizon}); \item  region B into T
(\defn{triangle}) and O (\defn{overlapping}: whole plane covered).
\end{itemize}
The IH regime can be further subdivided into
regimes with different classes of possible infinite trajectories:
\begin{itemize}
\item IH1: both vertical/horizontal and diagonal infinite
trajectories exist; \item IH2: only diagonal ones exist; and \item
IH3: only vertical/horizontal ones. \end{itemize} The region IH1
can itself be subdivided into infinitely many regions where
increasingly many corridors (at diagonal angles other than
$45\degrees$) become available.  Bleher \cite[Sec.~8]{Bleher}
gives a concise method to find the boundaries of these different
IH regimes in the case $b=0$.
\begin{figure}[p]
\placefignocap[0.8]{0.23}{gen_unit_cell.1}{IH1} \hfill
\placefignocap[0.8]{0.23}{gen_unit_cell.2}{IH2} \hfill
\placefignocap[0.8]{0.23}{gen_unit_cell.3}{IH3} \hfill
\placefignocap[0.8]{0.23}{gen_unit_cell.33}{FH}
 \caption{Non-localised configurations}
 \label{fig:non-localised-configs}
\end{figure}
\begin{figure}[p]
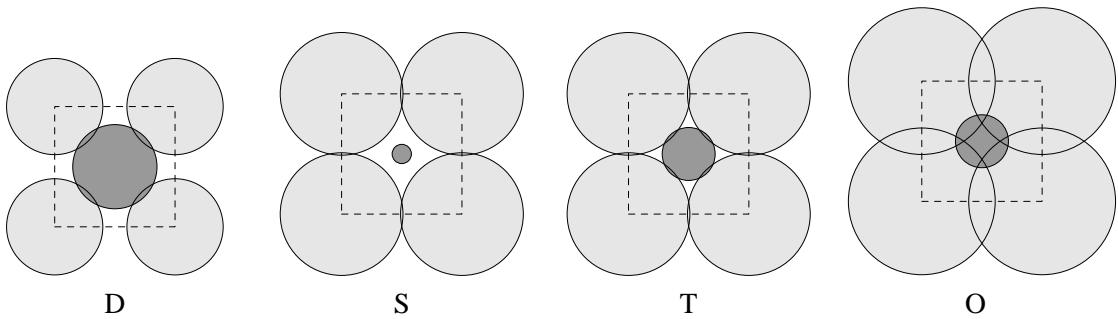

\placefignocap[0.8]{0.23}{gen_unit_cell.4}{D} \hfill
\placefignocap[0.8]{0.23}{gen_unit_cell.5}{S} \hfill
\placefignocap[0.8]{0.23}{gen_unit_cell.6}{T} \hfill
\placefignocap[0.8]{0.23}{gen_unit_cell.7}{O} \caption{Localised
configurations} \label{fig:localised-configs}
\end{figure}

To have a finite horizon we must block horizontal, vertical and
diagonal corridors.  (The necessity to block diagonal corridors
was overlooked in \cite{Garrido}.) Horizontal and vertical
corridors are blocked by requiring the central disc to be
sufficiently large, namely $b \ge \frac{1}{2}-a$, and by symmetry
it suffices to block diagonal trajectories at $45\degrees$ between
mid-points of adjacent sides; for this the $a$ discs must be large
enough. (If we instead use $b$ discs to block these diagonal
trajectories, we necessarily have $b>a$; by our convention, we
thus reduce to the case considered.) Hence for a finite horizon we
supplement Garrido's condition
\begin{equation}\label{eq:finite-horiz-condition-1}
    \texthalf - a \, \le \, b \, \le \, \textfrac{1}{\sqrt{2}}-a
\end{equation}
 with\footnote{Note that the critical case of a `grazing' collision (when
 a trajectory is tangent to a scatterer) is sufficient to have normal
diffusion.}
 \begin{equation}\label{eq:finite-horiz-condition-2}
    a \ge \frac{1}{2 \sqrt{2}}.
\end{equation}

The boundary between regions T and O occurs when the $b$ disc
covers the whole region left empty between the overlapping $a$
discs.  This occurs when the boundary of a $b$ disc reaches the
point of intersection of the boundaries of two $a$ discs, when
$b=\frac{1}{2} - \sqrt{a^2-\frac{1}{4}}$.

The complete phase diagram is shown in
\figref{fig:square-phase-diag}.  The dots give the parameter
values used to depict the various regimes in
\figref{fig:non-localised-configs}  (non-localised configurations)
and \figref{fig:localised-configs} (localised configurations);
white indicates the allowed region of motion of the particles.

\subsection{Channel geometry} \label{subsec:channel-geometry}

Diffusion in a 2D infinite lattice is described by a second order diffusion
tensor having
 $4$ components $D_{ij}$ with respect to a given orthonormal
basis, which we refer to as \defn{diffusion coefficients}.  If the
system is sufficiently symmetric then the diffusion tensor reduces
to a scalar multiple of the identity tensor. Square symmetry, as
in the model presented above, and hexagonal symmetry, as in the
standard triangular periodic Lorentz gas \cite{KlagesD00} are both
sufficient for this reduction to occur; see also
\secref{sec:reducing-symmetry}.

We study \emph{individually} the components of the diffusion
tensor, defined by
\begin{equation}\label{}
D_{ij} = \lim_{t \to \infty} \frac{1}{2t} \mean{\Dx_i \Dx_j}_t.
\end{equation}
This enables us to check that $D_{xx} = D_{yy} \eqdef D$ and
$D_{xy}=D_{yx}=0$ in fully symmetric systems. For those systems
the diffusion coefficient can then be evaluated as an average over
the multidimensional distribution via
\begin{equation}\label{}
\mean{\D\x^2}_t = \mean{\Dx^2}_t+\mean{\Delta y^2}_t \sim 4Dt,
\end{equation}
as used e.g.\ in \cite{KlagesD00}, but this cannot be applied to
systems where the diffusion tensor is not a multiple of the
identity tensor.

\begin{figure}
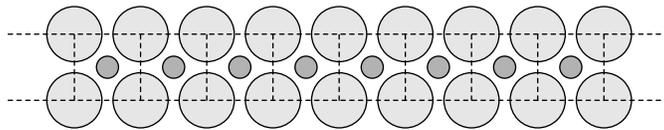

 \centrefig[1.1]{unfolding-torus.eps}
  \caption{One-dimensional channel created by unfolding torus only
in $x$-direction.}
  \label{fig:1dlattice}
\end{figure}

We can evaluate the diffusion coefficient $D_{xx}$ by looking at
the dynamics only in the $x$-direction, which corresponds to
studying the billiard dynamics in a $1$-dimensional
\defn{channel} extended in the $x$-direction; see
\figref{fig:1dlattice}.  Correspondingly, we restrict attention to
1D marginal distributions, which are easier to analyse.

\begin{figure}
\centering \subfigure[]{\label{fig:lorentz-channel}
\includegraphics{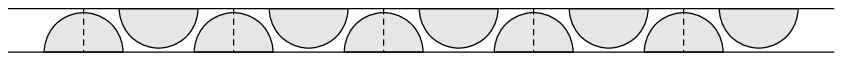}}

\subfigure[]{\label{fig:unfolded-lorentz-channel}
\includegraphics{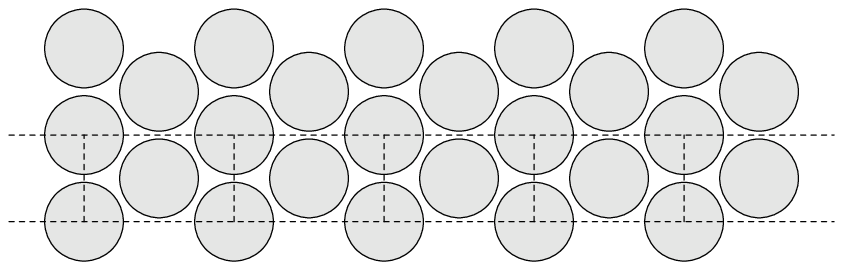}} \caption{ (a) Lorentz channel
studied in \cite{GaspChaotScattering,
  AlonsoLorentzChannel} with hard upper and lower boundaries; dotted lines
indicate unit cells. (b) Fully unfolded triangular Lorentz gas.
  Dotted lines indicate unit cells forming a channel with periodic upper and
  lower boundaries, created by reflecting the channel in
  (a) in the upper boundary.}
\end{figure}

A channel geometry, with hard horizontal boundaries, corresponding
to the triangular Lorentz gas was studied in
\cite{GaspChaotScattering, AlonsoLorentzChannel}
(\figref{fig:lorentz-channel}).
  This is equivalent  to
 a channel with twice the original height and
\emph{periodic} boundaries, by reflecting once in the hard boundary.
This new channel is shown in
\figref{fig:unfolded-lorentz-channel} as part of the whole
triangular lattice obtained by unfolding completely in the
vertical direction.
We can then view the triangular  lattice as consisting of
rectangular unit cells (\figref{fig:unfolded-lorentz-channel})
which are
 stretched versions of the square unit cell considered
above, with the extra condition $a=b$.

\subsection{Effect of length scale on diffusion
coefficient}

Since we have two choices of length scale
(\secref{subsec:length-scale}), we need to know how to convert
calculated diffusion coefficients between the two length scales.
Recall the convention that $\ta = a/r$, $\tb = b/r$; we also
introduce for clarity $\tr \defeq r/r = 1$.

Let $x(t)$ be the position at time $t$ in the system with $a=1$
fixed and speed $1$, with radii $r$ and $b$; and let
$\tilde{x}(t)$ be the position at time $t$ in the system with
$\tr=1$ fixed and speed $1$, with radii $\ta$ and $\tb$.

The motion in the tilde system is then equivalent to that in the
$a=1$ system, but with speed $r$; however, the size of the system
is also shrunk by a factor $r$. Thus we have
\begin{equation}\label{}
\tilde{x}(t) = \frac{x(rt)}{r},
\end{equation}
and so
\begin{equation}\label{}
\mean{\tilde{x}(t)^2} =
\left\langle\left(\frac{x(rt)}{r}\right)^2\right\rangle \sim
\frac{1}{r^2} 2 D rt.
\end{equation}
Hence
\begin{equation}\label{}
\tilde{D} = \frac{D}{r},
\end{equation}
where $\tilde{D}$ is the diffusion coefficient in the tilde
system.  (This result can also be obtained by dimensional
analysis.)

\section{Estimation of diffusion coefficient}
\label{sec:estimation-of-diffn-coeffs}

Having set up the model, we now consider the statistical question
of how to estimate the diffusion coefficient from numerical data.

\subsection{Estimation of moments}

A first approach to characterise the  distribution of a random
variable is to look at its moments. The $r$th \defn{raw population
moment} of a random variable $X$ is given by\footnote{We follow
the notation of \cite{RoseSmithMathstatica}.}
\begin{equation}\label{}
\rawmom_r(X) \defeq \mean{X^r} \defeq \E{X^r} =
\int_{-\infty}^\infty x\,^r \rd F_X(x) = \int_{-\infty}^\infty
x\,^r f_X(x) \rd x,
\end{equation}
where $F_X$ is the distribution function of the random variable
$X$; the integral with respect to $F_X$ is a Lebesgue--Stieltjes
integral, and the last equality holds if the random variable has a
density $f_X$.

If we take a sample $(X_i)_{i=1,\ldots,N}$ of independent and
identically distributed random variables from the distribution of
$X$, then the $m$th \defn{sample raw moment} is
\begin{equation}\label{}
\samprawmom_r \defeq \frac{1}{N} \sum_{i=1}^N X_i^r.
\end{equation}
This is an unbiased estimator of the $r$th population raw moment
$\rawmom_r$, i.e.\ we have $\E{\samprawmom_r}=\rawmom_r$. Halmos
\cite{HalmosUnbiased} showed that in fact it is the unique
unbiased estimator of $\rawmom_r(X)$ which is a symmetric function
of the $X_i$, and further that this estimator has the smallest
variance of all unbiased estimators;  in this sense it is the
\defn{best} estimator of $\rawmom_m(X)$ given the sample.

We are mainly interested in the distribution of position and
displacement in billiard models starting from an initial
distribution which is uniform in a unit cell. By symmetry, the
mean displacement $\mean{\Dx}_t$ then always vanishes, so that the
simplest non-trivial
characteristic of the distribution is the \emph{mean squared displacement}%
\footnote{The notation $\msd_t$ emphasises that we are averaging
over the distribution at time $t$, but we have $\msd_t =
\mean{[\Dx(t)]^2}$, where now we are thinking of averaging over
the distribution of initial conditions. This distinction is the
same as that between the Schr\"odinger and Heisenberg pictures in
quantum mechanics.} $\msd_t$. We can view $\msd_t$ either as the
2nd raw moment of the random variable $\Dx_t$, or as the 1st
moment of the distribution of the new random variable $Y_t \defeq
\Dx_t^2$, as follows:
\begin{equation}\label{}
\msd_t = \rawmom_2(\Dx_t) = \rawmom_1(Y_t),
\end{equation}
The best estimator of the mean squared displacement is thus
\begin{equation}\label{eq:msd-estimator}
\samprawmom_1 \defeq \samprawmom_1(Y_t) = \samprawmom_2(\Dx_t) =
\frac{1}{N} \sum_{i=1}^N (\Dx_t\up{i})^2;
\end{equation}
this can be regarded as a simple Monte Carlo estimator (see
\secref{sec:monte-carlo}).
 In our numerical experiments we restrict ourselves to a fixed number
of initial conditions $N$, although in principle we could add more
data until some pre-defined error tolerance was reached.

\subsection{Distribution of mean squared displacement at time $t$}

To establish how good an estimator \eqref{eq:msd-estimator} is, we
need to determine the width of the distribution of
$\samprawmom_1$. The $\Dx_t\up{i}$ are independent and identically
distributed (i.i.d.) random variables with mean $0$; since
particles with speed $v$ satisfy $\modulus{\Dx_t} \le vt$, they
also have finite variance. It follows that the $Y\up{i}
\defeq (\Dx_t\up{i})^2$ are also \iid, with positive mean and finite
variance. Hence the standard central limit theorem
(\secref{subsec:clt-iid}) applies, so that for large $N$ the
distribution of $\samprawmom_1$ is very close to normal, with
unknown mean and variance
\begin{equation}\label{}
\var{\samprawmom_1} = \frac{1}{N^2} \sum_{i=1}^N \var{\Dx_t^2} =
\frac{1}{N} \var{\Dx_t^2}.
\end{equation}
For fixed $t$, we thus arrive at the standard result that the
width of the distribution of $\samprawmom_1$, as measured by the
standard deviation (the square root of the variance) of
$\samprawmom_1$, is $\bigO{N^{-1/2}}$.

The best estimator for this variance can be found using the known
result for the variance of a sample mean, giving
\begin{equation}\label{}
S^2 \defeq \frac{1}{N-1} \sum_{i=1}^N
(Y_t\up{i}-\samprawmom_1(Y_t))^2.
\end{equation}


\subsection{Calculation of error bars: confidence interval for $\msd_t$}

We now establish a confidence interval for $\msd_t$.
For a normal random variable $Z$ with mean $\mu$ and variance
$\sigma$ we have
\begin{equation}\label{}
\prob{\frac{Z - \mu}{\sigma} \in [\confintsize,\confintsize]} =
\confintval,
\end{equation}
i.e.\ the probability that a normal random variable lies within
$\confintsize$ standard deviations of the mean is $99\%$. Since
neither $\mu$ nor $\sigma$ are known for the distribution of
$\rawmom_1(Y_t)$, we estimate them via $\samprawmom_1$ and $s$,
respectively, where $s$ is the value of $\sqrt{S^2}$ for the data.

\begin{figure}
\centering \subfigure[]{\includegraphics{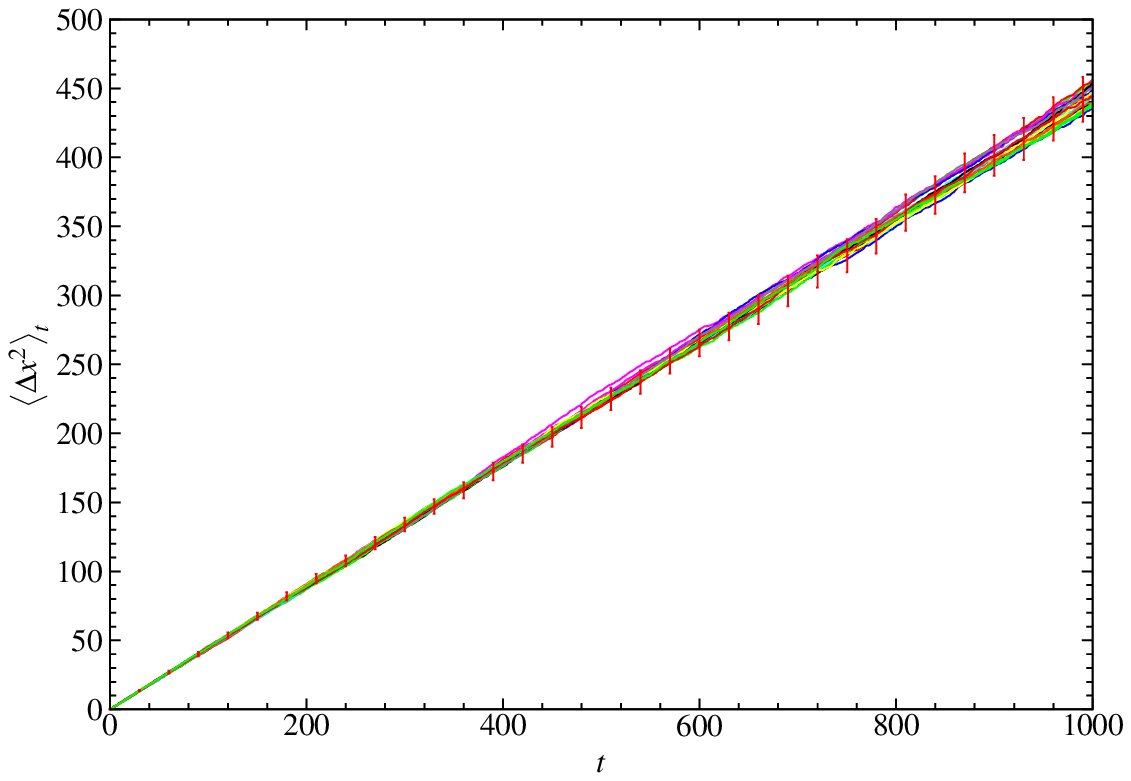}}

\subfigure[]{\includegraphics{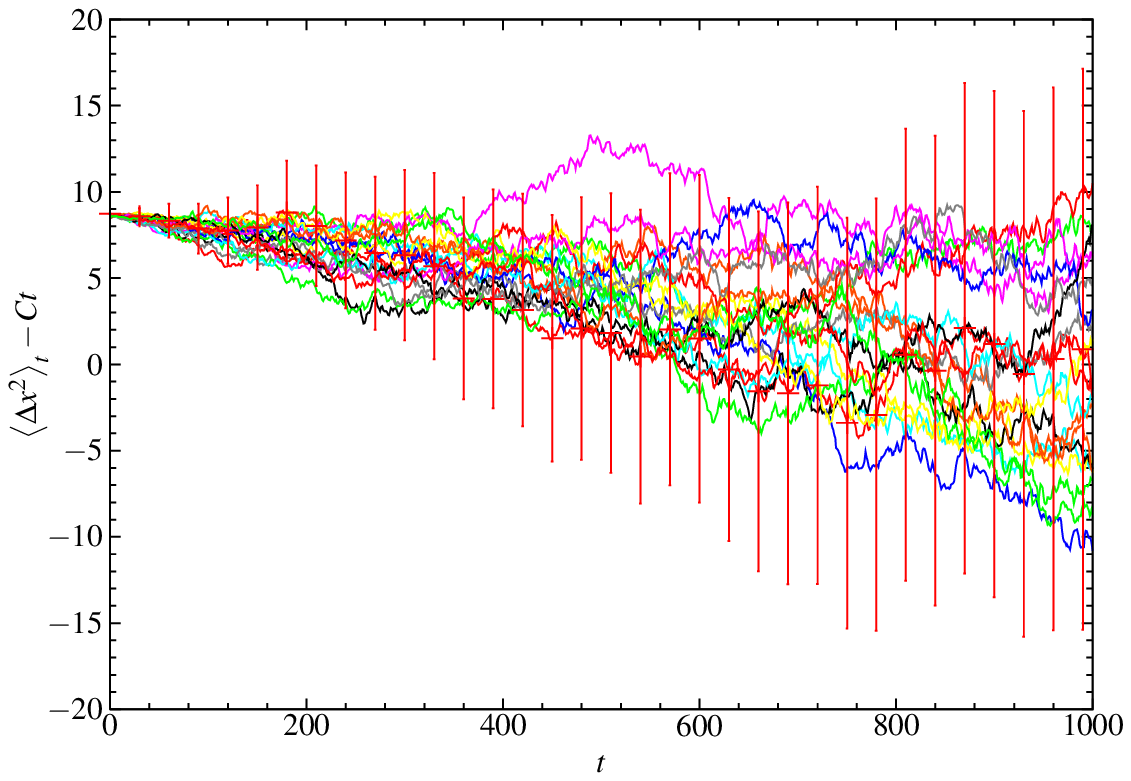}}
 \caption{\label{fig:msd-paths} (a) Mean
squared displacement $\msd_t$ as a function of $t$ for $20$
samples, each averaged over $N=10000$ data points.  Also shown are
error bars showing $99\%$ confidence intervals for one of the
samples. The other data mostly lie inside the error bars. (b)
Deviations from a linear fit to the data set whose error bars are
shown.}
\end{figure}

We will then estimate a $99\%$ confidence interval for
$\rawmom(Y_t)$ by the interval estimator
\begin{equation}\label{}
[\samprawmom_1-\confintsize s, \, \samprawmom_1+\confintsize s].
\end{equation}
\bfigref{fig:msd-paths} shows the estimate $\samprawmom_1(Y_t)$ as
a function of $t$ for several samples. Error bars for one sample
are also shown at $\samprawmom_1 \pm \confintsize s$; the error
bars contain most of the data for each sample.

\subsection{Estimation of diffusion coefficient}

The most efficient method to estimate the diffusion coefficient  is
to use the fact that
 the asymptotic growth rate of the mean squared
displacement in the linear regime is $2D$ and to calculate this by
fitting a straight line to the data in the presumed linear regime
\cite{KlagesD00}.  We must thus consider the evolution of $\msd_t$
over time. In practice, the linear regime is attained very
rapidly: see \figref{fig:msd-paths}(a). To find the slope in the
asymptotic regime  we apply \defn{linear regression} (see e.g.\
\cite{NR}), performing a least squares fit of a straight line to
the data in the region of large $t$, making sure that the
correlation coefficient of the fit is very close to $1$.

The linear regression procedure gives an estimate of the diffusion
coefficient as the slope of the fitted line.  We also need to
estimate the possible error that this estimate has made compared
to the true value.  This question is treated in statistics
textbooks mostly for data whose errors are independent.  This is
however \emph{not} the case here: if at a given time $t$ the
estimated mean squared displacement has deviated from the
`correct' value, then at a small time later, there is a large
probability that the mean squared displacement deviates away from
the `correct' value \emph{in the same direction} as in the
previous step.

\subsection{Growth of width of distribution of $\Dx^2_t$ with $t$}

To find the growth rate of the width of the $\Dx_t^2$
distribution, we argue as follows.  Since the distribution of $\Dx_t^2$ at a
fixed time $t$ is
(approximately)
normal, we only need to consider the standard deviation.  We have
\begin{equation}\label{}
\var{\Dx_t^2} = \E{\Dx_t^4} - (\E{\Dx_t^2})^2 = \rawmom_4(t) -
\rawmom_2(t)^2,
\end{equation}
setting $\rawmom_n(t) \defeq \rawmom_n(\Dx_t)$.

Chernov \& Dettmann \cite{CherDett} showed that for a periodic
Lorentz gas with finite horizon, the 4th-order
\defn{Burnett coefficient} $B$ exists, where
\begin{equation}\label{eq:def-burnett-coeff}
B \defeq \lim_{t\to\infty} \frac{1}{4!\,t} \, \cumulant_4(t),
\end{equation}
and
\begin{equation}\label{eq:def-cumulant}
\cumulant_4(t) \defeq
\rawmom_4(t) - 3 \rawmom_2(t)^2
\end{equation}
is the 4th-order \defn{cumulant}\footnote{The set of cumulants of
a distribution is an alternative  set of moment-type descriptors
of a distribution which have certain advantages over the raw
moments: see e.g.\ \cite{RoseSmithMathstatica}. For example,
$\cumulant_n(X+Y)=\cumulant_n(X)+\cumulant_n(Y)$ for independent
random variables $X$ and $Y$; this does not hold in general for
$\samprawmom_n$.}. This cumulant is related to the non-dimensional
\defn{kurtosis excess} $\gamma_2(t)$ via
\begin{equation}\label{}
\gamma_2(t) = \frac{\rawmom_4(t)}{\rawmom_2(t)^2} - 3 =
\frac{\cumulant_4(t)}{\rawmom_2(t)^2};
\end{equation}
the kurtosis excess $\gamma_2(t)$ is a common measure of how far a
distribution is from a Gaussian, the difference being exactly $0$
for a Gaussian \cite{MathworldKurtosis}.

\begin{figure}
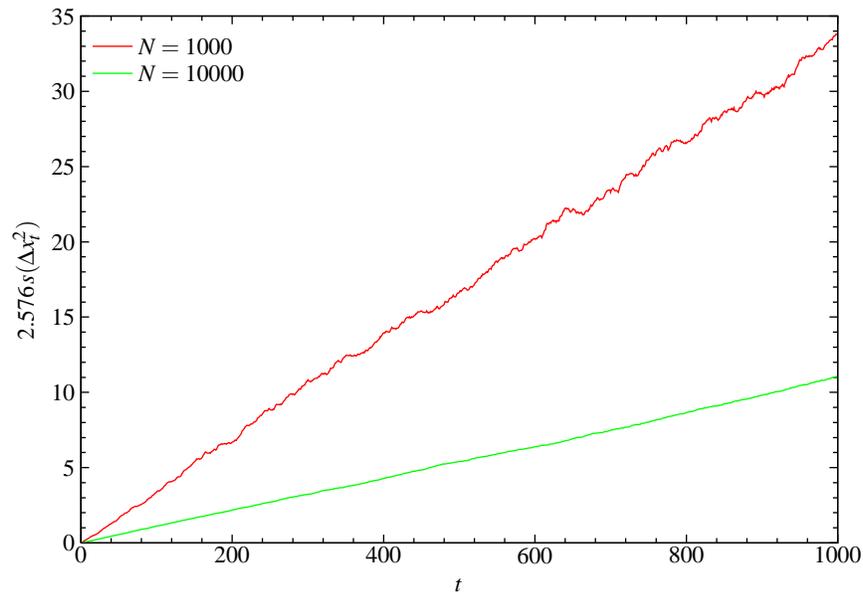

\centrefig{width-dx2-distn.eps} \caption{\label{fig:error-growth}
Growth of half-width of $99\%$ confidence interval for $\msd_t$
for two different sample sizes $N$.}
\end{figure}

We thus have
\begin{equation}\label{}
\var{\Dx_t^2} = \cumulant_4(t) + 2\rawmom_2(t)^2 \sim t+t^2 \sim
t^2 \quad \text{as }t \to \infty,
\end{equation}
since ${\rawmom_2(t)}^2$  grows like $t^2$ whereas
$\cumulant_4(t)$ grows like $t$. Hence the width of the
distribution of $\Dx_t^2$, as measured by the standard deviation,
grows like $t$. This does not seem to have been previously
remarked; it can be seen in \figref{fig:error-growth}, where the
half-width $2.576\,s(\Delta x_t^2)$ of a $99\%$ confidence
interval for $\msd_t$ is plotted as a function of $t$.

We note that Garrido \& Gallavotti \cite{GarrGall} considered  the
question of estimating the error in the diffusion coefficient by
an expansion assuming that the errors are small. However, if we
could continue the simulation indefinitely, the above calculation
shows that the errors would grow arbitrarily large.  Nonetheless
we can still estimate an error in the diffusion coefficient, as
follows.

\subsection{Sampling distribution of diffusion coefficient}

By repeating the above estimation procedure for $M$ different sets
of random initial conditions, we obtain a sample set
$(\hat{D}\up{i})_{i=1}^M$ of estimated values of the diffusion
coefficient.  From these we can estimate the
\defn{sampling distribution} of $D$; this will indicate how close
to the true underlying value of $D$ the estimate $\hat{D}\up{i}$
is expected to be.

Suppose that the confidence intervals for the mean squared
displacement enclose \emph{all} of the data.  Then the growth rate
of the data must lie between the rates of growth of the upper and
lower limits of the error bars.  In fact we can never guarantee
that all the data is enclosed, but nonetheless we estimate the
width of the  $D$ distribution by fitting straight lines to the
error bars.  For $99\%$ confidence intervals on the mean squared
displacement error bars, we would naively hope to obtain
 a $99\%$ confidence interval on $D$.  However, a confidence interval for each
 point does not necessarily correspond to a
 confidence \defn{band} for the whole curve \cite{RiceStatistics}.

\begin{figure}
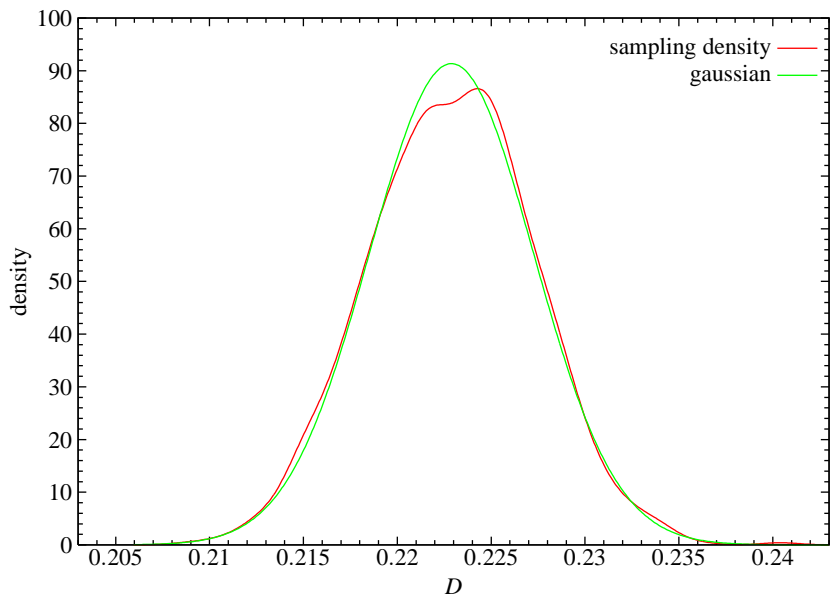

\centrefig{sampling-density.eps}
\caption{\label{fig:sampling-density-D}Sampling density of $D$ for
$M=1000$ samples of size $N=10000$, $r=2.5$, $b=0.5$, obtained by
a kernel density estimate.  A Gaussian with the same mean and
variance is shown for comparison.}
\end{figure}

We investigate the sampling distribution of $D$ by constructing a
kernel density estimate \cite{Silverman} of the sampling density.
\figref{fig:sampling-density-D} shows the sampling density
obtained for particular geometrical parameters for $M=1000$
samples of size $N=10000$.  For this case, the standard deviation
of the sampling distribution is $\sigma=0.044$. The supposed
$99\%$ confidence interval found by the above method (fitting
upper and lower error bar limits) is $0.08\pm0.012$. (This value
varies depending on the sample.)  The estimated error bars from
$D$ hence correspond to a width of $1.8\sigma$, which is
approximately a $90\%$ confidence interval.  Our method thus
slightly underestimates the width of the sampling distribution. We
have repeated the calculation for larger values of $N$, and have
found empirically that our method of estimating the width of the
distribution improves as $N$ increases.

\section{Geometry dependence of diffusion coefficients}
 \label{sec:geom-dependence}

We now consider the geometry-dependence of the diffusion
coefficient in our model as a function of the geometrical
parameters.  This question has been studied in \cite{KlagesD00}
for the triangular Lorentz gas, and in \cite{HarayamaKG02} in a
billiard with `flower-shaped' scatterers. As emphasised above, our
model has the advantage that we can vary \emph{independently} two
physical factors which we expect to influence the diffusion
coefficient: (i) the size of trap exits; and (ii) the available
area in each unit cell. We remark that striking results have also
been obtained for lifted circle maps on the real line
\cite{KlagesD95, GroeneveldK02}, where a fractal parameter
dependence of the diffusion coefficient was found.  This motivated
a conjecture of Klages \cite{KlagesD00} that the diffusion
coefficient in low-dimensional dynamical systems could in general
be a highly non-trivial function  of parameter\footnote{\revision{
Rigorous results on this question have recently been obtained for simple maps
\cite{KellerKlagesContinuityPropsTranspCoeffsSimpleMapsNonlin2008} and 
Lorentz gases
\cite[chapter~5]{ChernovDolgopyatBrownianBrownianMotionAMSMemoirs}.}}. We have
not
attempted to address this question: although we find that the
diffusion coefficient in our model is a reasonably smooth function
of parameter, we have not studied the fine-scale dependence in
detail.

\subsection{Constant trap exit size}

In this section we work in the tilde scaling, i.e.\ fixing the
side of the unit cell to be of length $1$.  The parameters varied
are then $\ta$ and $\tb$.

\begin{figure}
\centering \subfigure[]{\includegraphics{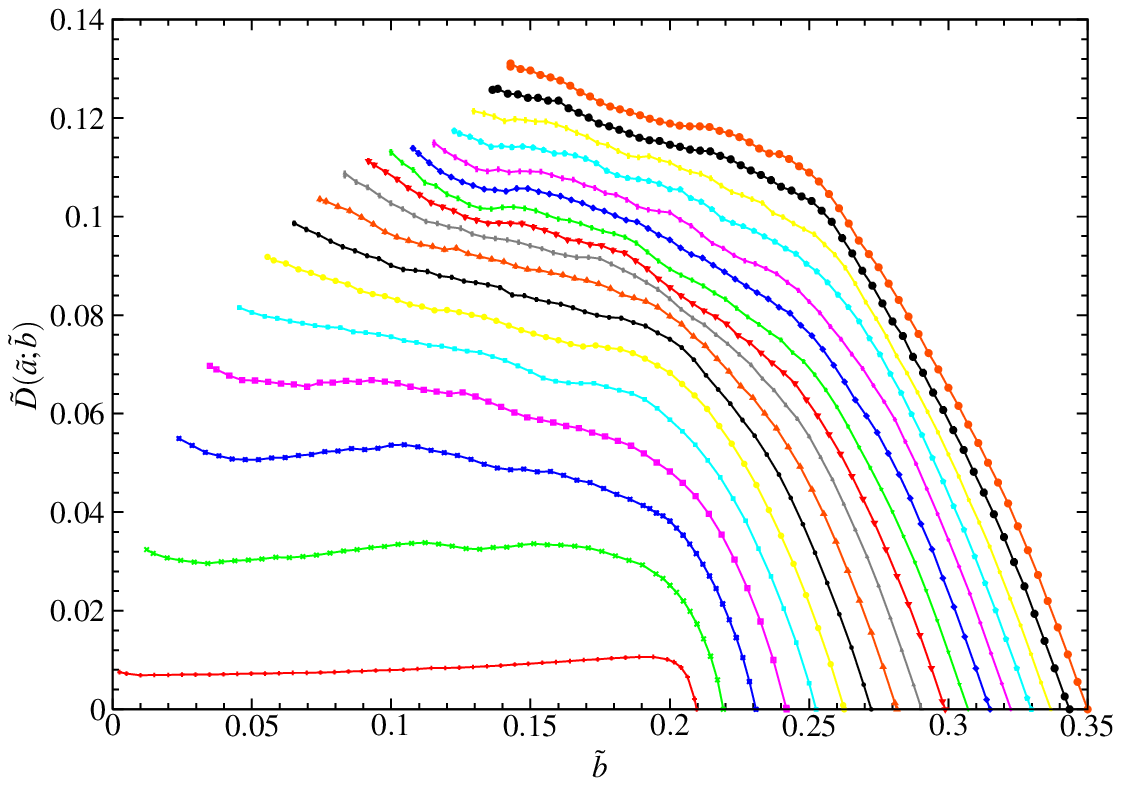}}

\subfigure[]{\includegraphics{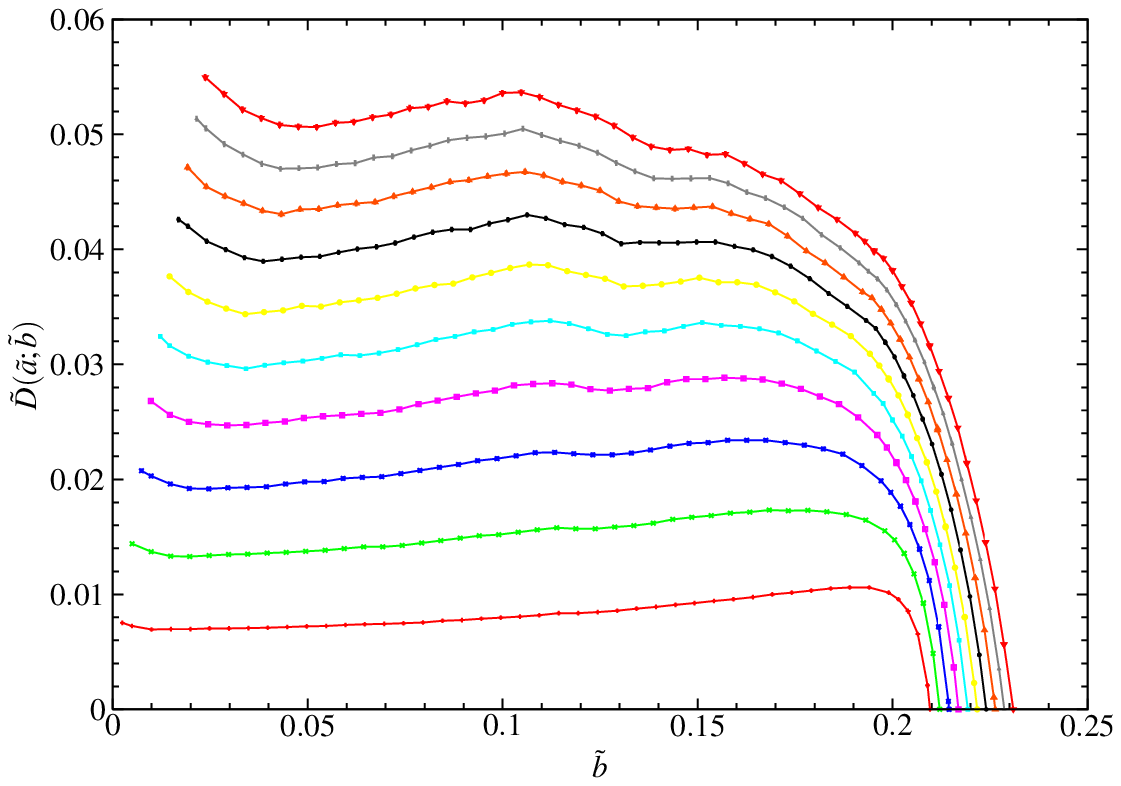}}
\caption{\label{fig:geom-dependence}(a) Geometry dependence of
diffusion coefficients over  whole finite-horizon regime.  Each
curve is for a fixed value of $\ta = 1/r$, where  $2.05 \le r \le
2.8$ and $r$ changes in increments of $0.05$; $r=2.01$ is also
shown. (b) Detailed view of the region $2.01 \le r \le 2.1$ in
steps of $0.01$. In both (a) and (b) $r$ increases from bottom to
top.}
\end{figure}

We fix the radius $\ta$ of the non-central discs and vary the
central disc radius $\tb$ over the allowed portion of the finite
horizon regime for that value of $\ta$.  We repeat this for
different values of $\ta$ covering the finite-horizon regime. Each
curve thus corresponds to fixing the size of the exits of the unit
cell, while the trap area varies.
\bfigref{fig:geom-dependence}(a) shows the diffusion coefficient
$\tilde{D}(\tb;\ta)$  over the whole finite-horizon regime.
  The curves are plotted over the allowed
range
\begin{equation}\label{eq:finite-horizon-regime}
    \frac{1}{2} -\ta \, \le \tb \, < \, \frac{1}{\sqrt{2}}-\ta \eqdef
\bmax(\ta);
\end{equation}
 $\tilde{D}$ vanishes for $\tb \ge 1/\sqrt{2}-\ta$, since the particle
is then completely trapped in a bounded region, so the rightmost
symbol on each curve is plotted at $(\bmax(\ta)$,0). Error bars,
estimated as in the previous section, are plotted, and are
indistinguishable from the data symbols (except possibly for the
largest values of $\ta$).

We note the following features of these graphs.
\begin{enumerate}

\item The curves seem to be continuous as a function of $\tb$, with
some degree of regularity.

\item They possess features such as local minima and
maxima which are reproduced in nearby curves, indicating that
there is also continuity of $\tilde{D}$ as a function of $\ta$
with $\tb$ fixed (see also below).

\item \label{enum:increasing-r} Decreasing $\ta$ with $\tb$ fixed results in
faster
diffusion.

\item \label{enum:critical-value} There is a critical value $\ta=\ta_c \simeq
0.48$,
below which the curves acquire a
non-trivial maximum away from the left end of the curve.
\bfigref{fig:geom-dependence}(b) shows in detail the region where
this transition occurs, i.e.\ the lower left portion of
\figref{fig:geom-dependence}.

\item For $\ta$ close to its upper bound $1/2$, i.e.\ when the
$a$-discs are very close together, the maximum diffusion rate
occurs for a value of $\tb$ close to $\bmax$.

\end{enumerate}

Point \itemref{enum:increasing-r} corresponds to the findings with
the triangular periodic Lorentz gas \cite{KlagesK02}, since there
both trap area and trap exit size increase when the parameter
is increased.  Intuitively, it is `easier' for the particle to
move through the lattice.

Point \itemref{enum:critical-value} is reminiscent of the
behaviour near to a critical point in the context of phase
transitions.  We have not found a physical explanation of this,
although \secref{sec:machta-zwanzig} contains some related
comments.

\paragraph{Diffusion coefficient as function of gap size}

Since the finite-horizon regime is a parallelogram, it is
interesting to consider the diffusion coefficient in terms of a
new variable $\tc \defeq \tb - (\frac{1}{2} - \ta)$ giving the
distance above the least allowed value for $\tb$.  In fact, we
also have
\begin{equation}\label{}
\tc = (\frac{1}{\sqrt{2}} - \frac{1}{2}) - \tilde{w_2},
\end{equation}
where $w_2$ is the minimum distance from the boundary of the
$b$-disc to the boundary of an $a$-disc (see also the next
section).  This is shown in \ref{fig:new-geom-dependence}. We can thus
view the graph as plotting the diffusion
coefficient as a function of this distance.

\begin{figure} 
\centrefig{all-data-new-new.eps} \caption{Geometry dependence of
diffusion coefficients over the whole finite-horizon regime, as a function of
$\tc \defeq \tb - (\frac{1}{2} - \ta$). The curves are as in
\figref{fig:geom-dependence}.}
\label{fig:new-geom-dependence}
\end{figure}

We see that the graphs approximately collapse for values of $w_2$
close to its maximum, i.e.\ when the $a$- and $b$-discs are close
to touching.  In particular, there seems to be a universal
approach of the diffusion coefficient to $0$ when $w_2 \to 0$.  We
remark that it was shown in \cite{BunDiff} that viewing
$\tilde{D}$ as a function of $w_2$, we have
\begin{equation}\label{}
\frac{\tilde{D}(w_2)}{w_2} \to \text{const},
\end{equation}
so that this approach occurs in a linear way as $\tilde{D} \sim
w_2$; this should be contrasted, for example, with the square
root-type behaviour $D \sim \sqrt{w_2}$ often found in phase
transitions.


\subsection{Diffusion coefficient variation with constant trap area}

We now fix the available area per unit cell and vary the trap exit
size; again we work in the tilde system.
\bfigref{fig:param-space-data} shows the variation of the
diffusion coefficient along circular curves in the parameter space
of \figref{fig:square-phase-diag} with constant radius $\alpha =
\sqrt{\ta^2 + \tb^2}$, and hence constant billiard domain area
$\modulus{Q} = 1 - \pi \alpha^2$.  The diffusion coefficient is
plotted as a function of $\theta \defeq \tan^{-1}(\tb/\ta)$, the
angle anti-clockwise from the $\tb$-axis along the circle of
radius $\alpha$.  When we move along these circles the radius
$\tb$ of the central disc increases, while the radius $\ta$ of the
other discs decreases, and hence so does the trap exit size. We
find the following:

\begin{figure}
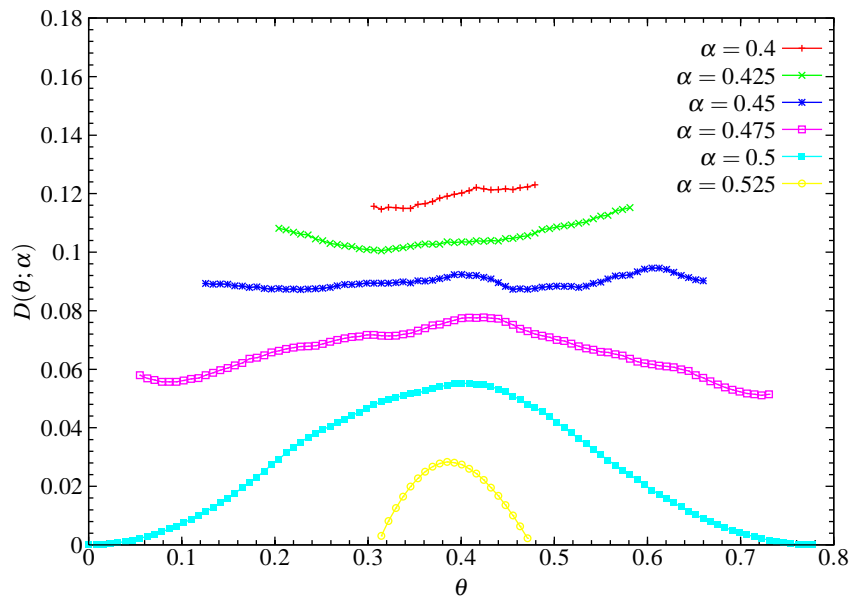

\centrefig{param_space_data.eps}
\caption{\label{fig:param-space-data} Diffusion coefficient as a
function of angle $\theta$ around circles of radius $\alpha$ in
phase space restricted to the finite-horizon regime.}
\end{figure}

\begin{enumerate}

 \item As $\alpha$ decreases, and so the available area increases, the
rate of diffusion increases.

\item For larger values of $\alpha$, the curves lie mainly within
two regions where a random walk model gives a good approximation
to the diffusion coefficient, as we show in
\secref{sec:machta-zwanzig}.

\item For $\alpha\approx0.45$,
the main effect seems to be due to variation of the trap area.
However, there is a still some variation along the curves.  This
provides some evidence against the application of an approximation
originating in the Boltzmann equation proposed in
\cite{KlagesD00}, since that approximation depends on density
alone.  In that paper the application was to the triangular
periodic Lorentz gas, but the lack of variable parameters in that
model did not allow the investigation of this question.
\end{enumerate}


\section{Machta--Zwanzig random walk model}
\label{sec:machta-zwanzig}

We wish to understand the dependence of the diffusion coefficient
$D$ on the geometry of the periodic Lorentz gas: can
we predict the gross structure of curves such as
those in \secref{sec:geom-dependence} via analytical arguments? In
this section we study the idea of Machta \& Zwanzig \cite{MZ} to
approximate the deterministic motion by a random walk.

\subsection{Derivation of Machta--Zwanzig random walk
approximation}

The idea of Machta \& Zwanzig \cite{MZ} is as follows. If the
lattice spacing is such that the discs are very close, then the
particle will on average be trapped for a long time in each unit
cell.  We refer to such a cell as a
\defn{trap}, where the exits of a trap should have a length which
is small compared to its total perimeter. Due to the scattering
nature of the boundaries, the velocity autocorrelation function
decays fast, in fact exponentially in the number of collisions
\cite{CherYoung}. Thus when the particle leaves a trap, its motion
will be almost uncorrelated with its motion on entering the trap.
We thus try to approximate the deterministic motion by a
completely uncorrelated
\defn{random walk} between traps.


Machta \& Zwanzig gave a physical argument which enabled them to
calculate  the mean residence time $\meanrestime$ in a trap.  The
result agrees with a rigorous calculation discussed below. Once
the mean residence time is known, we can derive the diffusion
coefficient of the random walk via
\begin{equation}\label{}
D = \frac{l^2}{4\meanrestime}
\end{equation}
for a 2D random walk on an isotropic lattice with lattice spacing
$l$ and  residence time $\meanrestime$.

We remark that the same idea of approximating irregular
deterministic dynamics by a stochastic process has also been
explored extensively in the context of transport in
area-preserving maps, where irregular motion  occurs only in part
of the phase space: see \cite{MacKayTransportHamilSysPhysD,
Wiggins, LichtenbergLieberman}.

\subsection{Application to our model}

Let $\trap$ be the trap region and $\trapboundary$ the part of its
boundary (lying on the edge of the torus viewed as a square) which
particles can cross, and denote by $\modulus{A}$ the
$m$-dimensional Lebesgue measure of the $m$-dimensional set $A$.
Then by an argument detailed in
\secref{subsec:torus-boundary-map}, together with a method of
Chernov (see \cite{CherMark, CherMFT}), we have
\begin{equation}\label{eq:mean-free-time-chernov-type}
\meanrestime = \frac{c_{\nu'}}{c_\mu} = \frac{\modulus{\trap}.
\modulus{S^{d-1}}}{\modulus{\trapboundary} . \modulus{B^{d-1}}}.
\end{equation}
Here $c_{\nu'} \defeq (\modulus{\trapboundary} .
\modulus{B^{d-1}})^{-1}$ is the normalising constant for the
measure $\nu'$ introduced in \secref{subsec:torus-boundary-map}
and $c_{\mu}
\defeq (\modulus{Q}.\modulus{S^{d-1}})^{-1}$ is the normalising constant
for the measure $\mu$.  Further, $B^{d-1}
\defeq \theset{\x \in \R^{d-1}: \norm{\x} \le 1}$ is the
$(d-1)$-dimensional unit ball and $S^{d-1}
\defeq \theset{\x \in \R^d: \norm{\x} = 1}$ is its boundary, the
$(d-1)$-dimensional unit sphere. In 2D, we have $\modulus{S^{d-1}}
= 2\pi$ and $\modulus{B^{d-1}} = 2$, so that
\begin{equation}\label{}
\meanrestime = \frac{\pi
\modulus{\trap}}{\modulus{\trapboundary}}.
\end{equation}
The exact expression \eqref{eq:mean-free-time-chernov-type} for
the mean residence time is precisely analogous to the exact
expression for the mean free path discussed in \cite{CherMFT}; it
also agrees with the more physical argument of \cite{MZ}.

\begin{figure}
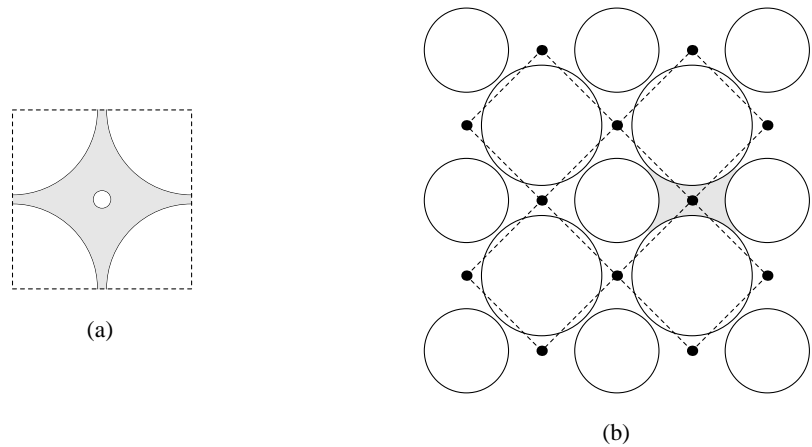

\centering
\subfigure[]{\begin{minipage}[c]{0.4\textwidth}
     \centering \includegraphics{mz.1}
      \vspace*{10pt}
    \end{minipage}}
    \qquad
    \subfigure[]{\begin{minipage}[c]{0.4\textwidth}
    \centering \includegraphics{mz.2}
    \vspace*{10pt}
    \end{minipage}}

\caption{\label{fig:mz-regimes}Two regimes where we expect the
Machta--Zwanzig approximation to be valid: (a) $r$ close to $2$,
so that the $a$-discs are close to touching; (b) $b$ close to
$r/\sqrt{2} - a$, so that the $b$- and $a$-discs are close to
touching. In each case, the shaded region indicates the trap shape
used in the Machta--Zwanzig argument.  In (b), the dashed lines
show the square lattice of traps whose centres are shown by dots.}
\end{figure}

For the square Lorentz gas that we consider, there are two
different regimes in which we can expect the Machta--Zwanzig
method to be valid: see \figref{fig:mz-regimes}. In (a), the
$a$-discs are almost touching, so that the $a$--$a$ gaps control
the escape process from each trap.  In (b), the $b$-disc is so
large that the $a$--$b$ gap will now control the dynamics.

In regime (a) we have
\begin{equation}\label{}
\modulus{\trap\up{1}} = r^2-\pi(a^2+b^2); \qquad
\modulus{\trapboundary\up{1}} = 4(r-2a); \qquad l = r
\end{equation}
and hence
\begin{equation}\label{}
{\DMZ}\up{1} = \frac{r^2}{\pi} \frac{r-2a}{r^2-\pi(a^2+b^2)},
\end{equation}
whilst in regime (b), we have
\begin{equation}\label{}
\modulus{\trap\up{2}} = \textfrac{1}{2} [r^2-\pi(a^2+b^2)]; \qquad
\modulus{\trapboundary\up{2}} = 4\lt[\frac{r}{\sqrt{2}}-(a+b)\rt];
\qquad l = \frac{r}{\sqrt{2}},
\end{equation}
and hence
\begin{equation}\label{}
{\DMZ}\up{2} = \frac{r^2}{\pi} \frac{
\frac{r}{\sqrt{2}}-(a+b)}{r^2-\pi(a^2+b^2)}.
\end{equation}
Introducing the quantities
\begin{equation}\label{}
w_1 \defeq r-2a; \qquad w_2
\defeq \frac{r}{\sqrt{2}}-(a+b),
\end{equation}
which are the the $a$--$a$ and $a$--$b$ gap sizes, respectively,
we can summarize the above as
\begin{equation}\label{}
{\DMZ}\up{i} = \frac{r^2}{\pi\lt[r^2-\pi(a^2+b^2)\rt]} \, w_i,
\quad i=1,2,
\end{equation}
where the $i$th approximation is expected to be valid when $w_i$
is small relative to the perimeter of a cell.

\subsection{Comparison of Machta--Zwanzig approximation with data}

\begin{figure}[p]
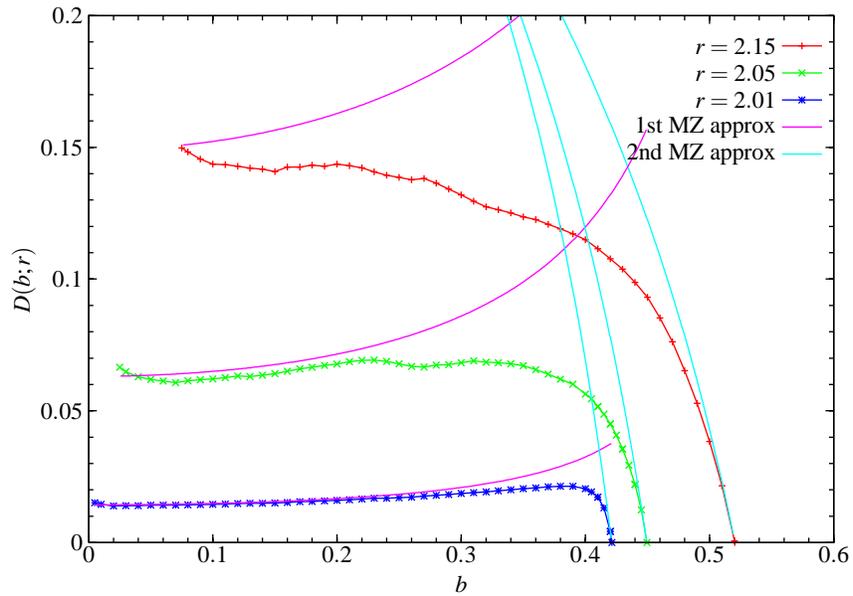

\centrefig{all-mz.eps}
\caption{\label{fig:comparison-mz-all-data}Comparison of
Machta--Zwanzig approximations with numerical diffusion
coefficient as a function of $b$ for different values of $r$.}
\end{figure}

\begin{figure}[p]
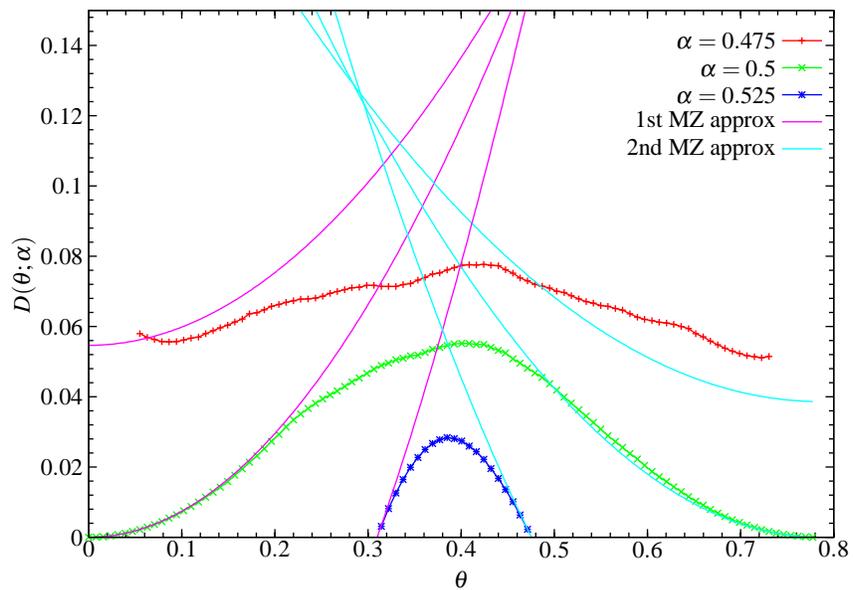

\centrefig{param-mz.eps}
\caption{\label{fig:comparison-mz-param-space-data}Comparison of
Machta--Zwanzig approximations with numerical diffusion
coefficient as a function of $\theta$ for different values of the
radius $\alpha$ in parameter space.}
\end{figure}

\bfigref{fig:comparison-mz-all-data} and
\figref{fig:comparison-mz-param-space-data} show comparisons of
the two Machta--Zwanzig approximations with numerical data for the
diffusion coefficient as a function of $b$ and of $\theta$,
respectively, where $\theta \defeq \tan^{-1}(\tb/\ta)$ is again
the angle of the point in parameter space from the line $\tb=0$.
We see that the Machta--Zwanzig approximations are good when the
respective $w_i$ are small, but fail elsewhere.

To some extent, these approximations explain the observation from
\secref{sec:geom-dependence}
 that there is a non-trivial maximum for values of $r$ near
$2$, since in that case the first approximation is valid for small
$b$ and the second is valid for large $b$, and they combine to
predict a maximum at an intermediate value of $b$.  It is clear,
however, that other effects are important in this intermediate
regime.

For larger values of $r$, decreasing $b$ has the effect of
allowing the particles freer passage through the system, so
intuition would again predict the observed increase in the
diffusion coefficient on decreasing $b$.


\subsection{Trap residence time distribution}
\label{subsec:trap-res-time-distn}

Since the above arguments involve the residence time $\restime$,
we here study numerically its distribution as a random variable,
where $\restime \from \tmapdomain \to \R$. (The space
$\tmapdomain$ and measure $\nu'$ are discussed in more detail in
\secref{sec:generalised-random-walk}.) To do this, we need to
distribute particles uniformly with respect to the invariant
measure $\nu'$ on $\tmapdomain$, i.e.\ on the trap exits with
inward-pointing velocities.  It is sufficient to distribute them
uniformly with respect to Liouville measure in the billiard domain
and evolve them forward until the first intersection with the trap
exit boundary.  By definition of the measure $\nu'$, they will
then be correctly distributed.

\begin{figure}
\subfig[0.7]{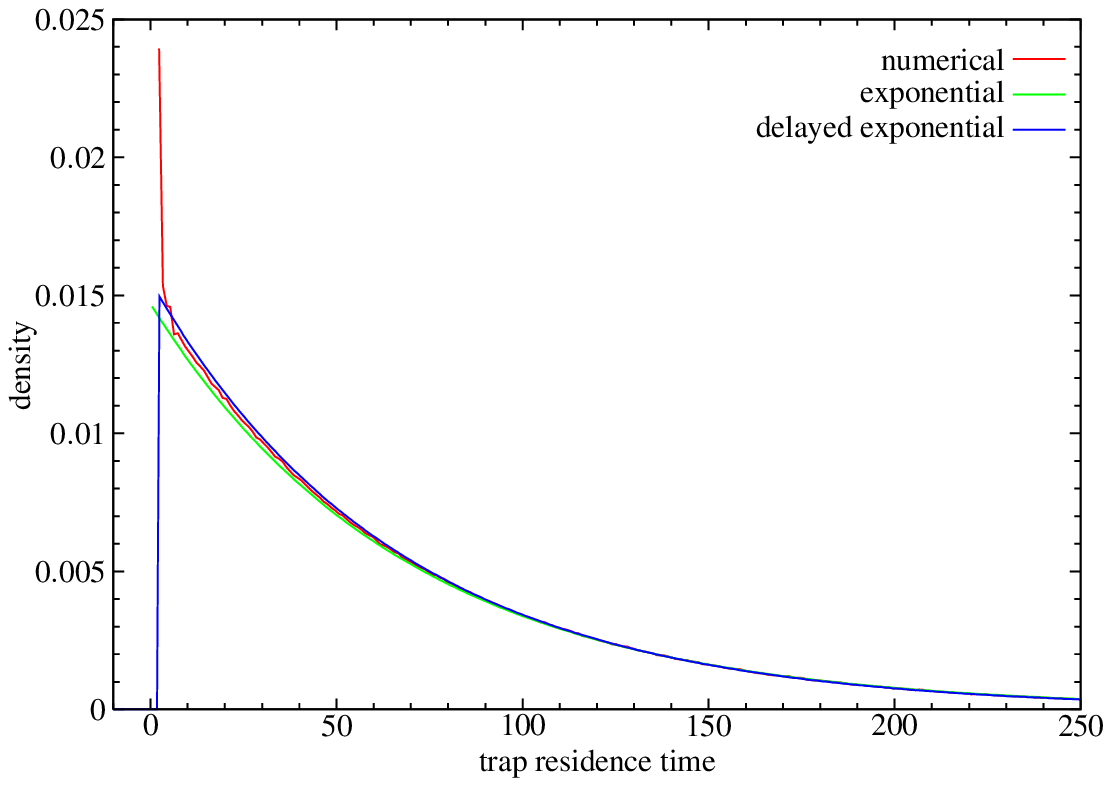} \hfill
\subfig[0.7]{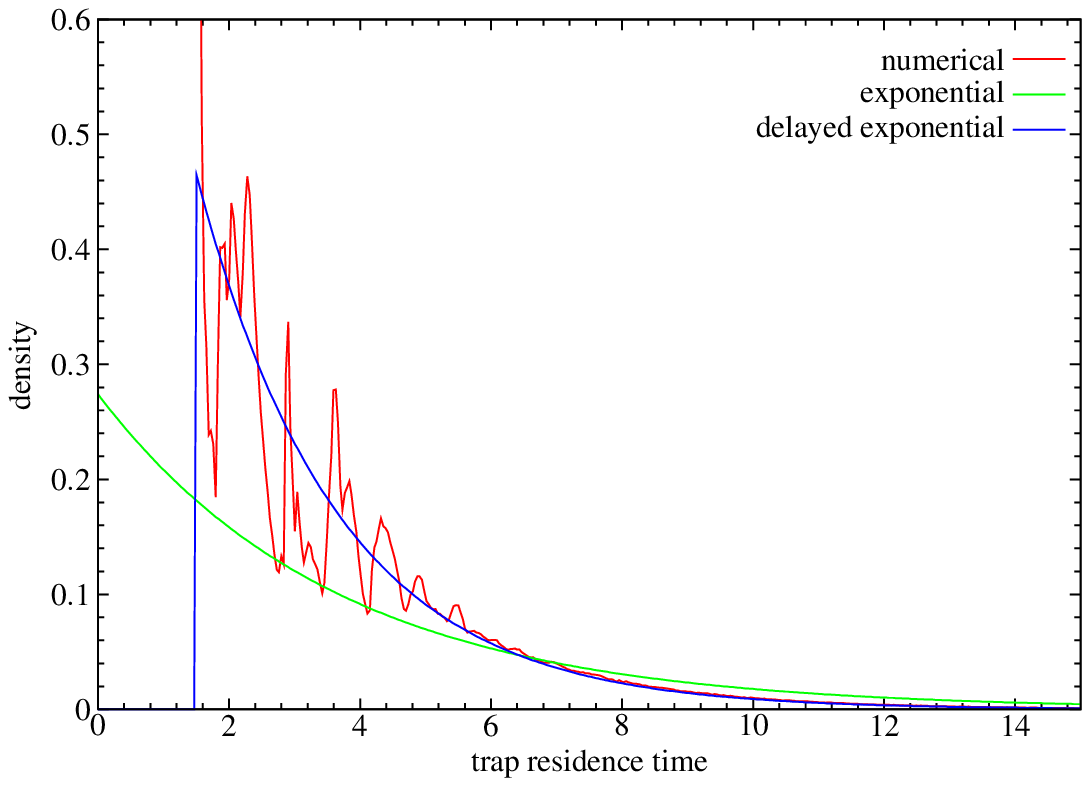}

\subfig[0.7]{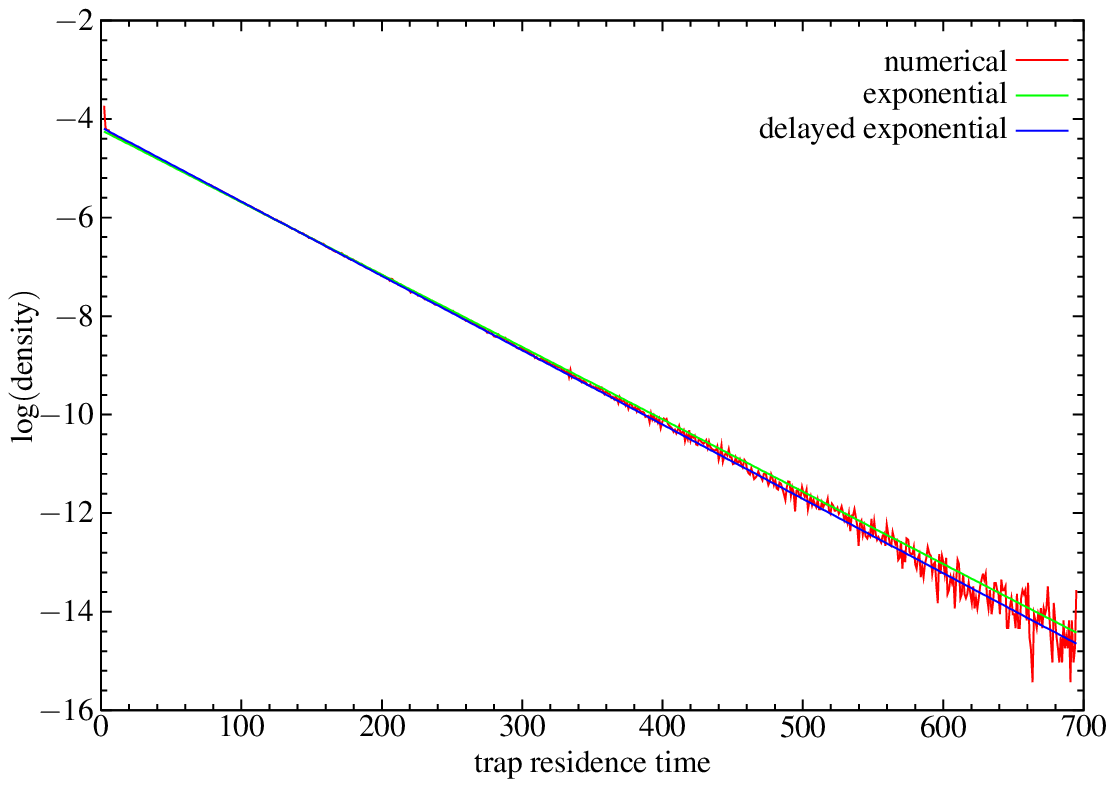} \hfill
\subfig[0.7]{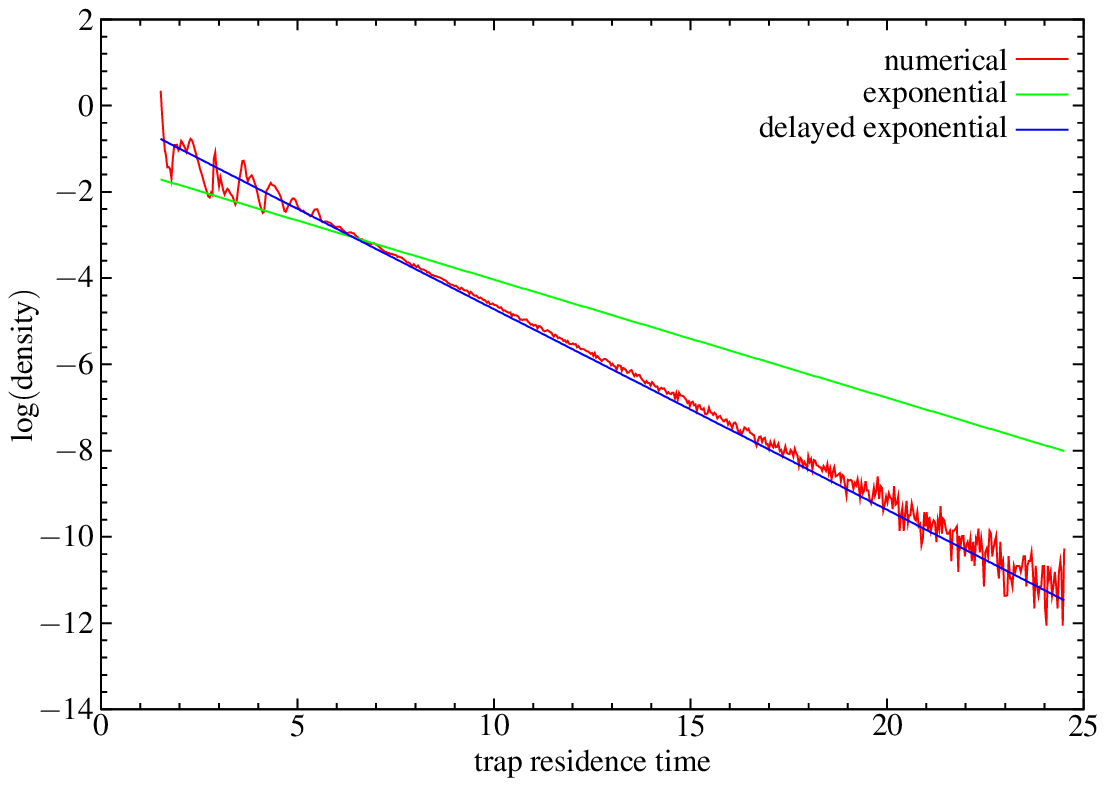}
\caption{\label{fig:trap-res-times} Trap residence time
distribution.  (a) is for $r=2.01$, $b=0.1$ and (b) for $r=2.5$,
$b-0.5$. (c) and (d) are the same as (a) and (b), respectively,
but plotted on semi-log scales. The continuous curves show the
delayed exponential density given by
\eqref{eq:delayed-exponential}.}
\end{figure}

We simulate the dynamics until the first exit from this trap and
record the trap residence time, for an ensemble of $N$ particles.
We then estimate the density of this distribution using a
histogram.  The results are shown in \figref{fig:trap-res-times}
for two different geometries.

An argument similar to that of Machta \& Zwanzig, looking at the
area available to escape in a small time $\Delta t$, would imply
an exponential distribution.  However, there is a minimum trap
residence time given by the time it takes to enter a trap in the
centre of one of its sides with a velocity perpendicular to that
side, collide with a $b$-disc, and exit along the same path.  This
minimum time is thus $2(r/2 - b) = r - 2b$.  In
\figref{fig:trap-res-times} we thus compare the numerical
distribution with an exponential distribution with density
\begin{equation}\label{}
f_{\text{exp}}(x) \defeq \frac{1}{\meanrestime}
\exp\paren{x/\meanrestime}
\end{equation}
with mean $1/\lambda \defeq \meanrestime$, and a \defn{delayed
exponential distribution} given by
\begin{equation}
f_{\text{delayed exp}}(x; \alpha) \defeq
\begin{cases}\label{eq:delayed-exponential}
\frac{1}{\meanrestime-\alpha} \exp
\paren{\frac{x-\alpha}{\meanrestime-\alpha}} &\text{if } x \ge
\alpha, \\
0,  &\text{if} x < \alpha,
\end{cases}
\end{equation}
where $\alpha \defeq r=2b$, both of which have mean
$\meanrestime$.
We see that the delayed exponential captures well the correct
asymptotic behaviour, whereas the standard exponential does not.

We remark that we can replace the simple Machta--Zwanzig approximation with an a priori
 more general one
where we use a \defn{continuous-time random walk} formulation with
a (delayed) exponential residence time distribution.  However, it
turns out that we obtain the same result for the diffusion
coefficient from any residence time distribution with a finite
mean, as was already found in \cite{ShlesingerAsymptotic}; see
also \cite[Chap.~3]{WeissRandomWalk}, and \secref{sec:CTRW-model}
for further discussion of continuous-time random walk models.

\section{Generalised random walk models and Green--Kubo approaches}
\label{sec:generalised-random-walk}

Klages \& Dellago \cite{KlagesD00} extended the \MZ\ approach in a
heuristic way, by including the probability to follow specified
sequences of traps. Klages \& Korabel \cite{KlagesK02} then found a way to include
such corrections in an exact Green--Kubo-type expansion.  We
believe, however, that while the result was correct, and the method was mostly
correct, there is a gap in the justification, since the relationship between the
discrete-time and continuous-time approaches studied above was not
treated correctly. Here we
 show how to close the gap in the argument.

\subsection{Original argument of Klages \& Korabel \cite{KlagesK02}}

In the appendix of \cite{KlagesK02}, Klages \& Korabel argued as
follows.  We wish to generalise the idea of a hopping process
between traps, for which we must pass from the continuous-time
expression for the diffusion coefficient $D$ in terms of the
Einstein formula for a symmetric system,
\begin{equation}\label{eq:cont-time-Einstein}
D = \lim_{t\to\infty} \frac{1}{4t} \mean{\lt[ \x(t) - \x(0)
\rt]^2},
\end{equation}
to a formula in terms of some discrete time $n$.  The mean $\mean{\cdot}$ here is
with respect to the invariant measure $\mu$ on the phase space $\M$ of the flow.

To do this, Klages \& Korabel \cite{KlagesK02} set\footnote{In
\cite{MZ} and \cite{KlagesK02},  $\tau$ denotes the average
residence time in a trap.  We follow the convention in the
mathematical literature (e.g.\ \cite{CherMark}), where $\tau$ is
used for the free path function, with mean $\bar{\tau}$. Denoting
by $\restime$ the trap residence time function, we replace the
$\tau$ of \cite{KlagesK02} by $\meanrestime$.}
\begin{equation}\label{}
\x_n \defeq \x(n \meanrestime),
\end{equation}
where $\x(t)$ is the random variable giving the position at time
$t$ and $\x_n$ is the position \emph{at the $n$th step}.  Note
that here the `$n$th step' simply means $n$ times the (so far arbitrarily
chosen) time step $\meanrestime$.  Then
\eqref{eq:cont-time-Einstein} implies that
\begin{equation}\label{eq:D-in-terms-of-discrete-time}
D = \lim_{n \to \infty} \frac{1}{4n\meanrestime} \mean{\x_n^2},
\end{equation}
where the mean is still with respect to $\mu$.

Since we want an expression in terms of traps, we must show that
we can replace $\x_n$ by $\X_n$, the centre of the trap in which
$\x_n$ lies.  Set $\tx_n \defeq \x_n - \X_n$, the displacement of
the position $\x_n$ at step $n$ from the centre of the respective
trap. Then \eqref{eq:cont-time-Einstein} implies that
\begin{equation}\label{}
D = \lim_{n \to \infty} \frac{1}{4n\meanrestime} \mean{\lt( \X_n +
\D\tx_n\rt)^2},
\end{equation}
where $\D\tx_n \defeq \tx_n - \tx_0$, assuming that all initial
conditions are in the trap at $\vect{0}$. We expand
\begin{equation}\label{}
\mean{
\paren{\X_n + \D\tx_n} ^2} = \mean{ X_n^2 + 2 \X_n \cdot \D\tx_n +
\D\tx_n^2}
\end{equation} and note that the last term is bounded by $r^2$, where
$r$ is the diameter of a unit cell. The Cauchy--Schwarz inequality
for the second term then gives
\begin{equation}\label{}
\modulus{\mean{2 \X_n \cdot \D\tx_n}} \le 2 \mean{\X_n^2}^{1/2}
\mean{\D\tx_n^2}^{1/2} \le r \,\mean{\X_n^2}^{1/2},
\end{equation}
so that the second term grows only as fast as the square root of
the first term.  Hence we obtain
\begin{equation}\label{eq:D-in-discrete-time}
D = \lim_{n \to \infty} \frac{1}{4n\meanrestime} \mean{\X_n^2},
\end{equation}
solely in terms of the trap at time $n$.

The next step is to obtain a Green--Kubo formula in terms of the
\defn{jump vector}
$\j_n \defeq \X_{n+1} - \X_n$ using the telescoping sum
\begin{equation}\label{}
\X_n = \X_0 + \sum_{k=0}^{n-1} \j_k = \sum_{k=0}^{n-1} \j_k,
\end{equation}
where the equality follows since $\X_0 = \vect{0}$. This reduces
$\mean{\X_n^2}$ exactly to  the variance of a sum of stationary
random variables, giving
\begin{equation}\label{}
\mean{\X_n^2} =\sum_{k,l=0}^{n-1} \mean{\j_k \cdot \j_l} = n\,
\mean{\j_0^2} + 2 \sum_{m=1}^{n-1} (n-m) \mean{\j_0 \cdot \j_m},
\end{equation}
so that
\begin{equation}\label{eq:mean-xn2-over-n}
\frac{\mean{\X_n^2}}{n} = \mean{\j_0^2} + \sum_{m=1}^{n-1}
\paren{1 - \frac{m}{n}} \mean{\j_0 \cdot \j_m} =
{\j_0^2} + \sum_{m=1}^{n-1} \mean{\j_0 \cdot \j_m} + \frac{1}{n}
\sum_{m=1}^{n-1} m \, \mean{\j_0 \cdot \j_m}.
\end{equation}
Thus if $\mean{\j_0 \cdot \j_m}$ decays sufficiently fast as $m
\to \infty$, then we can take the limit of
\eqref{eq:mean-xn2-over-n} as $n \to \infty$ to obtain the
Green--Kubo relation
\begin{equation}\label{eq:D-discrete-time-green-kubo}
D = \frac{1}{4\meanrestime} \mean{\j_0^2} +
\frac{1}{2\meanrestime} \sum_{m=1}^{\infty} \mean{\j_0 \cdot
\j_m}.
\end{equation}

Klages \& Korabel then use this result to try to extend the
Machta--Zwanzig approximation, as follows.  They follow each
trajectory and record the \emph{sequence of traps visited} via a
symbol sequence.  They numerically calculate the probability of
occurrence of each symbol sequence and use sequences of length up
to $n$ as an $n$th-order approximation to the Green--Kubo
expression, where the Machta--Zwanzig approximation can be
regarded as the $0$th-order approximation.  They obtain results
which numerically converge exactly to the (numerically) exact
diffusion coefficient over the whole finite-horizon regime of the
triangular Lorentz gas.  The same method was applied in a
different model in \cite{HarayamaKG02}.

\subsection{Re-examination of Klages--Korabel derivation}

We argue that the derivation given is not directly related to the
numerical method subsequently employed, although  both are
independently correct. The reason for this is confusion in the
definition of the \emph{$n$th time step}:
\begin{enumerate}
\item \label{item:orig}
In the  derivation, the $n$th time step was taken as looking
stroboscopically at the flow at the start of each period of length
$\meanrestime$.

\item \label{item:numerical}
 In the numerical method, however, the $n$th `time step'
instead refers to the time when the phase space point crosses into
the $n$th trap.
\end{enumerate}

Both interpretations correspond to valid reductions from the
continuous-time flow to a discrete-time map, but they are
\emph{not} equivalent: (\ref{item:orig}) corresponds to looking at
statistical properties of the
\defn{time-$\meanrestime$ map} of the flow, while we will show that
(\ref{item:numerical}) corresponds instead to looking at
statistical properties of a particular Poincar\'e map, which
differs from the usual billiard map and which we call the
\defn{torus-boundary map}.

The part of the derivation after the definition of the position
$\x_n$ `at the $n$th time step', is valid in each case, and shows
that in any of these cases it is equivalent to study the
asymptotic behaviour either of $\x_n$ or of the corresponding trap
centre $\X_n$, or of the jump vectors $\j_n$ defined as above,
provided their correlations decay sufficiently fast. As far as we
are aware, no rate of decay of correlations is yet available for
the time-$\meanrestime$ map (which would be related to the
still-unsolved problem of the correlation decay rate for the
billiard flow), nor for the torus-boundary map (which does not
seem to have been studied previously).

\subsection{Torus-boundary map}
\label{subsec:torus-boundary-map}

Analogously to the billiard map $\map$ (which takes one collision
with the scatterers to the next), we introduce a map $\torusmap$,
which we call the
\defn{\boundarymap}, mapping one intersection with the
trap/torus boundary $\trapboundary$ to the next.  Thus the domain
of $\torusmap$ is
\begin{equation}\label{}
M \defeq \theset{(\q,\v) \in \M \colon \q \in \trapboundary,
\thickspace \v \cdot \n(\q) > 0},
\end{equation}
and we follow the trajectory up to the edge of the torus and then
continue to the point on the opposite edge with which that point
is identified.

The  \boundarymap\ will be undefined for
 phase points whose trajectories  get trapped bouncing between the
scatterers, i.e.\ for points on the stable manifold of the fractal
repeller.  However, this set has measure zero \cite{GaspBook}, so that $\torusmap$ is
defined almost everywhere.

We can now view the billiard flow as a suspension flow over the
\boundarymap\ $\torusmap$, under the trap residence time function
$\restime$, since specifying a phase space point in $M'$ and a
value for the residence time function $\restime$ specifies a
unique point of $\M$. The invariant measure $\nu'$ on the space
$M'$ is defined analogously to the invariant measure $\nu$ for the
billiard map $\map$, i.e.\ by projecting to $M'$, and then by
\appref{app:suspension-flows}, $\torusmap$ is ergodic. Hence, by
the Birkhoff ergodic theorem, for almost every $x \in M'$, we have
that
\begin{equation}\label{}
\lim_{n \to \infty} \frac{1}{n} \sum_{i=0}^{n-1} \rho \comp
\torusmap^i(x) = \meanrestime,
\end{equation}
i.e.\ time averages of the residence time function $\rho$ are
equal to $\meanrestime$, so that $\meanrestime$ is indeed the mean
trap residence time.


We can now use the relation between the variance of the suspension
flow and the variance of the base transformation (see
\appref{app:suspension-flows}) to get an expression for $D$ in
terms of statistical properties of the base transformation.  This
reduces exactly to \eqref{eq:D-in-terms-of-discrete-time}, where
$\x_n$ is the position at the $n$th crossing of a trap boundary
and the average $\mean{\cdot}$ is with respect to the invariant
measure $\nu'$ for the \boundarymap.

The regularity  of the roof function required for this result (see
\appref{app:suspension-flows}) is satisfied, since the roof
function, which is here given by the trap residence time function,
appears to decay exponentially; this was shown numerically in
\secref{subsec:trap-res-time-distn}, although we are not aware of
any rigorous results.

The remainder of the above derivation of Klages--Korabel now goes
through to give a Green--Kubo formula in terms of the
\boundarymap, \emph{provided} that correlations decay sufficiently
fast.  As far as we are aware, the only results on decay of
correlations for billiards are for the billiard map, where
exponential decay of correlations is proved \cite{Young, Cher99,
CherYoung}. We expect that correlations for the \boundarymap\ also
decay exponentially fast.  However, the proofs for the billiard
map rely on hyperbolicity due to the scattering nature of billiard
boundaries.  In the case of the \boundarymap, the boundaries are
not scattering: the scattering occurs between, rather than at,
intersections with the boundary. It is thus possible that a
rigorous proof could be even harder to obtain than for the
billiard map. If we nonetheless assume that these correlations do
decay sufficiently fast, we do have a Green--Kubo expansion such
that the Machta--Zwanzig result occurs as the $0$th order
approximation.  This completes the justification of the
Klages--Korabel method.

We remark that this method extends to  any other map such that we
can express  the billiard flow as a suspension over that map.
However, we are not aware of other physically interesting
maps for which that is the case.

\subsection{Low-order approximations}

If the details of the above justification can be made rigorous, we
would then be able to rigorously assert that the Machta--Zwanzig
approximation is indeed a zeroth-order truncation of the infinite
Green--Kubo sum.  In \cite{KlagesK02}, it was observed that a
certain simple low-order expression  involving collisionless
flights across traps gave a very good approximation of the
numerical diffusion coefficient over the whole finite horizon
regime for the triangular Lorentz gas.  This was an anomaly from
the point of view of the expansion used in that paper, since this
truncation did not appear explicitly in the Green--Kubo sum used
there.

We believe that this anomaly can be explained by noting that their
expression was instead a natural low-order truncation of the
discrete-time Green--Kubo formula obtained from the
time-$\meanrestime$ map, rather than from the torus-boundary map:
there we must include in the first term a sum over all accessible
traps after a time $\meanrestime$, which includes at least part of
the probability of
 collisionless flights, together with part of the backscattering
 probability
 which also appears in their best
 low-order truncation.
Their observation then suggests that the Green--Kubo expression
discussed above converges more rapidly as the truncation order $n$
increases than that used in their paper.  This faster rate of
convergence depends on correlations for the two respective maps;
as such, we would not expect to be able to predict which of the
two expansions would give better low-order approximations without
a detailed knowledge of these correlations.

\section{Reducing the geometrical symmetry}
\label{sec:reducing-symmetry}

In this section we study the effect of altering the geometry of
the system to reduce the symmetry of the unit cells; again our
model is a good candidate for this, whereas the triangular model
would perhaps be less natural.

\subsection{Symmetry of diffusion tensor}

We first consider the effect of symmetry on
the diffusion tensor.
Suppose we measure the diffusion coefficients in the original
coordinate system and in a new orthonormal coordinate system $\x'$
based at the same origin, given by $\x' = \Rm \cdot \x$,
where $\Rm$ is the orthonormal change of basis matrix (\ie a
combination of rotations and reflections).  Then the diffusion
coefficients with respect to the new axes are given by
\begin{equation}\label{eq:diff_coeffs_new_axes}
  D_{ij}' = \lim_{t \rightarrow \infty} \frac{1}{2t}
  \mean{x_i'(t) \, x_j'(t)}
%
  %
  = R_{ik} R_{jl} \lim_{t \rightarrow \infty} \frac{1}{2t}\mean{x_k(t) \,
  x_l(t)}
  = R_{ik} \, R_{jl} \, D_{kl}
  = (\Rm \tD \Rm \transp)_{ij},
\end{equation}
where we have used the Einstein summation convention (sum over
repeated indices). If the system is symmetric under some symmetry
element $\Rm$, then for that particular $\Rm$ we have $D'_{ij} =
D_{ij}$, since the system looks the same after the transformation.
We remark that this idea is known as \defn{Neumann's principle} in
the literature on crystal properties \cite{NyePhysicalProperties},
where it is an empirically based fact.  In our setting, however,
it is a theorem, since Liouville measure is invariant under the
point group of the lattice.

Thus symmetry under the transformation $\Rm$ implies that the
diffusion tensor satisfies
\begin{equation}\label{eq:symmetry_diff_tensor}
  \Rm \tD \Rm \transp = \tD.
\end{equation}
This equation in general restricts the allowed values of the
diffusion coefficient. For example, if the system is unchanged
under reflection in the $x$-axis, given by the transformation $y
\mapsto -y$ having matrix
\begin{equation}\label{eq:reflection}
  \Rm = \begin{pmatrix} 1 & 0 \\ 0 & -1 \end{pmatrix},
\end{equation}
then we find that \eqref{eq:symmetry_diff_tensor} implies that
$D_{xy} = D_{yx} = 0$ for a 2D system, so that the diffusion
tensor is diagonal in this coordinate system.  In the case of 3D
systems (which we have not investigated), tables of crystal
symmetries and their consequences for possible forms of the
2nd-order diffusion tensor can be found e.g.\ in
\cite{CarslawJaeger, NyePhysicalProperties}.

\subsection{Displacing central disc}

Here we consider the effect of displacing the $b$-discs away from
the centre of the cell%
\footnote{I would like to thank Leonid Bunimovich for suggesting
this, with the idea that it could help to understand the validity
of the Machta--Zwanzig approximation.}, parallel to the $x$-axis.
Then the system remains symmetric under reflection in the
$x$-axis, but is now asymmetric under reflection in the $y$-axis.
By the above, the diffusion tensor must remain diagonal, but we
may now have $D_{xx} \neq D_{yy}$, which is not possible for the
square model.

Starting from geometrical parameters $(a, b)$ lying in the finite
horizon regime, we displace the $b$-disc by a distance $g$ along
the $x$-axis in the direction of increasing $x$. To continue to
block the central vertical channel, we cannot move the $b$-disc
too far: we require $\frac{r}{2} + g - b < a$, i.e.\
\begin{equation}\label{}
g < a + b - \frac{r}{2}.
\end{equation}
The $b$-disc touches the $a$-discs when
\begin{equation}\label{}
(a+b)^2 = \paren{\frac{r}{2} - g}^2 + \paren{\frac{r}{2}}^2,
\end{equation}
so that the discs are not touching when
\begin{equation}\label{}
g < \frac{r}{2} - \sqrt{(a+b)^2 - \paren{\frac{r}{2}}^2}.
\end{equation}
It is possible that there is a finite horizon even if the $b$- and
$a$-discs touch.  In this case, there is no possibility for the
particle to escape from the vertical channel in which it begins,
so that $D_{xx} = 0$; however, there may still be a non-trivial
diffusion coefficient $D_{yy}$ in the $y$-direction.  Note,
however, that the analysis of \cite{BS,BSC} does \emph{not}
immediately apply to this case, since the (convex) scatterers are no longer
disjoint.

\begin{figure}
\centering

 \subfig[0.7]{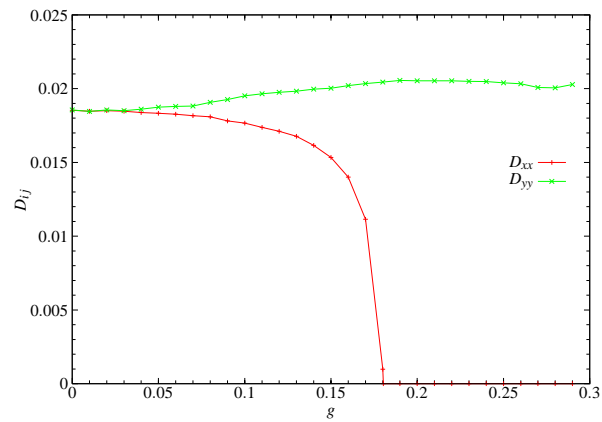}
 \hfill
\subfig[0.7]{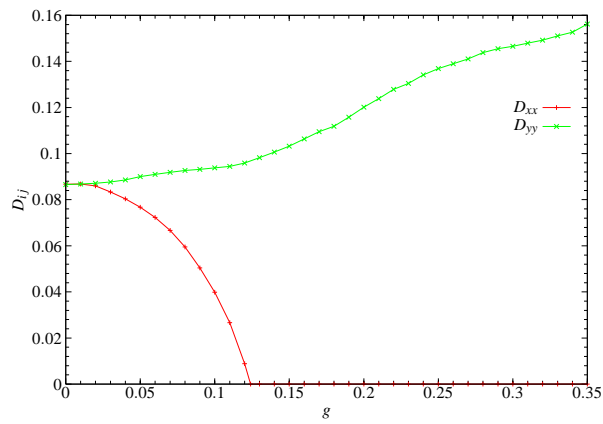}

\subfig[0.7]{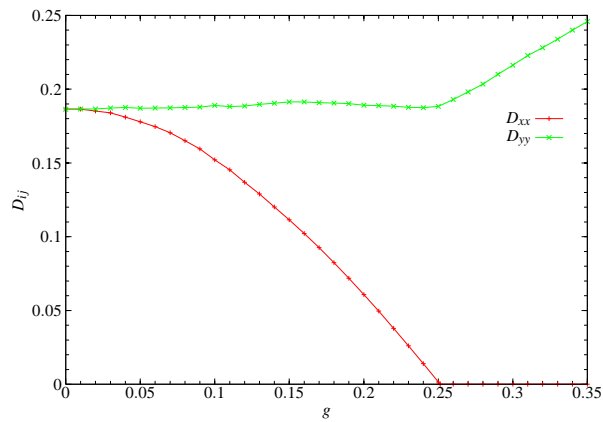}

\caption{\label{fig:asymmetric} The diffusion coefficients
$D_{xx}$ and $D_{yy}$ in the $x$- and $y$-directions,
respectively, as a function of the displacement $g$ of the
$b$-disc from the centre of the unit cell in the $x$-direction.
(a) $r=2.01$, $b=0.3$.  (b) $r=2.1$, $b=0.4$. (c) $r=2.5$,
$b=0.6$.}
\end{figure}

\bfigref{fig:asymmetric} shows numerical data for the diffusion
coefficients $D_{xx}$ and $D_{yy}$ for this system as a function
of the displacement $g$.  For $g=0$, we have $D_{xx}=D_{yy}$ by
the symmetry results above.  For non-zero $g$, however, $D_{xx}$
decreases and $D_{yy}$ increases.

Increasing $g$ past the point where the discs touch, the area
available in each cell increases; from the figure we see that
$D_{yy}$ suddenly increases its growth rate for two of the three
sets of geometrical parameters studied.

On the contrary, for $r=2.01$ the value of $D_{yy}$ seems to
stabilise for larger values of $g$, indicating some kind of
balance between the increased phase space volume available and the
control of the dynamics by the small trap exits.

\subsection{Rectangular model with less symmetry}

We could also stretch the unit cell to a rectangular one, imposing
different lattice spacings in the horizontal and vertical
directions; this introduces two new geometrical parameters.  In
particular, the standard triangular periodic Lorentz gas can be
considered as a particular case of such a stretched model.  Again
the symmetry of the diffusion tensor will be reduced in general,
although not in the special case of the triangular periodic
Lorentz gas.



\startonright




\graphicspath{{figs/}}

\chapter[Fine structure of distributions and central limit theorem]
{Fine structure of distributions and the central limit theorem in
2D periodic Lorentz gases} \label{chap:fine-structure}

In the previous chapter we studied the statistical properties of
the displacement as a function of time in terms of means. Those
means are taken over the probability distribution of the
observable, whose shape we investigate in this chapter.

As described in \chapref{chap:stat-props}, it was proved in
\cite{BS, BSC} that 2D periodic Lorentz gases with finite horizon
and disjoint scatterers (with sufficient smoothness of the
scatterer boundaries, namely piecewise $C^3$) satisfy a
\defn{central limit theorem}: the rescaled displacement
distribution converges in distribution to a limiting normal
distribution:
\begin{equation}\label{eq:recap-clt}
\frac{\x_t - \x_0}{\sqrt{t}} \distconv \z \quad \text{as } t \to
\infty.
\end{equation}


In this chapter we study the structure of position and
displacement distributions at \emph{finite} time $t$.  We show
that they possess a
\defn{fine structure}, consisting of a periodic oscillation
superimposed on the Gaussian shape that we expect from a diffusive
process.  We will find an analytical description of this fine
structure in terms of the geometry of the billiard domain, and
provide extensive numerical evidence that this is indeed the main
influence on the fine structure. This gives
 a physical picture of the \defn{weak}
type of convergence occurring in \eqref{eq:recap-clt}, and leads
to a conjecture on a possible stronger result; we can also give an
intuitive estimate of the rate of convergence to the limiting
distribution.

The results in \cite{BS, BSC} show how we can smooth away the fine
structure to obtain rigorous proofs of convergence. Our analysis
reinstates the fine structure to give a picture of how this
convergence occurs, making explicit
 the obstruction that prevents a stronger form of
convergence to the limiting normal distribution by showing how
density functions fail to converge pointwise to Gaussian
densities.

\section{Structure of 2D position and displacement
distributions} \label{sec:fine-structure}

\subsection{Statistical properties of position
and displacement distribution}

The displacement distribution occurs naturally in the central
limit theorem (\secref{sec:prob-limit-thms}) and in Green--Kubo
relations \cite{DorfBook, GaspBook}, whereas the position
distribution is more natural if we are unable to track the paths
of individual particles.  Their statistical properties are very
closely related, as shown by the following discussion for the
$x$-component.

Expanding the mean squared displacement as
\begin{equation}\label{eq:expansion-msd}
\mean{\Dx_t^2} = \mean{x_t^2} - 2 \mean{x_t x_0} + \mean{x_0^2},
\end{equation}
and applying the Cauchy--Schwarz inequality to the second term on
the right hand side, as in \cite{KlagesK02}, which gives
\begin{equation}\label{}
\modulus{\mean{x_t x_0}} \le \sqrt{\mean{x_t^2} \, \mean{x_0^2}},
\end{equation}
we see that $\mean{\Dx_t^2}$ and $\mean{x_t^2}$ have the same
asymptotic growth rate, so that they both grow linearly if one of
them does.

Further, since $\modulus{x_0} \le \frac{r}{2}$, we can show that
the $\sqrt{t}$-rescaled position distribution also satisfies the
central limit theorem if the displacement distribution does, with
the same limiting normal distribution.  From the point of view of
statistical properties it is hence equivalent to study either the
position or the displacement distribution.

\subsection{Shape of 2D  distributions}
\label{subsec:posn-disp-distns}

\begin{figure}[hpt]
\centering
\includegraphics{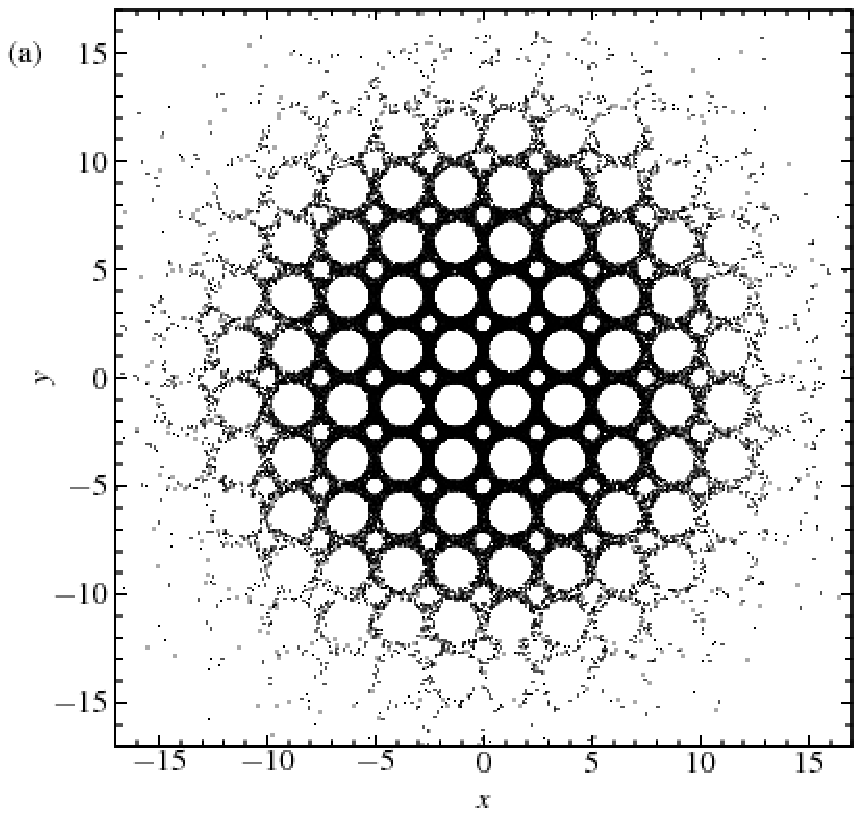}
\hfill
\includegraphics{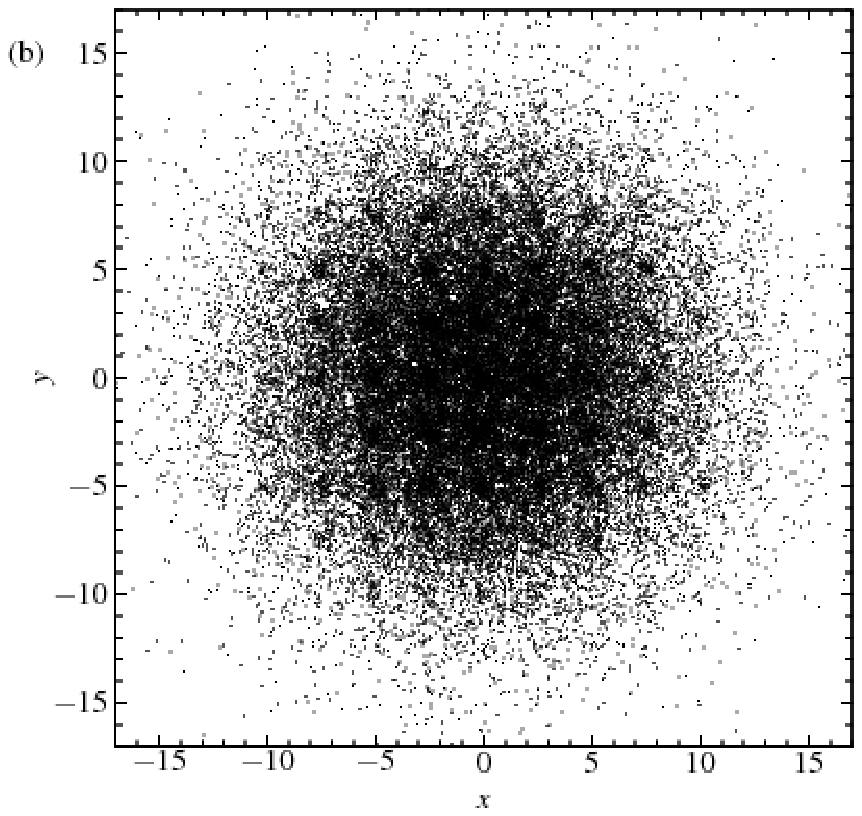}
\caption{\label{fig:2d-distns}
(a) 2D position distribution; (b) 2D displacement
distribution. $r=2.5$; $b=0.4$; $t=50$; $N=5 \times 10^4$ initial conditions.
}
\end{figure}

\bfigref{fig:2d-distns} shows scatterplots representing 2D
position and displacement distributions for a representative
choice of geometrical parameters. Each dot represents one initial
condition started in the central unit cell and evolved for time
$t=50$; $N=5 \times 10^4$ samples are shown, started from random
initial conditions distributed uniformly with respect to Liouville
measure in the central unit cell. Both distributions show decay
away from a maximum in the central cell, an overall circular
shape, and the occurrence of a periodic fine structure.

These figures are projections to the unfolded billiard domain
$\Qunf \subset \R^2$ of the density in the unfolded phase space
$\Qunf \times S^1$. Since the dynamics on the torus is mixing
\cite{CherMark}, the phase space density converges \emph{weakly}
\cite{LasotaMackey} to a uniform density on phase space
 corresponding to the invariant Liouville
measure: see \appref{app:conv-proj-densities}. Physically, the
phase space density on the torus develops a complicated layer
structure in the stable direction of the dynamics: see e.g.\
\cite{DorfBook}. Projecting corresponds to integrating over the
velocities; we expect this to eliminate this complicated structure
and result in some degree of smoothness of the projected densities
on the torus, and so presumably also of the unfolded projected
densities. However, we are not aware of any rigorous results in
 this direction, even
for relatively well-understood systems such as the Arnold cat map
\cite{DorfBook}.

These 2D distributions are difficult to work with, and we instead
restrict attention to one-dimensional marginal distributions,
i.e.\ projections onto the $x$-axis, which will also have some
degree of smoothness.
 We denote the 1D position
density at time $t$ and position $x \in \R$ by $f_t(x)$ and the
displacement density for displacement $x$ by $g_t(x)$. We let
their respective (cumulative) distribution
functions\footnote{Henceforth we use `distribution function',
eliminating the redundant `cumulative': this seems to be the usual
terminology  in probability theory, e.g.\
\cite{Grimmett,FellerII}.} be $F_t(x)$ and $G_t(x)$, respectively,
so that
\begin{equation}\label{}
F_t(x) \defeq \P(x_t \le x) = \int_{-\infty}^x f_t(s) \rd s,
\end{equation}
and similarly for $G_t$.  (When necessary, we will instead denote
displacements by $\xi$.)
The densities show the
structure of the distributions more clearly, while the distribution
functions are
more directly related to
analytical considerations.

\section{Numerical estimation of 1D distribution
functions and densities} \label{subsec:num-estimation-densities}

We wish to estimate numerically the above densities and
distribution functions at time $t$ from the $N$  data points
$(x_t\up{1}, \ldots, x_t\up{N})$. The most widely used method in
the physics community for estimating density functions from
numerical data  is the histogram \cite{NR}. However, histograms
are not always appropriate, due to their non-smoothness and
dependence on bin width and position of bin origin
\cite{Silverman}. In \cite{AlonsoPolyg}, for example, the choice
of a coarse  bin width obscured the fine structure of the
distributions that we describe in \chapref{chap:polygonal}.

We have chosen the following alternative method, which seems to
work well in our situation, since it is able to deal with strongly
peaked densities more easily, although we do not have any rigorous
results to justify this. We have also checked that histograms and
kernel density estimates (a generalization of the histogram
\cite{Silverman}) give similar results, provided that sufficient
care is taken with bin widths.

We first calculate the empirical cumulative distribution function
\cite{Scott, Silverman}, defined by $F_t^{\emp}(x) \defeq
\#\{i:x_t\up{i} \le x\}$ for the position distribution, and
analogously for the displacement distribution.
The estimator $F_t\emp$ is the optimal one for the distribution
function $F_t$ given the data, in the sense that there are no
other unbiased estimators with smaller variance \cite[p.\
34]{Scott}. We find that the distribution functions in our models
are smooth on a scale larger that that of individual data points,
where statistical noise dominates. (Here we use `smooth' in a
visual, nontechnical sense; this corresponds to some degree of
differentiability). We verify
that adding more data does not qualitatively change this
larger-scale structure: with $N=10^7$ samples we
seem to capture the fine structure.

We now wish to construct the density function $f_t = \partial F_t
/ \partial x$.  Since the direct numerical derivative of $F_t\emp$ is
useless due to
statistical noise, our procedure is
to fit an (interpolating)
\defn{cubic spline} to an evenly-spread sample of points from $F_t\emp$,
and differentiate the cubic spline to obtain the density function
at as many points as required. Sampling evenly from $F_t\emp$
automatically uses more samples where the data are more highly
concentrated, i.e.\ where the density is larger.

 We must confirm
(visually or in a suitable norm) that our spline approximation
reproduces the fine structure of the distribution function
sufficiently well, whilst ignoring the variation due to noise on a
very small scale. As with any density estimation method, we have
thus made an assumption of smoothness \cite{Silverman}. The
analysis of the fine structure in \secref{sec:fine-structure}
justifies this to some extent.

\section{Time evolution of 1D distributions}
\label{subsec:time-evolution-distns}

\begin{figure}[tp]
\centering
\includegraphics{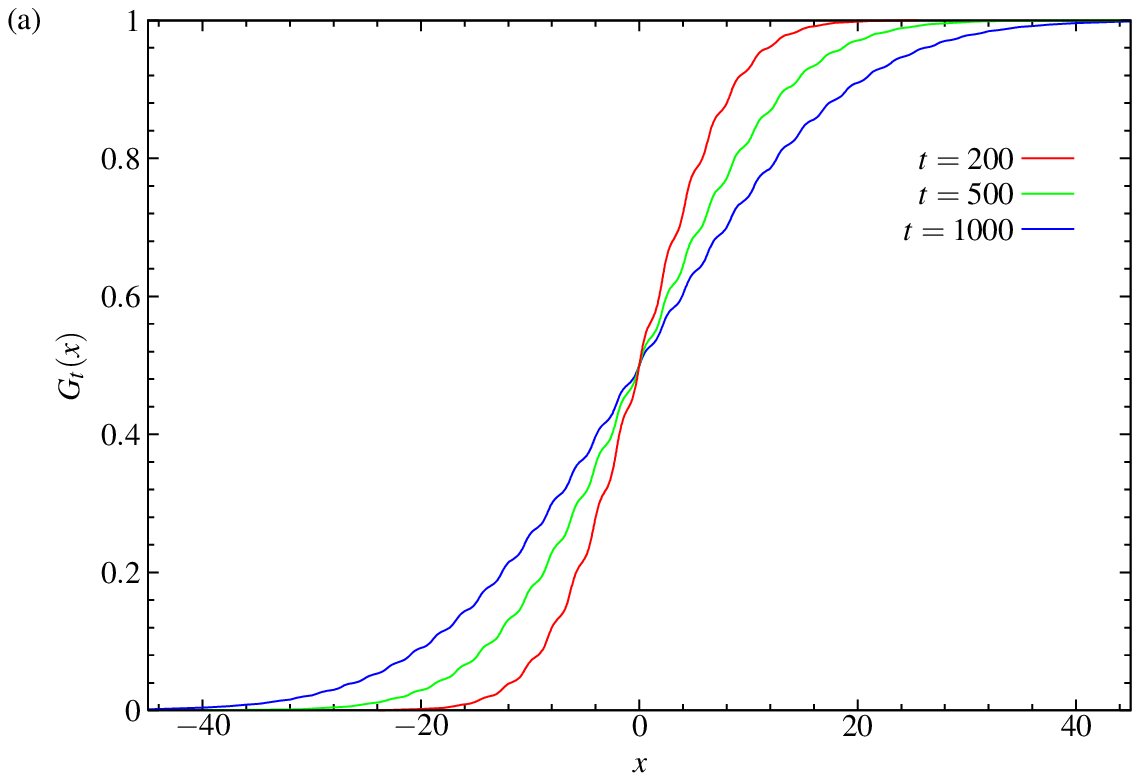}
\hfill
\includegraphics{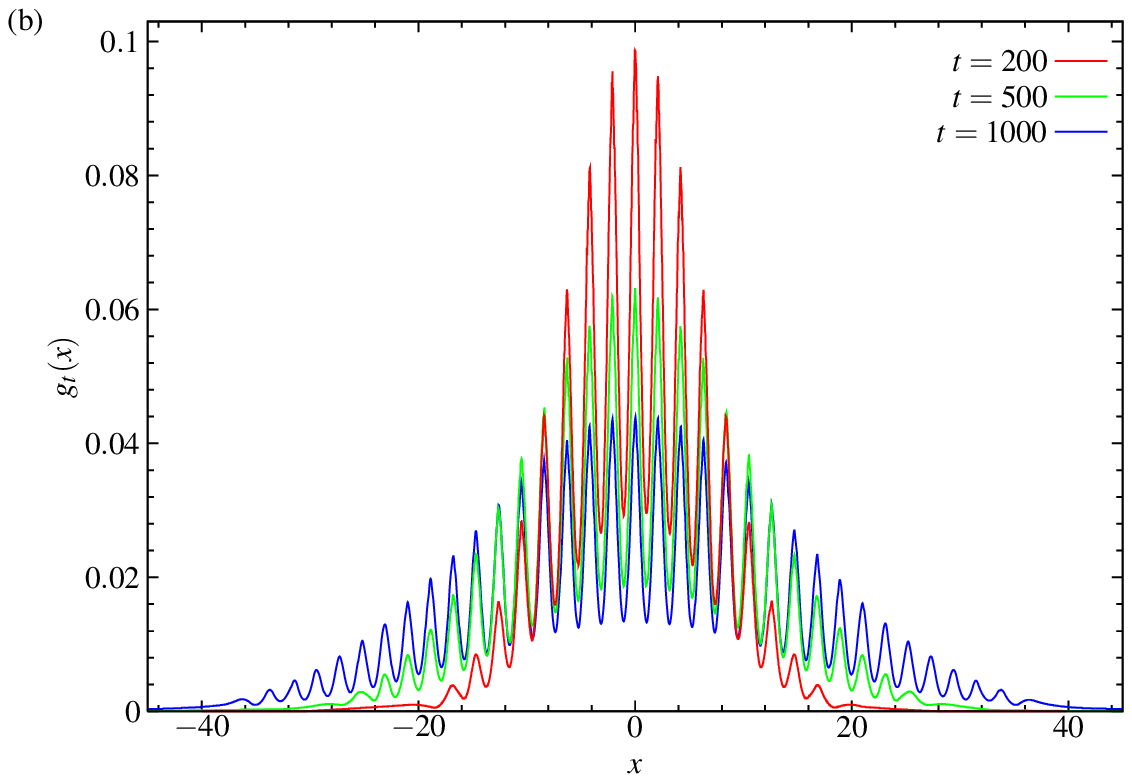}
\caption{\label{fig:time-evol} (a) Time evolution of displacement
distribution functions. (b) Time evolution of displacement
densities, calculated by numerically
  differentiating a cubic spline approximation to distribution functions.
$r=2.1$; $b=0.2$.}
\end{figure}

\bfigref{fig:time-evol} shows the time evolution of 1D
displacement distribution functions and densities for certain
geometrical parameters, chosen to emphasise the oscillatory
structure. Other parameters within the finite horizon regime give
qualitatively similar behaviour.

The distribution functions are smooth, but have a
step-like structure. Differentiating the spline
approximations to these distribution functions gives
 densities which have an
underlying Gaussian-like shape, modulated by a \emph{pronounced
fine structure} which persists at all times
(\figref{fig:time-evol}(b)). This fine structure is just
noticeable in Figs.\ 4 and 5 of \cite{AlonsoPolyg}, but otherwise
does not seem to have been reported previously, although in the
context of iterated 1D maps a  fine structure was found, the
origin of which is related to pruning effects: see
 e.g.\ Fig.\ 3.1 of
\cite{KlagesPhD}.  We will show
that in billiards this fine structure can be understood
by considering the
geometry of the billiard domain.

\section{Fine structure of position density}
\label{sec:fine-structure-posn-density}

\begin{figure}
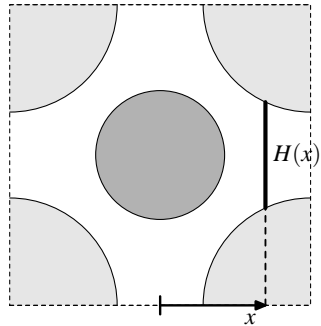

\centrefig{unfolding-torus.4} \caption{\label{fig:H-of-x}
Definition of the set $H(x)$.}
\end{figure}

Since Liouville measure on the torus is invariant, if
the initial distribution is uniform with respect to
Liouville measure, then the distribution at any time $t$ is still uniform.
Integrating over the velocities, the position distribution at time
$t$ is hence always uniform with respect to Lebesgue measure in
the billiard domain $Q$, which
 we normalize such that the measure of $Q$ is $1$.
Denote the two-dimensional position density on the torus at
$(x,y) \in [0,1)^2$ by
$\rtor(x,y)$.  Then
\begin{equation}\label{}
\rtor(x,y) = \frac{1}{\modulus{Q}} \indic{Q}(x,y) =
\frac{1}{\modulus{Q}} \indic{H(x)}(y).
\end{equation}
Here, $H(x) \defeq \{y: (x,y) \in Q\}$ is the set of allowed $y$
values for particles with horizontal coordinate $x$ (see
\figref{fig:H-of-x}),
and $\indic{B}$ is the indicator function of the (one- or
two-dimensional) set $B$, given by
\begin{equation}\label{}
\indic{B}(b) = \begin{cases} 1,\quad \text{if } b \in B \\
0, \quad \text{otherwise}.
\end{cases}
\end{equation}
Thus for fixed $x$, $\rtor(x,y)$ is
independent of $y$ within the available space $H(x)$.

Now unfold the dynamics onto a 1-dimensional channel in the
$x$-direction, as in \figref{fig:1dlattice}, and consider the torus as the
distinguished unit cell at the origin. Fix a vertical line with
horizontal coordinate $x$
in this cell,
 and consider its periodic translates $x+n$ along the channel,
where $n \in \Z$. Denoting the density there by $\rchan_t(x+n, y)$,
we  have that for all $t$ and for all $x$ and $y$,
\begin{equation}\label{eq:reduce_density_torus}
 \sum_{n \in \Z} \rchan_t(x+n, y) = \rtor(x,y).
\end{equation}

We expect that after a sufficiently long time,
the distribution within cell  $n$  will look
like the distribution on the torus, modulated by a slowly-varying
function of $x$.
In particular, we expect that the 2D position density will become
asymptotically
uniform in $y$ within
 $H(x)$ at long times.

\begin{figure}[p]
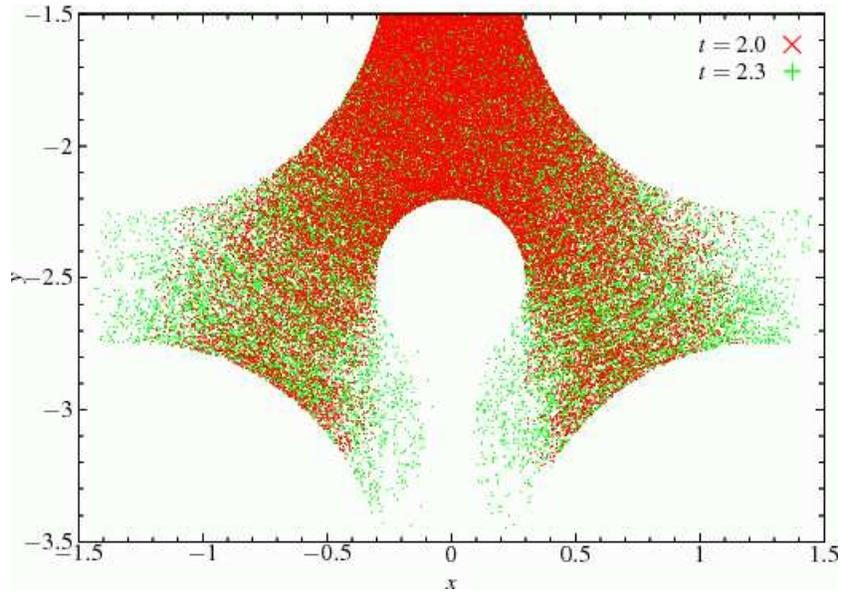

\centrefig{dynamics.eps} \caption{\label{fig:dynamics}Leading edge
of 2D position density for $r=2.5$; $b=0.3$; $a=1$, at times
$t=2.0$ and $t=2.3$.}
\end{figure}

We have not been able to prove this, but we have checked by
constructing 2D kernel density estimates \cite{Silverman} that it
seems to be true. A `sufficiently long' time would be one which is
much longer than the time scale for the diffusion process to cross
one unit cell. It does not, however, hold for short times: for
example, in \figref{fig:dynamics} we see the development of the 2D
density at the leading edge.  At the earlier time, we see that the
`fluid' has streamed past the small disc without reaching the
region below it; at the later time, this region is beginning to be
occupied. At long times, the mixing property of the dynamics will
have filled the region approximately uniformly.

Thus we have approximately
\begin{equation}\label{eq:first-def-rho-bar}
\rchan_t(x,y) \simeq \rtor(x,y) \, \rbar_t(x) = \rbar_t(x)
\frac{1}{\modulus{Q}} \indic{H(x)}(y),
\end{equation}
where $\rbar_t(x)$ is the
\defn{shape} of
the two-dimensional density distribution as a function of $x \in
\R$; we expect this to be a slowly-varying function. We use
`$\simeq$' to denote that this relationship holds in the long-time
limit, for values of $x$ which do not lie in the tails of the
distribution. Although this breaks down in the tails,
 the density is in any case small there.

The 1D marginal density that we measure will then be given  by
\begin{equation}\label{eq:1d-posn-density-from-2d}
f_t(x) = \int_{y=0}^1 \rchan_t(x,y) \rd y \simeq \rbar_t(x) \,
h(x),
\end{equation}
where $h(x) \defeq \modulus{H(x)}/\modulus{Q}$ is the normalized
height (Lebesgue measure) of the set $H(x)$ at position $x$ (see
\figref{fig:H-of-x}).  Note that $H(x)$ is not necessarily a
connected set.

 Thus the measured density $f_t(x)$ is
given by the shape $\rbar_t(x)$ of the 2D density,
\emph{modulated} by  fine-scale oscillations due to the geometry
of the lattice and described by $h(x)$, which we call the
\defn{fine structure function}.

The above argument motivates the \emph{(re-)definition} of
$\rbar_t(x)$ so that $f_t(x) =  h(x) \rbar_t(x)$ with strict
equality and for all times,  by setting
\begin{equation}\label{}
\rbar_t(x) \defeq \frac{f_t(x)}{h(x)}.
\end{equation}
  We can now view
$\rbar_t(x)$ as the density with respect to a \emph{new underlying
measure} $h \, \lambda$, where $\lambda$ is $1$-dimensional
Lebesgue measure; this measure takes into account the available
space, and is hence more natural in this problem (see also
\secref{subsec:conv-1d-distns-billiards}). We expect that
$\rbar_t$ will now describe the large-scale shape of the density,
at least for long times and $x$ comparatively small.

The conjecture underlying the above heuristic argument is then
that for all $x \in \R$, we have
\begin{equation}\label{}
\frac{\rchan_t(x,y)}{f_t(x)} = \frac{\rchan_t(x,y)}{\int_{y=0}^1
\rchan_t(x,y') \rd y' } \stackrel{t \to \infty}{\longrightarrow}
h(x) \quad \text{for all } y \in H(x),
\end{equation}
or equivalently
\begin{equation}\label{}
\frac{\rchan_t(x,y)}{\rbar_t(x)} \stackrel{t \to
\infty}{\longrightarrow} 1 \quad \text{for all } y \in H(x),
\end{equation}
where the ratio converges to something which is \emph{independent}
of $y$.  We must use this expression since the density at any
fixed $x$ tends to $0$ as $t\to\infty$. In fact, regarding both
sides as a function of $y$, we could even expect uniform
convergence of the form
\begin{equation}\label{}
\frac{\rchan_t(x,\cdot)}{\int_{y=0}^1 \rchan_t(x,y') \rd y' }
\stackrel{\text{unif}}{\longrightarrow} h(x) \indic{H(x)}(\cdot),
\end{equation}
where $x$ is still fixed.

\begin{figure}[p]
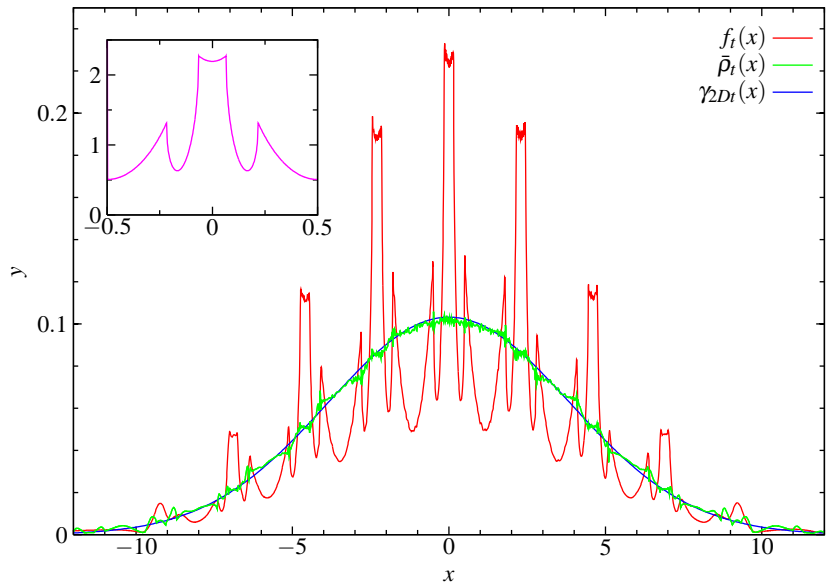

\centrefig{demod_ttr2.3b0.5.eps}
\caption{\label{fig:demodulate-posn-distn}
 Position density $f_t$ exhibiting a pronounced fine structure,
together with the demodulated slowly-varying function $\rbar_t =
f_t(x)/h(x)$ and a Gaussian with variance $2Dt$. The inset shows
one period of the demodulating fine structure function $h$.
$r=2.3$; $b=0.5$; $t=50$.}
\end{figure}

\bfigref{fig:demodulate-posn-distn} shows the original and
demodulated densities $f_t$ and $\rbar_t$ for a representative
choice of geometrical parameters.  The fine structure in $f_t$ is
very pronounced, but is eliminated nearly completely when
demodulated by dividing by the fine structure $h$, leaving a
demodulated density $\rbar_t$ which is close to the Gaussian
density with variance $2Dt$ (also shown).

 Note that although the density has non-smooth points, which affects the
 smoothness assumption in our density estimation procedure described in
 \secref{subsec:num-estimation-densities}, in practice these points are
 still handled reasonably well.  If necessary, we could treat these points
     more carefully, by suitable choices of partition points in that method.

We estimated the diffusion coefficient $D$ as follows. For $r=2.3$
and $b=0.5$, using $N=10^7$ particles evolved to $t=1000$, the
best fit line for $\log \msd_t$ against $\log t$ in the region $t
\in [500,1000]$ gives $\msd \sim t^{1.00003}$, which we regard as
confirmation of asymptotic linear growth. Following
\cite{KlagesD00}, we use the slope of $\log \msd_t$ against $t$ in
that region as an estimate of $2D$, giving $D = 0.1494 \pm
0.0002$; the error analysis is as in
\chapref{chap:geom-dependence}. (Throughout, we denote by
$\gauss{\sigma^2}$ the Gaussian density with mean $0$ and variance
$\sigma^2$, and by
 $\normal{\sigma^2}$ the corresponding normal distribution function.)

\section{Fine structure of displacement density}
\label{subsec:unfold-disp-density}

We can treat the displacement density similarly, as follows.
Let $\eta_t(x,y)$ be the 2D displacement density function at time
$t$, so that
\begin{equation}\label{}
\int_{-\infty}^x \int_{\infty}^y \eta_t(x,y) \rd x \rd y =
\prob{\D x_t \le x, \D y_t \le y}, \quad x,y \in \R.
\end{equation}
(Recall that $\D x_t \defeq x_t - x_0$.)
 We
\emph{define} the projected versions $\etachan$ and $\etator$ as
follows:
\begin{gather}\label{}
\etachan_t(x,y) \defeq \sum_{n \in \Z} \eta_t(x, y+n),
\quad x \in \R, y \in [0,1), \\
\etator_t(x,y) \defeq \sum_{n \in \Z} \etachan_t(x+n, y), \quad
x,y \in [0,1).
\end{gather}
Again we view the torus as the unit cell at the origin where all
initial conditions are placed.
Note that
 projecting the displacement distribution on $\R^2$
 to the channel or torus gives the same result as
first
 projecting and then obtaining the
displacement distribution in the reduced geometry.
Hence the designations as being associated with the
channel or torus are appropriate.

Unlike $\rtor$ in the previous section, $\etator_t$ is not
independent of $t$: for example, for small enough $t$, all displacements
 increase with time.  However, we show that $\etator_t$ rapidly approaches a
distribution which \emph{is} stationary in time.

Consider a small ball of initial conditions of positive Liouville
measure around a point $(\x,\v)$.  Since the system is mixing on
the torus, the position distribution at time $t$ corresponding to
those initial conditions converges as $t \to \infty$ to a
distribution which is uniform with respect to Lebesgue measure in
the billiard domain $Q$. The
 corresponding limiting displacement distribution is hence obtained by
averaging the displacement of
$\x$ from all points on the torus.

Extending this to an initial distribution which is uniform with
respect to Liouville measure over the whole phase space, we see
that the limiting displacement distribution is given by averaging
displacements of two points in $Q$, with both points distributed
uniformly with respect to Lebesgue measure on $Q$. This limiting
distribution we denote by $\etator(x,y)$, with no $t$ subscript.

\begin{figure}[tp]
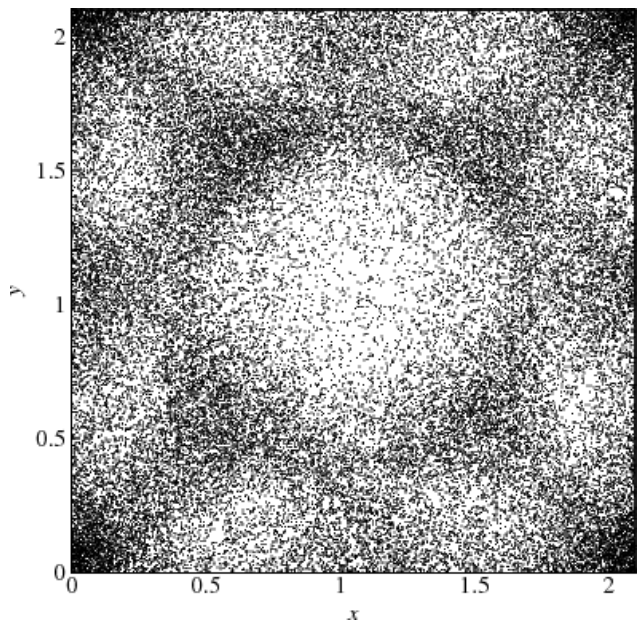

\centrefig{r2.1b0.4n50k_redt50.eps}
\caption{\label{fig:reduced-2d-displacement-distn} Scatterplot of
the 2D displacement density $\etator_t(x,y)$ on the torus for
$r=2.1$, $b=0.4$ and $t=50$. This corresponds to the projection of
\figref{fig:2d-distns}(b) to one unit cell (although the
geometrical parameters used are different). This density is  close
to the limiting displacement density $\etator$, since the
relaxation is fast.}
\end{figure}

As in the previous section, we expect the $y$-dependence of
$\etachan_t(x+n,\cdot)$ to be the same, for large enough $t$, as
that of $\etator(x,\cdot)$ for $x \in [0,1)$. However,
$\etator(x,\cdot)$ is not independent of $y$, as can be seen from
\figref{fig:reduced-2d-displacement-distn}, which is a projected
version of \figref{fig:2d-distns}(b) to the torus (with different
geometrical parameters). We thus expect
\begin{equation}\label{eq:def-eta-chan}
\etachan_t(x,y) \simeq \etator(x,y) \, \etabar_t(x).
\end{equation}

To obtain the 1D marginal density $g_t(x)$, we
integrate with respect to $y$:
\begin{equation}\label{}
g_t(x) = \int_{y=0}^1 \etachan_t(x,y) \rd y \simeq \phi(x)
\etabar_t(x),
\end{equation}
where
\begin{equation}\label{}
\phi(x) \defeq \int_{y=0}^1 \etator(x,y) \rd y.
\end{equation}
Again we now redefine $\etabar$ so that $g_t(x) = \phi(x)
\etabar_t(x)$, with the
 fine structure of $g_t(x)$ being described by $\phi$ and the
large-scale variation by $\etabar(x)$,  which can be regarded as
the density with respect to the new measure $\phi \, \lambda$
taking account of the excluded volume. In the next section we
evaluate $\phi(x)$ explicitly.

\subsection{Calculation of $x$-displacement density $\phi(x)$ on torus }
\label{subsec:x-disp-density-torus}

Let $(X_1, Y_1)$ and $(X_2, Y_2)$ be independent random variables,
distributed uniformly with respect to Lebesgue measure in the
billiard domain $Q$, and let $\D X \defeq \fracpart{X_2-X_1} \in
[0,1)$ be their $x$-displacement, where $\fracpart{\cdot}$ denotes
the fractional part of its argument. Then $\D X$ is the sum of two
independent random variables, so that  its density $\phi$ is given
by the following convolution, which correctly takes account of the
periodicity of $h$ and $\phi$ with period $1$:
\begin{equation} \label{eq:convolution}
  \phi(\xi) = \int_0^1 h(x) \, h(x+\xi) \rd x.
\end{equation}
This form leads us to expand in
Fourier series:
\begin{equation}\label{eq:fourierseries}
h(x) =  \sum_{k \in \Z} \hh{k} \, \e^{2 \pi \i k x} = \hh{0} + 2
\sum_{k \in \Nats} \hh{k} \, \cos 2 \pi k x,
\end{equation}
and similarly for $\phi$, where the Fourier coefficients are
defined by
\begin{equation}\label{eq:fouriercoeffs}
  \hh{k} \defeq \int_0^1 h(x) \, \e^{-2 \pi \i k x} \rd x
          =   \int_0^1 h(x) \, \cos(2 \pi k x) \rd x.
\end{equation}
 The last equality follows from the evenness of $h$, and shows
 that $\hh{k} = \hh{-k}$, from which the second
 equality in \eqref{eq:fourierseries} follows.
 Fourier transforming \eqref{eq:convolution} then gives
\begin{equation}\label{eq:fourierconv}
  \fh{k} =  \hh{k} \, \hh{-k} =  \hh{k}^2.
\end{equation}

Taking the origin in the centre of the disc of radius $b$ (see
\figref{fig:H-of-x}), the available space function $h$ is given by
\begin{equation}\label{eq:defn_of_h}
h(x) = \frac{1}{\modulus{Q}} \left( 1 - 2 \sqrt{b^2 - x^2}
-2 \sqrt{a^2-(\textstyle \frac{1}{2}-x)^2} \right)
\end{equation}
for $x \in [0,1/2)$, and is even and periodic with period $1$.
(Here we adopt the convention that $\sqrt{\alpha} = 0$ if
$\alpha<0$ to avoid writing indicator functions explicitly.) The
evaluation of the Fourier coefficients of $h$ thus involves
integrals of the form
\begin{equation}\label{eq:bessel}
  \int_0^a \cos zt \, \sqrt{a^2-t^2} \rd t
    = \frac{\pi a}{2z} \, J_1(za),  \qquad (z\neq 0)
\end{equation}
where $J_1$ is the first order Bessel function; this equality
follows from equation (9.1.20) of \cite{AbramowitzS} after a
change of variables.

The Fourier coefficients of $h$ are thus $\hh{0}=\int_0^1 h(x) =
1$ and, for integer $k\neq 0$,
\begin{equation}\label{eq:fouriercoeffofh}
\ \hh{k} = -\frac{1}{\modulus{Q} . \modulus{k}}
\left[ (-1)^k \, a \, J_1(2 \pi a \modulus{k})
+ b \, J_1(2 \pi b \modulus{k}) \right].
\end{equation}
Note that  $\int_0^1 \phi(x) \rd x = \fh{0} = \hh{0}^2 = 1$, so
that $\phi$ is correctly normalized as a density function on the
torus.

\begin{figure}[p]
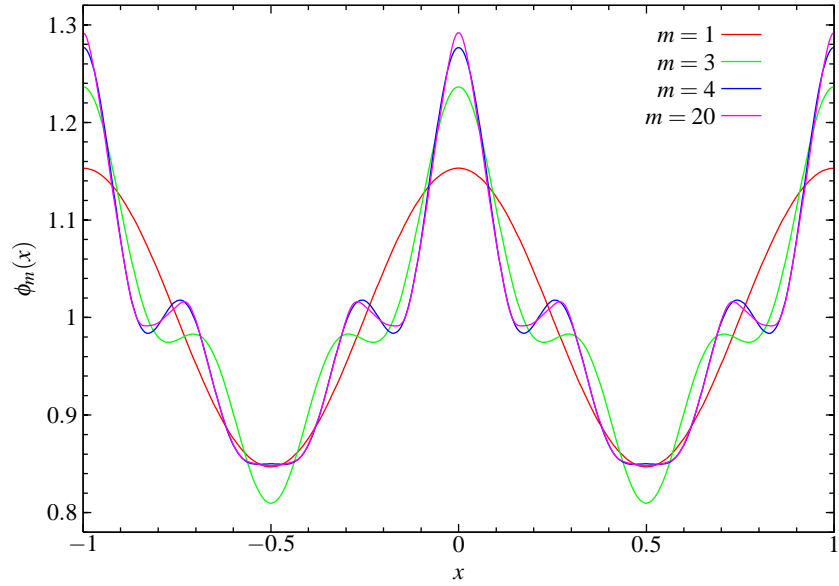

 \centrefig{fourier-of-phi.eps} \caption{\label{fig:partial-sums}
Partial sums $\phi_m$ up to $m$ terms of the Fourier series for
$\phi$, with $r=2.3$ and $b=0.5$.}
\end{figure}

In \figref{fig:partial-sums} we plot partial sums $\phi_m$ up to $m$ terms
of the Fourier
series for $\phi$ analogous to \eqref{eq:fourierseries}.
We can determine the degree of smoothness of
$\phi$, and hence presumably of $g_t$, as follows. The
asymptotic expansion of $J_1(z)$ for large real $z$ (equation
(9.2.1) of \cite{AbramowitzS}),
\begin{equation}\label{eq:asympexpansionforJ1}
J_1(z) \sim \sqrt{\frac{2}{\pi z}} \, \cos \left( {
\frac{3\pi}{4}} - z \right) = \bigO{z^{-1/2}},
\end{equation}
shows that $\hh{k} = \bigO{k^{-3/2}}$ and hence
$\fh{k}=\bigO{k^{-3}}$. From the theory of Fourier series (see
e.g.\ \cite[Chap.\ 1]{Katznelson}), we hence have that $\phi$ is
at least $C^1$ (once continuously differentiable).  Thus the
convolution of $h$ with itself is smoother than the original
function, despite the non-differentiable points of $h$.

We have
checked numerically  the approach of $\int \etator_t(x,y)
\rd y$ to $\phi(x)$, and it appears to be fast, although
the rate is difficult to evaluate,
since a large number of initial conditions are required for
the numerically calculated distribution function to approach
closely the limiting distribution.

\subsection{Structure of displacement distribution}

\begin{figure}[p]
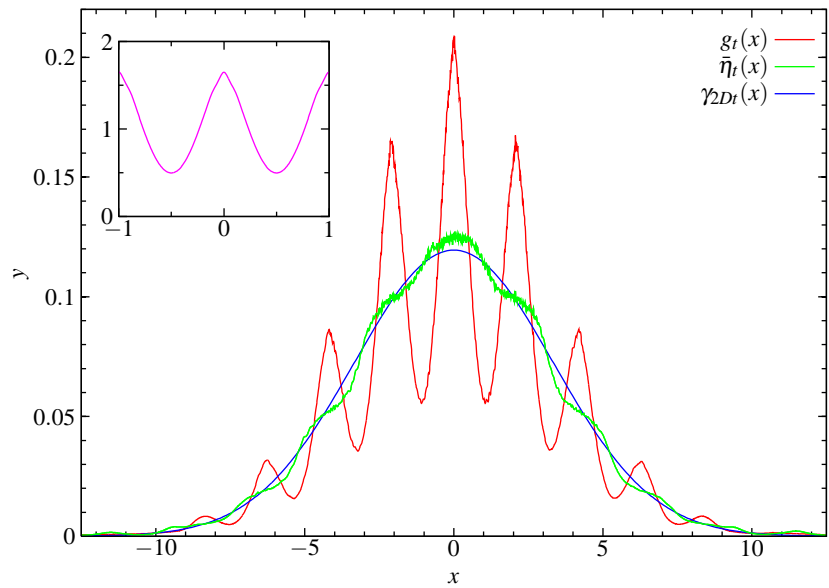

 \centrefig{demod_disp_r2.1b0.2.eps}
\caption{\label{fig:demodulate-disp-distn}%
 Displacement density $g_t$, with demodulated $\etabar_t$
compared to a Gaussian of variance $2D$.  The inset in (a) shows
the fine structure function $\phi$ for these geometrical
parameters. $r=2.1$; $b=0.2$; $t=50$. }
\end{figure}

In \figref{fig:demodulate-disp-distn} we plot the
numerically-obtained displacement density $g_t(x)$, the fine
structure function $\phi$ calculated above, and their ratio
$\etabar_t(x)$, for a certain choice of geometrical parameters.
Again the ratio is approximately Gaussian, which confirms that the
densities can be regarded as a Gaussian shape modulated by the
fine structure $\phi$.

\begin{figure}[p]
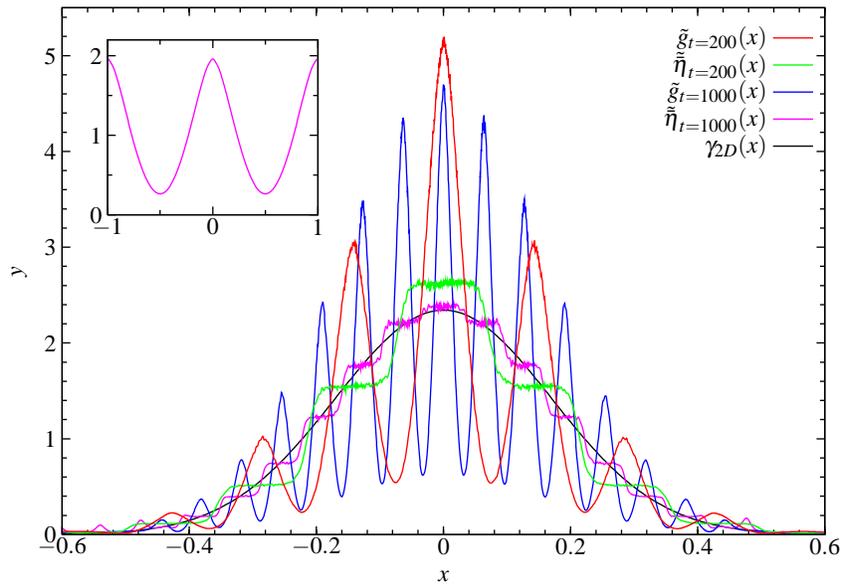

 \centrefig{demod_r2.01.eps}
\caption{\label{fig:demodulate-disp-flat}%
 Displacement density $g_t$ and demodulated $\etabar_t$, both
rescaled by $\sqrt{t}$, at $t=200$ and $t=1000$, compared to a
Gaussian of variance $2D$. The inset again shows the fine
structure function $\phi$. $r=2.01$; $b=0.1$. }
\end{figure}

However, if $r$ is close to $2a$, then $\etabar_t$ itself develops a
type of fine structure: it is nearly constant over each unit cell.
 This is shown in \figref{fig:demodulate-disp-flat} for two different times.
We plot both $g_t$ and
 $\etabar_t$, rescaled by $\sqrt{t}$ and compared to a Gaussian of
 variance $2D$.  (This scaling is discussed in \secref{sec:clt}.)

 This step-like structure of $\etabar_t$ is related to the validity of
the Machta--Zwanzig random walk approximation discussed in
\secref{sec:machta-zwanzig}.
   Having $\etabar_t$ constant across each cell
indicates that  the distribution of particles within the
billiard domain in each cell is uniform, as is needed for the Machta--Zwanzig
approximation to work.

As $r$ increases away from $2a$, the exit size of the traps
increases, and the Machta--Zwanzig argument ceases to give a good
approximation. The distribution then ceases to be uniform in each
cell: see \figref{fig:demodulate-posn-distn}. This may be related
to the crossover to a Boltzmann regime described in
\cite{KlagesD00}.

\subsection{Comparison of position and displacement distributions}

\begin{figure}[p]
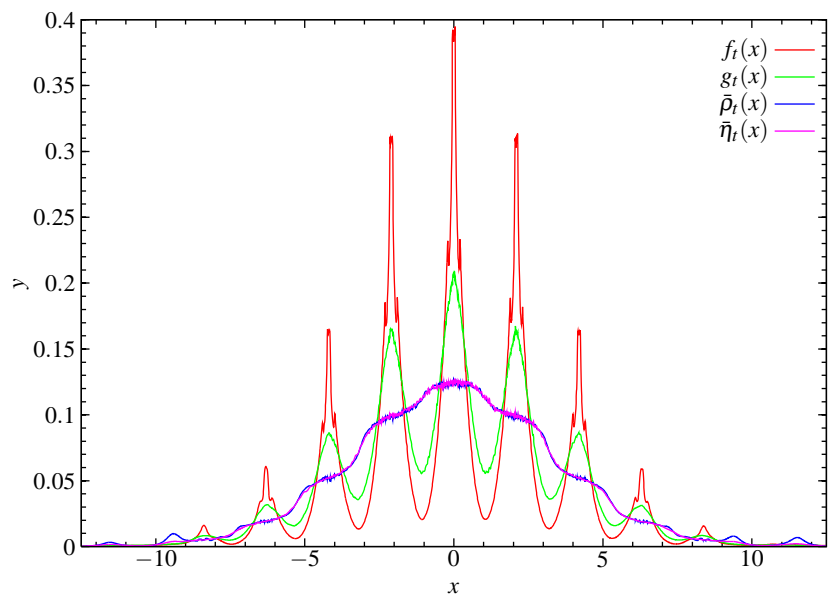

\centrefig{comparison.eps} \caption{\label{fig:comparison}
Comparison of the position density $f_t$, the displacement density
$g_t$, the demodulated position density $\bar{\rho}_t$ and the
demodulated displacement density $\bar{\eta}_t$, for $r=2.1$,
$b=0.2$ and $t=50$.}
\end{figure}

For long times, both position and displacement distributions
converge to the same limiting normal distribution, so we expect
the demodulated position and displacement distributions at a large
but finite time $t$ to be close.  This is confirmed in
\figref{fig:comparison}.

We can ask if there is in general any advantage in studying one
or the other of the position and displacement distributions. Since
$\fh{k} = \hh{k}^2$, the Fourier  coefficients of $\phi$ decay
faster as $k \to \infty$ than those of $h$.  This translates to
better \defn{regularity} (smoothness) of $\phi$ than of $h$, as we
have seen in the particular example above, where\footnote{$f$ is
\defn{H\"older continuous with exponent $0<\alpha < 1$}, denoted
$f \in C^{\alpha}$, if and only if $\modulus{f(x)-f(y)} \le K
\modulus{x-y}^{\alpha}$.} $h \in C^{1/2}$ but $\phi \in C^{1}$.

Further, the convolution reduces the  amplitude of the
oscillations of the fine structure.  We can see this by taking the
first Fourier coefficient alone as an initial indication of the
amplitude, putting
\begin{equation}
  h(x) = 1 + A \cos 2\pi x,
\end{equation}
although there may be significant contributions to the amplitude
from higher Fourier coefficients, as can be seen in
\figref{fig:partial-sums}.
 The first term ensures that  $h$ is a density (per unit length), i.e.\ that $h$
integrates to $1$, and $A$ is the amplitude of the oscillation.
Then
\begin{equation}
  \phi(\xi) = \int_{x=0}^1 h(x) \, h(x+\xi) \rd x = 1 + \texthalf A^2 \, \cos 2\pi \,
  \xi,
\end{equation}
so that the amplitude of the oscillations in (the lowest-order
Fourier approximation of) $\phi$ is $\texthalf A^2$. If $A\ge 1$
then the channel would have height $0$ at some point and no
particle could pass.  So the ratio of the amplitude of $\phi$ to
that of $h$ is $\texthalf A < 0.5$.  This argument leads us to
expect that indeed the amplitude of fine structure oscillations
for the displacement density should be significantly less than
that for the position density.

The above two results indicate that in numerical  investigations
it is more useful to concentrate on the displacement density,
since we expect better performance of density estimation methods
 when the densities are smoother and less oscillatory.

\section{Central limit theorem and rate of convergence}
\label{sec:clt}

We now discuss the central limit theorem as $t \to \infty$ in
terms of the fine structure described in the previous section.

\subsection{Central limit theorem: weak convergence to normal
distribution} \label{subsec:clt-weak-conv}

\begin{figure}[p]
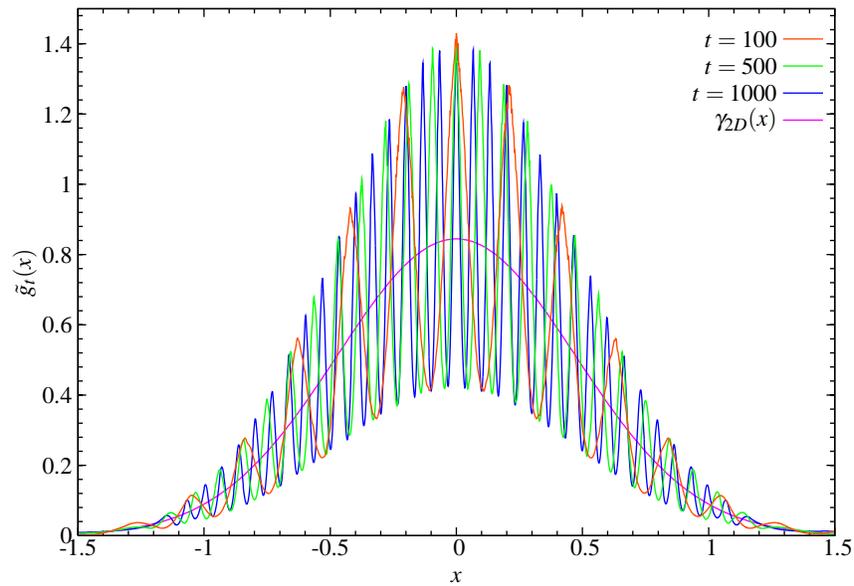

\centrefig{rescaled_pdfsr2.1b0.2.eps}
\caption{\label{fig:rescaled-densities} Displacement densities as
in \figref{fig:time-evol}(b) after  rescaling by $\sqrt{t}$,
compared to a Gaussian density with mean $0$ and variance $2D$.
$r=2.1$; $b=0.2$.}
\end{figure}

The central limit theorem requires us to consider the densities
scaled by $\sqrt{t}$, so we define the rescaled densities
\begin{equation}\label{}
\tg_t(x) \defeq \sqrt{t} \, g_t(x \sqrt{t}),
\end{equation}
where the first factor of $\sqrt{t}$ normalizes the integral of
$\tg_t$ to $1$, giving a probability density.
\bfigref{fig:rescaled-densities} shows the densities of
\figref{fig:time-evol}(a) rescaled in this way, compared to a
Gaussian density with mean $0$ and variance $2D$. We see that the
rescaled densities oscillate within an envelope which remains
approximately constant, but with an increasing frequency as $t \to
\infty$; they are oscillating around the limiting Gaussian, but do
not converge to it pointwise. See also
\figref{fig:demodulate-disp-flat}.

\begin{figure}[p]
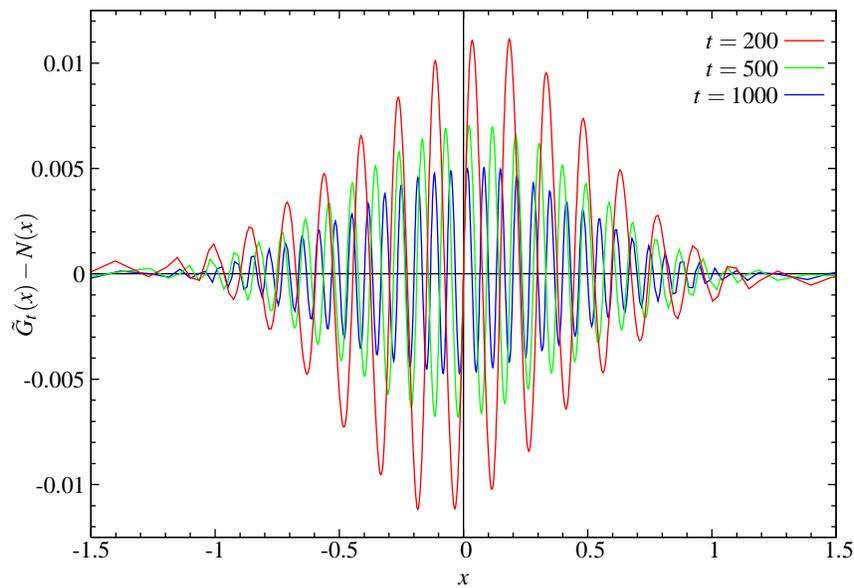

\centrefig{osc_rescaled_cdfsr2.1b0.2.eps}
\caption{\label{fig:rescaled-distn-fns} Difference between
rescaled distribution functions and
    limiting normal distribution with variance $2D$. $r=2.1$; $b=0.2$.}
\end{figure}

The increasingly rapid oscillations do, however, cancel out when
we consider the rescaled distribution functions, given by the integral of
the rescaled density functions:
\begin{equation}\label{}
\tG_t(x) \defeq \int_{s=-\infty}^x \tg_t(s) \rd s = G_t(x \,
\sqrt{t}).
\end{equation}
\bfigref{fig:rescaled-distn-fns} shows the difference between the rescaled
distribution
functions and the limiting normal distribution with mean $0$ and
variance $2D$.  We see that
 the rescaled distribution functions do converge
 to the limiting normal, in fact uniformly, as $t \to \infty$; we thus
have only \defn{weak} convergence of the distributions.

Although this is the strongest kind of convergence we can obtain
for the densities $\tilde{g}_t$ with respect to Lebesgue measure,
\figref{fig:demodulate-disp-flat} provides evidence for the
following conjecture: the rescaled densities $\tilde{\etabar}_t$
with respect to the new measure $h\lambda$ converge
\emph{uniformly} to a Gaussian \emph{density}.  This characterizes
the asymptotic behaviour more precisely than the standard central
limit theorem.

\subsection{Rate of convergence} \label{subsec:rate-conv}

Since the $\tG_t$ converge uniformly to the limiting normal
distribution, we can consider
the distance $\supnorm{\tG_t - \normal{2D}}$, where we define the
\defn{uniform norm} by
\begin{equation}\label{}
\supnorm{F} \defeq \sup_{x \in \R} \modulus{F(x)}.
\end{equation}
We denote by $\normal{\sigma^2}$  the
normal distribution function with variance $\sigma^2$, given by
\begin{equation}\label{}
\normal{\sigma^2}(x) \defeq \frac{1}{\sigma \sqrt{2 \pi}}
\int_{s=-\infty}^x \e^{-s^2/2\sigma^2} \rd s,
\end{equation}

\begin{figure}[tp]
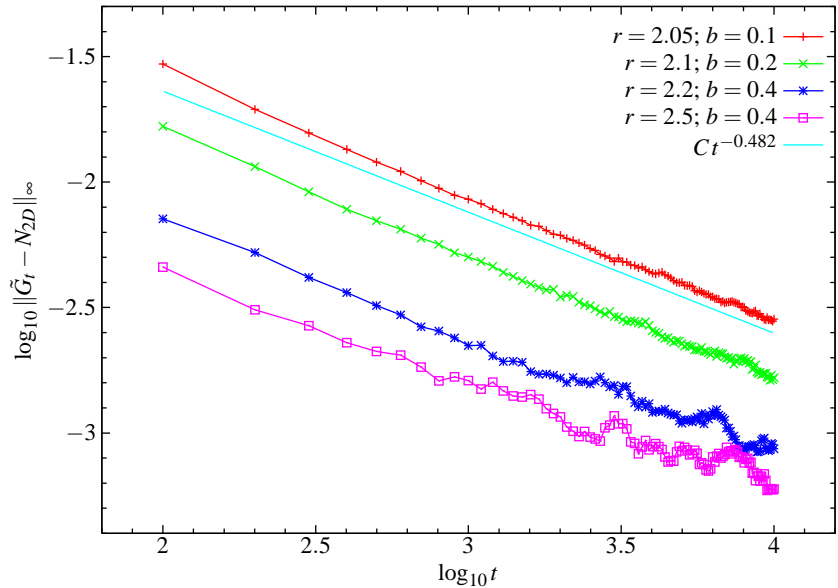

\centrefig{lorentz_clt.eps} \caption{ Distance of rescaled
distribution functions $\tilde{G}_t$ from limiting normal
distribution $N_{2D}$ in log--log plot.  The straight line is a
fit to the large-time decay of the data for $r=2.05$.}
  \label{fig:rate-conv}
\end{figure}

\bfigref{fig:rate-conv} shows a log--log plot of this distance
against time, calculated numerically from the full distribution
functions.  We see that the convergence follows a power law
\begin{equation}\label{}
\supnorm{\tG_t - \normal{2\Dest}} \sim t^{-\alpha},
\end{equation}
with a fit to the data for $r=2.05$ giving a slope $\alpha \simeq 0.482$.
The same decay rate is obtained for
a range of other geometrical parameters, although
the quality of the data deteriorates for larger $r$, reflecting  the fact
that diffusion is faster, so that the distribution spreads further
in the same time.  Since we use the same number
$N=10^7$ of
initial conditions, there is  a lower resolution near $x=0$ where, as shown
in the next section, the maximum
is obtained.

In \cite{Pene} it
was proved rigorously  that $\alpha \ge \frac{1}{6}
\simeq 0.167$ for \emph{any} H\"older continuous observable $f$.
Here we
have considered only the particular H\"older observable
$v$, but for this function we see that the rate of convergence is
much faster than the lower bound proved in \cite{Pene}.

\subsection{Analytical estimate of rate of convergence}
\label{subsec:qual-est-rate-conv}

We now show how to obtain a simple analytical estimate of the rate
of convergence using the  fine structure calculations in
\secref{sec:fine-structure-posn-density}.

Since the displacement distribution is symmetric, we have
$\tG_t(x=0) = 1/2$ for all $t$. The maximum deviation of $\tG_t$
from $\normal{2\Dest}$ occurs near to $x=0$, where the density
function is furthest from a Gaussian, while the fine structure of
the density $\tg_t$ means
 that $\tG_t$ is increasing there (\figref{fig:rescaled-distn-fns}).
 Due to the
oscillatory nature of the fine structure, this maximum thus
occurs at a distance of $1/4$ of the period of oscillation
 from $x=0$.

Since the displacement density has the form $g_t(x) = \phi(x) \etabar_t(x)$,
after rescaling
we have
\begin{equation}\label{}
\tg_t(x) = \phi(x \sqrt{t}) \tilde{\etabar}(x),
\end{equation}
where $\tilde{\etabar}_t(x) \defeq \sqrt{t} \, \etabar_t(x \sqrt{t})$ is the
rescaled slowly-varying part
of $g_t$, and the fine structure at time $t$ is given by
\begin{equation}\label{}
\phi(x \sqrt{t}) = 1 + 2 \sum_{k \in \Nats} \fh{k} \, \cos(2 \pi k
x \sqrt{t}).
\end{equation}
The maximum deviation occurs at $1/4$ of the period of
$\phi(x\sqrt{t})$, i.e.\ at $x=\frac{1}{4\sqrt{t}}$, so that
\begin{align}\label{}
\supnorm{G_t - N} &\simeq \int_0^{1/4 \sqrt{t}} \sum_{k \in \Nats}
\fh{k} \,
 \cos(2 \pi k x \sqrt{t})
\rd x \\
&= \frac{1}{\sqrt{t}} \sum_{k \in \Nats, \, k \text{ odd}} \fh{k}
\frac{(-1)^{(k-1)/2}}{2 \pi k}.
\end{align}
The correction due to the curvature of the underlying Gaussian
converges to $0$ as $t \to \infty$, since this Gaussian is flat at
$x=0$. Hence $\supnorm{G_t - N} = \bigO{t^{-1/2}}$.

This calculation shows that the fastest possible convergence is a
power law with exponent $\alpha = 1/2$, and provides an intuitive
reason why this is the case. This should be compared to the rate
$t^{-1/2}$ for convergence of rescaled distribution functions
corresponding to solutions of the diffusion equation in
\appref{app:conv-solns-diffn-eqn}.  If the rescaled shape function
$\tilde{\etabar}_t$ converges to a Gaussian shape at a rate slower
than $t^{-1/2}$, then the overall rate of convergence $\alpha$
could
 be slower than $1/2$.  However, the numerical results in
\secref{subsec:rate-conv} show that the rate is close to $1/2$.
We remark that for an observable
which is not so intimately related to the geometrical structure of
the lattice, the fine structure will in general be more
complicated, and the above argument may no longer hold.

\section{Maxwellian velocity distribution}
\label{sec:maxwell-vel-distn}

In this section we consider the effect of a non-constant
distribution of particle speeds\footnote{I would like to thank
Hern\'an Larralde for posing this question, and for the
observation that the resulting position distribution may no longer
be Gaussian.}. A Maxwellian (Gaussian) velocity distribution was
used in polygonal and Lorentz channels in
\cite{LiHeatLinearMixing} and \cite{AlonsoLorentzChannel},
respectively, in connection with heat conduction studies. The
  mean squared displacement was observed to grow
asymptotically linearly, but  the relationship with the unit speed
situation was not discussed.

  We show that the mean squared displacement grows asymptotically
linearly in time with the same diffusion coefficient as for the
unit speed case, but that the limiting position distribution may
be \emph{non-Gaussian}. For brevity we refer only to the position
distribution throughout this section; the displacement
distribution is similar.

\subsection{Mean squared displacement}

Consider a particle located initially at $(\x_0,\v_0)$, where
$\v_0$ has unit speed.  Changing the speed of the particle does
not change the path it follows, but  only  the distance along the
path traveled in a given time.  Denoting by $\flow_v^t(\x_0,\v_0)$
the billiard flow with speed $v$ starting from $\x_0$ and with
initial velocity in the direction of the unit vector $\v_0$, we
have
\begin{equation}
\flow_v^t(\x_0,\v_0) = \flow^{vt}(\x_0, \v_0),
\end{equation}
where the flow on the right hand side is the original unit-speed
flow. If all speeds are equal to $v$, then the radially symmetric
2D position probability density after a long time $t$ is thus
\begin{equation}
\conddensity{p_t}{x,y}{v} = \frac{1}{4\pi Dvt} \exp \left(
\frac{-(x^2+y^2)}{4 D v t} \right),
\end{equation}
giving a radial density
\begin{equation}
\conddensity{p_t}{r}{v} = \frac{r}{2 D v t} \, \exp \left( \frac{-r^2}
{4 D v t} \right).
\end{equation}
(Throughout this calculation we neglect any fine structure.)

If we now have a distribution of velocities with density $p_V(v)$,
then the radial position density at distance $r$ is
\begin{equation} \label{eq:maxwell-posn-density}
\frad_t(r) = \int_{v=0}^\infty \conddensity{p_t}{r}{v} \, p_V(v) \rd v.
\end{equation}
The variance of the position distribution is then given by
\begin{align}
\mean{\x^2} &= \int_{r=0}^\infty r^2 \, \frad_t(r) \rd r \\
&= 4Dt \int_0^\infty v \, p_V(v) \rd v \eqdef 4Dt \bar{v},
\end{align}
where $\bar{v}$ is the mean speed, after interchanging the integrals over
$r$ and $v$.

We thus see that for any speed distribution having a finite mean,
the variance of the position distribution, and hence the mean
squared displacement, grows asymptotically linearly with the same
diffusion coefficient as for the uniform speed distribution,
having normalized such that $\bar{v}=1$. We have verified this
numerically with a Gaussian velocity distribution: the mean
squared displacement is indistinguishable from the unit speed case
even after very short times.

\subsection{Gaussian velocity distribution}

Henceforth attention is restricted to the case of a Gaussian
velocity distribution. For each initial condition, we generate two
independent normally-distributed random variables $v_1$ and $v_2$
with mean $0$ and variance $1$ using the standard Box--Muller
algorithm \cite{NR}, and then multiply by $\sigma$, which is a
standard deviation calculated below. We use $v_1$ and $v_2$ as the
components of the velocity vector $\v$, whose probability density
is hence given by
\begin{equation}\label{}
p(\v)=p(v_1,v_2) = \frac{\e^{-v_1^2/2\sigma^2}}{\sigma \sqrt{2\pi}} \,
\, \frac{\e^{-v_2^2/2\sigma^2}}{\sigma \sqrt{2\pi}} \,
= \frac{\e^{-v^2/2\sigma^2}}{2 \pi \sigma^2} \, ,
\end{equation}
where $v \defeq \modulus{\v} = \sqrt{v_1^2+v_2^2}$ is the speed of
the particle.  The speed $v$ thus has density
\begin{equation}\label{}
p_V(v) = \frac{v}{\sigma^2}  \, \e^{-v^2/ 2\sigma^2}
\end{equation}
and mean
$\bar{v} = \sigma \sqrt{\pi/2}$.
 To compare
with the unit speed distribution we require $\bar{v} = 1$, and
hence $\sigma =  \sqrt{2/\pi}$. As before, we distribute the
initial positions uniformly with respect to Lebesgue measure in
the billiard domain $Q$.

\subsection{Shape of limiting distribution}

The position density \eqref{eq:maxwell-posn-density} is a function
of time. However, the Gaussian assumption  used to derive that
equation is  valid in the limit when $t \to \infty$, so the
central limit
 theorem
rescaling
\begin{equation}
\ftrad_t(r) \defeq \sqrt{t} \, \frad_t(r \sqrt{t})
\end{equation}
eliminates the time dependence in \eqref{eq:maxwell-posn-density},
giving the following shape
for the limiting radial density:
\begin{equation}\label{eq:maxwell-reduced-integral}
\ftrad(r) = \frac{\pi r}{4D} \int_{v=0}^\infty
\exp \left(-\frac{r^2}{4Dv} - \frac{\pi v^2}{4} \right) \rd v \eqdef
\frac{\pi r}{4D} \, I,
\end{equation}
denoting the integral by $I$. We can evaluate this integral
explicitly using \texttt{Mathematica} \cite{MathematicaBook} in
terms of the \defn{Meijer $G$-function}\footnote{The Meijer
$G$-function is a generalisation of the classical Gauss
hypergeometric function, defined as the following
Mellin--Barnes-type integral; for more details see
\cite{Erdelyi,MathaiSaxena,MetzlerKlafter}:
\begin{gather}\label{}
G^{m,n}_{p,q} \biggl(\, z \, \biggm|
\begin{matrix} a_1,\ldots,a_p \\ b_1,\ldots,b_q \end{matrix} \biggr)
\defeq \int_C \frac{1}{2\pi \i } \chi(s) z^{-s} \rd s; \\
\chi(s) \defeq \frac{\prod_{j=1}^m \Gamma(b_j+s) \prod_{j=1}^n
\Gamma(1-a_j-s)}{\prod_{j=m+1}^q \Gamma(1-b_j-s) \prod_{j=n+1}^p
\Gamma(a_j+s)}.
\end{gather}
$C$ is a certain contour in the complex plane and there are
 restrictions on the
$a_i$ and the $b_j$.}
 \cite{Erdelyi,MathaiSaxena}:
\begin{equation}\label{eq:exact-meijer-G}
I = G^{3,0}_{0,3}\biggl(\frac{\pi r^4}{256 D^2} \biggm|
\begin{matrix} \text{---} \\ -\textstyle \frac{1}{2},0,0 \end{matrix} \biggr).
\end{equation}
See \cite{MetzlerKlafter} and references therein for a review of the use
of such special functions in anomalous diffusion.

We can, however, obtain an asymptotic approximation to $I$ from
its definition as an integral, without using any properties of
special functions, as follows.
Define $K(v) \defeq \frac{r^2}{4Dv} + \frac{\pi v^2}{4}$,  the negative
of the argument of the exponential in \eqref{eq:maxwell-reduced-integral}.
Then $K$ has a unique minimum at $\vmin \defeq (r^2/(2 \pi D))^{1/3}$
and we expect the integral to be dominated by the neighborhood of this
minimum.
However, the use of standard asymptotic methods is complicated by the fact
that
as $r \to 0$, $\vmin$ tends to $0$, a boundary
of the integration domain.

To overcome this, we change variables to fix the minimum away from
the domain boundaries, setting $w \defeq v / \vmin$.  Then
\begin{equation}
I = \vmin \int_{w=0}^\infty e^{-\alpha \, L(w)} \rd w,
\end{equation}
where $\alpha \defeq \frac{\pi \vmin^2}{2}$ and $L(w) \defeq \frac{1}{w} +
\frac{w^2}{2}$, with a minimum at $\wmin=1$.
Laplace's method (see e.g.\ \cite{CarrierKrookPearson}) can now be applied,
giving the asymptotic approximation
\begin{equation}\label{eq:asymptotic-expansion-integral}
I \sim \vmin \, e^{-\alpha L(\wmin)} \frac{\sqrt{2\pi}}{\sqrt{\alpha \,
L''(\wmin)}} = \frac{2}{\sqrt{3}} e^{-3\alpha/2},
\end{equation}
valid for large $\alpha$, i.e.\ for large $r$.

Hence
\begin{equation}\label{eq:asymptotic}
\ftrad(r) \stackrel{r \to \infty}{\sim} C \, r \, e^{-\beta \, r^{4/3}},
\end{equation}
where
\begin{equation}\label{eq:defs-of-asymp-consts}
C \defeq \frac{\pi}{2D \sqrt{3}}; \quad
\beta \defeq \frac{3}{2} \left(\frac{\pi}{32 D^2} \right)^{1/3}.
\end{equation}

\subsection{Comparison with numerical results}

\begin{figure}[p]
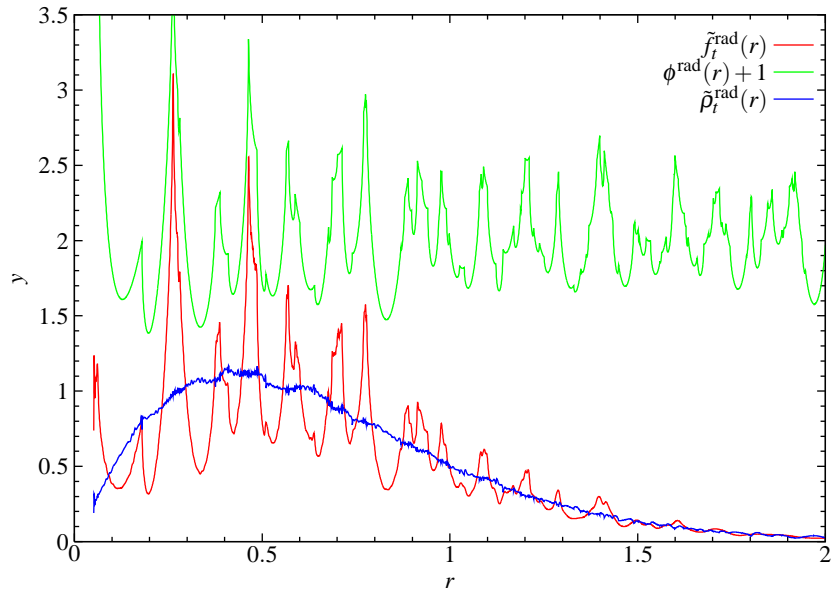

\centrefig{maxwellr2.3b0.5-radial.eps}
\caption{\label{fig:maxwell} The radial density function
$\ftrad_t$ compared to the numerically calculated  radial fine
structure function $\phirad$, rescaled to converge to $1$ and then
vertically shifted for clarity.  The demodulated radial density
$\rhorad_t$ is also shown. $r=2.3$; $b=0.5$; $t=100$.}
\end{figure}

\bfigref{fig:maxwell} shows the numerical radial position density
$\ftrad_t(r)$ for a particular choice of geometrical parameters.
We wish to demodulate
this as in \secref{sec:fine-structure} to extract the slowly-varying
shape function, which we can then compare to the analytical calculation.

The radial fine structure function $\phirad(r)$ must be calculated
numerically, since no analytical expression is available.  We do
this by distributing $10^5$ points uniformly on a circle of radius
$r$ and calculating the proportion of points not falling inside
any scatterer.  This we normalize so that $\phirad(r)
\to 1$ as $r \to \infty$, using the fact that when $r$ is large,
the density inside the circle of radius $r$ converges to the ratio
$[r^2 - \pi(a^2+b^2)]/r^2$ of available area per unit cell to
total area per unit cell. We can then demodulate $\ftrad_t$ by
$\phirad$, setting
\begin{equation}
\rhorad_t(r) \defeq \frac{\ftrad_t(r)}{\phirad(r \, \sqrt{t})}.
\end{equation}


\bfigref{fig:maxwell} shows the demodulated radial density
$\rhorad_t(r)$ at two times compared to the exact solution
\eqref{eq:maxwell-reduced-integral}--\eqref{eq:exact-meijer-G},
the asymptotic approximation
\eqref{eq:asymptotic}--\eqref{eq:defs-of-asymp-consts}, and the
radial Gaussian $\frac{r}{2D} e^{-r^2/2D}$.  The asymptotic
approximation agrees well with the exact solution except at the
peak, while the numerically determined demodulated densities agree
with the exact long-time solution over the whole range of $r$. All
three differ significantly from the Gaussian, even in the tails.
We conclude that the radial position distribution is
\defn{non-Gaussian}. A similar calculation could be done for the
radial displacement distribution, but a numerical integration
would be required to evaluate the relevant fine structure
function.

An explanation of the non-Gaussian shape comes by considering slow
particles which remain close to the origin for a long time, and
fast particles which can travel further than those with unit
speed. The combined effect skews the resulting distribution in a
way which depends on the relative weights of slow and fast
particles.


\begin{figure}[p]
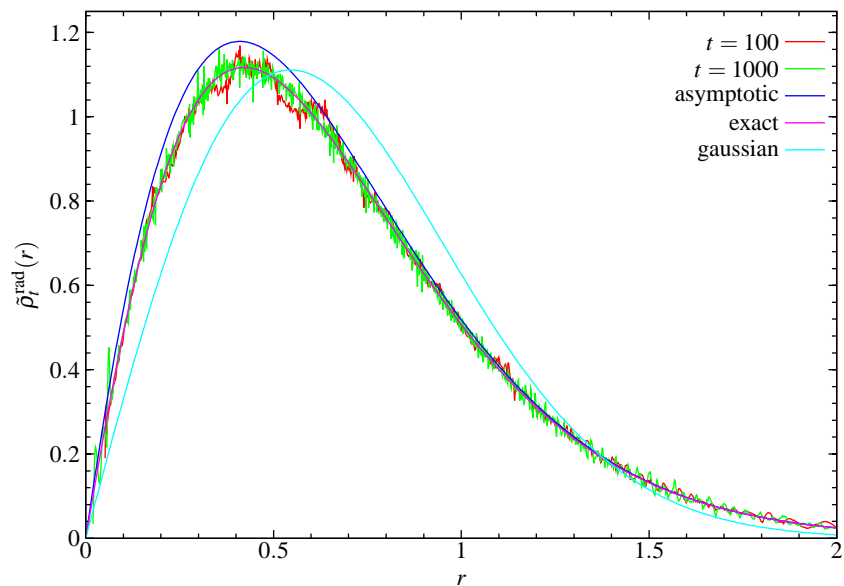

\centrefig{maxwellr2.3b0.5-radial-demod.eps}
\caption{\label{fig:maxwell-demod} Comparison of the demodulated
radial density $\rhorad_t$ with the exact Meijer-$G$
representation, the large-$r$ asymptotic approximation, and the
radial Gaussian with variance $2D$.}
\end{figure}


\subsection{1D marginal}

\begin{figure}[tp]
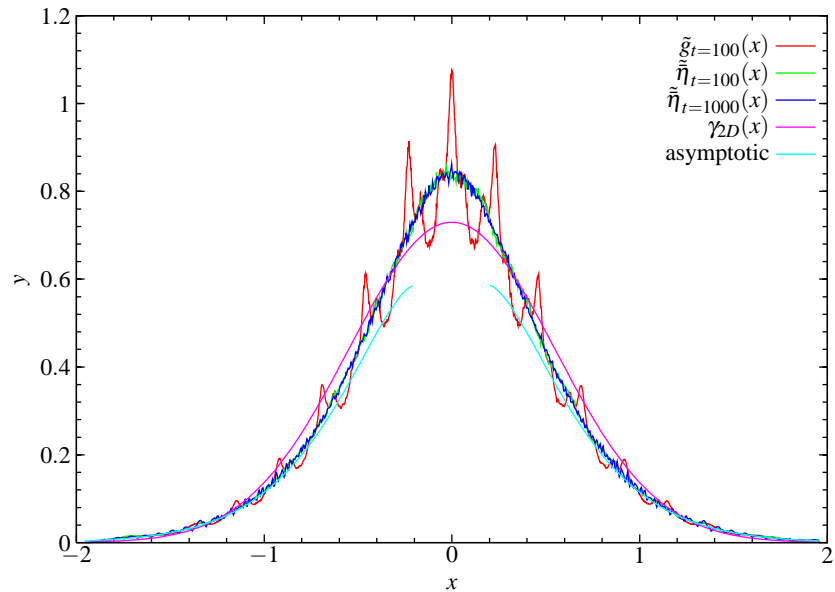

\centrefig{maxwell-r2.3b0.5.eps}
\caption{\label{fig:maxwell-marginal}Rescaled 1D marginal of the
displacement density $\tilde{g}$ and the demodulated version
$\tilde{\etabar}$ compared to the Gaussian with variance $2D$ and
to the asymptotic expression.  The latter is not shown close to
$x=0$, where it drops to $0$. $r=2.3$; $b=0.5$.}
\end{figure}

The 1D marginal in the $x$-direction is shown in
\figref{fig:maxwell-marginal}.  Again there is a significant
deviation of the demodulated density from a Gaussian. From
\eqref{eq:asymptotic}, the 2D density at $(x,y)$ is asymptotically
\begin{equation}\label{eq:2d-asymptotic}
\tilde{f}(x,y) \sim \frac{C}{2\pi} \exp\left[-\beta (x^2+y^2)^{2/3}\right],
\end{equation}
from which the 1D marginal $\tilde{f}(x)$ is obtained by
\begin{equation}\label{eq:asymptotic-marginal}
\tilde{f}(x) = \int_{y=-\infty}^\infty \tilde{f}(x,y) \rd y.
\end{equation}
It does not seem to be possible to perform this integration
explicitly for either the asymptotic expression
\eqref{eq:2d-asymptotic} or the corresponding exact solution in
terms of the Meijer $G$-function.  Instead we perform another
asymptotic approximation starting, from the  asymptotic expression
\eqref{eq:2d-asymptotic}. Changing variables in
\eqref{eq:asymptotic-marginal} to $z
\defeq y/x$ and using the evenness in $y$ gives
\begin{equation}
\tilde{f}(x) \sim \frac{C \modulus{x}}{2\pi} \int_{z=-\infty}^\infty
\exp \left[ -\kappa(1+z^2)^{2/3} \right] \rd z,
\end{equation}
where $\kappa \defeq \beta \modulus{x}^{4/3}$.
 Laplace's method  then gives
\begin{equation}
\tilde{f}(x) \sim \frac{C\sqrt{3}}{\sqrt{8\pi\beta}} \modulus{x}^{1/3}
e^{-\beta \modulus{x}^{4/3}},
\end{equation}
valid for large $x$. This is also shown in
\figref{fig:maxwell-marginal}.  Due to the $\modulus{x}^{1/3}$
factor, the behaviour near $x=0$ is wrong, but in the tails there
is reasonably good agreement with the numerical results.

\startonright


\graphicspath{{figs/}}

\chapter{Normal and anomalous diffusion in polygonal billiard channels}

\label{chap:polygonal}

\section{Introduction: ergodic and statistical properties in polygonal billiards}

\label{sec:poly-intro}

The rigorous results on statistical properties of Lorentz gases
discussed in \chapref{chap:stat-props}
 depend on
the hyperbolicity of the dynamics due to the defocusing nature of
the scatterers.  Recently the question has been asked to what extent
 hyperbolicity is actually \emph{necessary} (rather than
just sufficient) for these strong statistical properties to hold
\cite{DettmanNature,DettCoh1}.

The weak nature of the scattering from flat obstacles implies that
in polygonal billiards the Kolmogorov--Sinai entropy and all
Lyapunov exponents are $0$, so that infinitesimally separated
trajectories separate slower than exponentially; in fact they
separate linearly, as can be seen by unfolding the polygonal
billard and considering two initial conditions with slightly
different angles. Nonetheless, trajectories separated by a finite
distance may fall on different sides of a polygonal vertex,
causing them to separate faster.  Characterising this effect is
difficult: see e.g.\ \cite{vanBeijerenWeaklyChaoticEntropies} for
a recent attempt.

If all angles are \defn{rational}, i.e.\ rational multiples of
$\pi$, then the dynamics reduces to directional flows on invariant
surfaces and is well-understood; we do not consider this case.
Rigorous results on ergodic properties of more general polygonal
billiards are reviewed in \cite{Gutkin86review,GutkinJSPReview,
GutkinRegChaoticDynReview}. There are known rigorous examples of
ergodic polygonal billiards, but no examples known to be mixing,
although mixing has not been \emph{dis}proved in general. Recently
numerical evidence has been given that the billiard dynamics in
right-angled triangles with irrational acute angles is ergodic and
weak-mixing but not mixing \cite{ArtusoCasatiNumericalErgTriBill},
while in triangles with all angles irrational the dynamics
appears to be mixing \cite{CasatiProsenMixingTriBilliard}.

Despite these uncertainties, various polygonal billiard models
have been found numerically to exhibit normal diffusion, in the
sense that the mean squared displacement grows asymptotically
linearly in time (property (a) from \chapref{chap:stat-props})
\cite{DettCoh1,AlonsoPolyg,LiHeatLinearMixing}. In this chapter we
further explore statistical properties of polygonal billiards,
attempting to understand when we can hope to see normal diffusion
and to what extent stronger properties, such as the central limit
theorem, are satisfied.  We also investigate when normal diffusion
fails and characterise the anomalous diffusion that results.

Recently there has also been much interest in finding simple model
systems which show normal \emph{heat conduction}, in the sense
that Fourier's law holds.  Several simple periodic billiard models
were found which show this behaviour when placed between thermal
reservoirs at different temperatures
\cite{AlonsoLorentzChannel,AlonsoPolyg,LiHeatLinearMixing}.
Evidence has been given that the heat conduction properties of 1D
can be predicted once the diffusive properties are known
\cite{LiNormAndAnomDiffn, DenisovDynamHeatChannels}, so that it is
important to characterise the diffusive behaviour.

We remark that  the weak instability in polygonal billiards means
that numerical simulations should be \emph{more reliable} than in
scattering billiards, in the sense that computed trajectories
should lie close to true trajectories for a long time; see e.g.\
\cite{ArtusoGuarneriAnomPolyBillChaos}.  For this reason we
believe that numerical experiments on statistical properties of
polygonal billiards can provide useful information.

 We begin by reviewing existing models which numerically
 exhibit asymptotic linear growth of the mean squared displacement.  We then
 attempt to identify geometrical features of polygonal billiards which allow  or
 prevent the occurrence of normal diffusion, and construct two more classes of
 models which  have normal diffusion except when a particular
 geometrical condition is satisfied, namely that \emph{parallel scatterers} exist,
 when we find anomalous diffusion.


\section{Polygonal models exhibiting normal diffusion}

\subsection{Previous models}

\paragraph{Model of Alonso \etal}

\begin{figure}
\centering \subfigure[]{\label{fig:poly-bill-geom}
\includegraphics{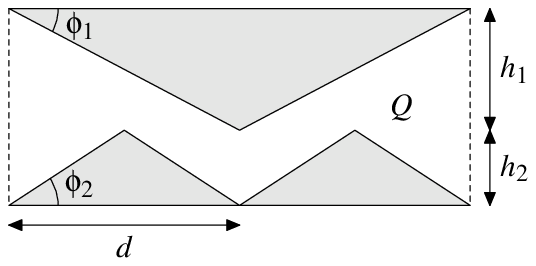}}

\subfigure[]{\label{fig:poly-channel}
\includegraphics{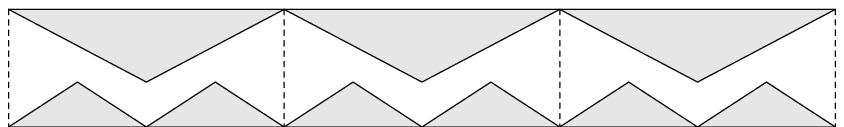}}
  \caption{(a) Geometry of the polygonal billiard unit cell of \cite{AlonsoPolyg}, shown to
scale with $\phi_2=\pi/(2e)$.  (b) Part of the polygonal channel
with the same parameters.}
\end{figure}

Alonso \etal\ \cite{AlonsoPolyg} studied the geometry shown in
\figref{fig:poly-bill-geom} and \figref{fig:poly-channel}. We fix
the angles $\phi_1$ and $\phi_2$ and choose $d$ such that the
bottom triangles are half the width of the top triangle.  This
determines the ratio of $h_1$ to $h_2$ in terms of the angles
$\phi_1$ and $\phi_2$.  We then
  require the inward-pointing vertices of each
triangle to lie on the same horizontal line to prevent infinite
horizon trajectories.  Taking $h_1+h_2=h=1$ then gives
\begin{equation}\label{eq:definition-of-d}
d = \frac{h}{\tan \phi_1 + \textstyle \frac{1}{2} \tan \phi_2},
\end{equation}
with $h_1 = d \tan \phi_1$ and $h_2 = (d/2) \tan \phi_2$. We
remark that in \cite{AlonsoPolyg} it was stated that the area
$\modulus{Q}=dh$ of the billiard domain is independent of $\phi_2$
when $\phi_1$ is fixed.  This is not correct, however, since by
\eqref{eq:definition-of-d}, $d$ depends on $\phi_2$.

In \cite{AlonsoPolyg}, the parameters $\phi_1 = \pi(\sqrt{5} -
1)/8$ and $\phi_2 = \pi/q$, $q \in \Nats$, $3 \leq q \leq 9$ were
used. For $q \ge 5$, the mean squared displacement was found to
grow like $t^\alpha$ with $\alpha$ in the range $1$ to $1.08$,
indicating normal diffusion.

For $q=3,4$, however, anomalous diffusion was found, for which
$\msd_t \sim t^\alpha$ with $\alpha \neq 1$. As far as we are
aware, there is as yet no explanation for this observed anomalous
behaviour, although presumably number-theoretic properties of the
angles are relevant; in a second paper \cite{AlonsoPolyII}, Alonso
\etal\ state that in these cases they find large classes of
periodic orbits which are either trapped in one cell in the case
$q=4$, resulting in sub-diffusion, or are propagating orbits,
resulting in super-diffusion for $q=3$.

 These results appear to be related to the rational angles used.
 We have instead used a value of $\phi_2$ which is not
rationally related to $\phi_1$, namely $\phi_2 = \pi / (2\e)
\simeq \pi/5.44$, where $\e$ is the base of natural logarithms. In
this case we find that $\msd_t \sim t^{1.008}$, which we regard as
asymptotically linear, with $D = 0.3796 \pm 0.0009$.

\paragraph{Model of Li \etal}

\begin{figure}[htbp]
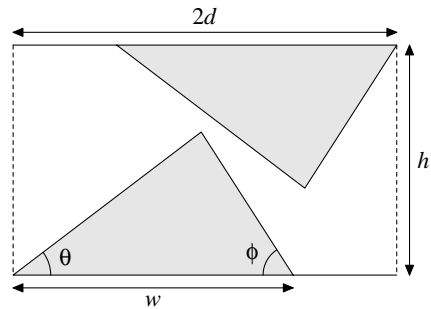

  \centrefig{li-billiard.1}
  \caption{ \label{fig:li-model} Unit cell of the model introduced in \cite{LiHeatLinearMixing},
  with $2d=3$, $h=1.8$, $w=2.19$, $\phi=1$ and
$\theta=(\sqrt{2}-1) \pi/2$.  Note that $\phi=1$ is a
transcendental multiple of $\pi$, whereas $\theta$ is only a
quadratic irrational multiple of $\pi$.}
\end{figure}

\bfigref{fig:li-model} shows the unit cell of the model introduced
by Li \etal\ \cite{LiHeatLinearMixing}. Unlike the other models we
study, this one generates a \emph{two-dimensional}  polygonal
billiard with finite horizon when unfolded: presumably this was
the reason for introducing this more subtle shape.

\subsection{Necessary conditions for normal diffusion}

Based on the features of the models presented above and those
introduced below, and on experiments with further models, we
arrive at the following heuristic ingredients for constructing
polygonal billiards with normal diffusion:
\begin{enumerate}

\item avoid vertex angles which are rationally related to $\pi$;

\item avoid infinite horizon trajectories; and

\item avoid parallel scatterers.

\end{enumerate}

As discussed above, point (i) has to some extent a rigorous
justification.  Point (ii) is related to heuristic arguments
discussed in \chapref{chap:3dmodel} which show that infinite
horizon trajectories lead to at least weak anomalous diffusion,
where the mean squared displacement grows like $t \log t$.  Point
(iii) is the main observation that we wish to emphasise.  As we
discuss in more detail in \secref{sec:poly-anom-diffn}, parallel
scatterers seem to cause a
\defn{channelling} effect, resulting in long laminar stretches and
thus
 anomalous diffusion.

\subsection{Constructing new models with normal diffusion}

Following the above heuristic rules, we now construct two classes
of models which seem to exhibit normal diffusion for most
parameter values, but  anomalous diffusion for certain special
geometrical configurations.  These new models provide clear
evidence of point (iii) in particular. We follow
\cite{AlonsoPolyg} in using strictly 1D channels, i.e.\ ones which
cannot be unfolded in the $y$ direction, which enables us more
easily to ensure a finite horizon.

\paragraph{Polygonal Lorentz model}

\begin{figure}
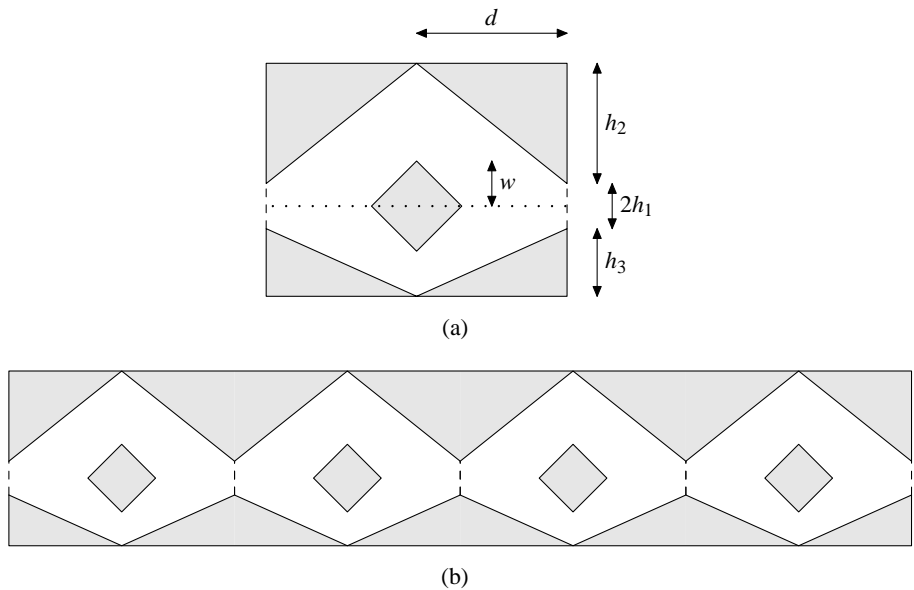

\centering \subfig{poly-billiard-mine.1}\hfill

\subfig{poly-billiard-mine.2}
\caption{\label{fig:poly-billiard-mine} (a) One unit cell, and (b)
several cells forming a 1D channel, for a polygonal model derived
from the Lorentz gas considered in \chapref{chap:geom-dependence}.
There is a crossover from normal to anomalous diffusion when
$h_2=h_3$.}
\end{figure}

Our first model is a polygonalised version of the model introduced
in \chapref{chap:geom-dependence},  restricted to be strictly
one-dimensional: we create a polygonal channel and then add an
extra scatterer at the centre of each unit cell to ensure a finite
horizon by blocking the central corridor: see
\figref{fig:poly-billiard-mine}.

We have defined this model in terms of heights of the scatterers.
The angles between the scatterers are then given as arctangents of
ratios of these heights. In general these angles will be
irrational multiples of $\pi$, although since in our calculations
we use simple heights, we will always end up with angles which are
nongeneric in the sense that they are arctangents of rationals.

To generate initial conditions uniformly distributed with respect
to Lebesgue measure in the available space inside the polygon, we
use the `rejection method' \cite{NR}, as follows.  We  generate
points in a rectangle which contains the polygonal billiard domain
and reject points which do not lie inside it, using the following
algorithm \cite{ORourke}: a point lies inside the polygon if  a
ray extending from the point to the exterior of the polygon
crosses the boundary of the polygon an odd number of times.

\paragraph{Simplified version: zigzag model}

\begin{figure}
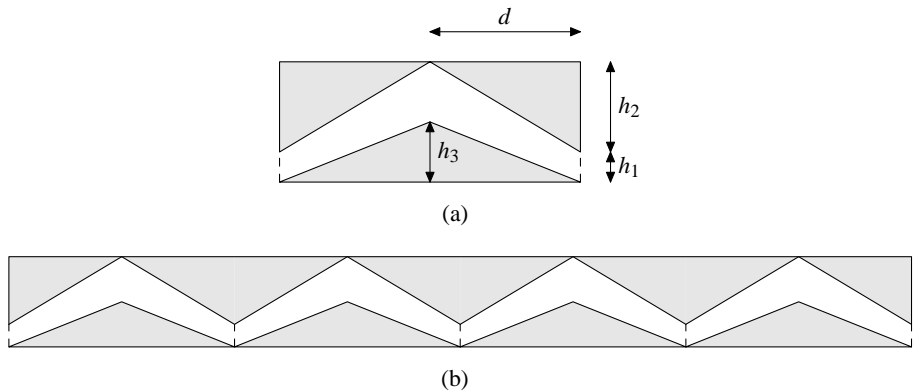


\centering \subfig{zigzag.1}

 \subfig{zigzag.2}
\caption{\label{fig:zigzag-channel} (a) One unit cell, and (b)
several unit cells forming a 1D channel, for a zigzag model which
is a modification of \figref{fig:poly-billiard-mine}.  This model
also shows a crossover from normal to anomalous diffusion when
$h_2=h_3$.}
\end{figure}

We can simplify the above model by eliminating the central
additional  scatterer.  In order  still to be able to block
horizontal trajectories in the centre of  the channel, we flip the
bottom line of scatterers to point up instead of down, resulting
in the model\footnote{\revision{The same model was studied independently in
\cite{JeppsRondoniSimpleTransportPhenomenaJPA2006} at around the same time,
with similar conclusions.}} shown in \figref{fig:zigzag-channel}.

\section{Normal diffusion}
\label{sec:polyg-normal-diffn}

We present evidence that the models discussed above do  indeed
exhibit normal diffusion.  It is difficult to
distinguish between
asymptotic $t$ and  $t \log t$ behaviour by examining only the
mean squared displacement\footnote{\revision{This was done in the
published version of this chapter
\cite{SandersLarraldeNormalAnomalousDiffusionPolygonalPRE2006}.}}, so that we
look at several different
indicators. (Of course, numerical evidence alone can never be
conclusive without a rigorous basis, especially as regards
asympotic behaviour; nonetheless, the available numerical evidence
presented below leads to this conclusion.)

\subsection{Moments}

\begin{figure}
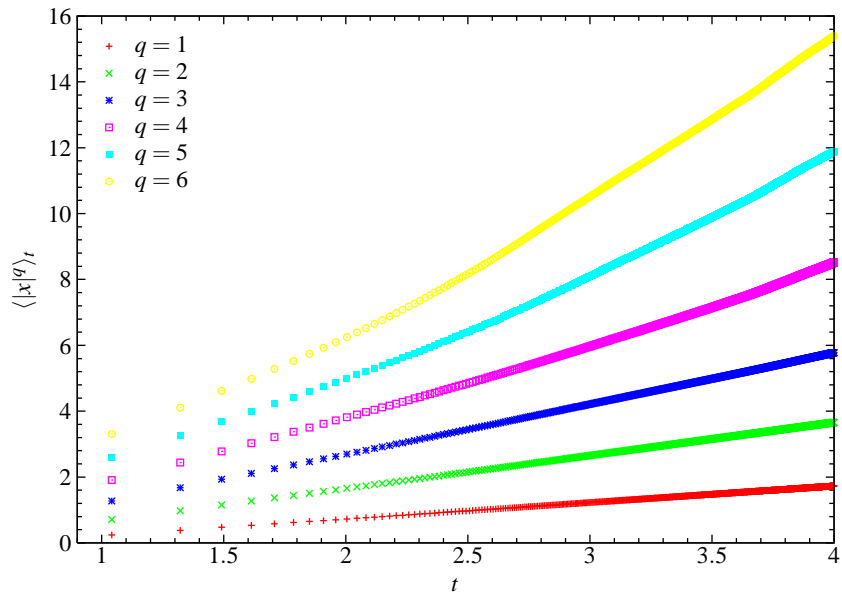

\centrefig{mine-normal-mom-fn-t.eps}
\caption{\label{fig:mine-normal-moments-as-fn-t} Moments
$\mean{\modulus{x}^q}_t$ as a function of $t$ for different values
of $q$.  Polygonal Lorentz channel with $h_1=0.1$, $h_3=0.45$,
$\Delta h=0.1$, $w=0.2$.}
\end{figure}

Following \cite{ArmsteadOtt}, we consider  the moments
$\mean{\modulus{x}^q}_t$. Several of these moments are shown as a
function of time in \figref{fig:mine-normal-moments-as-fn-t} for
the polygonal Lorentz channel with values of the geometrical
parameters for which we expect normal diffusion, i.e.\ without
parallel scatterers.

\begin{figure}
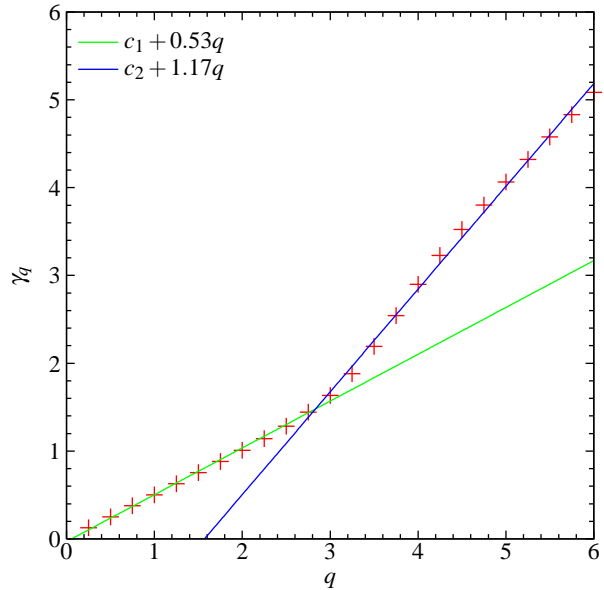

\centrefig{mine-normal-moments.eps}
\caption{\label{fig:mine-normal-moments} Growth rate  $\gamma_q$
of the $q$th moment $\mean{\modulus{x}^q}$ for the polygonal
Lorentz channel with $h_1=0.1$, $h_3=0.45$, $\Delta h=0.1$,
$w=0.2$.  The straight lines are best fits to the low- and
high-order moments.}
\end{figure}

Since $\modulus{x}^q<t^q$, we expect these moments to have power
law growth, possibly with e.g.\ logarithmic corrections.
 We thus define  the growth rates $\gamma_q$ by \cite{ArmsteadOtt}
\begin{equation}
  \gamma_q \defeq \lim_{t\to\infty} \frac{\log \mean{\modulus{x}^q}_t}{\log t}.
\end{equation}
Corrections to power law growth are ignored by this  definition.
Numerically we calculate $\gamma_q$ by fitting a straight line to
a $\log$--$\log$ plot of the $q$th moment in the long-time regime.
The growth rates are shown in \figref{fig:mine-normal-moments} for
the polygonal Lorentz channel in the  normal diffusion case. We
see a qualitative change at around $q=3$ between two approximately
linear regimes. Results of this type have been reported in a
variety of models in \cite{CastiglioneStrongAnomDiffn,
FerrariStrongSelfSimDiffn, ArtusoPerOrbTheoryStrongAnomTransp} and
are sometimes referred to as `phase transitions'.  Physically this
corresponds to a change in importance between ballistic
trajectories and diffusive trajectories
\cite{FerrariStrongSelfSimDiffn}.

We also remark that it was proved in \cite{ArmsteadOtt}
that $\gamma_q$ is a \defn{convex}
 function of $q$, assuming that the limit defining $\gamma_q$
 exists.  This considerably restricts the possible behaviour.
 Again the data points on the graph violate this and so cannot be
 correct; nonetheless we believe that the data gives an impression
 of the true behaviour.
In particular, the numerical accuracy of the high-order moments is
not good, since they become very large: the rate of increase in
the high-order regime cannot in fact exceed $1$, contradicting the
best-fit line\footnote{\revision{Higher moments are dominated by extreme
values, which must be eliminated to obtain reproducible results
\cite{SandersLarraldeNormalAnomalousDiffusionPolygonalPRE2006}.}}.

\subsection{Velocity autocorrelation function}

\begin{figure}
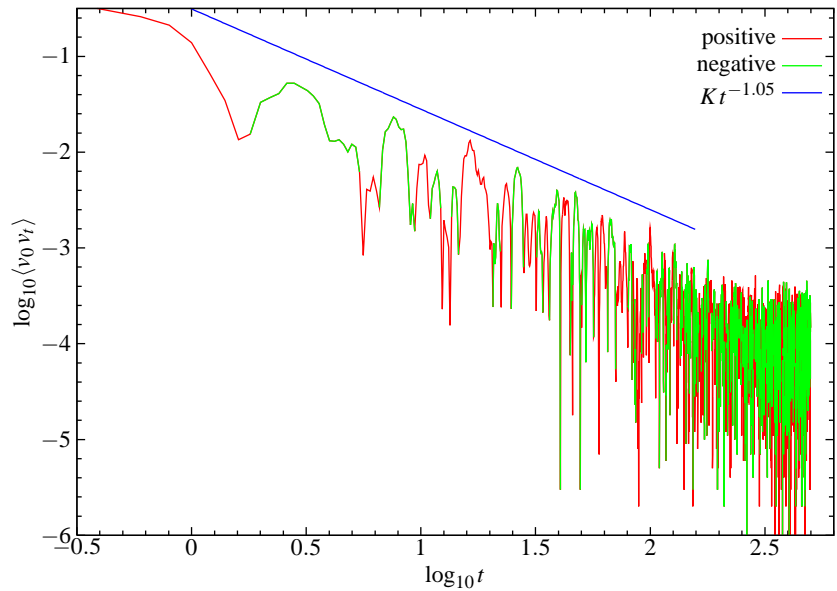

\centrefig{mine-vacf.eps} \caption{\label{fig:mine-vacf} Log--log
plot of  velocity autocorrelation function $\mean{v_0 \, v_t}$ as
a function of time.  The function oscillates about $0$, so
different colours are used to depict the positive and negative
parts.  A straight line with slope $-1.05$ is shown.}
\end{figure}

The above results provide some evidence of normal  diffusion: we
have $\mean{x^2}_t \sim t^{1.01}$, so that the growth of the
variance of the position distribution (and hence of the mean
squared displacement) is asymptotically linear.  We can attempt to
confirm this by studying the velocity autocorrelation function
$\mean{v_0 \, v_t}$, which must be integrable for the diffusion
coefficient to exist. As seen in \figref{fig:mine-vacf}, this is
an oscillatory function whose maxima seem to decay approximately
as $t^{-1.05}$. (Statistical errors (not shown) dominate for large
time.)  Since there is also cancellation due to the oscillation,
this gives weak evidence that the velocity autocorrelation
function is integrable.

\begin{figure}
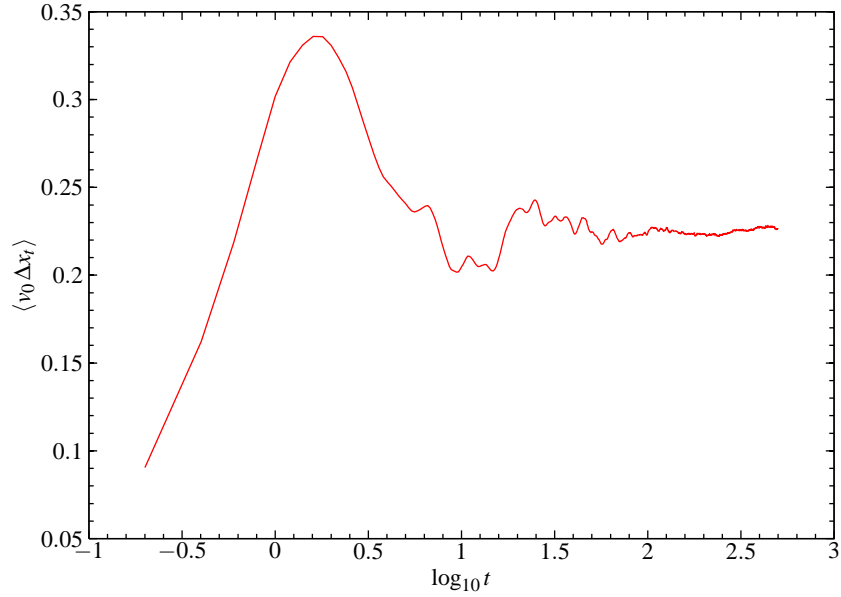

\centrefig{mine-intvacf.eps} \caption{\label{fig:mine-intvacf}
Integral $\mean{v_0 \,\Delta x_t}$  of the velocity
autocorrelation function as a function of $\log_{10} t$.}
\end{figure}

It was suggested in \cite{LoweMasters} to consider instead  the
integrated velocity autocorrelation function
\begin{equation}
  R(t) \defeq \int_0^t \mean{v_0 \, v_s} \rd s =
  \mean{v_0 \int_0^t v_s \rd s} = \mean{v_0 \, \Delta x_t},
\end{equation}
since the delicate cancellations in $C(t)$ will be seen  more
robustly in $R(t)$. If $C(t)$ is integrable then $R(t) \to 2D$ as
$t\to\infty$.  If $C(t)$ decays as, for example, $C(t) \sim
t^{-1}$ then $R(t)$ diverges logarithmically, so that plotting
$R(t)$ against $\log t$ should give a linear asymptotic growth.
\bfigref{fig:mine-intvacf} provides some evidence that $C(t)$ is
integrable, since an asymptotic limit seems to be attained as $t$
increases. Thus overall we have good evidence that the mean
squared displacement can grow asymptotically linearly, although we
have not ruled out a small logarithmic correction\footnote{\revision{This point
is again addressed in
\cite{SandersLarraldeNormalAnomalousDiffusionPolygonalPRE2006}.}}.

\subsection{Fine structure} \label{subsec:poly-fine-structure}

The shape of the displacement density was considered in
\cite{AlonsoPolyg} using histograms, but the results were not
conclusive. Here we use the more refined methods developed in
\chapref{chap:fine-structure} to describe the fine structure and
to show that the position and displacement distributions do seem
to be asymptotically normal.

\begin{figure}[p]
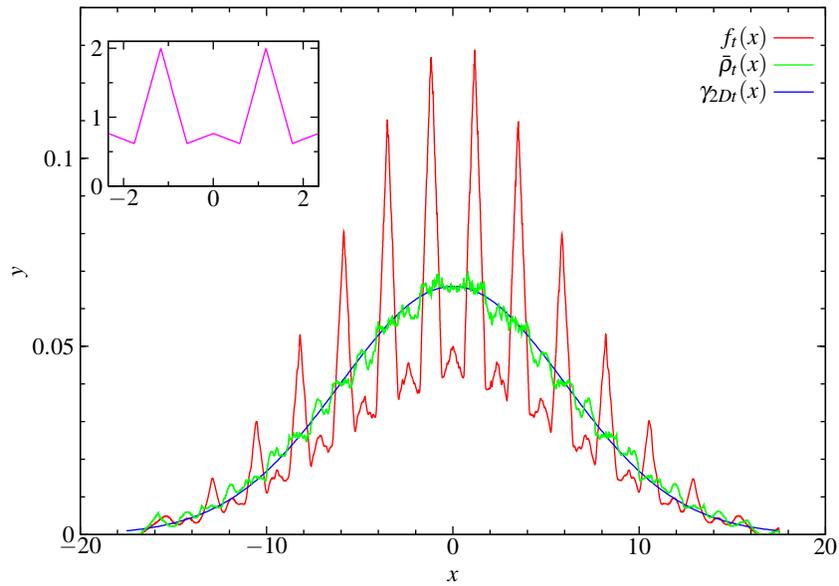

\centrefig{phi2_posn.eps} \caption{\label{fig:poly-posn-density}
Position density at $t=50$ in the polygonal model with $\phi_2 =
\pi/(2e)$.  The inset shows the function $h(x)$ over two periods.}
\end{figure}

\bfigref{fig:poly-posn-density} shows the position density
$f_t(x)$ in the Alonso model at a particular time. Following the
method of \secref{sec:fine-structure-posn-density}, we can
calculate the fine structure function $h(x)$ as the normalized
height of available space at position $x$. Taking the origin in
the centre of the unit cell in \figref{fig:poly-bill-geom}, we
have
\begin{equation}\label{}
h(x) = \frac{2d}{\modulus{Q}} \left( x \, \tan \phi_1 + \modulus{x
- \textstyle\frac{1}{2} d} \, \tan \phi_2 \right)
\end{equation}
for $0 \le x \le d$, with $h$ being an even function and having
period $2d$, and $\modulus{Q}=d\,h$. (The factor of $2d$ makes $h$
a density per unit length.) This fine structure function is shown
in the inset of \figref{fig:poly-posn-density}. We
 demodulate $f_t$ by dividing by $h$ to give $\rbar_t(x)\defeq f_t(x)/h(x)$, which
is also shown in the figure.  We see that it is close to the
Gaussian with variance $2Dt$.

With the same notation as in \secref{subsec:x-disp-density-torus},
 we can also calculate the fine
structure function $\phi$ of the displacement density for the
polygonal channel.
 The Fourier coefficients are $\hh{0} = 1$ and
\begin{equation}\label{}
\hh{k} = \frac{1}{2d} \int_{-d}^d h(x) \, \cos\left( \frac{\pi \,
k \, x}{d} \right) = \frac{1}{\modulus{Q}} \frac{d^2}{\pi^2 k^2}
\, l(k)
\end{equation}
for $k \neq 0$, where for $m\in\Z$ we have
\begin{equation}\label{}
l(k) = \begin{cases} 4 \tan(\phi_1), &
\text{if $k$ odd} \\
8 \tan (\phi_2), & \text{if }k = 4m + 2 \\
0, & \text{if }k = 4m.
\end{cases}
\end{equation}

For the polygonal Lorentz channel, we have for $x \in [0,d]$ that
\begin{equation}
h(x) = \frac{2d}{\modulus{Q}}
\left[ 2h_1 + h_2 + h_3 - \frac{h_2+h_3}{d} \, x - (w-x) \indic{[0,w]}(x)
\right],
\end{equation}
where
\begin{equation}
  \modulus{Q} = 2\left[ 2h_1\,d + \texthalf d(h_2+h_3) - w^2 \right].
\end{equation}
This gives
\begin{equation}
  \hh{k} = \frac{2}{\modulus{Q}} \frac{d^2}{\pi^2 k^2}
\left[\frac{h_2+h_3}{d} (1 - \cos \pi k) - \left(1 - \cos \frac{\pi k w}{d}\right)\right].
\end{equation}
In both cases we have $\hh{k} = \bigO{k^{-2}}$  and hence $\fh{k}
= \bigO{k^{-4}}$, so that $\phi$ is at least $C^2$, whereas $h$ is
Lipschitz continuous (i.e.\ H\"older with exponent $\alpha=1$).

\subsection{Central limit theorem} \label{poly-clt}

\begin{figure}[p]
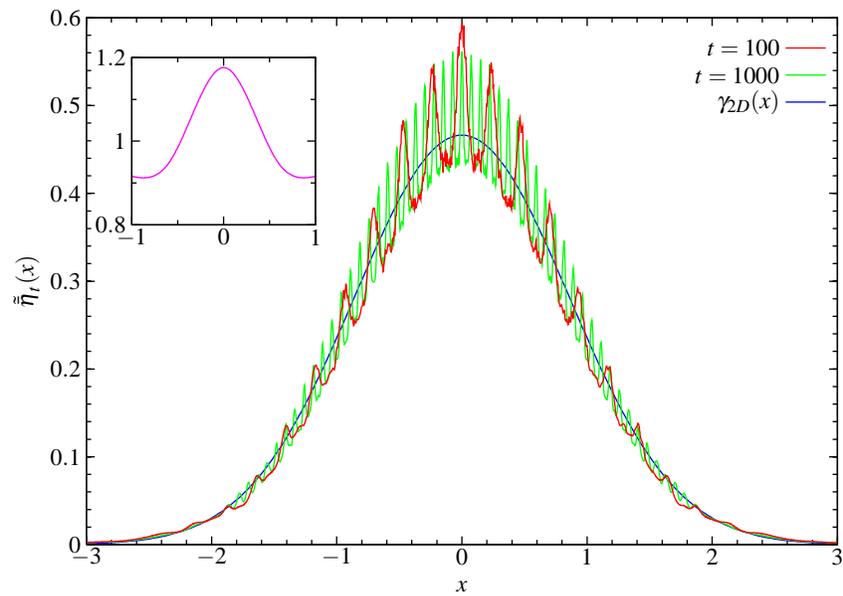

\centrefig{phi2-demod} 
\caption{
Rescaled displacement denstities compared to the Gaussian with
variance $2D$.  The inset shows the function $\phi$ for this
geometry.
\label{fig:phi2-demod}}
\end{figure}

\begin{figure}[p]
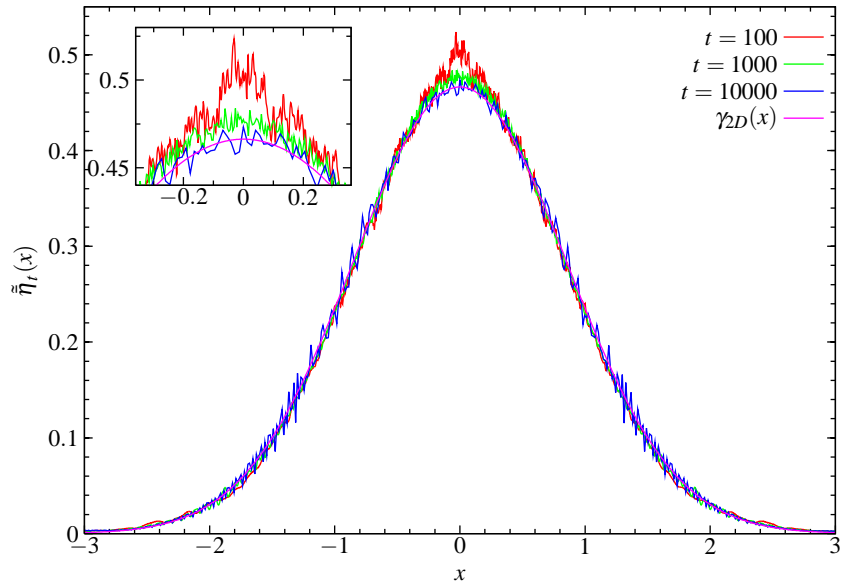

\centrefig{phi2-demod-rescaled}
\caption{\label{fig:poly-demod-rescaled} Demodulated densities
$\tilde{\etabar}_t$ for $t=100$, $t=1000$ and $t=10000$, compared
to a Gaussian with variance $2D$.  The inset shows a detailed view
of the peak near $x=0$.}
\end{figure}

As for the Lorentz gas, we rescale the densities and distribution
functions by $\sqrt{t}$ to study the convergence to a possible
limiting distribution. Again we find oscillation on a finer and
finer scale and weak convergence to a normal distribution: see
\figref{fig:phi2-demod}. \bfigref{fig:poly-demod-rescaled} shows
the time evolution of the demodulated densities
$\tilde{\etabar}_t$. There is an unexpected peak in the densities
near $x=0$ for small times, indicating some kind of trapping
effect; this appears to relax in the long time limit to a Gaussian
density.  Again we conjecture that we have uniform convergence of
these demodulated densities to a Gaussian density.

\begin{figure}[p]
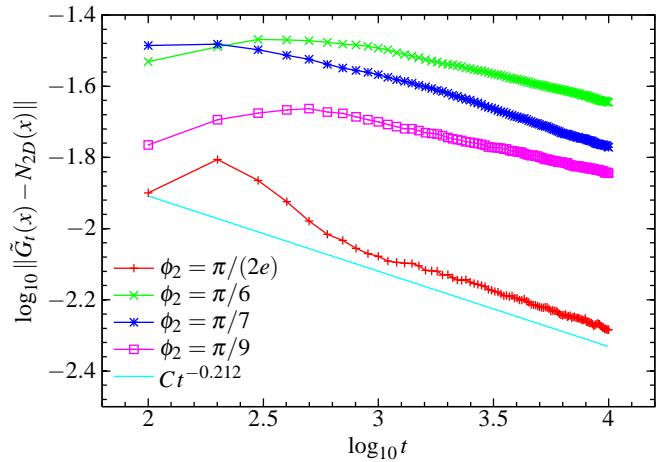

\centrefig{poly_clt} 
\caption{Distance of
the rescaled distribution functions from the limiting normal
distribution for the polygonal model with different values of
$\phi_2$. The straight line is a fit to the large-time decay of
the irrational case $\phi_2 = \pi/(2e)$.
\label{fig:poly-clt}}
\end{figure}

\bfigref{fig:poly-clt} shows the distance of the rescaled
distribution functions from the limiting normal distribution,
 analogously to \figref{fig:rate-conv}, for several values of $\phi_2$ for
which the mean squared displacement is asymptotically linear. The
straight line fitted to the graph for $\phi_2 = \pi/(2e)$ has
slope $-0.212$, so that the rate of convergence for this polygonal
model is substantially slower than that for the Lorentz gas,
presumably due to the much slower rate of mixing in this system. A
similar rate of decay is found for $\phi_2 = \pi/7$, while
$\phi_2 = 6$ and $\phi_2=9$ appear to have a slower decay rate.
Nonetheless, the distance does appear to converge to $0$ for all
these values of $\phi_2$, providing evidence that the
distributions are indeed asymptotically normal, i.e.\ that the
central limit theorem is indeed satisfied.  A similar calculation
for the polygonal Lorentz channel (for the parameters used above)
gives a convergence rate of $t^{-0.15}$,  of the same magnitude as
 for the Alonso model.

\section{Anomalous diffusion}
\label{sec:poly-anom-diffn}

In dispersing Lorentz gases, there is a weak form of anomalous
diffusion, with the mean squared displacement growing as $\msd_t
\sim t \log t$ for large $t$: see \chapref{chap:3dmodel}.  In
polygonal billiards of many types, however, a strong type of
anomalous diffusion has been observed, with  $\msd_t \sim
t^{\alpha}$ for some $\alpha
> 1$. Note that the  maximum possible rate  is $\alpha=2$, since
$\modulus{x}\le t$ due to the finite particle speed.

As stated above, we find anomalous diffusion in our models
precisely when there are \emph{parallel scatterers} in the unit
cell.  Previous observations of anomalous diffusion in billiard
models include \cite{ZaslavskyEdelmanMaxwell, ZaslavskyReview,
LiFiniteThermCond}, and \cite{ProsenAnomDiffn} in discrete time,
but they do not seem to have explicitly related the occurrence of
anomalous diffusion to the geometry of the system. We remark that
in \cite{ZwanzigCTRW} a model was presented for which the unit
cell is a rectangle with a small window in one edge.  It was
stated that the growth exponent of the mean squared displacement
depends on the number-theoretic properties of the aspect ratio of
the rectangle. It would be interesting to revisit this model in
the light of our results.

\subsection{Moments}

\begin{figure}
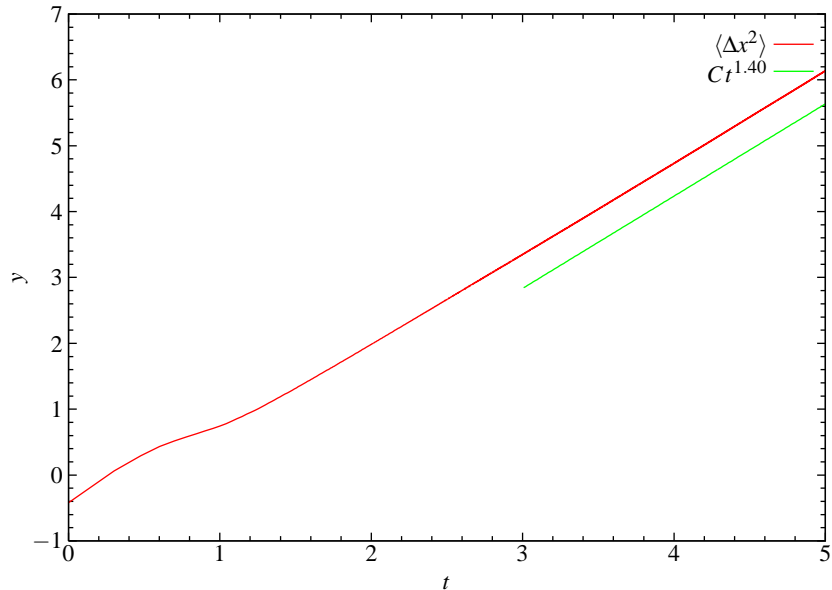

\centrefig{minee-anom-msd.eps} \caption{\label{fig:anom-msd} mean
squared displacement as a  function of time $t$, on a log--log
plot. $h_1=0.1$, $h_2=h_3=0.3$, $w=0.2$, $d=1$, for the polygonal
Lorentz channel.  The straight line is a fit to the long-time
slope.}
\end{figure}

\bfigref{fig:anom-msd} shows a typical plot of the mean squared
displacement when $h_2=h_3$.  The long-time behaviour is  well
described by
\begin{equation}
\mean{\Dx^2}_t \sim C \, t^\alpha,
\end{equation}
where $\alpha = 1.40$.

\begin{figure}
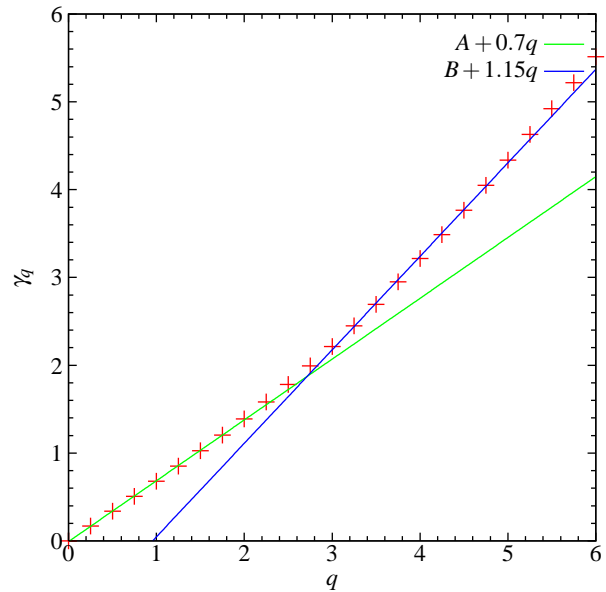

\centrefig{minee-anom-moments.eps}
\caption{\label{fig:minee-anom-moments}  Growth exponent of higher
order moments for the polygonal Lorentz channel with anomalous
diffusion. There is again a crossover between two different linear
regimes occurring close to $q=3$. }
\end{figure}

\begin{figure}
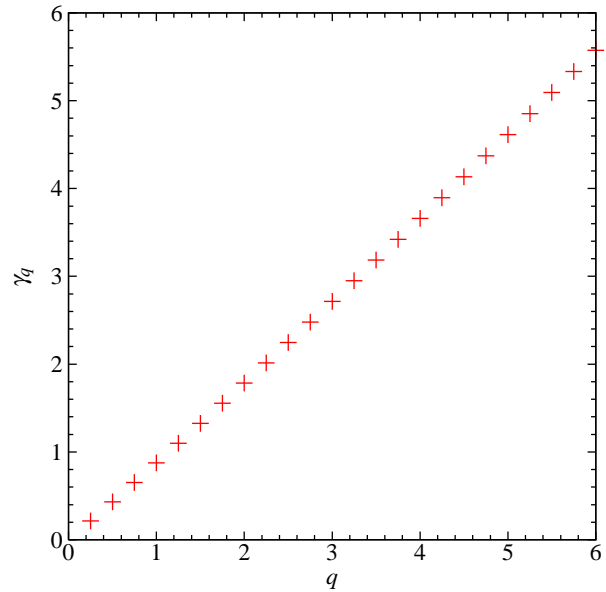

\centrefig{zigzag-anom-moments.eps}
\caption{\label{fig:zigzag-anom-moments}  Growth exponents
$\gamma_q$ of higher  order moments for zigzag channel.  The
growth exponent is close to a linear function of $q$, although
perhaps two different linear regimes can be detected for low and
high values of $q$.}
\end{figure}

\bfigref{fig:minee-anom-moments} shows the growth exponent of  the
higher order moments $\mean{\modulus{x}^q}_t$ for the polygonal
Lorentz channel.  There are two different linear regimes, as
previously, with a crossover occurring near $q=3$.
\bfigref{fig:zigzag-anom-moments} shows the  same for the zigzag
model, for which there does not seem to be a crossover, and we
have \defn{strong anomalous diffusion} under the definition of
\cite{CastiglioneStrongAnomDiffn}, where $\gamma_q = \nu \, q$ for
all $q$, with the constant $\nu$  given by $\nu \simeq 0.93$ for
these parameters.

\subsection{1D densities}

Since the mean squared displacement grows  like $\mean{\Dx^2}_t
\sim t^\alpha$ with $\alpha > 1$, the $\sqrt{t}$ scaling used in
the central limit theorem cannot give densities which converge in
any sense, since the variance of the rescaled densities would
still tend to infinity.  Instead we rescale to keep the variance
bounded, setting
\begin{equation}
\tg_t(x) \defeq t^{\alpha/2} \, g_t(x \, t^{\alpha/2}).
\end{equation}
This scaling was previously used in \cite{ProsenAnomDiffn} for the
Poincar\'e map of an anomalously diffusive billiard.

\begin{figure}
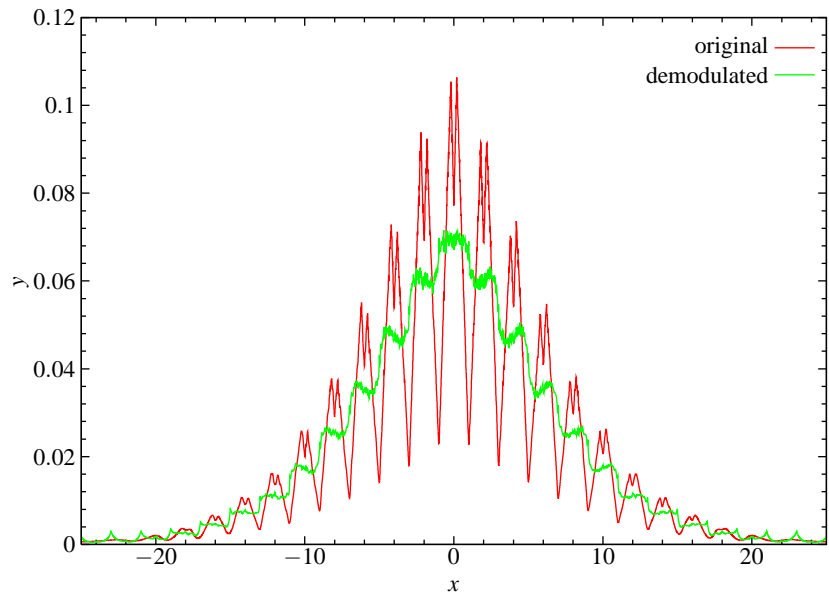

\centrefig{anom-demod.eps} \caption{\label{fig:anom-demod}
Position density and demodulated position density in the polygonal
Lorentz model with  $h_1=0.1$, $h_2=h_3=0.45$, $w=0.2$, $d=1$ and
$t=100$.}
\end{figure}

\bfigref{fig:anom-demod} shows the $x$-position density.   The
fine structure is again mostly removed by demodulating by the fine
structure function $h$ found above, although this demodulation
does not seem to be as effective as previously; this could be an
artifact of an insufficiently precise estimation of the original
density, or it could reflect weaker mixing properties of the
system.

\begin{figure}
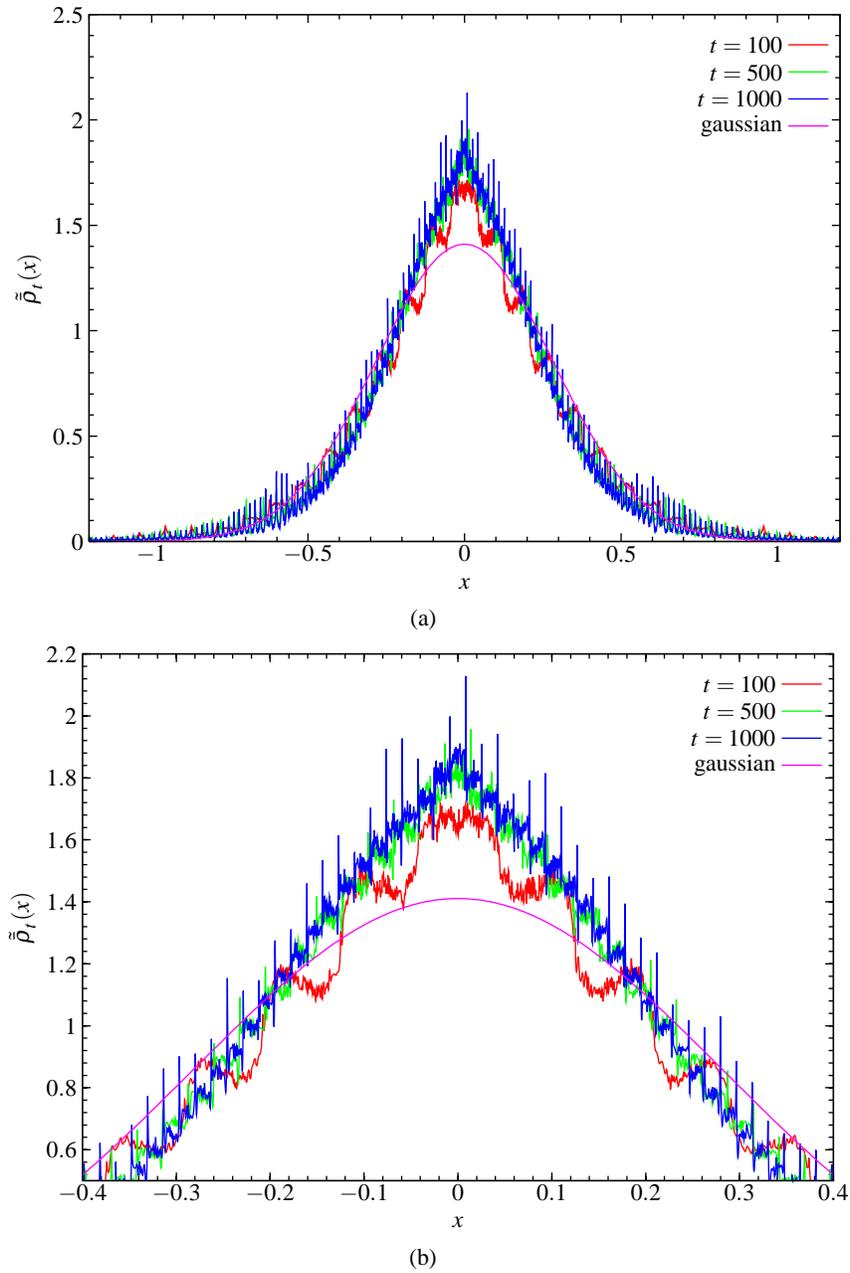

\centering \subfig{anom-rescaled.eps}
\subfig{anom-rescaled-zoom.eps} \caption{\label{fig:anom-rescaled}
(a) Rescaled demodulated densities in the polygonal Lorentz model
for several times, compared to a Gaussian with the same variance;
the central region is shown in more detail in (b). There seems to
be convergence at longer times to a limiting shape which is
non-Gaussian.}
\end{figure}

\bfigref{fig:anom-rescaled} shows a sequence of  demodulated
densities rescaled as above.  They appear to converge at long
times to a limiting shape which is non-Gaussian.  They are
compared to a Gaussian with variance $B$, where $B$ is the
generalisation of the diffusion coefficient given by
$\mean{\Dx^2}_t \sim B t^{\alpha}$.  The demodulated densities
are, however, rather noisy; those for the displacement
distribution are even more so.

\paragraph{Zigzag model}

Rescaled densities for the zigzag model are shown in
\figref{fig:zigzag-anom-pdfs};   for long time we again have
convergence to a limiting distribution.  The spikes in the tails
seem to correspond to long-lived propagating orbits. Note that
there is no fine structure to take into account for the zigzag
channel, since the channel width is a constant within the unit
cell.  The peak at $x=0$ has been observed previously in anomalous
diffusion: see e.g.\ \cite{FerrariStrongSelfSimDiffn}.
\begin{figure}
\centrefig{zigzag-anom.eps} \caption{\label{fig:zigzag-anom-pdfs}
Rescaled densities for the zigzag channel, where $\mean{\Dx^2}_t
\sim t^{1.81}$.}
\end{figure}

We remark that the unit cell of the zigzag model when the top and
bottom scatterers are parallel can be reduced to a parallelogram
with irrational angles.  The mixing properties of the billiard
flow inside such a geometry are not obvious, although we have
checked that velocity autocorrelations decay as a power law. There
may be an effect of \defn{slow ergodicity}, where the system takes
a long time to explore certain regions of phase space: see e.g.\
\cite{KaplanWeakQuantErg,ProsenAnomDiffn}.


\subsection{Maxwellian velocity distribution}

Just as for the Lorentz gas, we can consider the effect of a
different distribution of velocities; this has been considered for
heat conduction \cite{LiHeatLinearMixing, AlonsoPolyg}.

Recall from \secref{sec:maxwell-vel-distn} that if there is a
speed distribution $p_V(v)$, then the mean squared displacement at
time $t$ in 2D is given by
\begin{equation}\label{}
\mean{\Delta \x^2}= \int_v p_V(v) \, \mean{r^2(t)}_v.
\end{equation}
If a billiard system shows anomalous diffusion for particles with
unit speed $v_0=1$ which looks like
\begin{equation}\label{}
\mean{r^2(t)}_{v_0} \sim t^{\alpha},
\end{equation}
then
\begin{equation}\label{}
\mean{r^2(t)}_v = \mean{r^2(tv/v_0)}_{v_0} = (tv)^{\alpha}.
\end{equation}
Thus
\begin{equation}\label{}
\mean{\Delta \x^2} = t^{\alpha} \int_v p_V(v) \,v^{\alpha} \rd v.
\end{equation}
Hence there is now a correction factor of the $\alpha$th moment of
the speed distribution which enters, although the rate of growth
is still $t^{\alpha}$. We have observed such a correction
numerically.

\section{Continuous-time random walk model for anomalous diffusion}
\label{sec:CTRW-model}

A widely-used framework for understanding anomalous diffusion
processes is the theory of \defn{continuous-time random walks}
(CTRW): see e.g.\ \cite{WeissRandomWalk,
KlafterStochasticAnomDiffn, ZumofenKlafter}.   This is a
generalisation of standard discrete-time random walks where
individual steps are still independent, but are now described by a
density function $\psi(\r,t)$, where $\psi(\r,t) \d \r \d t$ is
the probability of having a step with distance in $(\r,\r+\rd \r)$
which takes time in $(t,t+\rd t)$.

To be able to model super-diffusion, $\psi$ must have a coupled
form \cite{ZumofenKlafter}, such as
\begin{equation}\label{fig:ctrw-delta-version}
  \psi(\r,t) = \texthalf \delta(\modulus{\r}-t) \, \psi(t).
\end{equation}
This form (called the \defn{velocity model} in
\cite{ZumofenKlafter}) models motion at a constant velocity for a
time $t$ in the direction $\r$; after each stretch the direction
is randomised and a new step is taken. It gives the following
long-time growth of the mean squared displacement $\sigma^2(t)$:
\begin{equation}\label{eq:ctrw-growth-msd}
\sigma^2(t) \sim \begin{cases}
t^2, & 0<\nu \le 1\\
t^{3-\nu}, & 1<\nu<2\\
t \ln t, &  \nu=2\\
t, &  \nu>2;
\end{cases}
\end{equation}
when $\psi(t) \sim t^{-1-\nu}$; see
\cite{GeiselReview,KlafterZumShlesReview} for reviews. More
generally we have
\begin{equation}\label{eq:ctrw-density-cond}
  \psi(\r,t) = p(\r \, | \, t) \, \psi(t),
\end{equation}
where $p(\r \, | \, t)$ is the conditional probability of a step
of length $\r$ if we know that it lasts for time $t$: see e.g.\
\cite[Sec.~2.5]{WeissRandomWalk} and \cite{ShlesingerKlafter1989}.
In general the rate of growth of the mean squared displacement
will then also depend on the properties of $p(\r \, | \, t)$.

\begin{sidewaysfigure}
\centering

\includegraphics{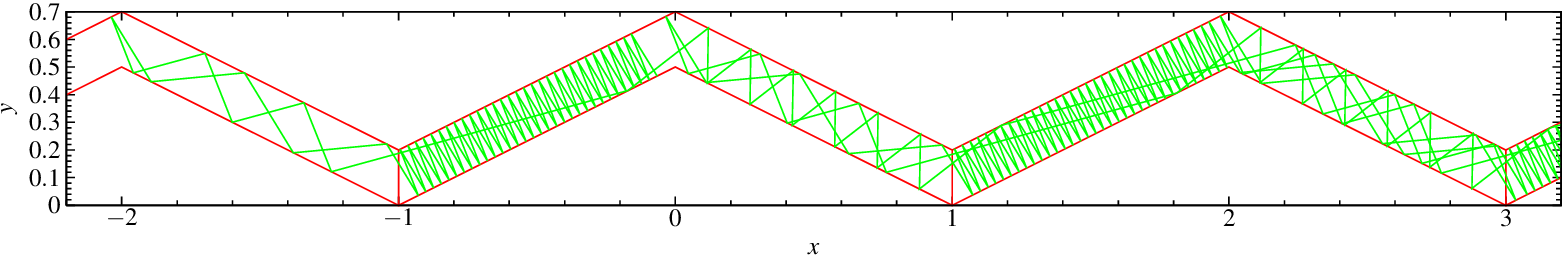}

\vspace*{40pt}

\includegraphics{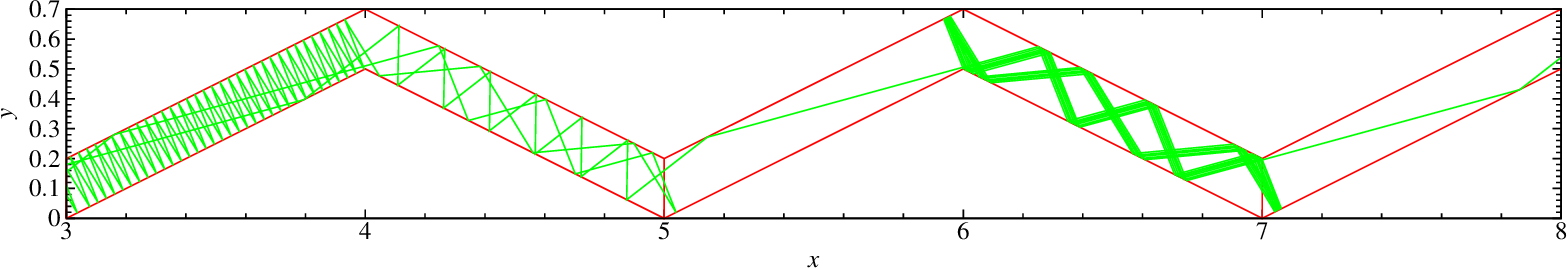}

\caption{\label{fig:zigzag-traj} Part of a representative single
trajectory in the zigzag model with anomalous diffusion for
$h_1=0.1$, $h_2=h_3=0.3$.  The upper figure shows the start of the
trajectory, which has initial condition $(-0.066, 0.47)$; the
lower figure shows the continuation of the trajectory.}
\end{sidewaysfigure}

\bfigref{fig:zigzag-traj} shows a representative trajectory in the
zigzag model with anomalous diffusion.  From the figure we can see
the importance of
\defn{laminar} motion, i.e.\ coherent motion in one direction
along the channel.  Referring to the above CTRW picture, we try to
model the motion using  random walks where the steps are stretches
of laminar motion.

\begin{figure}
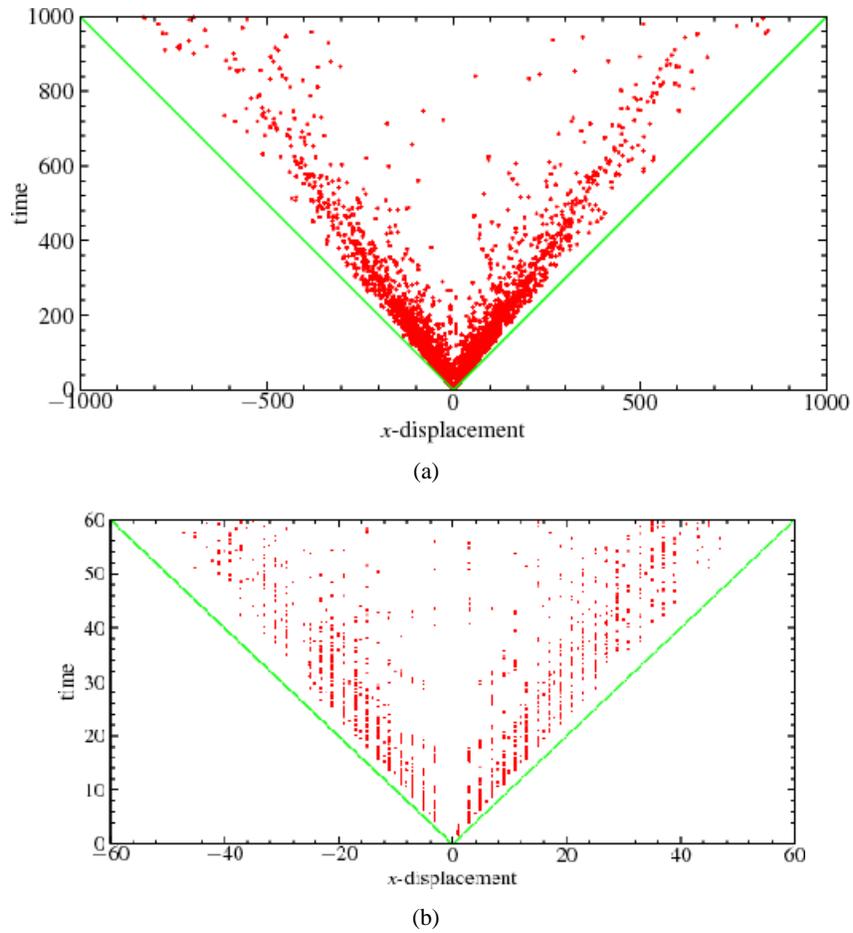

\centering \subfig{zigzag-laminar.eps}

\subfig{zigzag-laminar-zoom.eps}
\caption{\label{fig:zigzag-laminar-joint} Scatterplot representing
the joint density $\psi(\r,t)$ of laminar paths in the zigzag
model with anomalous diffusion ($h_1=0.1$, $h_2=h_3=0.3$). The
straight lines have slope $1$, representing the maximum possible
speed. (b) Shows the fine structure near the origin: the allowed
values of $x$ are restricted to be near integers.}
\end{figure}

We define laminar motion in the zigzag model as follows.  For each
section (half a unit cell) $i$ of the channel, we assign a vector
$\vect{L}_i$ parallel to the channel walls, with a consistent
$x$-direction along the channel.  We say that two consecutive
stretches of trajectory (between bounces) are part of the same
laminar motion if they have the same directions along the channel,
as measured by $\sgn(\v \cdot \vect{L}_i)$.
\bfigref{fig:zigzag-laminar-joint} shows a scatterplot
representing the joint distribution $\psi(x,t)$ of periods of
laminar motion with $x$-displacement $x$ along the channel and
taking time $t$.  Part (b) of that figure shows that allowed
values of $x$ are restricted to being near integers, since a
laminar motion cannot end in the middle of a unit cell.  (A
similar effect in infinite horizon 2D periodic Lorentz gases was
described in \cite{Bleher}, although no figure was given.)  We can
thus regard the walk as taking place on a one-dimensional lattice.

We see that the distribution $\psi(x,t)$ is non-trivial, and the
figure does not give much hope of finding an analytical expression
for it.  However we could hope that an approximation of the
 form $\psi(\r,t) = \texthalf \delta(\modulus{\r}-vt) \,
\psi(t)$ referred to above may be adequate, for some speed $v$,
since the distribution is somewhat concentrated along diagonal
lines. \bfigref{fig:zigzag-laminar-t-distn} shows the tail region
of the density $\psi(t)$ and its distribution function.  If we
assume the velocity model form, then the observed long-time power
law decay $\psi(t) \sim t^{-2.58}$, shown in
\figref{fig:zigzag-laminar-t-distn}, gives $\nu=1.58$ and hence a
mean squared displacement $\sigma^2(t) \sim t^{3-1.58}=t^{1.42}$.
This compares with the observed growth $\sigma^2(t) \sim
t^{1.81}$. The agreement leaves room for improvement, showing that
we must use a more general CTRW model incorporating information on
the complete $\psi(x,t)$, or perhaps reject any CTRW model which
omits important information on correlations between consecutive
laminar trajectories; such correlations can be seen in
\figref{fig:zigzag-traj}, for example.

\begin{figure}

\centrefig{zigzag-lam-pdf.eps}
\caption{\label{fig:zigzag-laminar-t-distn} (a) Density $\psi(t)$
of  times of laminar motion for the same model as in
\figref{fig:zigzag-laminar-joint}.}
\end{figure}

\begin{figure}
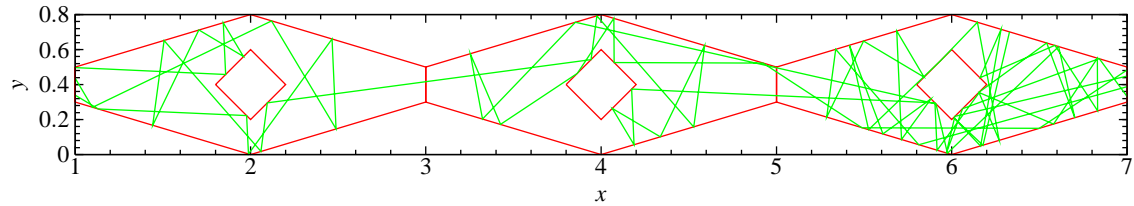

\centrefig{mine-traj.eps} \caption{\label{fig:mine-traj} Part of a
single trajectory of the polygonal Lorentz model with $h_1=0.1$,
$h_2=h_3=0.3$.}
\end{figure}

\bfigref{fig:mine-traj} shows a trajectory in the polygonal
Lorentz gas model.  Despite the fact that anomalous diffusion is
also found in this model when there are parallel scatterers, the
mechanism is much less clear, since it does not seem possible to
identify an obvious candidate `laminar motion' in this model by
looking at sample trajectories.

\section{Crossover from normal to anomalous diffusion}
\label{sec:crossover-normal-anom}

Since anomalous diffusion occurs under certain geometrical
conditions, whereas normal diffusion seems to be more general, it
is of interest to ask how the crossover from one to the other
occurs when the geometry is changed to approach one where
anomalous diffusion occurs. We study the two models introduced
above to show numerically how the crossover occurs.

In the following we fix all geometrical parameters except for
$h_2$, which we vary according to $h_2 = h_3 + \Delta h$.
Anomalous diffusion is found for $\Delta h = 0$, with
asymptotically normal diffusion found for any $\Delta h \neq 0$.
It is thus of interest to study how the statistical behaviour
changes from one asymptotic regime to the other as the symmetrical
configuration is approached when $\Delta h \to 0$.

\begin{figure}
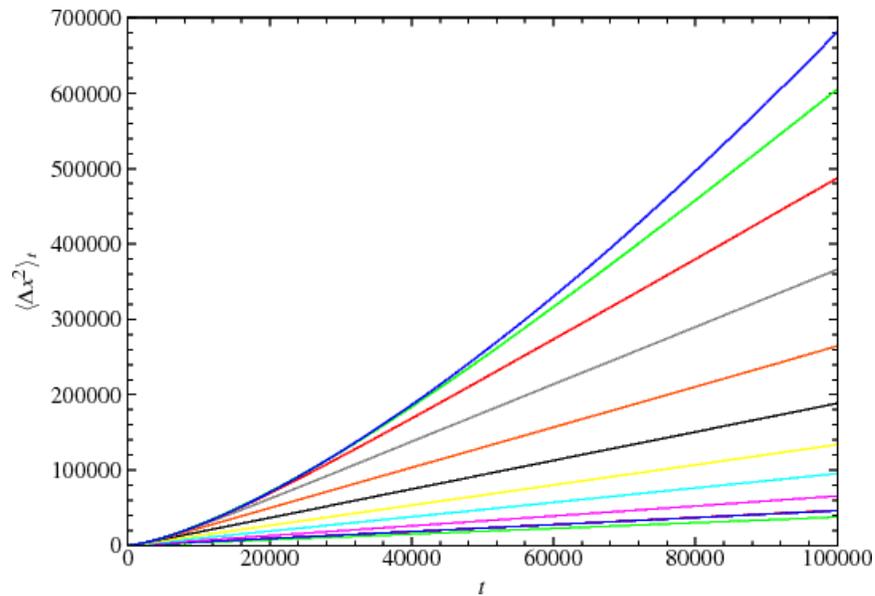

\centrefig{minee-msd.eps} \caption{\label{fig:minee-msd} Mean
squared displacement as a function of time for different values of
$\Delta h = h_2-h_3$ tending to $0$, for the polygonal Lorentz
model with $h_1=0.1$, $h_3=0.45$ and $w=0.2$.  Values of $\Delta
h$ shown are, from top to bottom, $\Delta h=0$, $10^{-11}$,
$10^{-10}$, $\ldots$, $10^{-1}$. }
\end{figure}

\begin{figure}
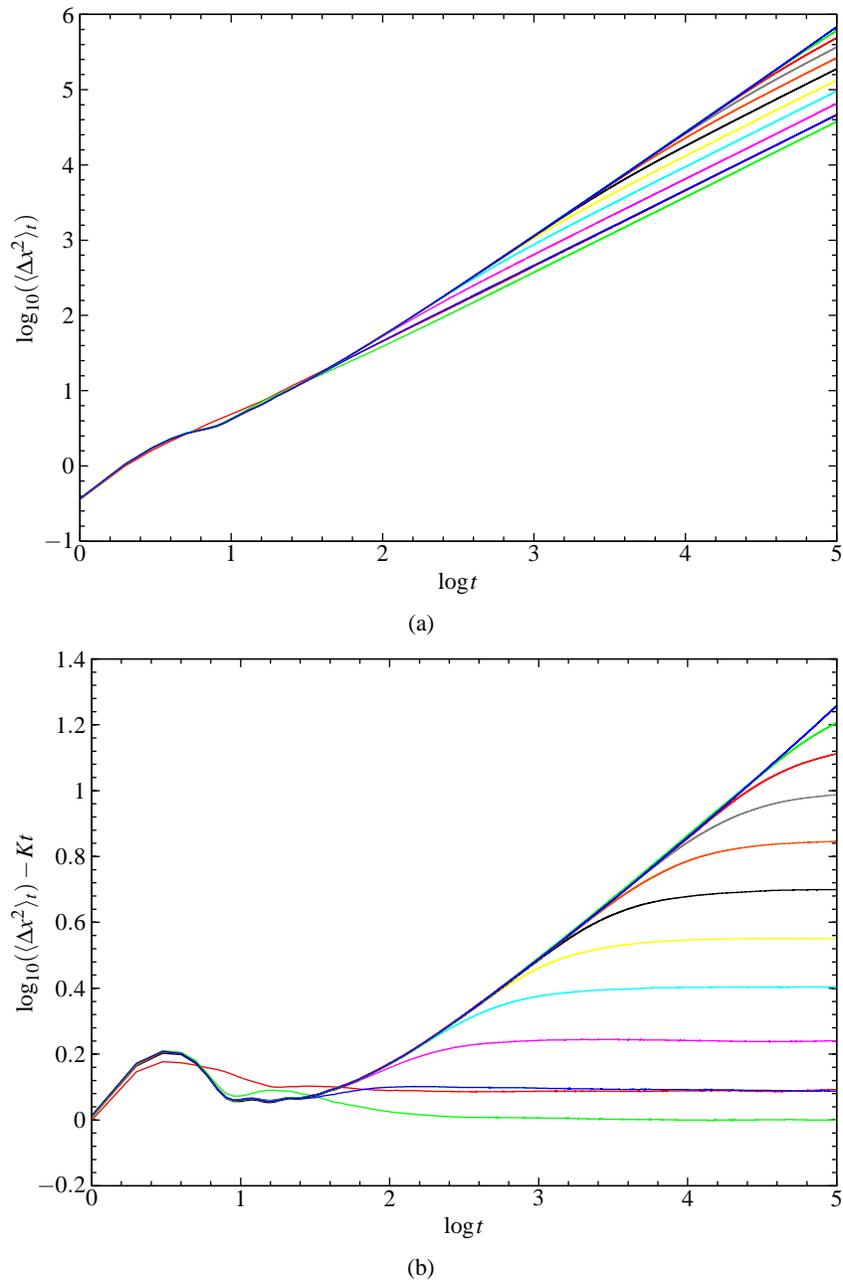

\centering
\subfig{minee-log-msd.eps} \subfig{minee-deviations.eps}
 \caption{\label{fig:minee-log-msd} (a) Log--log
plot of the mean squared displacement as a function of time. (b)
Deviations of the log--log plot from a straight line fitted to the
long-time part of the lowest curve.  As $\Delta h$ approaches $0$,
the mean squared displacement curve follows that for the anomalous
case $\Delta h=0$ for longer and longer times.  Values of $\Delta
h$ shown are as in the previous figure.}
\end{figure}

\bfigref{fig:minee-msd} shows the mean squared displacement as a
function of time for several values of $\Delta h$.  The same data
is plotted on log--log axes in \figref{fig:minee-log-msd}(a) and
as deviations from a linear fit to one of the curves in
\figref{fig:minee-log-msd}(b).  We see that as $\Delta h \to 0$,
the curves follow the anomalous diffusion curve (with slope $> 1$)
for longer times, before a crossover occurs to asymptotic linear
behaviour.

\begin{figure}
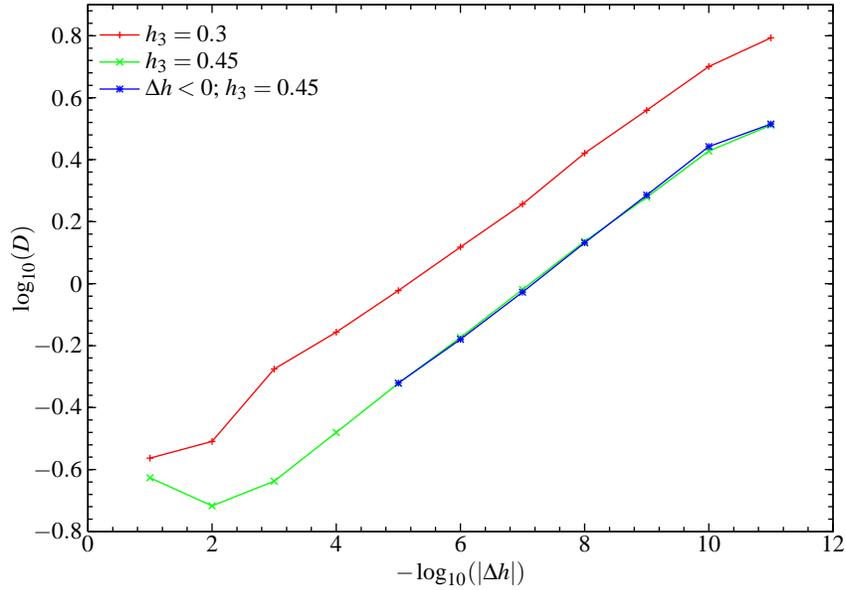

\centrefig{minee-diff-coeffs.eps}
\caption{\label{fig:minee-diff-coeffs} Log--log plot of the
diffusion coefficient $D$ as a function of $\Delta h$ for the
polygonal Lorentz model.  The linear growth indicates power law
behaviour.}
\end{figure}

The intercept of the asymptotic linear growth for $\Delta h \neq
0$ is related to the diffusion coefficient $D(\Delta h)$.  As
$\Delta h \to 0$, the intercept moves up, corresponding to an
increase in $D$.  \bfigref{fig:minee-diff-coeffs} shows the
diffusion coefficient in this asymptotic linear regime as a
function of $\Delta h$, obtained via the slope of the mean squared
displacement. Note that it is clear from
\figref{fig:minee-log-msd} that the asymptotic linear growth
regime  has not yet been reached for the smallest values of
$\Delta h$, so that the diffusion coefficients for those values
are expected to be underestimates.

 We obtain a straight line on a log--log plot, and a
fit in the region $3 \le -\log_{10}(\Delta h)$ gives
\begin{equation}\label{}
\log_{10} D(\Delta h) \sim -1.10 - 0.152 \log_{10}(\Delta h)).
\end{equation}
The diffusion coefficients obtained for negative values of $\Delta
h$ are also shown for $h=0.45$, and it is clear that the growth
rate is the same.  We hence obtain the power law behaviour
\begin{equation}\label{}
D(\Delta h) \sim 0.08 \, \modulus{\Delta h}^{-0.15} \quad \text{as
} \modulus{\Delta h} \to 0.
\end{equation}
Note that this  agrees with $D=\infty$ when $\Delta h = 0$.

We could also consider the \defn{crossover time}, i.e.\ the time
required to switch from  the anomalous diffusion regime to the
linear growth regime.  One possible definition of this time could
be  as the intersection of two straight line fits: a fit to the
initial anomalous growth and a fit to the asymptotic normal
growth. A construction of this type was used e.g.\ in
\cite{ClausRandomLorentz} in a different context.  From
\figref{fig:minee-log-msd} we see that this crossover time tends
to $\infty$ as $\Delta h \to 0$, and we could study\footnote{\revision{This 
was done in \cite{SandersLarraldeNormalAnomalousDiffusionPolygonalPRE2006},
where a simple scaling form was found allowing a data collapse of the data for
different $\Delta h$.}}
the dependence
of this time on $\Delta h$.

\subsection{Zigzag model}

\begin{figure}
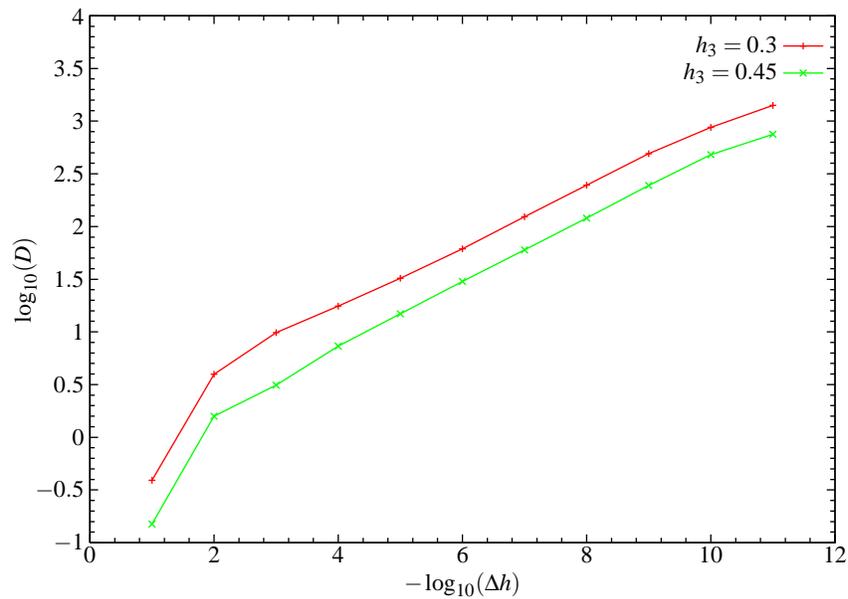

\centrefig{zigzag-diff-coeffs.eps}
\caption{\label{fig:zigzag-diff-coeffs} Log--log plot of the
diffusion coefficient as a function of $\Delta h$ for the zigzag
model.}
\end{figure}

There is a similar crossover to anomalous diffusion when $h_2=h_3$
in the zigzag model; the growth of the moments is very similar to
the polygonal Lorentz channel. \bfigref{fig:zigzag-diff-coeffs}
shows the growth of the diffusion coefficients as $\Delta h \to
0$.  A fit gives
\begin{equation}\label{}
\log_{10} D(\Delta h) = -0.320 \, \log_{10}(\Delta h) - 0.191,
\end{equation}
and hence
\begin{equation}\label{}
D(\Delta h) \sim 0.644 \, (\Delta h)^{-0.32} \quad \text{as }
\Delta h \to 0,
\end{equation}
so that we again obtain a power law, but with a different rate of
growth than in the polygonal Lorentz model.

\subsection{Qualitative explanation}

In the zigzag model we can find a qualitative explanation of the
above observations as follows. If we start a trajectory with the
same initial conditions in the cases $\Delta h=0$ and $\Delta h$
small, the latter trajectory will
\defn{shadow} (follow approximately) the former for a certain
length of time. However, the latter will gradually (linearly in
time) deviate from the first trajectory due to the (weak)
defocusing effect of the boundaries, eventually becoming
effectively decorrelated.  For smaller values of $\Delta h$, the
shadowing will persist for a longer period.

\begin{figure}
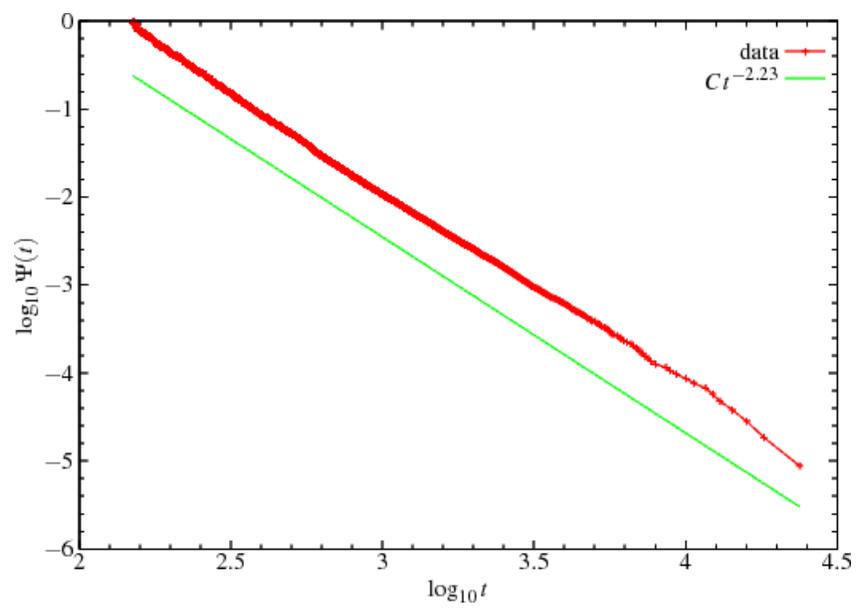

\centrefig{zigzag-lam-dh0.01-cdf.eps}
\caption{\label{fig:zig-lam-deltah-0.01-cdf} Tail of $\Psi(t)$ on
a log--log plot.}
\end{figure}

We refer again to the CTRW model described in
\secref{sec:CTRW-model}.  For $\Delta h = 0.01$, we find that the
density function $\psi(t)$ is oscillatory, so that it is difficult
to determine its decay rate; instead,
\figref{fig:zig-lam-deltah-0.01-cdf} shows $\Psi(t)$ for long
times, where
\begin{equation}\label{}
\Psi(t) \defeq \int_t^\infty \psi(t') \rd t'.
\end{equation}
It decays like $\Psi(t) \sim t^{-2.23}$, so that $\psi(t) \sim
t^{-1-2.23}$, giving $\nu=2.23$, which according to
\eqref{eq:ctrw-growth-msd} gives normal diffusion, as required.
However,  we may not  have managed to attain the asymptotic regime
in this calculation. For smaller $\Delta h$ we expect the
distribution to have a progressively longer tail, so that the
average laminar length will be longer, resulting in a larger
diffusion coefficient as seen in the numerical experiments.

Again we remark that although the same behaviour is found in the
polygonal Lorentz model at the level of the statistical
properties, it is less clear how to obtain a qualitative
understanding in that case without identifying the type of laminar
behaviour which is presumably responsible.


\startonright



\graphicspath{{figs/}}

\chapter{Three-dimensional periodic Lorentz gases} \label{chap:3dmodel}

We would like to extend the detailed knowledge available for 2D
systems to study more physically relevant 3D periodic Lorentz
gases. Relatively little work has been done on higher-dimensional
models, since computation time increases and technical
difficulties arise, some of which are discussed below. In this
chapter we succeed in addressing some of these issues and point
out where more work is still required\footnote{\revision{Some results from this
chapter are included in a submitted manuscript \cite{SandersNormalDiffusion3D},
which also contains updated
references.}}.

Previous work on 3D periodic Lorentz gases in a physical context
includes that of Bouchaud \& Le Doussal \cite{BouchaudDoussal85},
who studied the Lyapunov exponent and the decay of velocity
autocorrelation functions for simple (hyper-)cubic lattices in
$d\le 7$ dimensions, and Dettmann \etal\
\cite{DettmannMorrissConjPairing}, who studied a 3D hexagonal
close-packed lattice with an electric field and a
\defn{thermostat} (a device for keeping the kinetic energy of a particle
constant, despite the energy
input due to the electric field), showing that the Lyapunov
exponents satisfied a conjugate pairing rule.  Rigorous results,
discussed in more detail below, are proved in \cite{Chernov94} and
\cite{GolseFreePathsHigherDimensions}.

We first review results which show that higher-dimensional Lorentz
gases with finite horizon \emph{exist}.  We then construct a
particular 3D model with overlapping scatterers which has a finite
horizon regime, analogously to the model studied in
\chapref{chap:geom-dependence}; in this regime we show numerically
that our model is diffusive.  We then consider the effect on the
statistical properties of allowing an infinite horizon, and find
that there are two qualitatively different types of infinite
horizon regime.

\section{Existence of higher-dimensional Lorentz gases with finite
horizon}

Chernov \cite{Chernov94} extended the results of \cite{BS, BSC} to
higher-dimensional ($d \ge 3$) periodic Lorentz gases. He
proved\footnote{Some details of the proofs in \cite{Chernov94}
are corrected in 
\cite{BalintAnnHenriPoinc,BalintAsterisque}.} fast (at least stretched
exponential)
decay of correlations for the billiard map, and the central limit
theorem and functional central limit theorem for H\"older
continuous observables for the billiard flow, when the scatterers
are disjoint and have sufficiently smooth ($C^3$) boundaries, and
the finite horizon condition is satisfied\footnote{Recall from
\secref{sec:2d-lorentz-gas-model} that a periodic Lorentz gas
satisfies the finite horizon condition if no trajectory can be
extended arbitrarily far without hitting a scatterer.}. However,
no explicit example of a model satisfying the assumptions was
given in \cite{Chernov94}, even in the 3D case. In fact we are not
aware of \emph{any} examples of such models in the literature;
unfortunately we have also been unable to construct an explicit
example.

\subsection{Rigorous results from convex geometry}

Several results in convex geometry, which are seemingly unknown in
the physics community, bear light on the possibility of
constructing higher-dimensional periodic Lorentz gases with finite
horizon.

In \cite{Heppes} it was shown that any lattice packing of spheres
in 3D has an infinite horizon, and in fact a
\defn{cylindrical hole}, the three-dimensional version of the
corridors in infinite horizon 2D periodic Lorentz gases.  Such a
hole consists, in billiard language, of a collection of
parallel trajectories (forming a cylinder), none of which ever
collides with a scatterer.

We remark that the term `lattice' in this result refers to a set
of points of the form $\sum_{i=1}^n a_i \vect{e}_i$, where $a_i
\in \Z$; thus geometers' lattices are what physicists call
\defn{Bravais lattices} \cite{AshcroftMermin}. For example, the
hexagonal close-packed structure is not a lattice in this sense,
although the face-centred cubic is.  A `sphere packing' is
(roughly) a collection of touching, but non-overlapping, spheres.

 The result was later extended
to show the existence of cylindrical holes in lattice sphere
packings of any dimension: see references in \cite{HenkZong}. This
implies that to obtain a finite horizon it is necessary to have
more than one sphere per unit cell;  this was stated without proof
in \cite{CherDett}.

Recently it was proved in \cite{HenkZong} that this is no longer
true if we consider arbitrary periodic structures, rather than
just lattices, consisting of identical convex bodies.  They showed
that in any dimension $n$ there exist periodic arrays of
non-touching spheres with finite horizon (although they did not
use this terminology), and in fact with spheres replaced by any
convex body. The proof is constructive, but it is not easy to
convert it into the construction of an explicit
example\footnote{M. Henk, private communication.}, since it
involves finding sets of minimal cardinality satisfying certain
properties.

This result implies that the class of models considered in
\cite{Chernov94} is indeed non-empty. We remark that in any such
finite horizon model, the minimum number of spheres which can be
seen from a given sphere has been shown to be at least $30$: for
references on this and related results see the review
\cite[Sec.~5--6]{ZongReview}. Hence any such structure must be
quite complicated.

\subsection{Attempts to construct a finite-horizon model}

A common construction in solid state physics is to build lattices
as stacks of layers, with each layer being a 2D triangular lattice
of spheres (corresponding to the standard triangular 2D periodic
Lorentz gas). Consecutive layers are placed in one of three
possible  positions, labelled $A$, $B$ and $C$, relative to the
previous two layers \cite{AshcroftMermin}: a hexagonal
close-packed structure corresponds to the choice $\ldots
ABABAB\ldots$, and the face-centred cubic to $\ldots
ABCABC\ldots$; neither of these (somewhat surprisingly) has a
finite horizon. (The face-centred cubic structure is a Bravais
lattice, and so is covered by the theorem cited above, but the
hexagonal close-packed structure is not.) We could ask if it is
possible to produce a finite horizon by some other periodic
ordering of $A$s, $B$s and $C$s. However, in fact placing adjacent
layers on top of each other creates cylindrical holes
\emph{between the layers}, so that there is always an infinite
horizon.   If we also require disjoint (non-touching) scatterers,
then we would need to separate the scatterers slightly from their
touching positions, thereby introducing more holes.

\begin{figure}
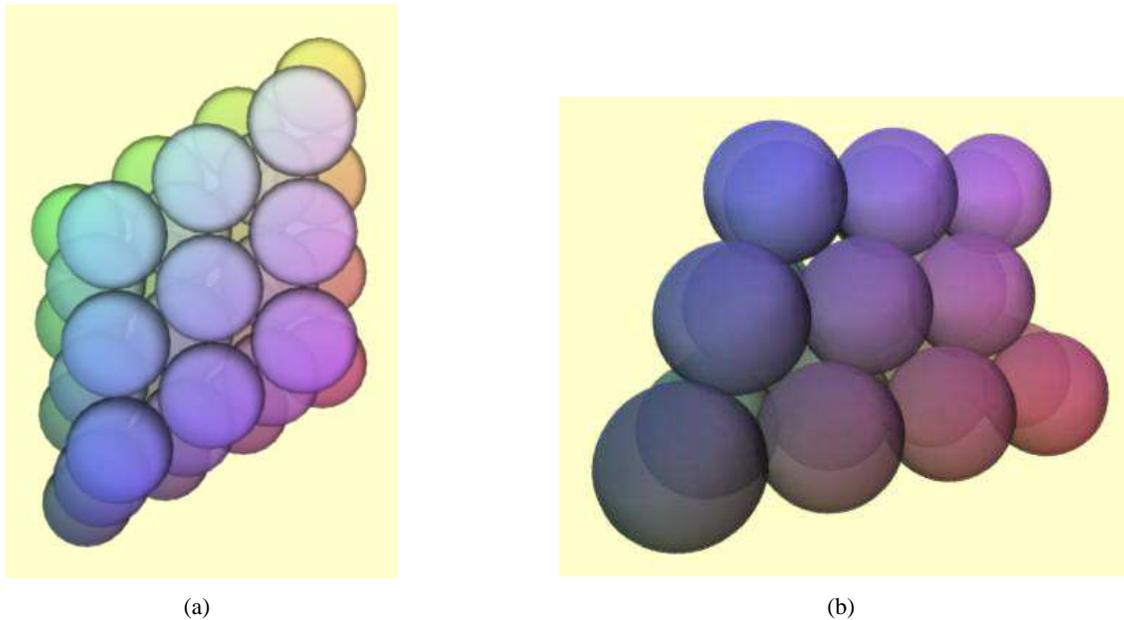

\centering \subfig{spheres1.eps} \hfill\subfig{spheres2.eps}
\caption{\label{fig:holes-3d-fcc} (a) Face-centred cubic lattice
as constructed from hexagonal close-packed layers of touching (but
non-overlapping) spheres. (b) Two small cylindrical holes (top
left) between two layers.}
\end{figure}

As an example, \figref{fig:holes-3d-fcc} shows the case of the
face-centred cubic lattice: \figref{fig:holes-3d-fcc}(a) shows the
structure as built up from stacks of hexagonal close-packed
layers, whilst \figref{fig:holes-3d-fcc}(b) shows a cylindrical
hole through the structure, as required by Heppes' theorem quoted
above.  In fact this hole lies between two adjacent hexagonal
layers, so that any structure built up of such layers would
contain a similar cylindrical hole.

Another possibility for constructing explicit 3D models with
finite horizon would be to make periodic layers with a lower
density of spheres, but still with finite horizon within the
layers.  If the density were low enough, the next layer could be
pushed down far enough to block the corridors between the layers
and possibly obtain a finite horizon.  However, we have not
managed to construct a model in this way.  We could also try
constructing a model with, for example, ellipsoids instead of
spheres, but again we have had no success.

\section{Construction of 3D periodic Lorentz gas with overlapping scatterers}

Here we present a 3D periodic Lorentz gas model analogous to the
2D model\footnote{In fact, we were led to the  2D model as a
cross-section of the 3D model.} of \chapref{chap:geom-dependence},
which has a finite horizon in certain regions of parameter space.
To accomplish this, however, we allow the spheres to
\emph{overlap}, which appears to be the only way of achieving a
finite horizon with a simple model, as discussed above.

Allowing the spheres to overlap is at first sight non-physical;
however, suppose that the moving particles are spheres of
\emph{non-zero} radius $s$. In this case, the dynamics is the same
as that of point particles moving through the same lattice, but
with the radius of the scatterers increased by $s$; this radius
could represent the effective radius of a particle interacting
with the crystal lattice via a more realistic potential. This is
the same construction originally used by Sinai \cite{Sinai70} to
reduce two discs moving on a torus to a periodic Lorentz gas; see
also the discussion in \chapref{chap:introduction}. We will find
conditions under which the overlapping model can be regarded as a
physical model in this way.

Note that allowing overlapping scatterers violates one of the key
requirements in Chernov's proof that the multi-dimensional Lorentz
gas is diffusive \cite{Chernov94}. Nonetheless, we expect
physically that the system still possesses strong ergodic and
statistical properties, and in particular is still diffusive. This
seems to be corroborated by numerical experiments, as discussed
below.

\subsection{Construction of finite horizon model by blocking corridors}

We consider a cubic unit cell of a 3D simple cubic lattice of
spheres of radius $a$. If the spheres do not touch, then there are
lots of holes in the structure.  In order to construct a finite
horizon model, we need to block all of these holes.  We first
allow the $a$ spheres to \emph{overlap}. The configuration in a
plane through the centres of  $4$ neighbouring $a$-spheres then
looks like \figref{fig:perp_geom}. There is no longer an infinite
horizon within this plane, and we have one corridor perpendicular
to the plane.

We now add a new scatterer, with (different) radius $b$, at the
centre of the $a$-unit cell.  If we make the $b$-sphere
sufficiently large that its \emph{projection} onto one of the
faces of the unit cell blocks the hole in that face
(\figref{fig:perp_geom}), then we have blocked all corridors
perpendicular to the faces.

\subsection{Phase diagram}

 The construction of the phase diagram follows
that of the 2D case: again we look for lines in parameter space
across which the geometry of the unit cell undergoes qualitative
changes.  However, the additional freedom in $3$ dimensions makes
the calculation more difficult.

\paragraph{Overlap of $a$-spheres}

The overlaps of the $a$-spheres change as follows when we vary
$a$.

\begin{itemize}
\item  The $a$-spheres overlap if and only if $a \ge \texthalf$;
this is a necessary condition for a finite horizon in our class of
models.

\item The space between the $a$-discs on a unit face closes when
two discs at opposite ends of a diagonal of the square touch,
which occurs when $a = \frac{1}{\sqrt{2}}$; for larger values of
$a$ it is impossible for a particle to move between different unit
cells, so the trajectory is localised and the diffusion
coefficient vanishes.

\item The $a$-spheres cover the entire unit cell when two spheres at
diametrically opposite vertices of the unit cube touch, which
occurs when $2a = \sqrt{3}$, \ie at $a = \frac{\sqrt{3}}{2}$.
\end{itemize}

\paragraph{Blocking vertical trajectories}

\begin{figure}
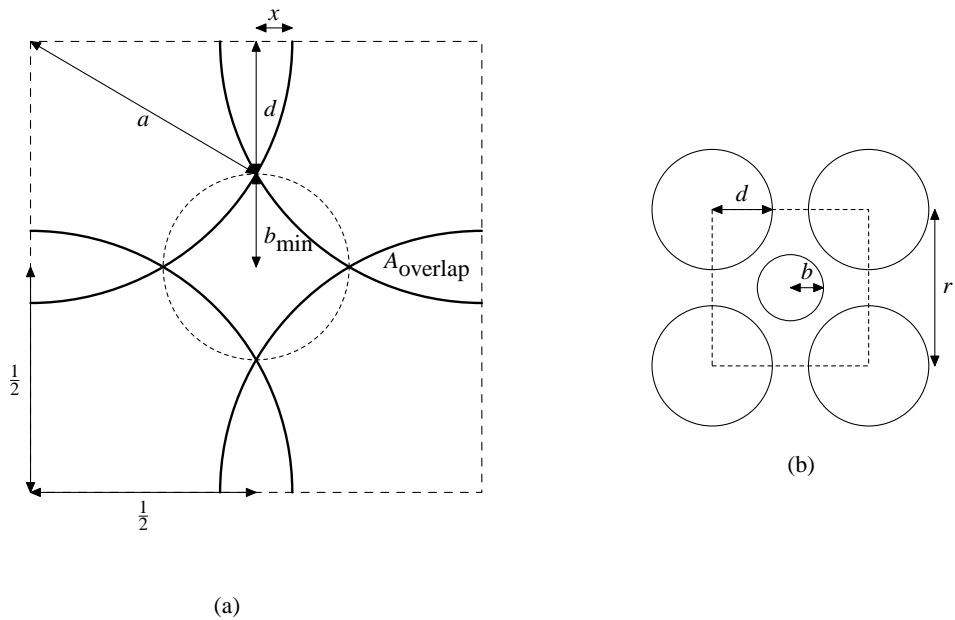

\centering
\subfigure[]{\begin{minipage}[c]{0.4\textwidth}
     \centering \includegraphics{overlap-geom.2}
     \vspace*{10pt}
    \end{minipage}\label{fig:perp_geom}}
    \qquad \qquad
\subfigure[]{\begin{minipage}[c]{0.4\textwidth}
    \centering \includegraphics{overlap-geom.1}
     \vspace*{10pt}
    \end{minipage}\label{fig:midplane}}

\caption{(a) Geometry of overlapping discs on one (and hence each)
face of the cubic unit cell. The dashed circle is the projection
of the $b$-disc at the centre of the cell. (b) Geometry in the
mid-plane of a unit cell.}
\end{figure}

Let $\bmin$ be the minimal value of the radius $b$ of the
additional scatterer required to block the vertical corridor
(i.e.\ such that its vertical projection covers the space on a
face between the $a$-sphere overlaps, as in
\figref{fig:perp_geom}), and let $d$ be the width of the overlap
of two $a$-discs on a face; this is also the radius of the disc of
intersection of two neighbouring $a$-spheres with the mid-plane
between them. The geometry is shown in figure \ref{fig:perp_geom}.

We have $\bmin + d = \texthalf$ and $d^2 + (\texthalf)^2 = a^2$,
so that $d = \sqrt{a^2 - \textstyle \frac{1}{4}}$ and a necessary
condition for finite horizon is
\begin{equation}\label{}
b \ge \bmin \defeq \texthalf - \sqrt{a^2 - \textstyle
\frac{1}{4}}.
\end{equation}

\paragraph{Blocking diagonal trajectories}
We now must now block any diagonal corridors (diagonal relative to
the lattice viewed as cubic cells). By symmetry, it suffices to
block trajectories in the mid-plane of the unit cell: there is
`most space' in this plane. (Compare this to blocking diagonal
angles at $45\degrees$ in 2D.)

The mid-plane is shown in \figref{fig:midplane}; its geometry is
 the same as that of the 2D model considered in \chapref{chap:geom-dependence},
  after replacing $a$ by $d$.
If $b \ge \bmin$, then horizontal and vertical trajectories within
this plane are already blocked, by definition of $\bmin$. Diagonal
trajectories at $45\degrees$ will be blocked if either $d \ge
\frac{1}{2 \surd 2}$, which reduces to $a \ge \frac{\sqrt{3}}{2
\surd{2}} \approx 0.612$, or if $b \ge \frac{1}{2 \surd 2}$.

If neither of these  conditions holds, then, by continuity, there
is an interval of non-zero size around the height of the midplane
such that planes at heights lying in that interval also have an
infinite horizon along trajectories at $45\degrees$, so that there
is a cylindrical hole of non-zero volume lying along this
diagonal, and hence the structure has an infinite horizon.

\paragraph{Conditions for localised trajectories}

If the $a$ or $b$ spheres are too large then they will overlap to
such an extent that trajectories will be localised, as in the S
and D regimes of the 2D model
(\secref{subsec:2d-parameter-space}).

Consider again the midplane shown in \figref{fig:midplane}.
Suppose that the $b$- and $d$-discs do not touch. Then any point
outside the discs in the midplane is accessible from any other. In
fact, since there are identical
 planes perpendicular to each of the coordinate directions,
 there is a connected network of empty space lying around the
grid of lines parallel to the coordinate directions which join the
centres of the $b$-spheres.  Hence for $b + d <
\frac{1}{\sqrt{2}}$ we do not have localised trajectories.  We now
show that in fact this is a necessary condition: all trajectories
are localised if $b \ge\frac{1}{\sqrt{2}} - d =\frac{1}{\surd 2} -
\sqrt{a^2 - \frac{1}{4}}$.

Consider cutting the unit cell by a horizontal plane which begins
at the mid-plane and moves upwards. Let  $h$ be the height of the
plane above the mid-plane, and let the radius at height $h$ of the
$a$- and $b$-cross-sections be $a(h)$ and $b(h)$, respectively.
Then $a(0) = d$, $b(0) = b$, $b(h) = \sqrt{b^2 - h^2}$ and $a(h) =
\sqrt{a^2 - (\texthalf - h)^2}$.

Trajectories will be localised if and only if there is a plane at
some height $h$ in which there is no available space, i.e.\ for
which the $a(h)$- and $b(h)$-discs fill this plane, for then there
are barriers in each direction blocking escape, whilst if there is
space in each plane $h$ then in fact this space must be connected
all the way from $h=0$ to $h=\texthalf$.  From
\secref{subsec:2d-parameter-space}, we know that this blocking
occurs exactly when
\begin{equation}\label{eq:b-of-h-condition}
b(h) \ge \texthalf - \sqrt{a(h)^2-\textfrac{1}{4}},
\end{equation}
since the geometry in this plane is the same as the 2D geometry
with $a$ and $b$ replaced by $a(h)$ and $b(h)$, respectively.
Considering when there exists an $h$ for which $b(h) = \texthalf -
\sqrt{a(h)^2-\textfrac{1}{4}}$ is equivalent to the following
slightly different approach.

Taking coordinates with the origin at the centre of the unit cell
(and hence at the centre of the $b$-sphere), if the intersection
point of the overlap of neighbouring $a$-spheres with the
$b$-sphere is at $(x,y,z)$ then we have $x=0$ by symmetry and thus
\begin{gather}
0^2 + y^2 + z^2 = b^2, \quad\text{and} \label{eq:on_b_sphere}\\
(0-\texthalf)^2 + (y-\texthalf)^2 + (z-\texthalf)^2 = a^2,
\label{eq:on_a_sphere}
\end{gather}
since the intersection point lies on each sphere.

Expanding \eqref{eq:on_a_sphere} and using \eqref{eq:on_b_sphere}
gives $y+z = \alpha \defeq b^2 - a^2 + \textstyle \frac{3}{4}$,
and substituting $z = \alpha - y$ into \eqref{eq:on_b_sphere} then
gives
\begin{equation}\label{}
y = \frac{1}{2} \lt[ \alpha \pm \sqrt{2 b^2 - \alpha^2} \rt],
\end{equation}
so that an intersection exists if and only if $2 b^2 - \alpha^2
\ge 0$, which is equivalent to $b \ge
\frac{\modulus{\alpha}}{\surd 2}$, since $b>0$.

We have $a<\textfrac{\sqrt{3}}{2}$, so that $a^2<\textfrac{3}{4}$,
and hence $\alpha=b^2-(a^2-\textfrac{3}{4}) > 0$. Thus the
condition for existence of the intersection is $b \sqrt{2} \ge b^2
- a^2 + \textfrac{3}{4}$, \ie $b^2 - b \sqrt{2} + (\textfrac{3}{4}
- a^2) \le 0$. Equality occurs when $b = \frac{1}{\surd 2} \pm
\sqrt{a^2 - \frac{1}{4}}$, so that there is an intersection if and
only if
\begin{equation}\label{}
\textfrac{1}{\sqrt{2}} - \sqrt{a^2 - \textfrac{1}{4}} \le b \le
\textfrac{1}{\sqrt 2} + \sqrt{a^2 - \textfrac{1}{4}}.
\end{equation}
The leftmost term is equal to $\textfrac{1}{\surd 2} - d$; the
rightmost term is greater than $\textfrac{1}{\surd 2}$, so that it
is never attained for non-localised trajectories for which we must
have $a<\textfrac{1}{\surd 2}$. Hence the condition for
intersection is that the $b$- and $d$-discs touch in the
mid-plane, as claimed.

\paragraph{Condition for all space to be covered}

To establish when the $b$-sphere covers all the space left by the
$a$-spheres,  consider first the case when $a < \textfrac{1}{\surd
2}$, so that the $a$-spheres leave holes on and near the unit cell
faces. These holes will be covered by the $b$-sphere once it is
big enough that its intersection with the unit cell face covers
the space on that face, \ie if
\begin{equation}\label{}
b(h) \ge \bmin \quad \text{for } h=\texthalf,
\end{equation}
which reduces to
\begin{equation}\label{}
\sqrt{b^2-(\texthalf)^2} \ge \texthalf - \sqrt{a^2 - \quarter},
\end{equation}
or finally
\begin{equation}\label{}
b \ge \sqrt{a^2 + \quarter - \sqrt{a^2 - \quarter}} =\texthalf -
\sqrt{a^2-\textfrac{1}{4}}.
\end{equation}

When $a > \textfrac{1}{\surd 2}$, we need $b$ to be large enough
for the $b$-sphere to cover the $3$-hole inside the overlap of the
$a$-spheres.  For this we need that the $b$-disc covers the
$2$-hole between the $d$-discs in the mid-plane.  The geometry is
exactly as in the 2D case, with $a$ replaced by $d$, so we can
quote the result:
\begin{equation}\label{}
b \ge \texthalf - \sqrt{d^2 - \quarter} = \texthalf - \sqrt{a^2 -
\texthalf}.
\end{equation}

\paragraph{Overlapping conditions}

Note that if $b > \sqrt{3}/2 - a$ then  the $b$-spheres meet the
$a$-spheres, whilst if $b>1/2$ then neighbouring $b$-spheres meet
each other.

\paragraph{Phase diagram}

\begin{figure}
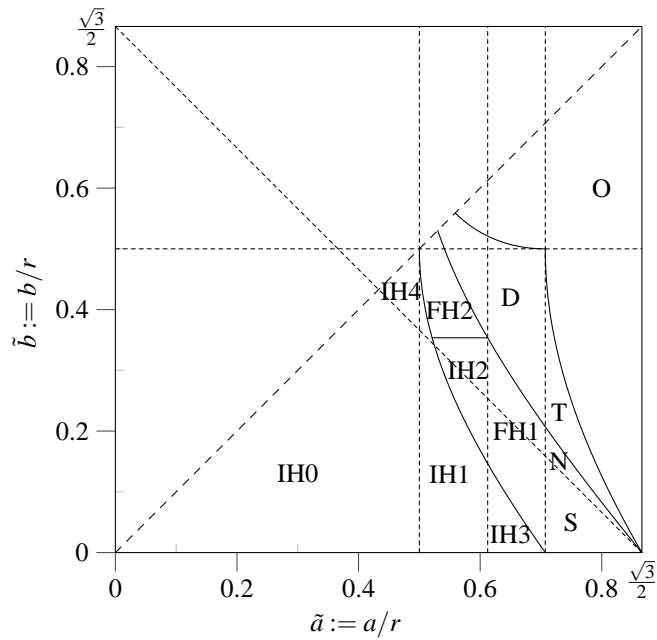

\centrefig{3d-phase-diag-new.1} \caption{\label{fig:3d-phase-diag}
Parameter space of the 3D model showing the different regimes
described in the text.}
\end{figure}

The parameter space of the 3D model is shown in
\figref{fig:3d-phase-diag} for $b<a$.  The regimes are named as
far as possible to agree with those of the 2D model in
\secref{subsec:2d-parameter-space}: if the name is the same as one
for the 2D model, then the geometrical features and statistical
properties are similar.  The diagonal from top left to bottom
right is the boundary line separating regimes where the
$b$-spheres do not touch the $a$-spheres (below) from those where
the $b$- and $a$-spheres overlap (above).  This cuts several of
the other regimes.

Several regimes do not occur in the 2D model. N has localised
motion in cells centred on the $b$-spheres, but with the
$b$-spheres overlapping the $a$-spheres. In IH4 the $a$-spheres
meet each other and also the $b$-spheres but still with infinite
horizon. In IH0 nothing is touching and we have a strongly
infinite horizon (see also \secref{sec:simple-cubic-lattices}).

Note that the finite horizon (FH) regime consists of two disjoint
pieces, corresponding to the two different ways of preventing
diagonal infinite horizon trajectories detailed above.  In fact,
the top of FH2, above the line $b=\texthalf$, could be regarded as
a third finite horizon regime.  Here, the $b$-spheres meet each
other (as well as the $a$-spheres), so that there is a single
connected scatterer network with an interconnected hole inside it.

\begin{figure}

\centering \subfig[1.2]{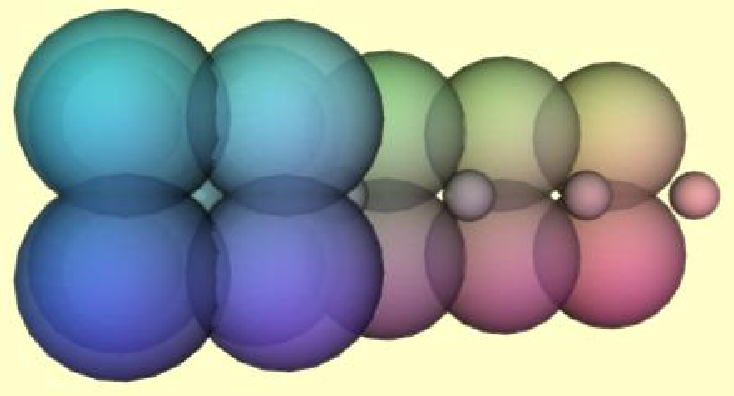}

\subfig[1.2]{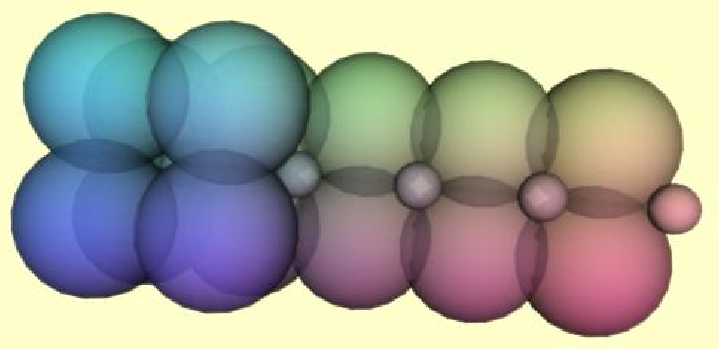}
\caption{\label{fig:overlapping-3d-lorentz} The 3D Lorentz gas
with finite horizon in a regime ($r=1.56$, $b=0.31$) where the
$b$- and $a$-spheres do not touch, as can be seen in (a); a
different view of the structure is shown in (b), which confirms
that there is a finite horizon, since no holes can be seen in the
complete unit cell.}
\end{figure}

\bfigref{fig:overlapping-3d-lorentz} shows the model in the finite
horizon regime with the $a$- and $b$-spheres not overlapping. We
can check by rotating the model that it does indeed have finite
horizon, since no holes can be seen in the complete
unit cell. 

\subsection{Physical realisation with a moving
particle of non-zero size}

The overlapping system with radii $a$ and $b$ studied above can be
regarded as physical if it is a reduction of a system with a
moving sphere of radius $s$ travelling within a lattice of spheres
with radii $a'$ and $b'$ with no overlaps. Regarding $(a, b)$ and
$(a', b')$ as position vectors in the phase diagram determined
above, the reduction condition is
\begin{equation}\label{}
(a', b') + (s,s) = (a, b),
\end{equation}
which corresponds geometrically to translating by the vector
$(s,s)$ from the initial point on the phase diagram.  The system
will be diffusive if this lands in the FH regime.

For the initial point to correspond to a system where no spheres
overlap, we must start in the region IH0.  In fact, any point in
IH0 corresponds to some point in FH via a translation by some
vector $(s,s)$, except for a line at $45\degrees$ passing through
the point where the FH1 and FH2 regions touch.

\section{Normal diffusion in the finite horizon regime}

We now use techniques developed in previous chapters to give
strong evidence that we have normal diffusion in the finite
horizon regime. We emphasise that the arguments of
\cite{Chernov94} require disjoint scatterers, and so do not
immediately apply to our model; nonetheless we expect that those
methods should be able to be refined to deal with our case and
prove normal diffusion, even in the strongest sense that the
functional central limit theorem is satisfied.

\subsection{Decay of velocity autocorrelations}


 From the discussion of the Green--Kubo relation in continuous
time of \secref{subsec:green-kubo-disc-cont}, a necessary
condition for the existence of the diffusion coefficient (which is
the weakest type of normal diffusion) is the integrability of the
velocity autocorrelation function $C(t)$.  Since this function is
highly oscillatory and rapidly becomes small, it is difficult to
study.  It was thus suggested in \cite{LoweMasters}
to look instead at the integrated velocity autocorrelation
function $R(t)$, given by
\begin{equation}\label{}
R(t) \defeq \int_0^t C(s) \rd s = \mean{\v_0 \cdot \Delta \x_t}.
\end{equation}
 The integration in the definition of $R(t)$ acts as a smoothing operation
 which deals efficiently with the highly oscillatory nature of $C(t)$.
 Note that in this definition we do not need to perform a direct
 numerical integration of the function $C(t)$; such an approach
 was taken, for example, in \cite{MatsuokaMartin}.

 If $C(t)$ is integrable, then $R(t) \to d\,D$ as $t \to \infty$,
where $d$ is the spatial dimension and $D$ is the diffusion
coefficient. We find numerically that $C(t)$ decays so fast that
we cannot extract any information about it, and we do not plot it.
 Similarly $R(t)$ reaches very rapidly a limiting value, which it then
 oscillates around in a random fashion. This provides evidence that $C(t)$ is
 integrable and hence that $D$ exists, giving normal diffusion.

\subsection{Growth of moments}

\begin{figure}
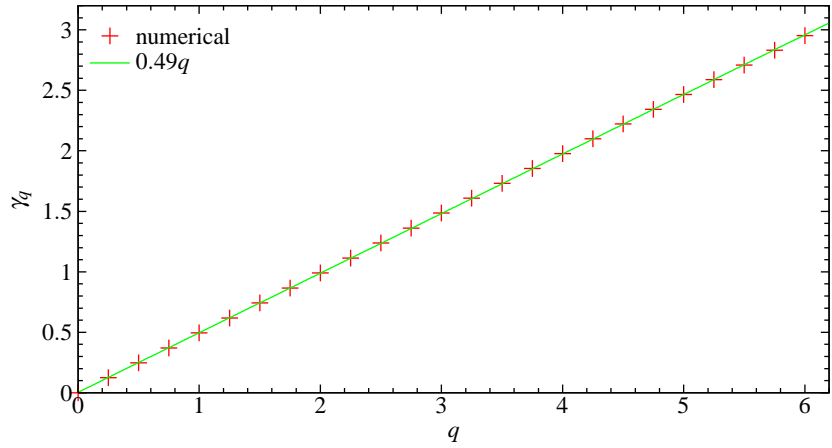

\centrefig{3d-fh-moments} \caption{\label{fig:3d-fh-moments}
Growth rate $\gamma_q$ of the moments $\mean{\modulus{r}^q}$ for
particular parameters in the finite horizon regime ($r=1.6$;
$b=0.3$), as a function of $q$. The growth rates obey well the
relation  $\gamma_q = q/2$, as for the 2D Lorentz gas with finite
horizon.}
\end{figure}

\bfigref{fig:3d-fh-moments} shows the growth rates $\gamma_q$ of
the moments $\mean{\modulus{r}^q}$ as a function of $q$.  As in
the 2D finite horizon case (not shown), they satisfy $\gamma_q =
q/2$, so that we have a strong form of normal diffusion; we
conjecture that the functional central limit theorem is satisfied
for our model in the finite horizon regime, as was proved in
\cite{Chernov94} in the case of disjoint scatterers.

\subsection{Shape of distributions and central limit theorem}

Knowing that the mean squared displacement grows asymptotically
linearly in the finite horizon regime, we can apply the techniques
of \chapref{chap:fine-structure} to investigate the shape of
densities and the central limit theorem.

Restricting attention to 1D projections of the position
distribution, we need to calculate the analogue for our model in
the finite horizon regime of the available height function $h(x)$
used in \chapref{chap:fine-structure}.  This analogue is the
available area, which we denote by $A(x)$, in a cross-section
perpendicular to the $x$-axis at distance $x$ from the centre of
the $b$-sphere;  we refer to this cross-section plane as the
$x$-plane. We denote the radii of the cross-sections of the $a$-
and $b$-spheres in that plane by $a(x)\defeq
\sqrt{a^2-(x-\texthalf)^2}$ and $b(x)
\defeq \sqrt{b^2-x^2}$, respectively, where we again use the
convention that $\sqrt{\alpha}=0$ when $\alpha<0$.

\bfigref{fig:perp_geom} can be thought of as depicting a generic
cross-section, after possibly adding  a small $b$-cross-section in
the centre. Denote by $\Aoverlap(x)$ the area of each overlap of
$a(x)$-discs in the $x$-plane.  Then
\begin{equation}\label{}
A(x) = \frac{1}{\modulus{Q}} \left[ 1 - \pi(a(x)^2+b(x)^2) + 4
\Aoverlap(x) \right],
\end{equation}
where if $a(x)>1/2$ then
\begin{equation}\label{}
\Aoverlap(x) = \int_{y=1/2}^{a(x)} 2\sqrt{a(x)^2-y^2} \rd y = 2
\left[ \textfrac{\pi}{4} a(x)^2 - F(\texthalf; x) \right],
\end{equation}
whilst there is no overlap if $a(x)\le 1/2$.
 Here
\begin{equation}\label{}
F(y; x) \defeq \texthalf y \sqrt{a(x)^2-y^2} + \texthalf a(x)^2
\tan^{-1} \left( \frac{y}{\sqrt{a(x)^2-y^2}} \right)
\end{equation}
is the anti-derivative of $\sqrt{a(x)^2-y^2}$ (with respect to
$y$).  Further, $\modulus{Q}$ is the volume of the available space
in a unit cell given by
\begin{equation}\label{}
\modulus{Q} = 1 - \textfrac{4}{3}\pi(a^3+b^3) + 3 \Voverlap,
\end{equation}
where
\begin{equation}\label{}
\Voverlap \defeq 2\int_{1/2}^a \pi (a^2-y^2) \rd y = 2 \pi \left[
\textfrac{2}{3} a^3 - \texthalf a^2 + \textfrac{1}{24} \right]
\end{equation}
is the volume of the intersection of two neighbouring spheres.
(Similar overlap calculations were used in \cite{DEAMemoire} in a
related context.) In the above calculations we have for simplicity
restricted the calculation to the case where the $b$- and
$a$-spheres do not overlap.

\begin{figure}
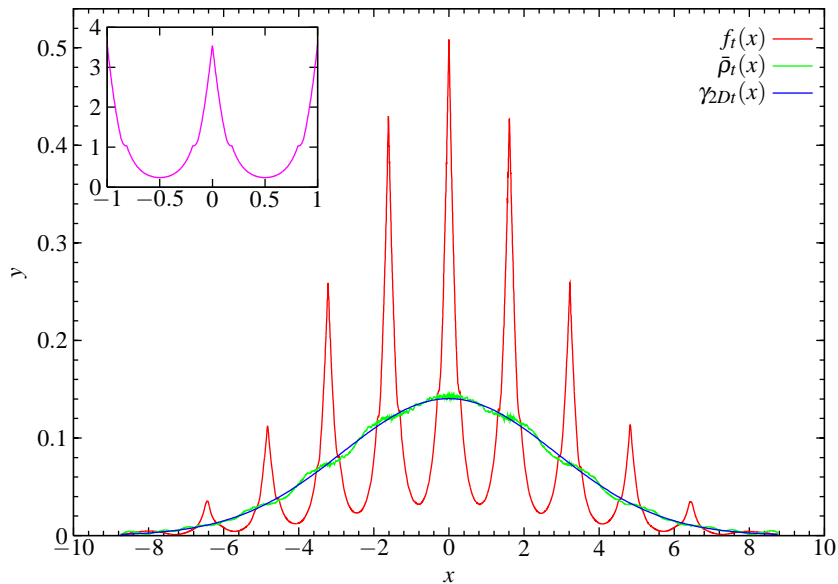

\centrefig{3d-demod.eps} \caption{\label{fig:3d-demod} 1D position
density $f_t(x)$ at time $t=50$ for the 3D Lorentz gas in the
finite horizon regime at $r=1.61$, $b=0.3$ (for which the $a$- and
$b$-spheres do not overlap). The inset shows the available space
function $A(x)$, and the main figure also shows the demodulated
density $\bar{\rho}_t(x)$, compared to a Gaussian with variance $2
D t$, where $D=0.081$ for this geometry.}
\end{figure}

\begin{figure}
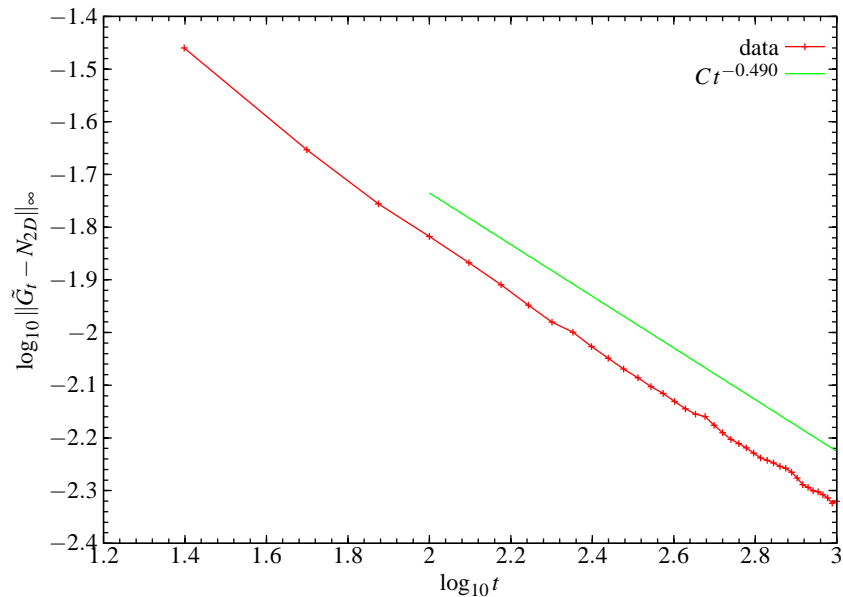

\centrefig{3d_clt.eps} \caption{\label{fig:3d-clt} Convergence to
limiting normal distribution with variance $2D$ for $r=1.61$,
$b=0.3$.  The rate for long times is $t^{-0.49}$.}
\end{figure}

The inset of \figref{fig:3d-demod} shows the available space
function $A(x)$ as calculated above, whilst the main part of that
figure shows the 1D position density $f_t(x)$ and the demodulated
version $\bar{\eta}_t(x) \defeq f_t(x)/A(x)$, for certain
geometrical parameters in the finite horizon regime which we
believe to be representative. Again we see that the demodulated
density is very close to Gaussian, and we have confirmed
numerically that the central limit theorem holds with rate of
convergence $t^{-0.49}$, close to the optimal $t^{-1/2}$ as
discussed in \secref{sec:clt}: see \figref{fig:3d-clt}.

\subsection{Geometry dependence of diffusion coefficients}

\begin{figure}
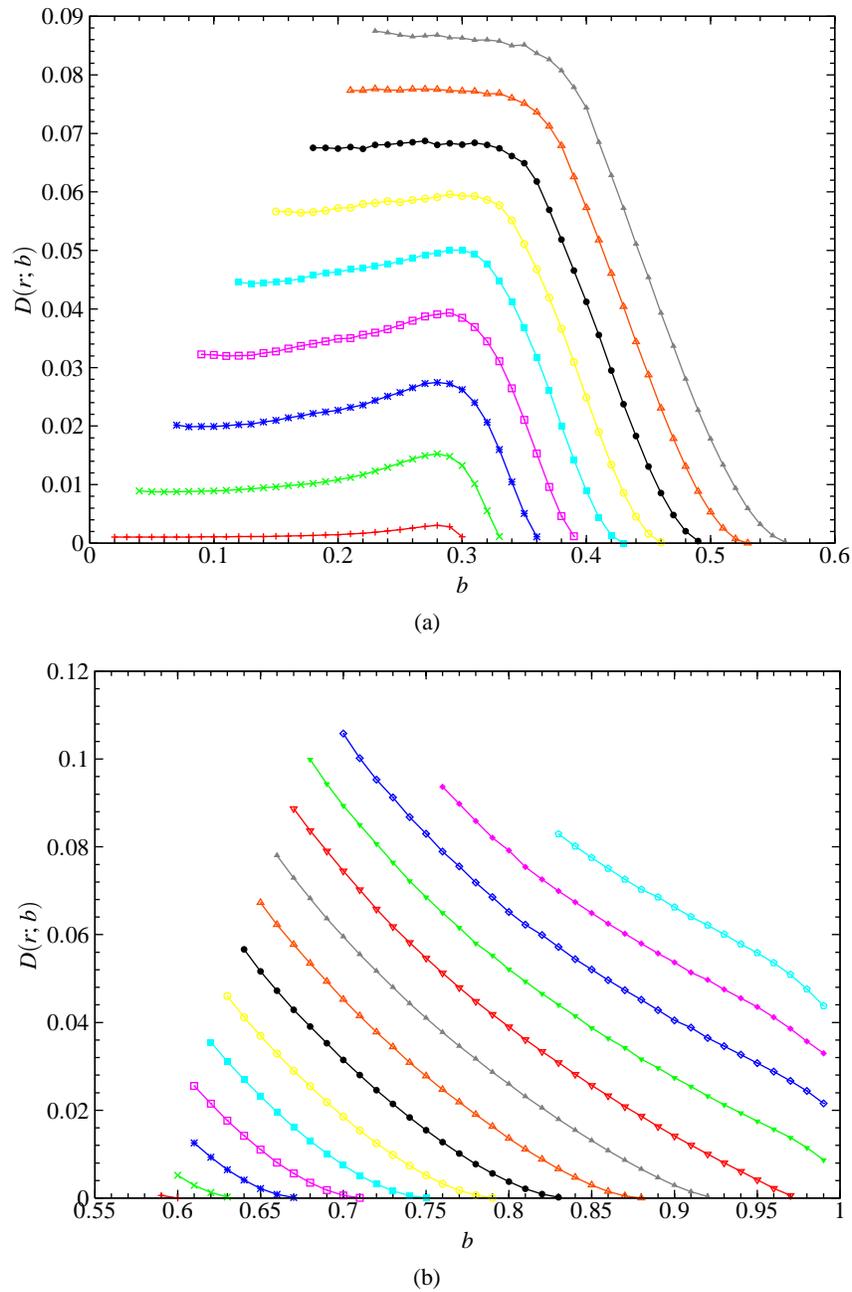

\centering \subfig{diff-coeffs-3d-FH1.eps}

\subfig{diff-coeffs-3d-FH2.eps}
\caption{\label{fig:diff-coeffs-3d} Diffusion coefficients for
parameter values in the finite horizon regimes: (a) in FH1; (b) in
FH2.  In (a), $r$ increases from $r=1.425$ to $r=1.625$ from
bottom to top, in steps of $0.025$.  In (b), $r$ increases from
$r=1.65$ to $r=1.975$ from bottom to top, in steps of $0.025$.}
\end{figure}

\bfigref{fig:diff-coeffs-3d} shows the geometry dependence of the
diffusion coefficient over the two finite horizon regimes.  In FH1
the behaviour resembles that in the 2D case, although the angle at
which the curves approach $0$ on the right hand side is different;
in particular, here too there is a qualitative change in behaviour
with a non-trivial maximum which disappears for larger values of
$r$. 

\section{Statistical behaviour in the infinite horizon regime}
\label{sec:stat-behaviour-inf-horiz}

\subsection{Shape of 2D distributions}

 How can normal
diffusion fail to hold in dispersing billiards with infinite
horizon?  \bfigref{fig:2d-distns-inf-horiz} shows scatterplots
representing the position and displacement distributions for
representative parameters in the infinite horizon regime of a 2D
square Lorentz gas.  These distributions have a distinctive shape
caused by the possibility of particles travelling arbitrarily far
without ever hitting a scatterer.  Bleher \cite{Bleher} showed,
assuming some natural conjectures\footnote{\revision{A complete, rigorous
proof has now been given
\cite{SzaszVarjuLimitLawsRecurrencePlanarLorentzInfiniteHorizonJSP2007}.}}, that
the mean squared
displacement in this case grows like $\msd_t \sim t \log t$ as $t
\to \infty$.  He also showed that
\begin{equation}\label{}
\frac{\x_t - \x_0}{\sqrt{t \log t}} \distconv \z,
\end{equation}
where $\z$ is a normal random variable; this is a central limit
theorem type result with a different normalisation constant.  The
faster growth rate of the variance corresponds to the tails of the
distribution visible in \figref{fig:2d-distns-inf-horiz}.

\begin{figure}
\centering \subfig{{r2.5b0.0n50ktt50.eps}}

\subfig{r2.5b0.0n50kt50.eps}
\caption{\label{fig:2d-distns-inf-horiz} (a) 2D position
distribution; (b) 2D displacement distribution. $r=2.5$; $b=0.0$;
$t=50$; $N=5 \times 10^4$ initial conditions. }
\end{figure}

\subsection{Discrete-time dynamics}

 Following \cite{Bleher} (who discussed the 2D case), we
write
\begin{equation}
  \x_n - \x_0 = \sum_{j=0}^{n-1} (\x_{j+1} - \x_j) = \sum_{j=0}^{n-1} \r_j,
\end{equation}
where $\x_n \defeq \x(t_n)$ is the position at the $n$th collision
(occurring at time $t_n$) and $\r_j \defeq \x_{j+1} - \x_j$ is the
free flight vector between the $j$th and $(j+1)$th collisions.

In the finite horizon regime the free path length is bounded
above,  so that $\r_j$ is a piecewise H\"older continuous function
defined on the collision phase space.  The central limit theorem
of \cite{BS, BSC} can then be applied to show that $(\x_n -
\x_0)/\sqrt{n}$ converges in distribution to a normal
distribution.

In the infinite horizon case, however, $\r_j$ is no longer bounded
and so  cannot be piecewise H\"older continuous, since a
continuous function on the compact phase space $M$ must be
bounded. The central limit theorem thus does not apply. Defining
the finite-time diffusion coefficient $D_n$ by
\begin{equation}
  D_n \defeq \frac{1}{4n} \mean{\modulus{\x_n-\x_0}^2}_{\nu},
\end{equation}
the limiting diffusion coefficient for the discrete time dynamics
takes the form \cite{Bleher}
\begin{equation}\label{eq:green-kubo-inf-horiz}
  D^0 \defeq \lim_{n\to\infty} D_n = \frac{1}{4} \mean{\modulus{\r_0}^2} +
\frac{1}{2} \sum_{j=1}^{\infty}\mean{\r_0\cdot\r_j},
\end{equation}
if it exists.

In \cite{Bleher} it was conjectured that (at least in 2D)
$\mean{\r_0\cdot\r_j}$ decays fast for large $j$, so that the
infinite sum in \eqref{eq:green-kubo-inf-horiz} exists.  However,
it was proved there that
$\mean{\modulus{\r_0}^2}_{\nu}=\mean{\freepath^2}_{\nu}$, the
second moment of the free path length for the billiard \emph{map},
diverges (logarithmically), so that the first term in
\eqref{eq:green-kubo-inf-horiz} is infinite and hence so are the
discrete-time and continuous-time diffusion coefficients.

\subsection{Obstructions to normal diffusion}

Generalising the above to higher dimensions, there are two
possible obstructions to normal diffusion outside the FH regime:
\begin{enumerate}
\item the continuous-time VACF $C(t)$ decays slowly as a function
of time $t$, so that the  Green--Kubo integral, giving $D$ as the
infinite-time integral of this VACF,  diverges; and
\item the second moment of the (discrete-time) free path length diverges.
\end{enumerate}
If neither of these obstructions is present, then we expect to
have normal diffusion: this is certainly the case in the finite
horizon regime treated above. We now discuss the situation in the
infinite horizon regime, i.e.\ when there exist trajectories which
never collide with the scatterers.

\subsection{Review of two-dimensional case}
\label{subsec:2D-inf-horiz}

\paragraph{Decay of velocity autocorrelations}

The following type of argument seems to have first been published
in \cite{FriedmanMartin84}. Consider a 2D lattice with
\emph{infinite} horizon (for example a square lattice of
scatterers of radius $a$). Due to the infinite horizon, there
exist
\defn{corridors} in the structure, i.e.\ empty regions of constant,
non-zero width, extending infinitely far in opposite directions.
Bleher \cite{Bleher} gives a detailed description of these
corridors.

Consider one particular corridor, of width $d>0$. Take coordinates
such that the $y$-axis is parallel to the centre-line of the
corridor and the $x$-axis is perpendicular to it, with the origin
at an arbitrary point on one edge of the corridor. Let $\theta$ be
the
 angle from the $y$-axis.

We consider initial conditions $(x_0, y_0)$ in $\fund$ whose
trajectories have no collisions within (continuous) time $t$, and
in fact remain inside the corridor at least until time $t$; thus
these trajectories are straight during this period, so that $\v_t
= \v_0$. A trajectory emanating from $(x_0, y_0)$ in the direction
$\theta$ from the $y$-axis will not leave the corridor within time
$t$ provided the line segment of length $t$ at angle $\theta$ ends
within the corridor, \ie provided $\theta_{-} \le \theta \le
\theta_{+}$, considering only one direction of flight down the
corridor initially. The geometry gives
\begin{equation}\label{}
  \sin \theta_+ = \frac{d-x}{t}\,; \qquad
  \sin \theta_- = -\frac{x}{t}\,.
\end{equation}

The Green--Kubo formula for the diffusion coefficient means that
we are interested in the rate of decay of the velocity
autocorrelation function (VACF) as $t \to \infty$. For large $t$,
$1/t$ is small, hence $\sin \theta_\pm$ is small, so that $\sin
\theta_\pm \approx \theta_\pm$.

For large enough $t$, denoting the average of the VACF solely over
those long trajectories which do not escape after time $t$ by
$\mean{\v_0 \cdot \v_t}_\txtr{long}$, we thus have
\begin{equation}\label{}
\mean{\v_0 \cdot \v_t}_\txtr{long} \approx K \, \int_{x=0}^d \d x
\, \int_{\theta=-x/t}^{(d-x)/t} \d \theta = \frac{K'}{t}.
\end{equation}
Here, $K$ and $K'$ are constants which are related to the area of
intersection of the corridor with $\fund$.

It is then argued in \cite{FriedmanMartin84} that if the rest of
the trajectories are more efficiently mixed, as we expect they are
due to collisions with the scatterers, then the VACF averaged over
those other trajectories will decay faster. Hence the rate of
decay averaged over the whole of $\fund$ will be dominated by the
slow $1/t$ decay of the long trajectories considered above. In
this case, the integral of the VACF will diverge, so that the
diffusion coefficient will not exist (or is infinite), implying
that the system is super-diffusive.

Unfortunately the above argument is not rigorous, and indeed
currently  there are no rigorous results on decay of correlations
for the billiard flow\footnote{\revision{Such results have now been obtained in
the finite-horizon case \cite{ChernovBoundsCorrelationsBilliardFlowsJSP2007}.}},
even in the finite-horizon case
\cite{CherYoung}, although the recent results of
\cite{SzaszVarjuII} prove that $D = \infty$, so that they show
indirectly that $C(t)$ must be non-integrable.

In \cite{GarrGall} the exact result of Bleher for the rate of
increase of the mean squared displacement was compared to
numerical results.  They pointed out, and we have confirmed, that
the coefficient of the $t \log t$ term is very small, so that it
is difficult to observe numerically. Nonetheless it is sometimes
possible to observe this effect by plotting $\msd_t / t$ against
$\log t$, as first suggested in the context of 1D maps in
\cite{GeiselThomaePRL1984}.  As argued in \cite{GarrGall}, if
$\msd_t \sim t \log t$ then the lower order terms presumably take
the form
\begin{equation}\label{}
\msd_t \sim A + B \, t + C \, t \, \log t.
\end{equation}
It then follows that
\begin{equation}\label{}
\frac{\msd_t}{t} \sim A \, t^{-1} + B + C \, \log t.
\end{equation}
Introducting the variable $z \defeq \log t$, we then have
\begin{equation}\label{}
\frac{\msd_t}{t} \sim A \, \e^{-z} + B + C \, z,
\end{equation}
so that for large enough values of $z$, i.e.\ for large enough
values of $t$, we should find asymptotically linear growth of
$\msd_t/t$.

\paragraph{Growth of higher-order moments}

In \cite{ArmsteadOtt}, a similar line of argument was used to give
a lower bound on the rate of increase of higher order moments,  as
follows.  A proportion $C/t$ of trajectories do not collide in
time $t$ (as described above), so that for those trajectories we
have the lower bound
\begin{equation}\label{}
\mean{\modulus{r}^q}_{\text{long}} \ge \frac{C}{t}(vt)^q = C \,
t^{q-1},
\end{equation}
where $v$ is the speed, so that $\gamma_q \ge q-1$.

They show that $\gamma_q$ is a convex function of $q$. We also
know that  $\gamma_0=0$ and $\gamma_2=1$ lie on the curve, that
$\gamma_q \ge q/2$, assuming that the process acts at least as
fast as a random walk, and that $\gamma_q \le q$. These together
lead to the conclusion that
\begin{equation}\label{}
\gamma_q = \begin{cases} q/2, \quad \text{when } q<2\\
q-1, \quad \text{when } q>2.
\end{cases}
\end{equation}

\paragraph{Tail of free path distribution}

As discussed in \secref{sec:stat-behaviour-inf-horiz}, the
existence of $\Exp{\nu}{\freepath^2}$ is crucial for the
possibility of having normal diffusion.  We introduce the
following functions:
\begin{gather}\label{}
\Psi(T) \defeq \prob[\nu]{\freepath > T} \defeq \nu\left(\theset{x
\in M \colon \freepath(x) > T}\right);\\
\tPsi(T) \defeq \prob[\mu]{\freepath > T} \defeq
\mu\left(\theset{x \in \M \colon \freepath(x) > T}\right).
\end{gather}
(Recall that the billiard map $T$ preserves the measure $\nu$ on
the phase space $M$, and the billiard flow $\Phi^t$ preserves the
measure $\mu$ on the phase space $\M$.)  These functions describe
the tail of the distribution of the free path length considered in
discrete time and continuous time, respectively. Note that in the
discrete time case $\modulus{\r}=\freepath$, where $\r$ is the
free flight vector.

In 2D, it was proved in \cite{Bleher} that
\begin{equation}\label{}
\Psi(T) \sim T^{-2} \quad \text{as } T \to \infty.
\end{equation}
By definition of expectation,
\begin{equation}\label{}
\E[\nu]{\freepath^2} = -\lim_{T'\to\infty} \int_0^{T'} T^2 \, \rd
\Psi(T),
\end{equation}
where the integral is a Lebesgue--Stieltjes integral.  Integration
by parts is valid for such integrals, so that
\begin{equation}\label{}
-\int_{T_0}^{T'} T^2 \, \rd \Psi(T) = \int_{T_0}^{T'} 2T\, \Psi(T)
\rd T + \bigO{1} \sim \log T',
\end{equation}
and hence the expectation diverges logarithmically.

The same result can be obtained using the following relation
proved in \cite{DumasGolseMeanFreePathII} (see also
\cite{GolseReview}):
\begin{equation}\label{}
c_{\nu} \, \E[\nu]{f(\freepath)} = c_{\mu} \E[\mu]{f'(\freepath)},
\end{equation}
for any $C^1$ function $f\from \Rplus \to \R$ such that $f(0)=0$;
here $f'$ denotes the derivative of $f$. For $f(z)=z$ we recover
the known expression (see \cite{CherMFT}) for the mean free path
in terms of $c_{\nu}$ and $c_{\mu}$ discussed in
\chapref{chap:geom-dependence}, while for $f(z)=z^2$ we obtain
\begin{equation}\label{eq:golse-free-path-moments}
\E[\mu]{\freepath} = \frac{c_{\nu}}{2 c_{\mu}}
\E[\nu]{\freepath^2}.
\end{equation}
 Here we extend the definition of
$\freepath$ to the whole of $\M$ as
\begin{equation}\label{}
\tau(\x,\v)
\defeq \inf \theset{t \in \Rplus \colon \x+t\v \in M},
\end{equation}
 the minimum time needed to collide with
a scatterer.

Thus the second moment of $\freepath$ with respect to $\nu$ exists
if and only if the mean of $\freepath$ with respect to $\mu$
exists.  We now use an argument similar to those above involving
the proportion of trajectories which have not collided within time
$t$: those trajectories have (continuous-time) free paths which
are at least $t$, so that integrating over initial conditions in
one unit cell we have
\begin{equation}\label{}
\tPsi(T) \ge C\, T^{-1}.
\end{equation}
The mean $\E[\mu]{\freepath}$ thus diverges logarithmically, and
hence so does  $\E[\nu]{\freepath^2}$.

\paragraph{Summary}
The above arguments all involve calculating the proportion of
trajectories starting from a single unit cell which undergo no
collisions up to time $t$.  For large $t$ these trajectories are
ones which remain inside the corridor and lie on a circle of
radius $t$.

We obtain rigorous lower bounds for the higher-order moments
$\mean{\modulus{\r(t)}^q}$ as $t\to \infty$ and for the size of
the tail $\tPsi(T)$ of the free path distribution function as $T
\to \infty$, whilst we obtain only a heuristic estimate of $C(t)$
as $t \to \infty$.  In the next sections we shall apply these
methods to our 3D Lorentz gas.

\section{Simple cubic lattices}
\label{sec:simple-cubic-lattices}

\subsection{Free path distribution}

We begin by discussing simple cubic lattices with $r>2$ and $b=0$.
In \cite{Chernov94} it was stated that point (ii) of the previous
section, i.e.\ divergence of $\E[\nu]{\freepath^2}$  holds in
higher dimensions, and that the proof is ``easy to verify'', but
it was not stated which configurations this applies to.  For
simple cubic lattices a proof was given in
\cite{GolseFreePathsHigherDimensions}. In the next section we show
that in certain other geometries the result is in fact no longer
true, so that the existence or otherwise of normal diffusion
reduces in that case to determining if (i) holds.

The method of \cite{GolseFreePathsHigherDimensions} to prove this
 result is to
 generalise the method discussed in the previous section, by using
the existence of
\defn{free planes} (called `sandwich layers' in
\cite{GolseFreePathsHigherDimensions}) in the simple cubic
lattice.
These
are infinite planes
 which do not intersect any
scatterer, and are the 3D analogue of corridors related to
infinite horizon trajectories (which we can think of as `free
lines') in 2D\footnote{\revision{Figures showing the different types of
holes in 3D, as well as an analytical argument for systems of any dimension,
are given in a submitted manuscript \cite{SandersNormalDiffusion3D}.}}.

\paragraph{Analytical argument}

Analogously to the calculations in the previous section, we need
to find the proportion of trajectories remaining inside a sandwich
layer and lying on a sphere of radius $t$. The intersection of
these two objects is approximately a circle with non-zero width,
and the proportion is then the ratio of the area of this thickened
circle, which is $2\pi t \Delta x$, divided by the area $4 \pi
t^2$ of the sphere of radius $t$.  The ratio is hence $C/t$.  A
more careful argument, taking account of the intersections of such
circles corresponding to different sandwich layers, shows that the
same result holds \cite{GolseFreePathsHigherDimensions}.  We
remark that this result was also stated in
\cite{FriedmanMartin84}.

The  argument of the previous section now gives a lower bound on
the size of the tail $\tPsi(T)$ of the free path distribution for
the billiard flow of $T^{-1}$, so that again the expectation
$\E[\mu]{\freepath}$ diverges.

\paragraph{Numerical calculation}

\begin{figure}
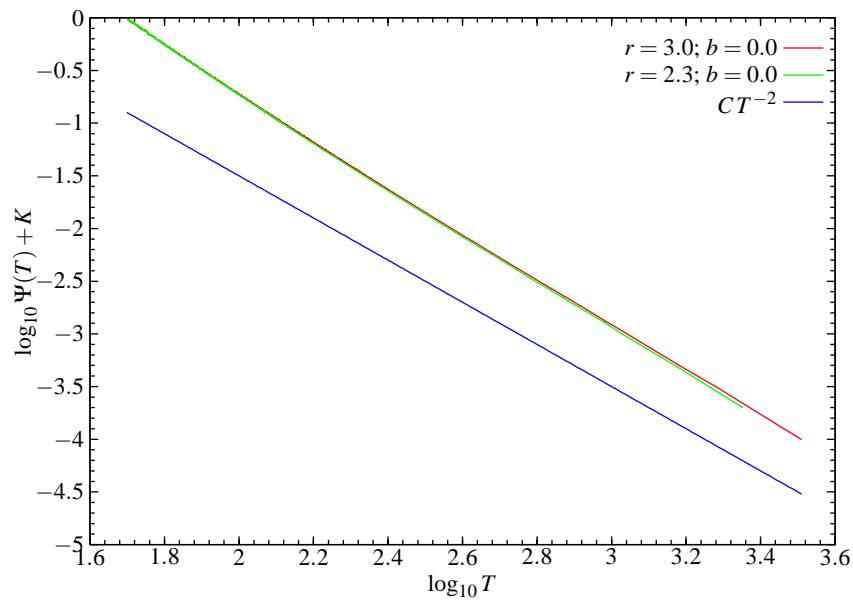

\centrefig{simple-cubic-freepath.eps}
\caption{\label{fig:simple-cubic-freepath} Decay of tail of the
free path distribution function for the billiard map for simple
cubic lattices ($b=0$) with $r=3.0$ and $r=2.3$, compared to the
predicted decay rate of $t^{-2}$.}
\end{figure}

\bfigref{fig:simple-cubic-freepath} shows the tail $\Psi(T)$ of
the free path distribution for the billiard \emph{map}, for two
simple cubic lattices.  This was obtained by recording only values
of the free path exceeding a certain lower threshold, since
otherwise the predominance of small values hides the information
about the tail. A straight line corresponding to a decay rate of
$T^{-2}$ is also shown. Although the data appear to decay slightly
faster than this, the decay rate is decreasing for large $T$; we
believe that the asymptotic behaviour does obey the analytical
prediction.

\subsection{Higher-order moments}

The growth rate $\gamma_q$ of the higher-order moments is shown in
\figref{fig:simple-cubic-moments}.  They are in good agreement
with the analytical prediction, which in this case is the same as
in the 2D case reviewed in the previous section:
\begin{equation}\label{}
\gamma_q = \begin{cases} q/2, \quad \text{when } q<2\\
q-1, \quad \text{when } q>2.
\end{cases}
\end{equation}

\begin{figure}
\centrefig{simple-cubic-moments.eps}
\caption{\label{fig:simple-cubic-moments} Growth rate $\gamma_q$
of moments $\mean{\modulus{r}^q}$ for the simple cubic lattice
with $r=3.0$.}
\end{figure}

\subsection{Decay of velocity autocorrelations}

\begin{figure}
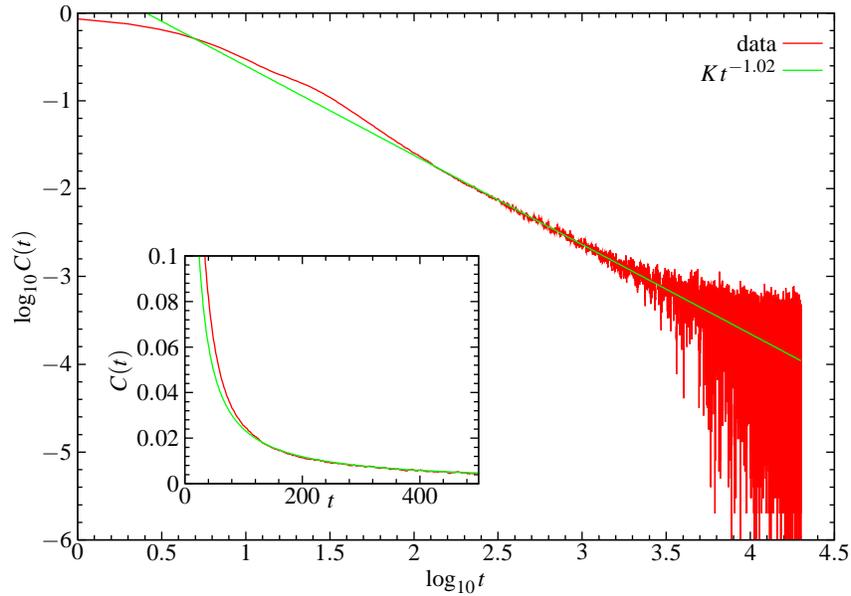

\centrefig{r3-vaf.eps} \caption{\label{fig:simple-cubic-vaf}
Velocity autocorrelation function $C(t)$ as a function of $t$
(inset) and on a log--log scale (main figure), for the simple
cubic lattice with $r=3.0$.  The data agree well with a fit to the
central less-noisy region, showing an approximate $t^{-1}$ decay.}
\end{figure}

\begin{figure}
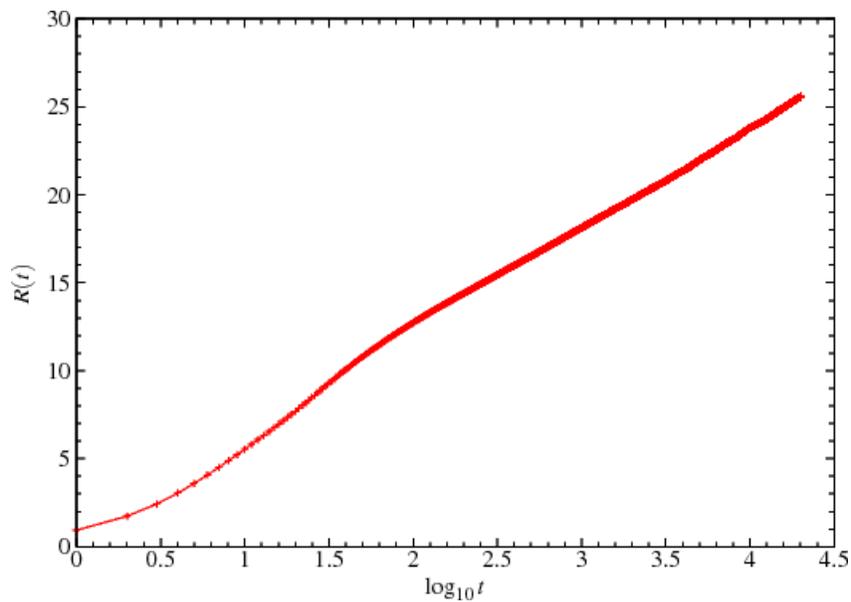

\centrefig{r3.0-intvaf.eps}
\caption{\label{fig:simple-cubic-intvaf} Integral $R(t)$ of the
velocity autocorrelation function as a function of $\log t$ for
the simple cubic lattice with $r=3.0$.}
\end{figure}

\bfigref{fig:simple-cubic-vaf} shows the velocity autocorrelation
function (VACF) $C(t)$ as a function of time $t$.  A fit to the
less-noisy central part of the graph (also shown) gives a power
law decay with an exponent which is close to $1$.  In
\figref{fig:simple-cubic-intvaf} we also show $R(t)$, the
integrated velocity autocorrelation function. If $C(t) \sim
t^{-1}$ as $t\to \infty$, then we expect that
\begin{equation}\label{}
R(t) \sim \log t,
\end{equation}
so that \figref{fig:simple-cubic-intvaf} plots $R(t)$ as a
function of $\log t$.  The linear growth for long time provides
confirmation of the non-integrability of $C$.  We note that $C(t)$
appears to converge to $0$ \emph{monotonically} from above.

\section{Lattices with cylindrical holes}
\label{sec:cylind-holes}

\subsection{Tail of free path distribution}

\paragraph{Analytical argument}

In the previous section we showed that the existence of free
planes in the structure implies the divergence of
$\E[\nu]{\freepath^2}$, and hence the non-existence of the
diffusion coefficient. However, for our model it is possible to
\emph{block} all free planes, either with the $a$-spheres
overlapping, or for large enough values of $b$ with the
$a$-spheres non-touching, leaving only holes of cylindrical type.
The previous argument proving the divergence of
$\E[\nu]{\freepath^2}$ now no longer holds, and in fact this
quantity is finite, as follows.


Again we calculate the proportion of non-colliding trajectories of
length $t$, now which remain inside a cylindrical hole of radius
$r$.  The set of these allowed directions along the cylinder is a
circle of area (approximately) $\pi r^2$, independent of $t$,
whilst the set of all possible directions is the surface of a
sphere of radius $t$, with total area $4 \pi t^2$.  (Actually
there are two such circles in opposite directions along the
cylinder.) The proportion of such trajectories is thus $C/t^2$,
compared to $C/t$ in two dimensions and in the case of simple
cubic lattices.

This  gives a  decay rate of $\tPsi(T) \sim T^{-2}$ for the
distribution function of the free path length for the billiard
flow, so that $\E[\mu]{\freepath}$ exists and hence by
\eqref{eq:golse-free-path-moments} we also have
$\E[\nu]{\freepath^2}<\infty$.  Thus point (ii) of
\secref{sec:stat-behaviour-inf-horiz} is \emph{no longer an
obstruction} to the possibility of normal diffusion. This
 does not appear to have been observed previously, although we later discovered
that it was stated in \cite{BourgainGolseFreePathDistn} (see also
\cite{GolseFreePathsHigherDimensions}) that there is a maximum
decay rate of $\tPsi(T)$ of $1/T^{d-1}$ in $d$ space dimensions.
No details of the derivation were given, although  presumably the
argument  was based on a similar idea.  We however give
\emph{explicit} examples of models where this optimum value is
\emph{attained}.




\subsection*{Numerical results} 
\revision{The numerical results in this section are \emph{incorrect}.  This is
due to the use of an incorrect method for generating the initial velocity
distribution.  The initial velocities should be generated uniformly on a unit
sphere (since the speed is fixed to be $1$).  The incorrect method used for the
numerical results depicted here was to distribute uniformly the spherical angles
$\theta$ and $\phi$, and then assign the components of the velocity based on
these angles.  This method, however, does not produce a uniform distribution
on the sphere, but rather produces a noticeable \emph{concentration} of
directions close to the poles of the sphere.  These concentrations align with
the holes in the structure, thus skewing the results sufficiently to make it
impossible to see the expected effect.\\[10pt]
\color{red}
Correct numerical results are given in a submitted
manuscript \cite{SandersNormalDiffusion3D}.
There, the  initial
velocities are correctly distributed uniformly on the sphere, by choosing
the velocity vector $\v$ uniformly in $[-1,1]^3$, rejecting those $\v$ with
$\modulus{\v} > 1$, and then normalising $\v$ to unit length. This generates
points $\v$ uniformly on the unit sphere.\\[10pt]
We find that diffusion is indeed normal, as the heuristic argument predicts, if
the only holes are cylindrical.
}

\begin{figure}
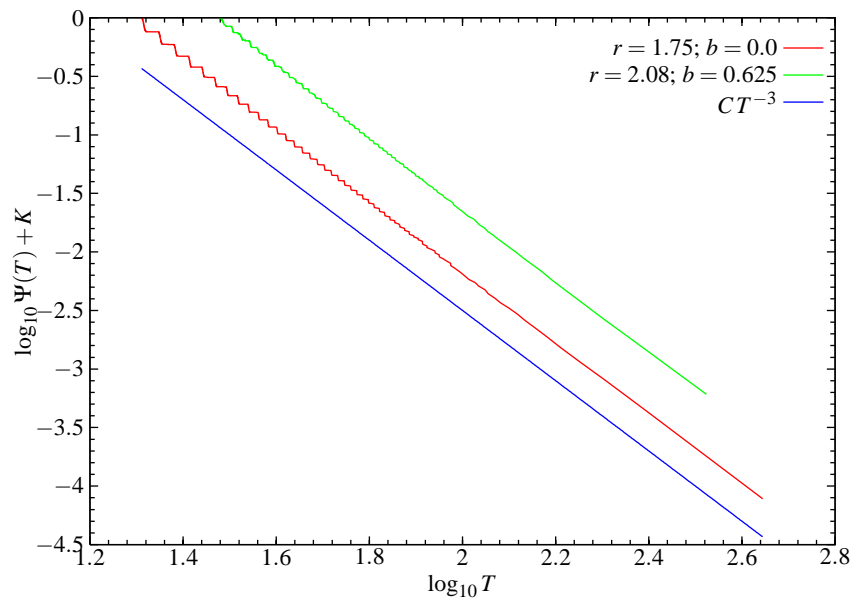

\centrefig{cylind-hole-freepath.eps}
\caption{\label{fig:cylind-freepath} Tail of free path
distributions for the billiard map, for two geometries in which
the only holes are cylindrical.  For $r=1.75$, the scatterers
overlap, but for $r=2.08$ and $b=0.625$ they are all disjoint.
There is good agreement with the analytical prediction of the
decay rate $T^{-3}$.}
\end{figure}

  \bfigref{fig:cylind-freepath} shows
the tail of the free path distribution for the billiard map for
two geometries for which \emph{the only holes are cylindrical}.
The tails decay as $T^{-3}$, so that
$\E[\nu]{\freepath^2}<\infty$, in agreement with the analytical
calculation. We remark that the step-like character visible in
\figref{fig:cylind-freepath} corresponds to the fact that the mass
of the distribution is localised near points of a periodic
lattice, as remarked for the 2D case in \cite{Bleher}. We do not
expect to see this effect in the continuous-time free path
distribution, and indeed it is absent in plots of that
distribution given in \cite{GolseFreePathsHigherDimensions}.

\subsection{Higher-order moments}

\begin{figure}
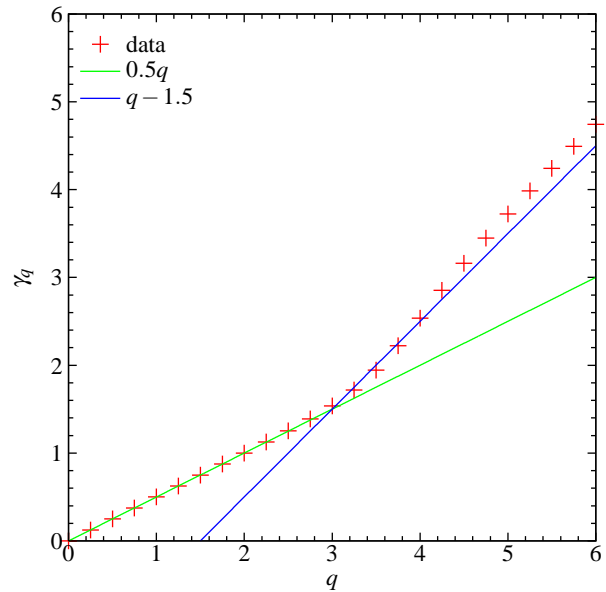

\centrefig{cylind-hole-moments.eps}
\caption{\label{fig:cylind-moments} Growth rate of moments for
$r=2.05$ and $b=0.6$. The crossover now appears to occur at
$q=3$.}
\end{figure}

The  generalisation to 3D  of the argument of \cite{ArmsteadOtt}
discussed above gives
\begin{equation}\label{}
\mean{\modulus{\r}^q} \ge K\,t^{q-2}.
\end{equation}
This does not give enough information to determine uniquely the
shape of the curve as it did in two dimensions, although it does
show that it must grow like $q$ for large $q$. Numerically we find
that the data, shown in \figref{fig:cylind-moments}, fit well the
function
\begin{equation}\label{}
\gamma_q = \begin{cases} q/2, \quad \text{when } q<3\\
q-3/2, \quad \text{when } q>3.
\end{cases}
\end{equation}
The occurrence of the number $3$ here is perhaps related to the
fact that we are now in $d=3$ space dimensions.

\subsection{Velocity autocorrelation function}

The decay of the velocity autocorrelation function at long times
has a lower bound of $C't^{-2}$, by the 3D version of the
Friedman--Martin argument discussed above.  It is thus now
\emph{possible that the VACF $C(t)$ is integrable} and thus that
the diffusion coefficient could exist, despite the infinite
horizon.

\begin{figure}
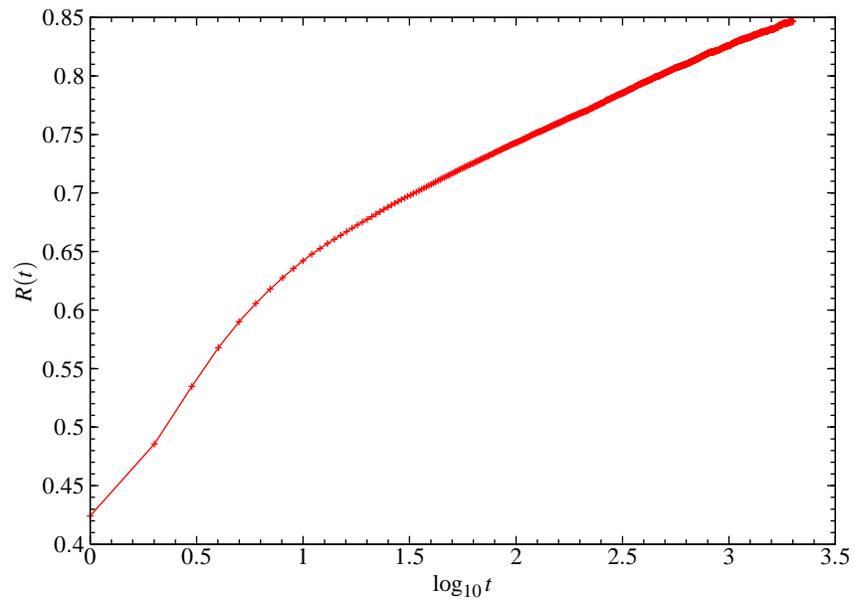

\centrefig{r1.7-intvaf.eps} \caption{\label{fig:r1.7-intvaf}
Integrated VACF $R(t)$ for $r=1.7$, $b=0.0$.}
\end{figure}

\begin{figure}
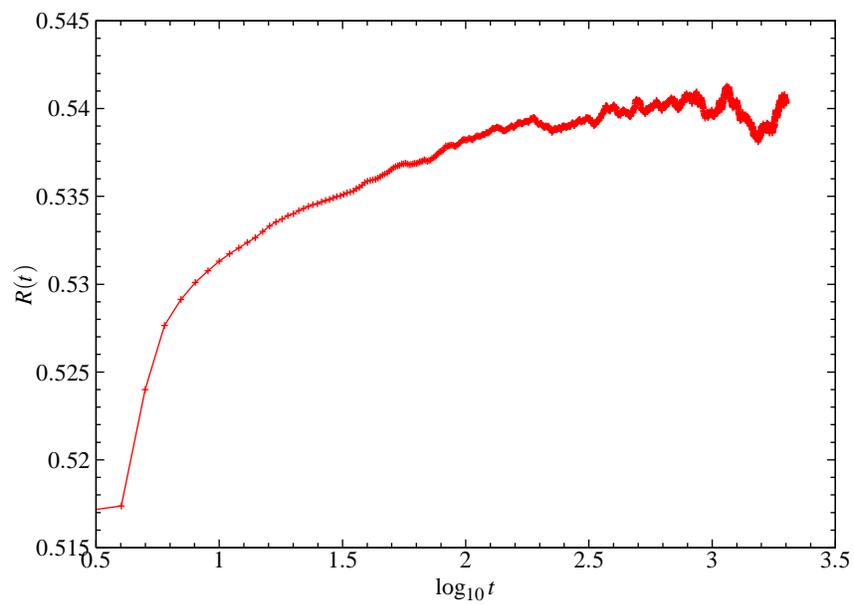

\centrefig{r1.8b0.3-intvaf.eps}
\caption{\label{fig:r1.8b0.3-intvaf} Integrated VACF $R(t)$ for
$r=1.8$, $b=0.3$.}
\end{figure}

\begin{figure}
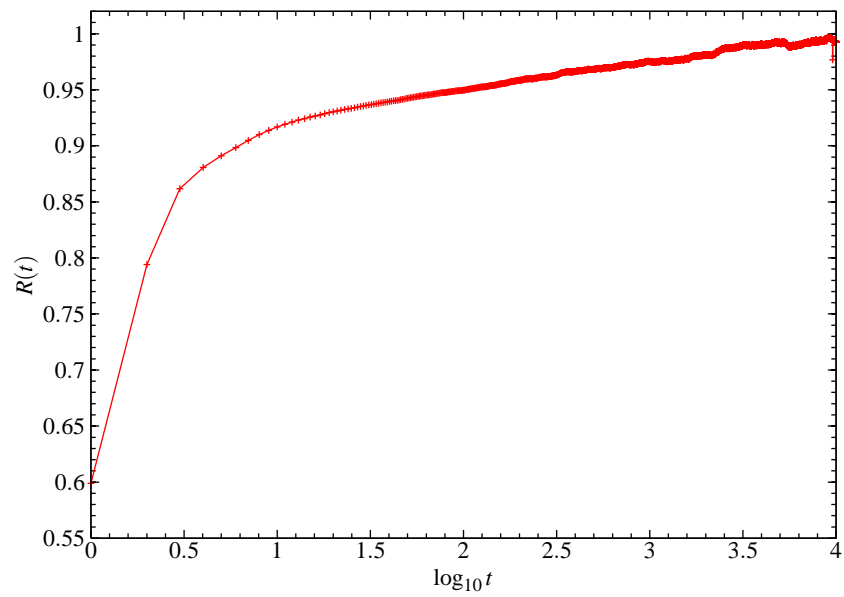

\centrefig{r2.07b0.6-intvaf.eps}
\caption{\label{fig:r2.07b0.6-intvaf} Integrated VACF $R(t)$ for
$r=2.07$, $b=0.6$.
 Computed using an average over $N=10^8$ initial conditions.}
\end{figure}

Figures \ref{fig:r1.7-intvaf}--\ref{fig:r2.07b0.6-intvaf} show the
growth of the integrated velocity autocorrelation function $R(t)
\defeq \int_0^t C(s) \rd s$ for several sets of geometrical
parameters.  As argued in the previous section, if $C(t) \sim
t^{-1}$ as $t \to \infty$ then we expect $R(t) \sim \log t$.  The
data in \figref{fig:r1.7-intvaf} provide strong evidence that for
$r=1.7$ and $b=0.0$, we do have such a $t^{-1}$ decay, and hence
that the diffusion coefficient does not exist in this case.

\bfigref{fig:r1.8b0.3-intvaf} shows $R(t)$ for $r=1.8$ and
$b=0.3$.  This differs from the previous plot in the presence of a
large central scatterer which severely restricts the size of the
cylindrical holes.  The figure is not conclusive, but it is
 possible to believe that $R(t)$ converges as $t \to
\infty$, implying that $D<\infty$.

The most interesting case is that for $r=2.07$, $b=0.6$, shown in
\figref{fig:r2.07b0.6-intvaf}, since here the scatterers are
\emph{disjoint}. The data here seem to be consistent with $R(t)
\sim \log t$, and hence $C(t) \sim t^{-1}$, although perhaps the
growth levels off for larger values of $t$.  We have also looked
directly at $C(t)$, but again this converges rapidly to zero.
There is some evidence that it oscillates around zero, which gives
more chance of having convergence.

These remarks suggest that there \emph{may} be a possibility of
having normal diffusion, in the sense of asymptotically linear
growth of the mean squared displacement, even in the presence of
an infinite horizon, provided that there are only cylindrical
holes; and this \emph{may} even be possible with \emph{disjoint}
scatterers.

 However, our numerical evidence also points to the
fact that many parameters for which there are `large' cylindrical
holes in fact have $C(t)\sim t^{-1}$, resulting in
super-diffusion.  (We remark that extending our numerical results
to longer times would be difficult, since the calculations for
\figref{fig:r2.07b0.6-intvaf} required a total of approximately 6
weeks of CPU time on modern workstations, distributed over 16
processors.) This slow decay is not predicted by the heuristic
arguments given above, and it is important to determine its
origin. One possibility could be that in fact there are strong
correlations $\mean{\r_0\cdot\r_j}$ between flights along the
cylindrical holes, so that the sum in
\eqref{eq:green-kubo-inf-horiz} does not converge.




\startonright


\chapter{Conclusions and future directions}
\label{chap:conclusions}
\section{Conclusions}

We have studied geometrical, statistical and physical aspects of
deterministic diffusion in three classes of billiard models.

We first discussed how best to estimate diffusion coefficients in
billiard models from numerical data, presenting a method which
estimates the width of the sampling distribution of the diffusion
coefficient in terms of the rate of growth of the width of the
distribution of the mean squared displacement.

We applied these numerical estimates in a 2D periodic Lorentz gas
model to study the geometry-dependence of the diffusion
coefficient, finding an unexpected qualitative change in behaviour
as one of the parameters is varied.  We extended a
previously-known random-walk approximation to our model, showing
that there are two regimes where it can be applied.  We then
considered a Green--Kubo formula, the zeroth-order term of which
is this random-walk approximation.  We showed how to improve the
justification of the derivation of this formula.  We also made
preliminary investigations on the variation of the diffusion
coefficient when the symmetry of the system is reduced.

In the same 2D periodic Lorentz gas model we then studied the
shape of position and distribution functions, which we know
converge to normal distributions at long times.  Using a method
which we believe is more appropriate than the usual histograms in
this case, we showed that at finite time the densities possess a
fine structure which prevents them from converging pointwise to
Gaussians.  We showed how  this fine structure could be understood
in terms of the geometry of the billiard domain, giving an
analytical expression for the fine structure function for both
position and displacement distributions.

Using these expressions, we showed that demodulating the densities
by the fine structure gives functions which describe the
two-dimensional density in a quasi-one-dimensional channel;
information about these 2D density functions is otherwise
difficult to obtain. This demodulation eliminates most of the very
fine-scale oscillations, resulting in much smoother functions
which seem to converge uniformly to Gaussian densities,
strengthening the standard central limit theorem. Nonetheless, in
certain parameter regimes these underlying densities  themselves
possess a type of fine structure, now corresponding to a
nearly-uniform distribution within traps. We also used the
knowledge of the fine structure to give a physical picture of the
rate of convergence in the central limit theorem.

 We considered the effect of imposing a non-constant distribution of
 particle speeds.  We showed that for a general speed distribution
 with finite mean this does not affect the mean squared displacement.
  In the special case of a Gaussian
 distribution of velocities, however, we showed that the
resulting limiting
 position distribution is skewed away from a Gaussian shape.  Our
  analytical calculations of the resulting shape matched numerical
  results very well once the densities had been demodulated by the
   relevant fine structure function.

We studied the extent to which similar statistical properties hold
in a polygonal billiard channel for which there are very few
rigorous results.  Although chaotic and mixing properties are much
weaker for this model, we showed that similar methods can be
applied, giving evidence that normal diffusion can occur, in the
sense that the mean squared displacement can grow asymptotically
linearly in time.  We also confirmed that the central limit
theorem can be satisfied, but with a slower rate of convergence
than for the Lorentz gas. We established in several particular
cases that  the existence of parallel scatterers in the structure
results in anomalous diffusion, and conjectured that this is
generally the case. We found a crossover from normal to anomalous
diffusion when this geometrical configuration is approached.  We
were able to understand this qualitatively in terms of a
continuous-time random walk model, although we found that the
quantitative prediction of a simple version of that formalism did
not match numerical data well.

We finally discussed to what extent results on two-dimensional
periodic Lorentz gases can be extended to the more physically
realistic three-dimensional case.  We constructed a 3D periodic
Lorentz gas with overlapping scatterers and showed that it has a
finite horizon regime, in which it exhibits normal diffusion. We
discussed how different types of holes in the structure affect the
statistical properties, giving evidence that it may be possible to
have normal diffusion  even when corridors exist, provided they
are small enough.

\section{Future directions}

Our results point in several directions:

\begin{itemize}

\item We hope that it is possible to  find a better physical model
of diffusion in the 2D Lorentz gas which can predict qualitative
features of the geometry dependence of the diffusion coefficient.

\item We would like to prove strong convergence of projected
densities, as discussed in \secref{sec:fine-structure} and
\appref{app:conv-proj-densities}, at least in a simple model such
as the Arnold cat map \cite{DorfBook}.

\item It may be possible to develop a more quantitative version of
the qualitative arguments given in \secref{sec:poly-anom-diffn} to
model anomalous diffusion in polygonal billiards.  In particular
it would be interesting to derive (approximations to) the step
length distribution in the continuous-time random walk model for
the zigzag model, directly from the shape of the unit cell.

\item Our analysis of the fine structure in
\chapref{chap:fine-structure} may have implications for the escape
rate formalism for calculating transport coefficients (see e.g.\
\cite{GaspBook}),
 where the diffusion equation
with absorbing boundary conditions is used as a phenomenological
model of the escape process from a finite length piece of a
Lorentz gas; analyzing the fine structure in this situation could
provide information about the validity of the use of the diffusion
equation in that context.

\item Further investigation is needed of the effect of cylindrical
holes in 3D models described in \secref{sec:cylind-holes}. It is
important to establish if (and to what extent) it is possible for
a 3D periodic Lorentz gas with non-overlapping scatterers to have
normal diffusion.

\item The arguments discussed in
\secref{sec:cylind-holes} relating to the possibility of faster
decay of correlations and free path distributions should extend to
billiards in higher dimensions; in particular, we should in
principle be able to obtain bounds on the decay of correlations
and the free path distribution of hard-sphere fluids, by treating
them as billiards in a high-dimensional phase space,
 as described in \secref{subsec:hard-spheres-as-billiards}.
An understanding of the shape of `free' regions in the
configuration space, together with arguments similar to those of
\secref{sec:cylind-holes}, should give bounds on and/or estimates
of the rate of
 decay of velocity autocorrelations and of the free path
distribution, corresponding to the \defn{trapping} effects
referred to
 in \cite{CherYoung}.  However,
 the complicated geometry of the high-dimensional phase space \cite{CherMFT} means that
 this would be difficult
 to implement.

\end{itemize}

We expect that billiard models will remain of interest to
mathematicians and to physicists in the future.



\appendix
\startonright


\chapter{Convergence of rescaled solutions of the diffusion equation to a Gaussian}
\label{app:conv-solns-diffn-eqn}

We  show that the rescaled solution of the diffusion equation
starting from an initial density which decays sufficiently rapidly
at infinity converges to a Gaussian shape as $t \to \infty$.

\section{Pointwise convergence}
\label{sec:pointwise-conv}

Let $\rho_t$ be the solution of the diffusion equation with
initial condition $\rho_0$ which is a density, i.e.\ which
satisfies $\rho_0 \ge 0$ and $\int \rho_0(y) \rd y = 1$.  We also
assume that $\rho_0 \from \R \to \Rplus$ is a piecewise continuous
function and decays sufficiently fast at infinity.  By translating
the coordinate origin if necessary, we further assume that the
centre of mass of the initial condition is fixed at the origin:
$\int y \, \rho_0(y) \d y = 0$.

Define the rescaled solution $\trho_t(x)
\defeq \sqrt{t}\, \rho_t(x\,\sqrt{t})$. Then
\begin{align}\label{}
\trho_t(x) &= \frac{\nsqrt{t}}{\nsqrt{4\pi D t}}
\int_{-\infty}^{\infty} \rz(y) \, \e^{-\lt(x\nsqrt{t}-y\rt)^2/4Dt}
\d y.
\\
&=\frac{1}{\nsqrt{4 \pi D}} \, \e^{-x^2/4D}
\int_{-\infty}^{\infty} \rz(y) \, \exp\lt[ \frac{xy}{2D\nsqrt{t}}
- \frac{y^2}{4Dt} \rt] \d y. \label{eq:rho-tilde-last-eqn}
\end{align}

For \emph{fixed} $x$, the argument of the exponential in the
integrand in \eqref{eq:rho-tilde-last-eqn} tends to $0$ as $t \to
\infty$.  Hence the integrand tends to $\rz(y)$ and it is bounded
above provided $\rz(y)$ decays sufficiently fast at infinity, for
example if it is exponentially localised in the sense that
\begin{equation}\label{}
\rz(y) \le \e^{-K\modulus{y}}
\end{equation}
for some constant $K>0$.

 The Lebesgue dominated convergence
theorem\footnote{If $(f_n)$ is a sequence of measurable functions
such that $f_n(x) \to f(x)$ almost everywhere as $n \to \infty$
and there exists an integrable function $g$ such that
$\modulus{f_n(x)} \le g(x)$ almost everywhere, then
$\lim_{n\to\infty} \int f_n = \int f$.}
\cite{RudinPrinciples}
then implies that for fixed $x$,
\begin{equation}\label{}
\trho_t(x) \stackrel{t \to \infty}{\longrightarrow}
\frac{1}{\nsqrt{4 \pi D}} \, \e^{-x^2/4D},
\end{equation}
\ie the rescaled solution of the diffusion equation tends
\emph{pointwise} to a Gaussian with variance $2D$. In fact  the
convergence is uniform, as follows.

\section{Uniform convergence of rescaled density
functions}

Consider the difference
\begin{equation}\label{}
\modulus{\trho_t(x) - \gauss{2D}(x)},
\end{equation}
where $\gauss{\sigma^2}$ is the Gaussian density with mean $0$ and
variance $\sigma^2$ given by
\begin{equation}\label{}
\gauss{\sigma^2}(x) \defeq \frac{1}{\sigma \sqrt{2 \pi}} \, \exp
\lt( \frac{-x^2}{2 \sigma^2} \rt).
\end{equation}
The difference is then given by
\begin{equation}\label{}
\modulus{\trho_t(x) - \gauss{2D}(x)} = \frac{1}{\nsqrt{4 \pi D}}
\, \e^{-x^2/4D} \lt[\int_{-\infty}^{\infty} \rz(y) \, \exp\lt[
\frac{xy}{2D\nsqrt{t}} - \frac{y^2}{4Dt} \rt] \d y - 1 \rt]. 
%
\end{equation}
We may bring the $-1$ term inside the integral, since $\int
\rho_0(y) \d y = 1$ by assumption.

 We  expand the exponential
in a Taylor series for $t$ large:
\begin{align}\label{eq:exp-expansion}
\exp\lt[ \frac{xy}{2D\sqrt{t}} - \frac{y^2}{4Dt} \rt] - 1 &=
\sum_{n=1}^\infty \frac{1}{n!} \lt[ \frac{xy}{2D\sqrt{t}} -
\frac{y^2}{4Dt} \rt]^n \\
&= \frac{xy}{2D\sqrt{t}} - \frac{y^2}{4Dt} + \frac{x^2 y^2}{8 D^2
t} - \frac{xy^3}{8 D^2 t^{3/2}} + \frac{t^4}{32 D^2 t^2} +
\sum_{n=3}^\infty {}.
\end{align}

 Assuming that it is permissible to integrate term-by-term, we have
\begin{align}\label{}
\modulus{\trho_t(x) - \gauss{2D}(x)} &= \gauss{2D}(x) \lt[
\frac{x}{2D\nsqrt{t}} \int y \, \rho_0(y) \d y + \lt( \frac{x^2}{8
D^2 t} - \frac{1}{4Dt} \rt) \int y^2 \, \rho_0(y) \d y \rt] \\
&= \frac{1}{\nsqrt{4 \pi D}} \, \e^{-x^2/4D}
\frac{\meanat{x^2}{0}}{t} \lt( \frac{x^2}{8 D^2} - \frac{1}{4D}
\rt) + \bigO{t^{-3/2}},
\end{align}
since $\int y \, \rho_0(y) \d y = 0$ by our choice of coordinates
and $\meanat{x^2}{0} = \int y^2 \rho_0(y) \, \d y$ by definition.

Since the decay of $\e^{-x^2/4D}$ is faster than the growth of any
polynomial in $x$, the terms in $x$ in the above are bounded.  If
also the term written as $\bigO{t^{-3/2}}$ is bounded, then we
have the following estimate for large $t$ which is \emph{uniform}
in $x$ (\ie \emph{independent} of $x$):
\begin{equation}\label{}
\modulus{\trho_t(x) - \gauss{2D}(x)} \le \frac{C}{t},
\end{equation}
for some constant $C$.  Thus $\supnorm{\trho_t - \gauss{2D}} \le
C/t$ for sufficiently large $t$, where the
\defn{uniform norm}  is defined by
\begin{equation}\label{}
\supnorm{f} \defeq \sup_{x \in \R} \modulus{f(x)}.
\end{equation}
Hence the rate of convergence of the rescaled density to the
limiting Gaussian is $\bigO{t^{-1}}$.

\section{Rigorous proof of uniform convergence}

We now give a rigorous proof of the result on uniform convergence
for which a heuristic argument was given in the previous section.
 We adapt a method from \cite{MillerRatesConvergenceThesis},
where convergence to a Gaussian was considered \emph{without}
rescaling; see also \cite{MillerRatesConvPaper} and references
therein.  We remark that a faster rate of convergence was obtained
in \cite{MillerRatesConvergenceThesis} by choosing a different
time origin for the Green function, but this does not work in our
situation due to the rescaling.  We set $D=1$ for simplicity
(e.g.\ by rescaling time).

\begin{theorem}
Let the density $\rho_0$ be piecewise continuous and such that the
first two moments $\int y \, \rho_0(y) \rd y$ and $\int y^2 \,
\rho_0(y) \rd y$ exist. Let $\rho_t$ be the solution of the
diffusion equation with diffusion coefficient $D=1$, starting from
the initial condition $\rho_0$. Then the rescaled density
$\trho_t(x)
\defeq \sqrt{t} \, \rho_t(x \, \sqrt{t})$ converges uniformly to
the limiting Gaussian $\tG^t(x) \defeq \frac{1}{\sqrt{4 \pi}} \,
\e^{-x^2/4}$ as $t \to \infty$, with rate of convergence
$\bigO{t^{-1}}$.
\end{theorem}


\begin{proof}
Let the initial condition be $h(x) \defeq \rho(0,x)$.  Then the
solution at time $t$ is given by the convolution $\rho_t = h \conv
G^t$, so that taking Fourier transforms gives $\rhat_t(k) = \hh{k}
\, \e^{-k^2\,t}$, where the second term is the Fourier transform
of the Green function $G^t(x)$.

Let $E^t(x) \defeq \rho_t(x) - G^t(x)$ be the error of  the
solution at time $t$ compared to the Green function, and let
\begin{equation}\label{}
\tE^t(x) \defeq \trho_t(x) - \tG^t(x) = \sqrt{t} \,
E^t(x\,\sqrt{t})
\end{equation}
be the error of the rescaled solution from the limiting Gaussian.

Then the Fourier transform of the error is
\begin{equation}\label{}
\Eh^t(k) = (\hh{k} - 1) \e^{-k^2\,t}.
\end{equation}

By Taylor's theorem with remainder we can expand $\hh{k}$ as
\begin{equation}\label{}
\hh{k} = \hhat(0) + k\, \hhat'(0) + \texthalf k^2 \hhat''(c),
\end{equation}
for some $c=c(k) \in [0,k]$.  But $\hhat(0)=1$ and $\hhat'(0) =
-\i \mean{x}_0 = 0$, by our choice of coordinates.  Furthermore,
\begin{equation}\label{}
\hhat''(c) = \int_{-\infty}^\infty (-\i x)^2 \e^{-\i cx} h(x) \rd
x,
\end{equation}
so that
\begin{equation}\label{}
\modulus{\hhat''(c)} \le \int_{-\infty}^\infty x^2 h(x) \rd x <
\infty,
\end{equation}
since we assumed that the second moment of $h$ exists. Thus
\begin{equation}\label{}
\modulus{\Eh^t(k)} \le C k^2 \e^{-k^2 t},
\end{equation}
for some constant $C$.

We now need to convert to the rescaled functions. We have
\begin{align}\label{}
\tEh^t(k) &= \int_{-\infty}^\infty \e^{-\i k x} \sqrt{t} E^t(x
\sqrt{t}) \rd x\\ & = \Eh^t(k/\sqrt{t})
\end{align}
after making the change of variables $y=x \sqrt{t}$.

Hence
\begin{equation}\label{}
\modulus{\tEh^t(k)} = \modulus{\Eh^t(k/\sqrt{t})} \le \frac{C}{t}
k^2 \e^{-k^2}.
\end{equation}
Reverting to real space using the inverse Fourier transform, we
have
\begin{align}\label{}
\modulus{\tE^t(x)} = \modulus{\frac{1}{2\pi}\int_{-\infty}^\infty
\e^{\i k x} \tEh^t(k) \rd k} \le
\frac{1}{2\pi}\int_{-\infty}^\infty \modulus{\tEh^t(k)} \rd k \le
\frac{C'}{2\pi t}\int_{-\infty}^\infty k^2 \e^{-k^2} \rd k=
\frac{C}{2\pi t},
\end{align}
where $C$ is a constant independent of $x$. Thus $\trho_t$
converges uniformly to $\tG^t$ as $t \to \infty$, and the size of
the error is
\begin{equation}\label{}
\norm{\tE^t}_\infty = \norm{\trho^t - \tG^t}_\infty  =
\bigO{t^{-1}}.
\end{equation}
\end{proof}

The numerical results reported in \secref{sec:diffn-eqn} provide
evidence that this upper bound $t^{-1}$ for the asymptotic rate is
in fact the actual decay rate.

\section{Convergence of distribution functions}

We have shown that the convergence of the rescaled density
functions is uniform in the case of the diffusion equation.
However, for the diffusive dynamical systems considered in this
thesis, rescaled \emph{density} functions do not usually converge
even pointwise to a limiting Gaussian distribution; rather, we
must consider convergence of the (cumulative) distribution
functions.

Let $F_t$ be the distribution function  at time $t$, given by
\begin{equation}\label{}
F_t(x) \defeq \int_{-\infty}^x \rho_t(x') \d x',
\end{equation}
and $\normal{}$ be the distribution function of the limiting
Gaussian distribution, so that
\begin{equation}\label{}
\normal{}(x) \defeq \int_{-\infty}^x \, \frac{1}{\sqrt{4\pi D}} \,
\e^{-x'^2/4D} \d x' = \int_{-\infty}^x \gauss{2D}(x') \d x'.
\end{equation}

Then $\rho_t(x) = F_t'(x)$ (where the prime denotes
differentiation), so that $\hat{\rho}_t(k) = \i k \hat{F_t}(k)$.
Following the same type of argument as for the density functions
gives
\begin{equation}\label{}
\norm{F_t - N}_{\infty} \le \frac{C'}{t^{1/2}} = \bigO{t^{-1/2}},
\end{equation}
for sufficiently large $t$ and some constant $C'$.

The rate $t^{-1/2}$ is known to be the fastest rate in the central
limit theorem for independent and identically distributed random
variables \cite{FellerII}, and is also the maximum rate that we
find in billiard models: see \secref{sec:clt}.


\startonright


\graphicspath{{figs/}}

\chapter{Suspension flows}
\label{app:suspension-flows}

Since rigorous results on the billiard flow $\flow^t$ are usually
proved by using the fact that it is a suspension over the billiard
map $\map$, under the free path function $\freepath$, we recall
the definition of suspension flows and some key properties. A
clear recent reference on limit theorems for suspension flows,
with strengthened versions of relevant theorems, is
\cite{MelbourneStatLimit}.

\section{Definition of suspension flows}

We follow closely the definitions in Cornfeld \etal\
\cite[Chap.~11]{Cornfeld}.

Let $(X, \Balg, \nu)$ be a measure space with an automorphism $T$,
and let $\roof \from X \to \Rplus$ be a function such that $\int_X
\roof \rd \nu < \infty$.

Define the \defn{space under the roof function $\roof$} by
\begin{equation}\label{}
Y  \defeq X^r \defeq \lt\{ (x,t): x \in X, \quad 0 \le t <
\roof(x) \rt\}.
\end{equation}
We assign a sigma-algebra $\Calg$ on $Y$ by taking as measurable
sets the measurable subsets of $X \times \R$ which belong to $Y$,
and we put
\begin{equation}\label{}
\mu \defeq \frac{1}{\bar{\roof}} \nu \times \ell,
\end{equation}
where
\begin{equation}\label{}
\bar{\roof} \defeq \Exp{\nu}{r} \defeq \int_X r(x) \rd \nu(x)
\end{equation}
and $\ell$ is Lebesgue measure on $\Rplus$, so that
\begin{equation}\label{}
\mu(A) = \frac{1}{\bar{\roof}} \iint_A \rd \nu(x) \rd t.
\end{equation}
This gives a measure space $(Y, \Calg, \nu)$ with the
normalisation $\mu(Y) = 1$.

We visualise the space $Y$ as a subset of the Cartesian product $X
\times \Rplus$, as in \figref{fig:invariant-sets-suspension}.  We
then define the flow under the roof function $\roof$ by flowing
vertically from $(x, 0)$ at unit speed for a time $\roof(x)$,
until we hit the roof function at $(x, \roof(x))$, when we
instantaneously jump to $(T(x), 0)$. This corresponds to
identifying the points $(x, \roof(x))$ and $(T(x), 0)$, in which
case we can write the flow as
\begin{equation}\label{}
V^t(x,s) = (x, s+t),
\end{equation}
computed using the identification:  see \cite{Cornfeld, GaspBook}
for explicit expressions.

 The following result shows that any flow
satisfying certain conditions can be viewed as a suspension flow,
which is technically simpler to study. For a proof, see \cite[pp.\
295ff]{Cornfeld}.

\begin{theorem}
Any flow $V^t$ without fixed points on a Lebesgue space $(M,
\Balg, \mu)$ is measure-theoretically isomorphic to a suspension
flow (also called a \defn{special flow}).
\end{theorem}

\section{Ergodicity of suspension flows}

The following theorem is stated in
\cite[Sec.~4]{ChernovInvMeasures}, but  I could not find a proof
in
the literature so I give one here%
\footnote{I would like to thank
Peter Walters for showing me the idea of this proof.}.

\begin{theorem}\label{thm:ergodicity-susp-flow}
Let $V^t$ be a suspended flow over the transformation $T \from X
\to X$, under the roof function $r\from X \to \Rplus$. Then $V^t$
is ergodic if and only if $T$ is ergodic.
\end{theorem}

\begin{proof}
Firstly suppose that the map $T$ is not ergodic.  Then there
exists an invariant set $A$ for the map which has non-trivial
measure, \ie\ $\mu(A) \neq 0$ and $\mu(A) \neq 1$.  Define the set
$B$ by
\begin{equation}\label{eq:set-over-A}
B \defeq \lt\{(x,t): x\in A, \thickspace 0\le t < \roof(x)\rt\} =
\bigcup_{x \in A} \{x\} \times [0,\roof(x));
\end{equation}
see \figref{fig:invariant-sets-suspension}. Then $B$ is an
invariant set for the flow $V^t$ with measure $\nu(B)$ which is
non-trivial, \ie\ $\nu(B) \neq 0$ and $\nu(B) \neq 1$.  Hence the
flow $V^t$ is not ergodic.

\begin{figure}
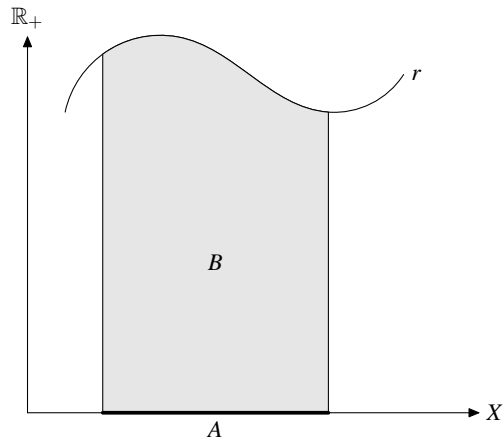

\centrefig[1]{suspension-flow.1}
\caption{\label{fig:invariant-sets-suspension}Invariant sets $A$
for the base transformation and $B$ for the suspended flow.  $A$
does not need to be connected in general.}
\end{figure}

Conversely, suppose that the flow is not ergodic, so that there is
an invariant set $B$ for the flow with non-trivial measure.  Then
the set $B$ \emph{must} be of the form \eqref{eq:set-over-A},
since otherwise $B$ would not be invariant.  Hence the set $A$
defined by projecting $B$ down to $X$ has non-trivial measure, so
that the map is not ergodic.
\end{proof}

Thus if a flow can be expressed as a suspension over some map,
then the ergodicity of either implies the ergodicity of the other.
We need the following application of this in
\secref{subsec:torus-boundary-map}.

\begin{corollary}
The \boundarymap, which maps one intersection with the torus
boundary into the next, is ergodic.
\end{corollary}

\begin{proof}
We consider the billiard dynamics on the torus.  Sinai
\cite{Sinai70} proved that the billiard map $\map$, which takes
one collision with a scatterer to the next, is ergodic. Hence the
billiard flow $\flow$ is ergodic, by Theorem
\ref{thm:ergodicity-susp-flow}, since it is a suspension over the
billiard map under the free path function $\freepath$ which gives
the time taken from one collision to the next.

However, the billiard flow can also be viewed as a suspension over
the \boundarymap, under the trap residence time function
$\restime$. Using Theorem \ref{thm:ergodicity-susp-flow} again, we
find that the \boundarymap\ is ergodic.
\end{proof}

\section{Central limit theorem for suspension flows}

The standard method to prove statistical properties for suspension
flows is via the above construction, relating them to statistical
properties of the Poincar\'e map.

Let $f_T \defeq \int_0^T f \comp \fl^t \rd t$. We say that the
central limit theorem (CLT) is satisfied for $f_T$ if
\begin{equation}\label{}
\frac{1}{\sqrt{T}} \left[ f_T - T \int_\M f \rd \mu \right]
\distconv Z,
\end{equation}
for some normal random variable $Z$.

Define $F(x) \defeq \int_0^{r(x)} f(x,u) \rd u$. The following
theorem is proved in \cite{MelbourneStatLimit} under certain
technical conditions.

\begin{theorem}
Suppose that $F$ and $r$ both satisfy the CLT. Then $f$ satisfies
the CLT.  If the CLT for $F$ has variance $\sigma_1^2 \ge 0$, then
the CLT for $f$ has variance $\sigma^2 = \sigma_1^2/\bar{r}$,
where $\bar{r} \defeq \frac{1}{\nu(M)} \int_M r \rd \nu$ is the
mean of the roof function.
\end{theorem}



\startonright


\chapter{Convergence of projected densities}
\label{app:conv-proj-densities}


\section{Densities and the Perron--Frobenius operator}

We consider a flow $\flow^t \from \M \to \M$ on the phase space
$\M$. Let $\mu_0$ be a measure describing the distribution of
initial conditions at time $t=0$.  After time $t$, this has
evolved to the pushed-forward measure $\mu_t$ defined in
\chapref{chap:stat-props}.

We say that  $\mu_0$ is
\defn{absolutely continuous} with respect to $\mu$, denoted
$\mu_0 \abscont \mu$, if $\mu(A)=0 \Rightarrow \mu_0(A) = 0$,
i.e.\ if $\mu_0$ is not concentrated on any sets of zero
$\mu$-measure. In this case, the Radon--Nikodym theorem
\cite{Royden} shows that there exists a unique non-negative
function $f_0 \in L^1(\M)$ such that
\begin{equation}\label{}
\mu_0(A) \defeq \int_A f_0 \rd \mu \defeq \int_A f_0(x) \rd \mu(x)
\quad \text{for all } A \in \Balg.
\end{equation}
We call $f_0$ the \defn{density} of $\mu_0$ with respect to $\mu$.

If $\mu$ is measure-preserving, so that $\mu_t(A) =
\mu_0(\fl^{-t}(A))$ for all $A \in \Balg$, then $\flow^t$ takes
sets of measure $0$ to sets of measure $0$.  Hence if $\mu_0
\abscont \mu$ then also $\mu_t \abscont \mu$, so that $\mu_t$ also
has a density, which we denote by $f_t$.  The map $P^t$ given by
\begin{equation}\label{}
P^t: f_0 \mapsto f_t
\end{equation}
which describes the time evolution of phase space densities is
called the
\defn{Perron--Frobenius operator} \cite{LasotaMackey,
KatokHasselblatt}.  In the case of invertible, measure-preserving
transformations, we can write an explicit formula for the
time-evolved density \cite[Chap.~5]{KatokHasselblatt}:
\begin{equation}\label{}
f_t(x) = f_0(\flow^{-t}(x)).
\end{equation}
This shows that in this case the density just gets `moved around';
nonetheless if the flow is \emph{mixing} then this `moving around'
occurs in such a way that the density gets spread out over phase
space, as follows.

\section{Mixing and weak convergence of densities}

Recall that the flow $\flow^t$ is \defn{mixing} with respect to
the invariant measure $\mu$ if
\begin{equation}\label{}
\mu(\flow^t(A) \cap B) \stackrel{t \to \infty}{\longrightarrow}
\mu(A) \, \mu(B).
\end{equation}
We can re-express this in terms of functions as follows:
\begin{equation}\label{}
 \int_\M \indic{\fl^t(A)}
\indic{B} \rd \mu \stackrel{t \to \infty}{\longrightarrow} \int_\M
\indic{A} \rd \mu \int_\M \indic{B} \rd \mu.
\end{equation}

By an approximation argument the following theorem relating mixing
to convergence of density functions can then be proved: see e.g.\
 \cite[p.~73]{LasotaMackey}.

\begin{theorem}
Let $(\M, \Balg, \mu)$ be a probability space (i.e.\ such that
$\mu(\M)=1$), $\flow^t \from \M \to \M$ a measure-preserving flow,
and $P^t$ the Perron--Frobenius operator corresponding to
$\flow^t$. Then $\flow^t$ is mixing if and only if $(P^t f)$ is
weakly convergent [see below] to $1_{\M}$ for all $f \in D$, where
$D
\defeq \{f \in L^1(\M): f \ge 0\}$ and $1_{\M}(x)=1$ for all $x
\in \M$.
\end{theorem}
(Note that $f \ge 0$ does not strictly make sense for an $L^1$
function, whose values can be changed on a set of measure $0$
without affecting the function.  The meaning is that it is
possible to find a representative of the equivalence class for
which $f \ge 0$, or equivalently that $f(x) \ge 0$ for almost all
$x \in \M$.)

We now define the notion of convergence appearing in the theorem.
\begin{definition}
A sequence $(f_n)_{n \in \Nats}$, $f_n \in L^p$,
\defn{converges weakly} to $f \in L^p$, denoted $f_n \weakconv f$, if and only if
\begin{equation}\label{}
\lim_{n \to \infty} \mean{f_n,g} = \mean{f,g} \quad \text{for all
} g \in L^{p'},
\end{equation}
where
\begin{equation}\label{}
\mean{f,g} \defeq \int_{\M} f \, g \rd \mu.
\end{equation}
\end{definition}
Here, $L^{p'}$ is the dual space of $L^p$ with
\begin{equation}\label{}
\frac{1}{p} + \frac{1}{p'} = 1 \quad \text{for } 1 < p \le \infty
\end{equation}
and $p' = \infty$ when $p = 1$, where $L^{\infty}$ is the space of
essentially bounded functions, i.e.\ functions which are bounded
except on a set of measure $0$.

\begin{corollary}
$\Phi^t$ is mixing if and only if
\begin{equation}\label{}
\mean{P^t f,g} \to \mean{1_{\M},g} = \int_\M g \rd \mu \quad
\text{as } t \to \infty,
\end{equation}
for all bounded functions $g \from \M \to \R$.
\end{corollary}

\section{Weak convergence of projected densities}
In general we cannot have stronger than weak convergence of
densities in phase space to the invariant density. For example, it
is proved in \cite{GoldsteinLebowitzSinai} that there cannot be
convergence in the $L^q$ norm for any $q \ge 1$, since  the
measure of the set with density in any interval is conserved.

However, we might expect that \emph{marginal} densities obtained
by projecting onto lower-dimensional subspaces of the phase space
may be able to converge more strongly \cite{DorfBook}.  Here we
prove that they converge weakly; in the next section we discuss
the question of strong convergence.

Consider for concreteness the 2D periodic Lorentz gas, with
coordinates $(x,y,\theta)$ in the phase space $\M = Q \times S^1$.
Consider an initial distribution in phase space given by the
density $f_0 \from \M \to \R$ with respect to normalised Liouville
measure $\rdd \mu \defeq \frac{1}{2 \pi \modulus{Q}} \rd x \rd y
\rd \theta$, where $\rdd x$ is the differential of Lebesgue
measure in the $x$-direction.  This density evolves in phase space
via the Perron--Frobenius operator $P^t$, defined as above.

We define the measure $\mu'$ by $\rdd \mu' \defeq
\frac{1}{\modulus{Q}} \rd x \rd y$, i.e.\ normalised Lebesgue
measure on $Q$, and  projected densities $\phi_t \from Q \to \R$
on the configuration space $Q$ by
\begin{equation}\label{}
\phi_t \defeq \int_{S^1} (P\,^t \! f_0) \rd \theta =
\int_{\theta=0}^{2\pi} f_t(x,y,\theta) \rd \theta,
\end{equation}
setting $f_t \defeq P^t f_0$. Then $\phi_t \in L^1(\mu')$ and
$\phi_t \ge 0$, so that $\phi_t$ is a density.

 We wish to show that the $(\phi_t)$ converge weakly to $1_Q$,
the constant invariant density on $Q$, with respect to the measure
$\mu'$. We distinguish when necessary the measure over which we
integrate by writing
\begin{equation}\label{}
\mean{f,g}_{\mu} \defeq \int_{\M} f \, g \rd \mu
\end{equation}
 for the inner product of $f$ and $g$ with respect to the measure
 $\mu$,
although in principle the measure is implicit in the domain of the
functions $f$ and $g$.

Let $\gamma \in \Linf(Q)$ be a bounded function on $Q$.  We want
to show that
\begin{equation}\label{}
\mean{\phi_t, \gamma}_{\mu'} \to \mean{1_Q, \gamma}_{\mu'} =
\int_Q \gamma \rd \mu' \quad \text{as } t \to \infty,
\end{equation}
by relating  the left  hand side to objects in phase space.

We have
\begin{align}\label{}
\mean{\phi_t, \gamma}_{\mu'} &= \frac{1}{\modulus{Q}} \int_Q
\phi_t(x,y) \, \gamma(x,y)
\rd x \rd y\\
=& \frac{1}{\modulus{Q}} \int_Q \left[ \frac{1}{2\pi} \int_{S_1}
f_t(x,y,\theta) \rd \theta \right] \, \gamma(x,y) \rd x \rd y\\
=&\frac{1}{2\pi \modulus{Q}} \int_{Q \times S^1} f_t(x,y,\theta)
\, g(x,y,\theta) \rd x \rd y \rd \theta,
\end{align}
where we define $g$ by
\begin{equation}\label{}
g(x,y,\theta) \defeq \gamma(x,y) \, 1_{S^1}(\theta),
\end{equation}
so that $g$ is constant on \defn{fibres} over $Q$.  Thus
\begin{align}
\mean{\phi_t, \gamma}_{\mu'} &= \mean{f_t, g}_{\mu}\\
&\stackrel{t \to \infty}{\longrightarrow} \int_{Q \times S^1} g
\rd \mu = \int_Q \gamma \rd \mu' = \mean{1_Q, \gamma}_{\mu'}.
\end{align}
But $\gamma \in \Linf(Q)$ was arbitrary. Hence $\phi_t \weakconv
1_Q$, as required.

\section{Convergence of 1D distributions in billiards}
\label{subsec:conv-1d-distns-billiards}

The above constitutes a general method for such proofs.  We now
consider the special case of projecting down from densities on $Q$
to densities in one coordinate direction, as in
\chapref{chap:fine-structure}.

We define the measure $\nu'$ on the $x$-space to be Lebesgue
measure. We then define 1D projected densities in the
$x$-direction by
\begin{equation}\label{}
\psi_t(x) \defeq \int_{y=0}^1 \phi_t(x,y) \rd y.
\end{equation}
(We could write $\psi_t \defeq \int_{S^1} \phi_t \rd y$, since in
fact the $x$ and $y$ coordinates are defined on a torus.)

 To
study weak convergence of the $\psi_t$, consider an arbitrary
bounded (i.e.\ $\Linf$) function $\rho$. To mimic the previous
proof, we wish to define $\gamma \from Q \to \R$ such that
\begin{equation}\label{}
\mean{\psi_t, \rho}_{\nu'} = \mean{\phi_t, \gamma}_{\mu'}.
\end{equation}
But
\begin{align}
\mean{\psi_t, \rho}_{\nu'} &= \int_{x=0}^1 \left[ \int_{y=0}^1
\phi_t(x,y) \rd y \right] \rho(x) \rd x\\
&= \iint_{x,y} \phi_t(x,y) \, \gamma(x,y) \rd x \rd y
\end{align}
if we set
\begin{equation}\label{}
\gamma(x,y) \defeq \rho(x) \, \indic{H(x)}(y).
\end{equation}
The subtlety here is that different fibres have a different amount
of associated measure.

With the above definition of $\gamma$, we have
\begin{align}\label{}
\mean{\psi_t, \rho}_{\nu'} &= \mean{\phi_t, \gamma}_{\mu'}\\
&\stackrel{t \to \infty}{\longrightarrow} \mean{1_Q,
\gamma}_{\mu'} = \int_Q \gamma \rd \mu' = \int_x \rho(x)
\left[\int_y \, \indic{H(x)}(y) \rd y \right] \rd x \\
&=\int_x \rho(x) \, h(x) \rd x = \mean{h, \rho}_{\nu'}.
\end{align}
Thus
\begin{equation}\label{}
\psi_t \rightharpoonup h
\end{equation}
with respect to the measure $\nu'$.

A different point of view is to consider the canonical projection
\begin{equation}\label{}
\pi \from Q \to S^1; \quad (x,y) \mapsto x.
\end{equation}
Denoting by $\nu$ the push-forward of the measure $\mu'$ under
this projection, we have
\begin{align}\label{}
\nu(A) \defeq [\pi\lowerstar(\mu')](A) \defeq \mu'(\pi\inv(A)) &=
\int_{Q}
\indic{x \in A} \indic {(x,y) \in Q} \rd \mu' \\
&= \int_{Q} \indic{\{x \in A\} \cup \{(x,y) \in Q\}} \rd \mu' \\
&= \int_{x \in A} \int_{y \in H(x)} \rd y \rd x = \int_{x \in A}
h(x) \rd x.
\end{align}
Thus the natural measure $\nu$ on the $x$-space has density $h$
with respect to Lebesgue measure, so that we could equally look at
the density $\psi'(x) \defeq \psi(x)/h(x)$ with respect to the
measure $\nu$ and say that
\begin{equation}\label{}
\mean{\psi', \rho}_{\nu} \to \mean{1, \rho}_{\nu},
\end{equation}
i.e.\ that $\psi'$ converges weakly to the constant density with
respect to the geometrical measure $\nu$.

\section{Stronger convergence of projected densities?}

We would like to prove that projected densities converge strongly,
e.g.\ in $L^2$ or even uniformly.  This cannot be true in general,
since if we project along the stable direction then we do not
obtain any smoothing effect: see the discussion of the baker and
cat maps in \cite{DorfBook}.  We expect, however, that if we avoid
this special direction then we should get strong convergence.  We
are not aware of any rigorous results on this, even for relatively
well-understood systems such as the cat map.  However, Sinai
proves in \cite[Chap.~18]{SinaiTopicsErgTheory} that densities
projected to unstable manifolds converge pointwise; see also
\cite{GoldsteinLebowitzSinai}.



\startonright


\chapter{Weak convergence of measures in path space}
\label{app:weak-conv-measures}

\section{Measures on path space}

We recall the definition of convergence in distribution of the
random path $\tilde{\x}_t$ to the Wiener process, a convergence of
measures on the space of paths.

Define the accumulation function by
\begin{equation}\label{}
S_t(\cdot) \defeq \int_0^t f \comp \Phi^s(\cdot) \rd s,
\end{equation}
so that $S_t \from \M \to \R$.  We denote elements of $\M$ by
$\omega$.

Define a  rescaled process $W_T$ by \cite{CherYoung}
\begin{equation}\label{}
W_T(s; \omega) \defeq \frac{S_t(\omega) - t\mean{f}}{\sqrt{T}},
\end{equation}
where $T>0$, $t=sT$, and the path is parametrised by $s\in[0,1]$.

 For fixed $T$,
$W_T(\cdot; \omega) \from [0,1] \to \R^2$ is a continuous path in
$\R^2$. In the case of diffusion in billiards, we have
$S_t(\omega) =  \int_0^t \v(s; \omega) \rd s =
\x_t(\omega)-\x_0(\omega)$, where $\omega=(q,v)$ denotes the
initial condition.

The continuous random function $W_T$ induces a probability measure
$P_T$ on the space $\mathcal{C}([0,1])$ of continuous paths from
$[0,1]$ to $\M$, via
\begin{equation}\label{}
P_T(A) \defeq \mu(\omega \in \M \colon W_T(\cdot; \omega) \in A),
\end{equation}
where $A \subset \mathcal{C}([0,1])$ is a subset of the space of
continuous paths which is measurable with respect to the Borel
$\sigma$-algebra on the metric space $\mathcal{C}([0,1])$ with
metric
\begin{equation}\label{}
d(f,g) \defeq \sup_{x \in [0,1]} \modulus{f(x) - g(x)}.
\end{equation}

\section{Weak convergence of measures on metric spaces}

The notion of convergence 
in the standard central limit
is convergence
\defn{in distribution} 
of the
rescaled distributions to a normal distribution,
which can be expressed in terms of pointwise (and 
uniform)
convergence of the rescaled distribution functions to a normal
distribution function on $\R^n$ \cite{Billingsley}.  This
definition in terms of distribution functions cannot be directly
generalised to convergence in path spaces, which we require here,
but it is
equivalent to the following notion of
\defn{weak convergence}\footnote{`Weak convergence' in probability theory
is close to \emph{weak-$\ast$} convergence in analysis
\cite[Section II.6]{WilliamsRogers}.} of measures, which does
generalise to arbitrary metric spaces \cite{Billingsley}.

Let $(X, \Balg)$ be a metric space together with the
$\sigma$-algebra of Borel sets on it.  Let $(P_n)_{n \in \Nats}$
and $P$ be probability measures on $(X, \Balg)$.  Then $P_n$
\defn{converges weakly} to $P$, written $P_n \weakconv P$, if and
only if 
\begin{equation}\label{}
\int_X f \rd P_n \to \int_X f \rd P,
\end{equation}
for all bounded, continuous functions $f\from X \to \R$ \cite{Billingsley}.
There is
an equivalent formulation in terms of sets, reflecting the duality
between measures thought of as the dual space of functions and
measures as set functions.  Namely, $P_n \weakconv P$ if and only
if
\begin{equation}\label{}
P_n(A) \to P(A) \quad \text{for all } A \text{ such that }
P(\boundary A)=0,
\end{equation}
where $\boundary A \defeq \bar{A}\setminus A^{\circ}$ is the
boundary of the set $A$, i.e.\ the set of points which are limits
of sequences of points in A and limits of sequences of points
outside A.  (Since $\boundary A$ is a closed set, it belongs to
the Borel $\sigma$-algebra $\Balg$, so that $P(\boundary A)$ is
defined.)


Sufficient conditions for this weak convergence are
\cite{Billingsley}: (i) the finite-dimensional distributions
converge; and (ii) this convergence is
\defn{tight}.
 Property (i),
 in the case of convergence to Brownian motion, is a
 \defn{multidimensional central limit theorem}
\cite{CherLimit}.  In
\secref{subsec:weak-conv-to-brownian-non-rigorous} we reformulate
it using the more intuitive notation \cite{Bleher}
\begin{equation}\label{}
\tx_t(s) \defeq \frac{\x(st)-\x(0)}{\sqrt{t}}, \quad s \in [0,1]
\end{equation}
for the rescaled process, where $\tx_t(s) = W_t(s, \cdot) \from \M
\to \R$.  Property (ii) means that for all $\epsilon
> 0$, there is a compact set $K=K(\epsilon)$ such that
$P_T(K)>1-\epsilon$ for all $T$; this prevents mass from escaping
to infinity \cite{Billingsley}.




\startonright

\addcontentsline{toc}{chapter}{Bibliography}
\bibliographystyle{my-style-latest}
\bibliography{billiards}            

\end{document}